\documentclass[11pt,aps,prd,preprintnumbers,nofootinbib,showkeys,superscriptaddress]{revtex4-1}

\pdfoutput=1

\usepackage{float}
\usepackage{cancel}
\usepackage{amsfonts}
\usepackage{amsmath}
\usepackage{amssymb}
\usepackage{caption}
\usepackage{color}
\usepackage{tikz}
\usepackage{graphicx}
\usepackage{epsfig}
\usepackage{hyperref}
\usepackage{dsfont}
\usepackage{bm}
\usepackage{soul}
\def\d{{\rm d}}
{}
{}
  \def\beq{\begin{eqnarray}}
\def\eeq{\end{eqnarray}}

\def \nchi0{\widetilde\chi^0}

\begin{document}

\title{Confronting electroweak MSSM through one-loop renormalized neutralino-Higgs interactions for dark matter direct detection and muon $(g-2)$}

\author{Subhadip Bisal}
\email{subhadip.b@iopb.res.in}
\affiliation{ Institute of Physics, Sachivalaya Marg, Bhubaneswar, 751 005, India}

\affiliation{Homi Bhabha National Institute, Training School Complex, Anushakti Nagar, Mumbai 400 094, India}

\author{Arindam Chatterjee}
\email{arindam.chatterjee@snu.edu.in}
\affiliation{Shiv Nadar University, Gautam Buddha Nagar, Uttar Pradesh, 201314, India}

\author{Debottam Das}
\email{debottam@iopb.res.in}
\affiliation{ Institute of Physics, Sachivalaya Marg, Bhubaneswar, 751 005, India}
\affiliation{Homi Bhabha National Institute, Training School Complex, Anushakti Nagar, Mumbai 400 094, India}

\author{Syed Adil Pasha}
\email{sp855@snu.edu.in}
\affiliation{Shiv Nadar University, Gautam Buddha Nagar, Uttar Pradesh, 201314, India}

\preprint{}

\begin{abstract} 
We compute the next-to-leading order (NLO) corrections to the vertices where a pair of the lightest neutralino couples to CP-even (light or heavy) Higgs scalars. 
In particular, the lightest neutralino is assumed to be a dominantly Bino-like mixed state, composed of Bino and Higgsino or Bino, Wino, and Higgsino. After computing all the three-point functions in the electroweak MSSM, we detail the contributions from the counterterms that arise in renormalizing these vertices in one-loop order. The amendment of the renormalized vertices impacts the spin-independent direct detection cross-sections of the scattering of nucleons with dark matter. We perform a comprehensive numerical scan over the parameter space where all the points satisfy the present B-physics constraints and accommodate the muon's anomalous magnetic moment. Finally, we exemplify a few benchmark points, which indulge the present searches of supersymmetric particles. After including the renormalized one-loop vertices, the spin-independent DM-nucleon cross-sections may be enhanced up to $20\%$ compared to its tree-level results. Finally, with the NLO cross-section, we use the recent LUX-ZEPLIN (LZ) results on the neutralino-nucleon scattering to display the relative rise in the lowest allowed band of the Higgsino mass parameter in the $M_1-\mu$ plane of the electroweak MSSM.
  \end{abstract}

\maketitle

\section{Introduction}
\label{sec:intro}

One of the main motivations of the Supersymmetric (SUSY) Standard Model (SM) with minimal field content or the Minimal Supersymmetric SM (MSSM) is the prediction of the Lightest Supersymmetric Particle (LSP) in the form of the lightest neutralino, which is neutral and weakly interacting with the SM particles. If $R$-parity is conserved, in most parts of the MSSM parameter space, the lightest neutralino ($\tilde \chi_1^0$) becomes stable, thus forming a good Dark Matter (DM) candidate (see, e.g.,~\cite{Jungman:1995df, Bertone:2004pz}). 
In the MSSM, $\tilde \chi_1^0$ can be dominated by one of the interaction states $-$ Bino, Wino, or Higgsino, or by any of their suitable admixtures. For instance, the LSP can be mixed Bino-Higgsino, Bino-Wino, or even Bino-Wino-Higgsino-like. Such mixed LSP scenarios are also known as ``well-tempered" neutralinos in Ref.~\cite{Arkani_Hamed_2006}. The Bino with mass $M_1$ carries no gauge charge and thus does not couple to gauge bosons. Over the parameter space of the MSSM, a dominantly Bino-like LSP results in an overabundance of dark matter except for a few fine-tuned strips characterized by, e.g., (a) slepton coannihilations ($\tilde\chi_1^0 \tilde{l} \xrightarrow{l} 
{l \gamma }$) and (b) resonant annihilation ($\tilde \chi_1^0 \tilde \chi_1^0 \xrightarrow
{A} {b \bar b, t \bar t, l^+ l^- }$). A
non-zero value of the Higgsino components will be necessary for the latter. Moreover, a
somewhat precise relation will be required between the masses of the annihilating LSP and the mediator for the $s$-channel resonance or between the co-annihilating supersymmetric state and the lightest neutralino for satisfying the observed relic abundance. 
The DM relic density of the Higgsino~\cite{Chan:1997bi,Chattopadhyay:2003xi,Chattopadhyay:2005mv,Akula:2011jx,
Baer:2011ab,Ellis:2012aa,Buchmueller:2013rsa,Baer:2015tva, He:2023lgi}
(Wino~\cite{Chattopadhyay:2006xb,Hisano:2006nn,Masiero_2005,
Cohen:2013ama,
Bhattacherjee:2014dya,Baumgart:2014saa,
Hryczuk:2014hpa,
Baer:2015tva,Ibe:2015tma,
Beneke:2016ync, Beneke:2016jpw, Beneke:2020vff, Beneke:2019qaa}) 
is primarily realized through the pair annihilation of $\tilde{H} \tilde{H} (\tilde{W} \tilde{W}) \to WW,f\bar f,..$.
With isospins =1/2 and 1, Higgsino (Wino)-like states can be observed to produce the correct abundance with mass term $\mu \simeq 1$ ($M_2=2$)~TeV respectively.  
Otherwise, in most of the MSSM parameter space, the relic density falls below the experimental value $\Omega_{DM}h^2 \sim 0.12$~\cite{WMAP:2012nax, Planck:2018vyg}. This is also supplemented by the fact that the second lightest neutralino $\tilde \chi_2^0$ and the lightest chargino $\tilde \chi_1^\pm$ can be degenerate with $\tilde \chi_1^0$, thus causing
too strong coannihilations to have too small DM relic density.
An exception may be observed in the unconstrained MSSM (pMSSM), with coannihilations may help to lower the effective thermally averaged annihilation cross-sections $\left\langle \sigma_{\rm eff} v \right\rangle$ thereby causing an increase in the DM relic density~\cite{Chakraborti:2017dpu}.
On the other hand, a well-tempered or a mixed LSP dominated by the Bino component is expected to give cosmologically compatible DM relic density  in an intermediate-mass (sub-TeV) range~\cite{Feng:1999zg, Feng:1999mn,
Feng:2000gh,
Chattopadhyay:2000fj,
Das:2007jn,
Chattopadhyay:2000qa,Baer_200522,
Chattopadhyay:2007di,
Chattopadhyay:2008hk,Chattopadhyay:2009fr,Chattopadhyay:2010vp,
Chakraborti:2017dpu,Chakraborti:2014fha}. It may be added here that a Higgsino or a mixed Bino-Higgsino DM naturally appears in most of the hyperbolic branch /focus point region of minimal supergravity (mSUGRA) inspired models~\cite{Chan:1997bi, Chattopadhyay:2003xi, Feng:1999zg, Feng:1999mn, Feng:2000gh, Feldman:2007fq, Feldman:2008jy} where the scalars may become considerably heavier (multi-TeV) satisfying the universal boundary conditions at the gauge coupling unification scale
($M_G \sim 2 \times 10^{16}$~ GeV)\footnote{A relatively small value of $\mu$ parameter $\leq 1$ TeV, is typically favored in most of the SUSY models guided by ``naturalness" ( for recent searches of the natural SUSY see~\cite{Drees:2015aeo, Barducci:2015ffa, Baer:2016usl, Chatterjee:2017nyx}).}.
Similarly, a Wino-like LSP arises naturally in the anomaly-mediated supersymmetry breaking (AMSB)
model~\cite{Randall:1998uk, Giudice:1998xp} where the gaugino and scalar masses are calculated from supergravity breaking in the hidden sector via super-Weyl anomaly contributions~\cite{Gherghetta:1999sw}.

In this era of LHC, with strongly interacting squarks and gluino heavier than a few TeV~\cite{ATLAS:2020syg, CMS:2019zmd}, a sub-TeV neutralino or chargino (will be referred to as Electroweakino) becomes the torchbearer for the TeV scale SUSY. On the one hand, unlike the colored sparticles, LHC constraints are much weaker for electroweak (EW) particles due to a smaller production cross-section~\cite{ATLAS:2020syg, CMS:2019zmd, Canepa:2019hph} (for heavier Higgsino searches at the LHC, see~\cite{Chakraborti:2015mra}). On the other hand, the pursuance of explaining muon anomalous magnetic moment $a_\mu= (g-2)_\mu/2$ through SUSY contributions is another instance where lighter electroweak sparticles are highly welcome.
The measured value (combining the BNL E821~\cite{Muong-2:2006rrc} and the Fermilab Muon $g$-2~\cite{Muong-2:2021ojo} experiments) deviates by 4.2$\sigma$ from the SM~\cite{Aoyama:2020ynm, Davier:2017zfy, Keshavarzi:2018mgv, Colangelo:2018mtw, Davier:2019can, Keshavarzi:2019abf, Kurz:2014wya, Hoferichter:2019mqg, Melnikov:2003xd, Masjuan:2017tvw, Colangelo:2017fiz, Hoferichter:2018kwz, Gerardin:2019vio, Bijnens:2019ghy, Colangelo:2019uex, Colangelo:2014qya, Blum:2019ugy, Aoyama:2012wk, atoms7010028, Czarnecki:2002nt, Gnendiger:2013pva}. The recent update released by Fermilab using Run-2 and Run-3 data, the new experimental average predicts  5.1$\sigma$ deviations from the SM~\cite{Muong-2:2023cdq}.
In the MSSM, lighter smuons and electroweakinos, e.g.,  $\tilde{\chi}^{-}-\tilde{\nu}_\mu$ and $\tilde{\mu}-\tilde{\chi}^{0}$ contribute to $\delta a_\mu= a_\mu^{\rm Exp}-a_\mu^{\rm SM}$ at one-loop level. 
A dominantly Bino-like light $\tilde \chi_1^0$ accompanying light sleptons seems to be favored by $\delta a_\mu$, especially if the observed DM abundance has to come entirely from the lightest neutralino in the R-parity conserving MSSM. 
For a mixed $\tilde\chi_1^0$, such as Bino-Higgsino DM, stringent limits from the direct dark matter detection
experiments (DD)~\cite{PandaX-II:2020oim, XENON:2023cxc, XENON:2018voc, LUX:2017ree, PandaX-II:2017hlx,XENON:2020kmp, PICO:2016kso,LUX:2016sci,PandaX-II:2016wea,LUX-ZEPLIN:2022xrq}
can be placed~\cite{Baer_2016, Badziak:2017the, Profumo:2017ntc,Abdughani:2019wai}. The Spin-Independent (SI) 
searches are particularly severe, as it directly curbs the gaugino-Higgsino-Higgs coupling in the $\tilde\chi_1^0 \tilde\chi_1^0 h(H)$ ($h$ and $H$ indicate the light and heavy Higgs bosons) vertex.
Following Ref.~\cite{Badziak:2017the}, one finds that even with the maximal mixing, a narrow strip is still viable for $\tan\beta\lesssim3$  (with stop mass $m_{\tilde{t}_1} \sim 25 ~\rm TeV$).
Otherwise, pockets exist in the parameter space where small DD cross-sections can be realized to comply with the DM-initiated recoils. In one example, ``Blind spots"  can be realized when the tree-level couplings of $\tilde \chi_1^0$ to $Z$ or the Higgs bosons may be highly suppressed or even zero identically~\cite{Cheung_2013, Cheung_2014} or
through the destructive interference between light and heavy CP-even Higgs bosons, as first noticed in~\cite{Huang:2014xua}. Another example follows when the SM-like Yukawa couplings of the light quarks are relaxed~\cite{Das:2020ozo}.
For a Bino-Wino scenario (i.e., with negligible Higgsino fraction), the Spin-Independent (SI) and Spin-Dependent (SD) DD rates vanish (see, e.g.,~\cite{Duan:2018rls}). This is because the Higgs coupling to the LSP pair is proportional to the product of their Higgsino and gaugino components, and the $Z$ boson coupling to the LSP pair is proportional to the square of its Higgsino components $\lvert\mathcal{N}_{13}\rvert^2-\lvert\mathcal{N}_{14}\rvert^2$ (but vanishes for pure Higgsinos).

More recently, Ref.~\cite{Abdughani:2019wai} zoom out the regions
of the allowed MSSM parameter space, compatible with the muon $g-2$ anomaly, DM relic density, DD limits, and the latest LHC Run-2 data. 
It turns out that a Bino-like light $\tilde \chi_1^0$ with minimal Higgsino contributions where sleptons are not far from the LSP or to a compressed scenario of Bino, Wino, and sleptons are still viable for future searches. The leading order (LO) process is only considered for evaluating DM observables, specifically for the DD cross-section. 

It is known that the next-to-leading order (NLO) corrections may lead to important effects in specific examples of DM phenomenology.
For instance, heavy quarks ($t,b$) and their superpartners can induce mass splitting between the Higgsino-like states~\cite{Giudice:1995np}, which in turn can influence the estimate of the LSP relic density~\cite{Drees:1996pk}. Latter follows from the fact that (i) the coannihilation rate is weighted by the exponential factors, thereby suppressed with the relative mass splitting of the Higgsino-like states, and (ii) gaugino and Higgsino components may get changed, which may affect the LSP couplings to the gauge and Higgs bosons. Ref.~\cite{Drees:1996pk, Drees:1993bu, Bisal:2023fgb} also presented the important SUSY corrections in the
cross-section of a Higgsino-like neutralino DM with the nucleon. In~\cite{Hisano:2004pv}, Wino/Higgsino-nucleon one-loop cross-sections generated by the gauge interactions were calculated. Ref.~\cite{Harz:2023llw, Klasen:2016qyz} considered the SUSY QCD corrections for the DD of neutralino DM. The DM-nucleon cross-section at one one-loop level for a general class of weakly interacting massive particles was considered in Ref.~\cite{Cirelli:2005uq, Hisano:2011cs}. At the same time, the interaction of gluon with the DM was noted in Ref.~\cite{Hisano:2010fy, Hisano:2010ct}. However, none of the analyses considers the renormalization of chargino/
neutralino sector, which we employ here to explicitly estimate the Higgs interactions with $\tilde \chi_1^0$ pairs. Adopting a suitable renormalization scheme, the vertex counterterms are calculated and added to the three-point vertex corrections. Here, we recall that in the limit of vanishing mixings among the different constituents in $\tilde \chi_1^0$, counterterms may be calculated to vanish. Since for a pure $\tilde \chi_1^0$, there is no tree-level interaction that an SM-like Higgs scalar can couple to $\tilde{\chi}_1^0$ pairs, the renormalization of $\tilde{\chi}_1^0\tilde{\chi}_1^0 h$ or $\tilde{\chi}_1^0\tilde{\chi}_1^0 H$ coupling at the NLO is neither needed nor possible\footnote{However, due to the off-diagonal terms in the neutralino mass matrix (generated after EWSB), a small admixture of the gauge eigenstates is inevitable even when the respective mass parameters are $\mathcal{O}$(1 TeV).}. For a general and dominantly Bino-like $\tilde \chi_1^0$, as in this case, tree-level coupling exists, and counterterms may boost the DM-nucleon scattering. 
Based on this lesson, we explore the MSSM regions through the muon $g-2$ anomaly, DM relic density, and the spin-independent DM direct detections (SI-DD) at the one-loop level. The latest LHC Run-2 data is also considered. The relic density constraint is not always respected in the analysis; thus, thermal relic abundance of the LSP may satisfy (i) the observed cosmological dark matter abundance, (ii) falls below the dark matter abundance (known as under-abundant neutralinos), (iii) overshoots the observed cosmological data (over-abundant neutralino). Since our primary interest is to find out the role of the NLO corrections to the neutralino-Higgs vertices and the SI-DD cross-section in compliance with $(g-2)_\mu$; we relax the relic density constraint in the first place\footnote{However, scenario (ii) and (iii) can be made viable in the presence of different DM components or through modifying the standard cosmological thermal history.}. 
In particular, two regions with a dominantly Bino-like $\tilde{\chi}_1^0$, but having (i) a minimal Higgsino ($M_1<<\mu$) component and (ii) a minimal Wino-Higgsino ($M_1<M_2\lesssim\mu$) component (assuming $M_1$, $M_2$, and $\mu$ to be real and positive in both the scenarios) will come out as interesting for future searches. Henceforth, we
refer them as $\tilde B_{\tilde H}$ and $\tilde B_{\tilde W\tilde H}$ 
zones of the MSSM parameter space.

The rest of the paper is organized as follows. In Sec.~\ref{neutralinocharginosector}, we briefly review the neutralino and chargino sectors of the MSSM. We fix the notation and convention for the masses and mixing matrices and then discuss the different possibilities of the mixed neutralino states. We present the effective Lagrangian for the neutralino-nucleon scattering in Sec.~\ref{sec:effectivelagrng}. 
In Sec.~\ref{subsec:ver}, we show the generic triangular topologies of the relevant Feynman diagrams and present analytical results, while Sec.~\ref{sec:Renormalization} covers the 
important aspects of renormalizations of the chargino and neutralino sectors. In Sec.~\ref{sec:amu} and Sec.~\ref{sec:collider}, we summarize the supersymmetric contributions to anomalous magnetic moments of muon ($\delta a_\mu$) and the limits from the SUSY searches at the collider experiments. Sec.~\ref{sec:methodology} illustrates the methodology adopted for the numerical calculations, followed by the evaluation of the neutralino-nucleon scattering cross-section in Sec.~\ref{sec:numeical}. Finally, we conclude in Sec.~\ref{sec:conclusion}.

\section{The Neutralino and Chargino sectors of the MSSM} 
\label{neutralinocharginosector}
In the MSSM, the supersymmetric partner of the neutral gauge bosons, known as Bino, $\tilde{B}$ (the supersymmetric partner of the $U(1)_Y$ gauge boson $B$) and Wino, $\tilde{W}^0$ (the supersymmetric partner of the $SU(2)_L$ neutral gauge boson $W^0$) mix with the supersymmetric partners of the two MSSM Higgs bosons, known as down and up type Higgsinos $\tilde{H}_d^0$ and $\tilde{H}_u^0$ respectively. The $4\times4$ neutralino mass matrix in the basis $(\tilde{B}, \tilde{W}^0, \tilde{H}^0_d, \tilde{H}^0_u)$ can be written as
 
\begin{align}
\overline{\mathbb{M}}_{\tilde{\chi}^0}=\left(\begin{array}{c c c c}
M_1 & 0 & -M_Zs_W c_\beta & M_Z s_W s_\beta \\
0 & M_2 & M_Zc_W c_\beta & -M_Z c_W s_\beta\\
-M_Zs_W c_\beta & M_Zc_W c_\beta & 0 & -\mu\\
M_Zs_W s_\beta & -M_Zc_W s_\beta & -\mu & 0\\
\end{array}\right),
\label{eq:nu_mass}
\end{align}
where $M_Z$ is the mass of the $Z$ boson, $\beta$ represents the mixing angle between the two Higgs VEVs, and $s_\beta=\sin\beta$, $c_\beta=\cos\beta$. Here, $c_W=\cos\theta_W$ and $s_W=\sin\theta_W$ are the cosine and sine of the Weinberg angle $\theta_W$, respectively. The mass parameters $M_1$, $M_2$, and $\mu$ can generally be complex, allowing the CP-violating interactions in MSSM. But we restrict to the case of CP-conserving interactions; hence, $M_1$, $M_2$, and $\mu$ are real in our scenario.

Besides, the mass matrix of charginos 
can be read from the mass eigenstates of the $2 \times 2$ complex mass matrix, $\overline{\mathbb{M}}_{\tilde{\chi}^\pm}$ in the Wino-Higgsino basis,
\begin{align}
\overline{\mathbb{M}}_{\tilde{\chi}^\pm}=\left(\begin{array}{c c }
M_2 & \sqrt{2} s_\beta M_W \\
 \sqrt{2} c_\beta M_W& \mu 
\end{array}\right),
\label{eq:char_mass}
\end{align}
which can be diagonalized by two unitary $2\times 2$ matrices $\mathbf{U}$ and $\mathbf{V}$.

The neutralino mass matrix in Eq.~\eqref{eq:nu_mass} can be diagonalized by a $4\times4$ unitary matrix $\mathbb{N}$ (in this case $\mathbb{N}$ is orthogonal).
\begin{align}
 \mathbb{N} \overline{\mathbb{M}}_{\tilde{\chi}^0} \mathbb{N}^{-1} = \mathbb{M}_{\tilde{\chi}^0} \hspace{2cm} [\rm since\,\,\, \mathbb{N}^{*} = \mathbb{N}]
\end{align}
where $\mathbb{M}_{\tilde{\chi}^0}=$ diag.$(m_{\tilde{\chi}_1^0}, m_{\tilde{\chi}_2^0}, m_{\tilde{\chi}_3^0}, m_{\tilde{\chi}_4^0})$ refer to the physical masses of the neutralinos with $\tilde{\chi}_1^0$ being the lightest, and $\mathcal{N}_{ij}$ are the elements of the neutralino mixing matrix $\mathbb{N}$. The Lagrangian for the neutralino-neutralino-scalar interaction is given by~\cite{Drees:2004jm},
\begin{align}
\mathcal{L}_{\tilde{\chi}_\ell^0\tilde{\chi}_n^0\phi} \supset& \,\,\frac{g_2}{2}h \bar{\tilde{\chi}}_n^0 \Big[\mathbf{P_L}\Big(Q_{\ell n}^{\prime\prime *}s_\alpha + S_{\ell n}^{\prime\prime *}c_\alpha\Big) + \mathbf{P_R}\Big(Q_{n\ell}^{\prime\prime}s_\alpha + S_{n\ell}^{\prime\prime}c_\alpha\Big)\Big]\tilde{\chi}_{\ell}^0 \nonumber\\
&- \frac{g_2}{2}H \bar{\tilde{\chi}}_n^0 \Big[\mathbf{P_L}\Big(Q_{\ell n}^{\prime\prime *}c_\alpha - S_{\ell n}^{\prime\prime *}s_\alpha\Big) + \mathbf{P_R}\Big(Q_{n\ell}^{\prime\prime}c_\alpha - S_{n\ell}^{\prime\prime}s_\alpha\Big)\Big]\tilde{\chi}_{\ell}^0\nonumber\\
&-i\frac{g_2}{2}A\bar{\tilde{\chi}}_n^0 \Big[\mathbf{P_L} \Big(S^{\prime\prime *}_{\ell n}c_\beta -Q^{\prime\prime *}_{\ell n}s_\beta\Big) \tilde{\chi}_{\ell}^0 + \mathbf{P_R} \Big(Q^{\prime\prime}_{n\ell}s_\beta -S^{\prime\prime}_{n\ell}c_\beta\Big)\Big] \tilde{\chi}_{\ell}^0
\label{Eq:lochichihi}
\end{align} 
Here $\phi=h_i,A$, with $h_i$, for $i=1$ and 2 refer to an SM-like 
 scalar $h$ and a heavy CP-even Higgs boson $H$, respectively. Similarly, $\mathbf{P_{L,R}}= \frac{1\mp \gamma 5}{2}$ as usual. The neutral CP-odd Higgs is denoted by $A$, $\alpha$ is the Higgs mixing angle, and $g_2$ is the $SU(2)_L$ gauge coupling strength. Couplings
$Q^{\prime\prime}_{n\ell}$ and $S^{\prime\prime}_{n\ell}$ are defined in the Appendix ~\ref{eq:couplings;}.

Similarly, the neutralino-neutralino-$Z$ interaction can be read from,
\begin{align}
\mathcal{L}_{\tilde{\chi}_\ell^0\tilde{\chi}_n^0 Z} \supset \frac{g_2}{2c_W} Z_\mu \bar{\tilde{\chi}}_\ell^0 \gamma^\mu \Big(N_{\ell n}^L \mathbf{P_L} + N_{\ell n}^R \mathbf{P_R}\Big)\tilde{\chi}_n^0\,,
\end{align}
where $N_{\ell n}^L$ and $N_{\ell n}^{R}$ are defined in the Appendix~\ref{eq:couplings;}.

A few interesting limits can now be observed.
{{A pure Higgsino-like $\tilde{\chi}_1^0$~}}: it refers to a limit $\mathcal{N}_{11}=\mathcal{N}_{12} =0$ or the soft masses $M_1$ and $M_2$ are large and decoupled.
{{A pure Gaugino-like $\tilde{\chi}_1^0$~}}: it
refers to a limit $\mathcal{N}_{13}=\mathcal{N}_{14} =0$.
In this pure limit, the coupling $\tilde{\chi}_1^0\tilde{\chi}_1^0\phi$ (with $\phi=h,H,A$) vanishes since $Q^{\prime\prime}_{11}= S^{\prime\prime}_{11}=0$. In the 
{{mixed LSP scenarios}}, as said before, we are interested in Bino-Higgsino ($\tilde B_{\tilde H}$) and Bino-Wino-Higgsino ($\tilde B_{\tilde W\tilde H}$) DM scenarios with a predominantly Bino component. A qualitative understanding of the neutralino masses and their mixings (mainly $\mathcal{N}_{1j}$) can be instructive here, which we detail in the Appendix~\ref{eq:appB} (see also~\cite{Gunion:1987yh, ElKheishen:1992yv, Barger:1993gh}). In fact, using the expressions derived for the mixing matrices in the Appendix~\ref{eq:appB}, we can rewrite the $\tilde{\chi}_1^0\tilde{\chi}_1^0h$ and $\tilde{\chi}_1^0\tilde{\chi}_1^0H$ couplings for $\tilde B_{\tilde W\tilde H}$ DM as
\begin{align}
\mathcal{L}_{\tilde{\chi}_1^0\tilde{\chi}_1^0\phi} = 
\phi \bar{\tilde{\chi}}_1^0  \Big[\mathbf{P_L} C_L^{\rm LO} + \mathbf{P_R} C_R^{\rm LO}\Big]\tilde{\chi}_{1}^0~.
\end{align}
For $\phi=h$,
\begin{align}
    C_L^{\rm LO} = C_R^{\rm LO} &= \frac{g_2}{2}\Big(Q_{11}^{\prime\prime}s_\alpha + S_{11}^{\prime\prime}c_\alpha\Big)
   =-\frac{g_2}{2}\frac{M_Z s_W}{\mu^2 - M_1^2}\big(\mu s_\alpha - M_1 c_\alpha\big)\Bigg[\frac{M_1 M_Z^2 s_{2W}}{2(M_2-M_1)}+t_W\Bigg]~,
   \label{eq:chichihHlo}
\end{align}
and for $\phi=H$,
\begin{align}
    C_L^{\rm LO} = C_R^{\rm LO} &= -\frac{g_2}{2}\Big(Q_{11}^{\prime\prime}c_\alpha - S_{11}^{\prime\prime}s_\alpha\Big)
    = \frac{g_2}{2}\frac{M_Z s_W}{\mu^2 - M_1^2}\big(\mu c_\alpha + M_1 s_\alpha\big)\Bigg[\frac{M_1 M_Z^2 s_{2W}}{2(M_2-M_1)}+t_W\Bigg]~.
\end{align}
Similarly, the couplings for $\tilde{B}_{\tilde{H}}$ DM are given as follows.\\
For $\phi=h$,
\begin{align}
    C_L^{\rm LO}=C_R^{\rm LO} = -\frac{g_2}{2}t_W \frac{M_Z s_W}{\mu^2-M_1^2}\big(\mu s_\alpha- M_1 c_\alpha\big)~,
    \label{Eq:LO_h}
\end{align}
and for $\phi=H$,
\begin{align}
     C_L^{\rm LO}=C_R^{\rm LO} = \frac{g_2}{2}t_W \frac{M_Z s_W}{\mu^2-M_1^2}\big(\mu c_\alpha+ M_1 s_\alpha\big)~.
     \label{Eq:LO_H}
\end{align}
The coefficients of $\mathbf{P_L}$ and $\mathbf{P_R}$ 
are equal due to the Majorana nature of the neutralinos. Recall that, in the above expressions, $s_\beta\to 1$ and $c_\beta\to 0$ are assumed. If instead we keep the $\tan\beta$ dependence, one can obtain the $\tilde{\chi}_1^0\tilde{\chi}_1^0h(H)$ coupling goes as $\propto \big[M_1+\mu s_{2\beta}\big](\mu c_{2\beta})$.
\section{Effective Lagrangian For Neutralino-Nucleon Scattering}
\label{sec:effectivelagrng}
This section presents the effective Lagrangian governing the neutralino-nucleon scattering process and provides the corresponding formulae for the cross-section (\cite{Jungman:1995df, Bertone:2004pz, Goodman:1984dc, Griest:1988ma, Ellis:1987sh, Barbieri:1988zs, Drees:1993bu, Ellis:2000ds, Vergados:2006sy, Oikonomou:2006mh, Ellis:2008hf}). In the realm of non-relativistic neutralinos, the effective interactions between $\tilde{\chi}_1^0$ and the light quarks and gluons, at the renormalization scale $\bar{\mu}_0\simeq m_p$, can be elegantly described as follows~\cite{Drees:1993bu, Nath:1994ci, Hisano:2004pv, Hisano:2017jmz, PhysRevD.59.055009}:
\begin{align}
\mathcal{L}^{\rm eff}=\sum_{q=u,d,s}\mathcal{L}_q^{\rm eff}+\mathcal{L}_g^{\rm eff}~,
\label{eq:L_eff}
\end{align}
where,
\begin{align}
\mathcal{L}_q^{\rm eff}&=\eta_q\bar{\tilde{\chi}}^0_1\gamma^\mu\gamma_5\tilde{\chi}^0_1\bar{q}\gamma_\mu\gamma_5 q + \lambda_qm_q\bar{\tilde{\chi}}^0_1\tilde{\chi}^0_1 \bar{q}q +\frac{g_q^{(1)}}{m_{\tilde{\chi}^0_1}}\bar{\tilde{\chi}}^0_1 i\partial^\mu\gamma^\nu\tilde{\chi}^0_1\mathcal{O}^q_{\mu\nu}+ \frac{g_q^{(2)}}{m^2_{\tilde{\chi}^0_1}}\bar{\tilde{\chi}}^0_1 (i\partial^\mu)(i\partial^\nu)\tilde{\chi}^0_1\mathcal{O}^q_{\mu\nu}~,\nonumber\\
\mathcal{L}_g^{\rm eff}&= \lambda_G\bar{\tilde{\chi}}^0_1\tilde{\chi}^0_1G^a_{\mu\nu}G^{a\,\mu\nu}+ \frac{g_G^{(1)}}{m_{\tilde{\chi}^0_1}}\bar{\tilde{\chi}}^0_1 i\partial^\mu\gamma^\nu\tilde{\chi}^0_1\mathcal{O}^g_{\mu\nu}+ \frac{g_G^{(2)}}{m^2_{\tilde{\chi}^0_1}}\bar{\tilde{\chi}}^0_1 (i\partial^\mu)(i\partial^\nu)\tilde{\chi}^0_1\mathcal{O}^g_{\mu\nu}~.
\label{eq:Lq_Lg}
\end{align}
The terms up to the second derivative of the neutralino field have been incorporated in the above. The spin-dependent interaction refers to the first term of $\mathcal{L}_q^{\rm eff}$ while the spin-independent ``coherent" contributions arising from the remaining terms in $\mathcal{L}_q^{\rm eff}$ and $\mathcal{L}_g^{\rm eff}$.
 The third and fourth terms in $\mathcal{L}_q^{\rm eff}$ and the second and third terms in $\mathcal{L}_g^{\rm eff}$ are governed by the twist-2 operators (traceless part of the energy-momentum tensor) for the quarks and gluons~\cite{Drees:1993bu, Hisano:2004pv}. Note that the contributions of the twist-2 operators of gluon are suppressed by the strong coupling constant
$\alpha_s$, thus not included in the subsequent sections. Finally, the SI scattering cross-section of the neutralino with target nuclei can be expressed as,
\begin{align}
\sigma_{\rm SI}=\frac{4}{\pi}\left(\frac{m_{\tilde{\chi}^0_1}M_A}{m_{\tilde{\chi}^0_1}+M_A}\right)^2\Bigg[\{Zf_p+(A-Z)f_n\}^2
\Bigg],
\label{eq:sigma_nucl}
\end{align} 
where 
$Z$ and $A$ represent its atomic and mass numbers, respectively. 

The spin-independent coupling of the neutralino with nucleon (of mass $m_N$), $f_N$ ($N=p,n$) in Eq.~\eqref{eq:sigma_nucl} can be expressed as 
(neglecting the contributions from twist-2 operators and also from squark loops)
\begin{align}
\frac{f_N}{m_N}=&\sum_{q=u,d,s}\lambda_qf^{(N)}_q
-\frac{8\pi}{9\alpha_s}\lambda_Gf^{(N)}_{G},
\label{Eq:fnSI}
\end{align}
where the matrix elements of nucleon are defined as 
\begin{align}
f_q^{(N)}&\equiv \frac{1}{m_N}\langle N|m_q\bar{q}q|N\rangle~.
\end{align}

The second term in Eq.~\eqref{Eq:fnSI} involves effective interactions between the WIMP, heavy quarks, and gluons, which can be evaluated utilizing the trace anomaly of the energy-momentum tensor in QCD~\cite{SHIFMAN1978443, Drees:1993bu}. Here, one finds heavy quark form factors are related to that of gluons.
\begin{align}
\langle N|m_Q\bar{Q}Q|N\rangle= -\frac{\alpha_s}{12\pi} c_Q
\langle N|G^a_{\mu\nu}G^{a\mu\nu}|N\rangle~, \nonumber \\
m_N f_G^{(N)}= -9 \frac{\alpha_s}{8\pi}\langle N|G^a_{\mu\nu}G^{a\mu\nu}|N\rangle,
\end{align}
with $\alpha_s=g_s^2/4\pi$ and the leading order QCD correction $c_Q=1+11\alpha_s(m_Q)/4\pi$ is considered. 
The coefficient 
$\lambda_G$ in Eq.~\eqref{Eq:fnSI} is related to heavy quarks.
\begin{align}
\lambda_G=-\frac{\alpha_s}{12\pi}\sum_{Q=c,b,t}c_Q\lambda_Q,
\end{align}
with $\lambda_Q$ often involves $\phi Q\bar Q$ 
vertex at the tree level. Here, $\lambda_{q}$ and $\lambda_{Q}$ contains all SUSY model-dependent information.

The parameters, $f^{(N)}_{q}$ ($q \in u,d,s$)               
can be determined from lattice QCD calculations~\cite{Thomas:2012tg}.
We use the following central values of $f^{(N)}_{q}$~\cite{Thomas:2012tg,Belanger:2013oya},
\begin{align}
f^{(p)}_{u} =0.0153,\quad f^{(p)}_{d} =0.0191,\quad f^{(p)}_{s} =0.0447, \nonumber\\
f^{(n)}_{u} =0.0110,\quad f^{(n)}_{d} =0.0273,\quad f^{(n)}_{s} =0.0447,
\label{eq:ff}
\end{align}
which leads to $f^{(N)}_G \sim 0.921$\footnote{{One gets slightly different values from chiral perturbation theory~\cite{Alarcon:2011zs,Crivellin:2013ipa,Hoferichter:2015dsa}.}}.
It should be noted that the above numerical values are subject to some uncertainties as
they are evaluated using the hadronic data~\cite{Ellis:2000ds,Hooper:2009zm}. 

The scalar cross-section depends on $t$-channel Higgs exchange ($h,H$) (neglecting squark contributions) $\Bigg(\sigma_{\rm SI}\propto\dfrac{1}{m_{h,H}^4}\Bigg)$. Apart from the masses of the Higgs scalar, the cross-section depends strongly on the $\tilde{\chi}_1^0\tilde{\chi}_1^0h(H)$ couplings (Eq.~\eqref{eq:chichihHlo}- \eqref{Eq:LO_H}) and also on the $q\bar{q}h(H)$ coupling (through $\lambda_{q}$ and $\lambda_{Q}$). Note that for down-type fermions $q\bar{q}h$ coupling goes as $\sim \tan\beta \cos(\beta-\alpha)$ while $q\bar{q}H$ coupling goes as $\sim \tan\beta \sin(\beta-\alpha)$. For $H$ scalar, $\tilde{\chi}_1^0\tilde{\chi}_1^0H$ and $q\bar{q}H$ couplings assume larger values compared to that of the SM-like Higgs scalar in the decoupling region ($M_A^2>>M_Z^2$) and with large $\tan\beta$. This makes the heavier Higgs boson contributions in the direct detection quite important. 

\section{Spin-Independent $\tilde{\chi}_1^0$-nucleon scattering at One-Loop: Theory and Implementation}

As already discussed, in general, a tree-level $\tilde{\chi}_1^0\tilde{\chi}_1^0 h_i$ coupling depends on the product of gaugino and Higgsino components. The one-loop correction to this vertex leads to a UV-divergent result. Therefore, one has to renormalize the vertex to get a UV-finite result. Here, we systematically analyze the vertex corrections' generic triangular topologies along with the renormalization procedure.
\subsection{Vertex corrections}
\label{subsec:ver}
\begin{figure}[H]
	\centering
	\includegraphics[width=0.75\linewidth]{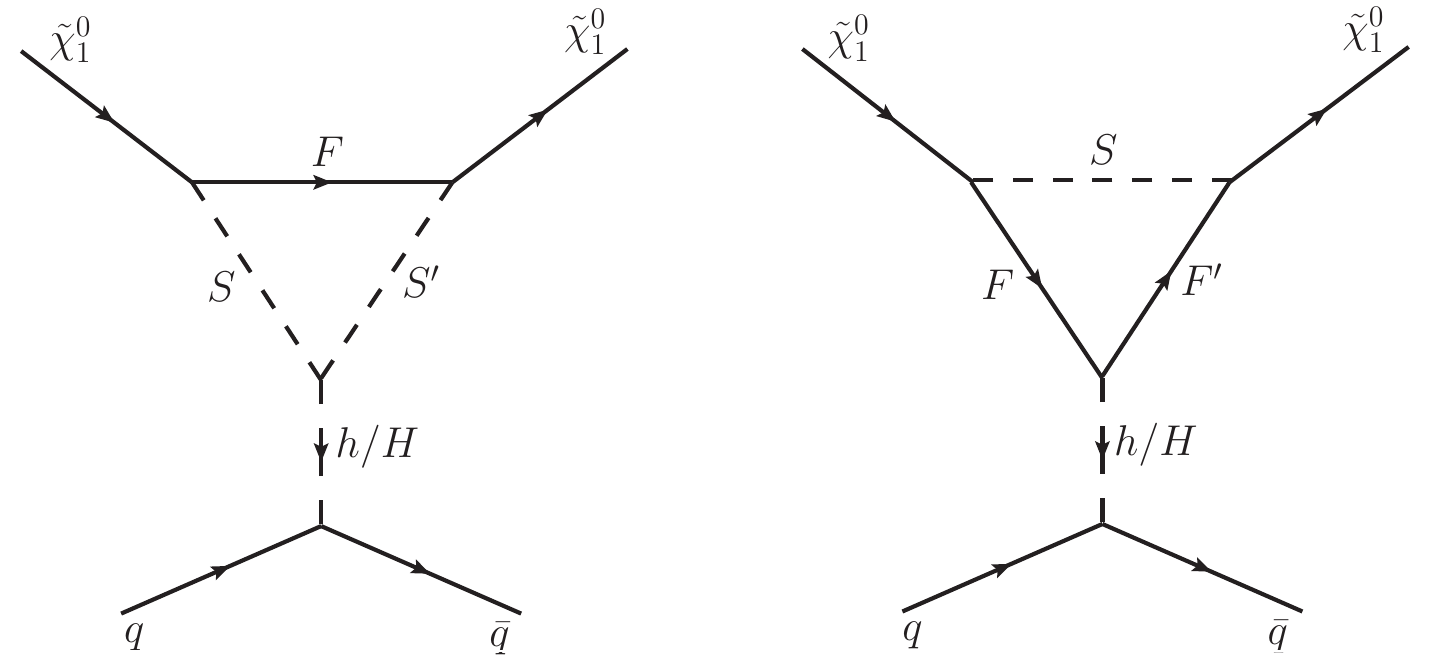}\\
	(a)\hspace{6.9cm}(b)
	\includegraphics[width=0.75\linewidth]{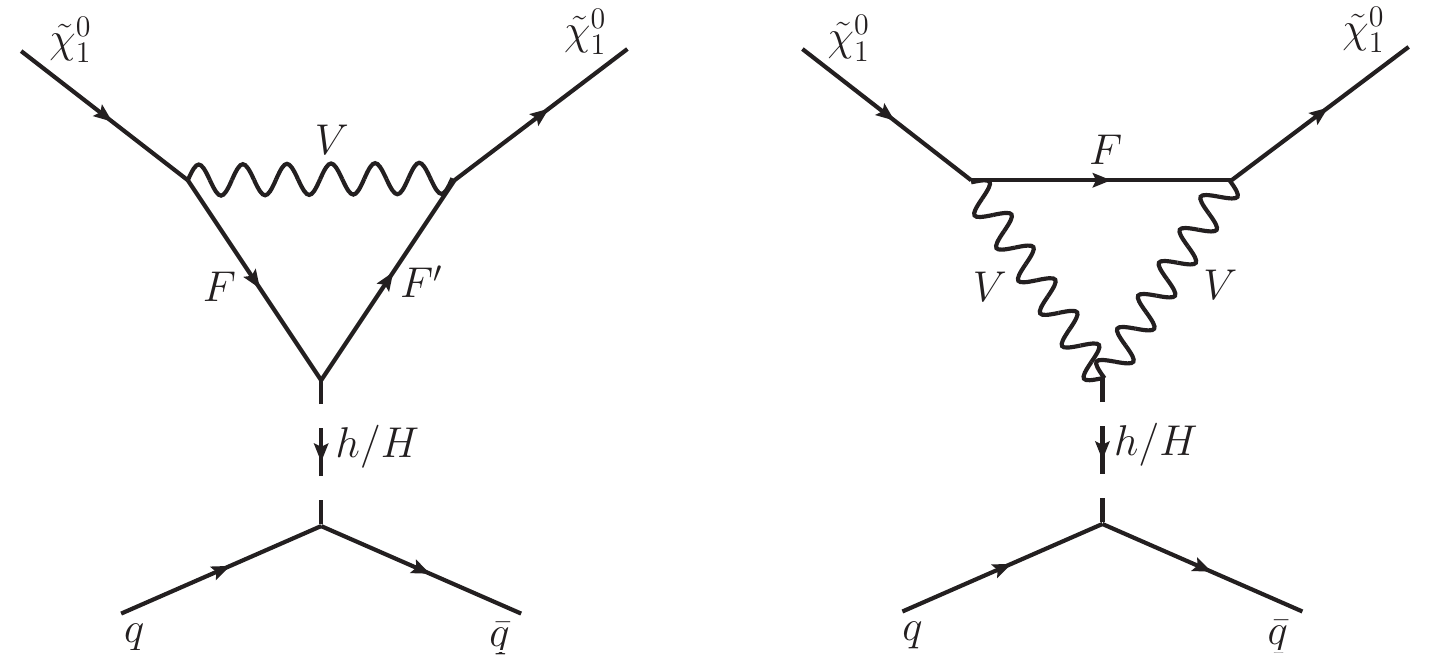}\\
	(c)\hspace{6.9cm}(d)
	\includegraphics[width=0.75\linewidth]{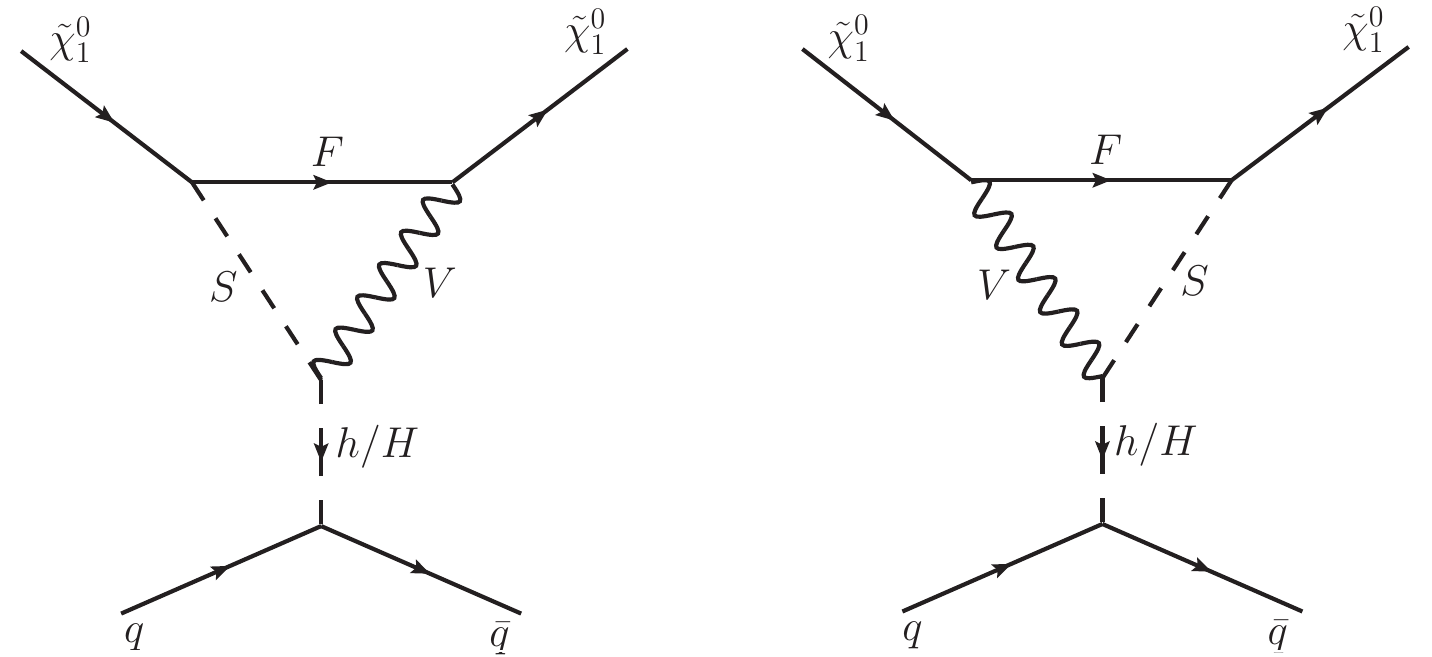}\\
	(e)\hspace{6.9cm}(f)
	\caption{Relevant topologies for the one-loop correction to the $\tilde{\chi}^0_1 \tilde{\chi}^0_1 h_i$ vertex which in turn yields the one-loop correction to the $\tilde{\chi}^0_1 \tilde{\chi}^0_1 q\bar{q}$ scattering. Here, $S,S^\prime \in$ \big\{$h$, $H$, $A$, $H^\pm$, $G^0$, $G^\pm$, $\tilde{\ell}$, $\tilde{\nu}_\ell$ (where $\ell = e$, $\mu$, $\tau$)\big\}; $F,F^\prime \in$ \big\{$\tilde{\chi}_n^0$, $\tilde{\chi}_k^\pm$ (where $n = 1,...,4$ and $k=1,2$), $\ell$, $\nu_\ell$\big\}; $V \in$ \big\{ $W^\pm$, $Z$\big\}. 
  }
	\label{fig:topology1}
\end{figure}

We start by classifying different triangular topologies for 
$\tilde\chi_1^0$-nucleon elastic scattering in Fig.~\ref{fig:topology1}
where the $\tilde{\chi}_1^0\tilde{\chi}_1^0 h_i$ vertex has been modified by the one-loop radiative corrections from the SM and SUSY particles. We adopt a general notation $S,S^\prime =h$, $H$, $A$, $H^\pm$, $G^0$, $G^\pm$, $\tilde{\ell}$, and $\tilde{\nu}_\ell$ (where $\ell = e$, $\mu$, $\tau$); $F,F^\prime =$ $\ell$, $\nu_\ell$, $\tilde{\chi}_n^0$ and $\tilde{\chi}_k^\pm$ (where $n = 1,...,4$ and $k=1,2$); $V = W^\pm$ and $Z$. The squark contributions can be ignored because we set them as heavy $\geq$ 4 TeV. For explicit calculations, we find a total of 468 diagrams where 234 diagrams for the $\tilde{\chi}_1^0\tilde{\chi}_1^0h$ vertex and another 234 diagrams for the $\tilde{\chi}_1^0\tilde{\chi}_1^0H$ vertex at the particle level.

The analytical expressions for the one-loop diagrams are calculated using $\mathtt{Package}$-X-2.1.1~\cite{Patel:2015tea, Patel:2016fam} and given in terms of the Passarino-Veltman functions. The Higgs propagator ($h/H$), which connects the quark line to the one-loop vertex, has been taken as off-shell with four-momentum $q$, known as the momentum transfer. 
The momentum transfer $q$ is generally very small ($q^2\sim \mathcal{O}(10^{-6})$ $\rm GeV^2$ for $v_{\chi_1^0} \sim 10^{-3}$) for the elastic scattering process. It may be noted here that $q^2\sim 0$ is assumed for numerical estimation. 
In the Appendix ~\ref{eq:couplings;}, we present the prefactors of different topologies for the MSSM.

\underline{{\bf Topology-(\ref{fig:topology1}a):}} 
\begin{align}
i\Gamma_{\tilde{\chi}_1^0\tilde{\chi}_1^0h_i}^{(a)} =& -\frac{i}{16\pi^2}\Big[\mathbf{P_L}\Bigl\{\xi_{LL}m_F\mathbf{C}_0 - \xi_{LR}m_{\tilde{\chi}_1^0}\mathbf{C}_1 - \xi_{RL}m_{\tilde{\chi}_1^0}\mathbf{C}_2\Bigr\}+ \mathbf{P_R}\Bigl\{\xi_{RR}m_F\mathbf{C}_0-\xi_{RL}m_{\tilde{\chi}_1^0}\mathbf{C}_1 \nonumber\\
&-\xi_{LR}m_{\tilde{\chi}_1^0}\mathbf{C}_2\Bigr\}\Big]~,
\label{eq1:topology2a}
\end{align}
where $\mathbf{C}_i=\mathbf{C}_i\big(m_{\tilde{\chi}_1^0}^2,q^2,m_{\tilde{\chi}_1^0}^2;m_F,m_S,m_{S^\prime}\big)$ and 
\begin{align*}
\xi_{LL} = \lambda_{h_iSS^\prime} \mathcal{G}_{\tilde{\chi}_1^0FS^\prime}^{L}\mathcal{G}_{\tilde{\chi}_1^0FS}^{L}~,\qquad\qquad \xi_{LR} = \lambda_{h_iSS^\prime} \mathcal{G}_{\tilde{\chi}_1^0FS^\prime}^{L}\mathcal{G}_{\tilde{\chi}_1^0FS}^{R}~,\\
\xi_{RL} = \lambda_{h_iSS^\prime} \mathcal{G}_{\tilde{\chi}_1^0FS^\prime}^{R}\mathcal{G}_{\tilde{\chi}_1^0FS}^{L}~,\qquad\qquad \xi_{RR} = \lambda_{h_iSS^\prime} \mathcal{G}_{\tilde{\chi}_1^0FS^\prime}^{R}\mathcal{G}_{\tilde{\chi}_1^0FS}^{R}~.
\end{align*}
\underline{{\bf Topology-(\ref{fig:topology1}b):}}
\begin{align}
i\Gamma_{\tilde{\chi}_1^0\tilde{\chi}_1^0h_i}^{(b)} =& -\frac{i}{16\pi^2}\Big[\mathbf{P_L}\Bigl\{\zeta_{LLL}m_{F}m_{F^\prime}\mathbf{C}_0 + \zeta_{LLR} m_{\tilde{\chi}_1^0}m_{F^\prime}\big(\mathbf{C}_0+\mathbf{C}_1\big) + \zeta_{LRL}\bigl\{\mathbf{B}_0+m_S^2\mathbf{C}_0+m_{\tilde{\chi}_1^0}^2\big(\mathbf{C}_1+\mathbf{C}_2\big)\bigr\}\nonumber\\
& + \zeta_{LRR}m_{\tilde{\chi}_1^0} m_{F}\mathbf{C}_1+ \zeta_{RLL} m_{\tilde{\chi}_1^0}m_{F}\big(\mathbf{C}_1+\mathbf{C}_2\big) + \zeta_{RLR} m_{\tilde{\chi}_1^0}^2\big(\mathbf{C}_0+ \mathbf{C}_1+ \mathbf{C}_2\big) + \zeta_{RRL} m_{\tilde{\chi}_1^0}m_{F^\prime}\mathbf{C}_2\Bigr\}\nonumber\\
& +\mathbf{P_R}\Bigl\{\zeta_{LLR} m_{\tilde{\chi}_1^0}m_{F^\prime}\mathbf{C}_2 + \zeta_{LRL} m_{\tilde{\chi}_1^0}^2\big(\mathbf{C}_0+ \mathbf{C}_1+\mathbf{C}_2\big) + \zeta_{LRR} m_{\tilde{\chi}_1^0}m_{F}\big(\mathbf{C}_1+ \mathbf{C}_2\big) + \zeta_{RLL} m_{\tilde{\chi}_1^0}m_{F}\mathbf{C}_1 \nonumber\\
&+ \zeta_{RLR}\bigl\{\mathbf{B}_0 + m_S^2\mathbf{C}_0 + m_{\tilde{\chi}_1^0}^2\big(\mathbf{C}_1+\mathbf{C}_2\big)\bigr\} + \zeta_{RRL} m_{\tilde{\chi}_1^0}m_{F^\prime}\big(\mathbf{C}_0 + \mathbf{C}_1\big) + \zeta_{RRR} m_{F}m_{F^\prime} \mathbf{C}_0\Bigr\}\Big]~,
\label{eq2:topology2b}
\end{align}
where $\mathbf{B}_0 = \mathbf{B}_0\big(q^2 ; m_F, m_{F^\prime}\big)$, $\mathbf{C}_i = \mathbf{C}_i\big(m_{\tilde{\chi}_1^0}^2, q^2, m_{\tilde{\chi}_1^0}^2 ; m_S,m_F,m_{F^\prime}\big)$ and 
\begin{align*}
\zeta_{LLL} = \mathcal{G}_{\tilde{\chi}_1^0 F^\prime S}^L \mathcal{G}_{FF^\prime h_i}^L \mathcal{G}_{\tilde{\chi}_1^0 FS}^L~, \qquad\qquad \zeta_{LLR} = \mathcal{G}_{\tilde{\chi}_1^0 F^\prime S}^L \mathcal{G}_{FF^\prime h_i}^L \mathcal{G}_{\tilde{\chi}_1^0 FS}^R~,\\
\zeta_{LRL} = \mathcal{G}_{\tilde{\chi}_1^0 F^\prime S}^L \mathcal{G}_{FF^\prime h_i}^R \mathcal{G}_{\tilde{\chi}_1^0 FS}^L ~,\qquad\qquad \zeta_{LRR} = \mathcal{G}_{\tilde{\chi}_1^0 F^\prime S}^L \mathcal{G}_{FF^\prime h_i}^R \mathcal{G}_{\tilde{\chi}_1^0 FS}^R~,\\
\zeta_{RLL} = \mathcal{G}_{\tilde{\chi}_1^0 F^\prime S}^R \mathcal{G}_{FF^\prime h_i}^L \mathcal{G}_{\tilde{\chi}_1^0 FS}^L ~,\qquad\qquad \zeta_{RLR} = \mathcal{G}_{\tilde{\chi}_1^0 F^\prime S}^R \mathcal{G}_{FF^\prime h_i}^L \mathcal{G}_{\tilde{\chi}_1^0 FS}^R~,\\
\zeta_{RRL} = \mathcal{G}_{\tilde{\chi}_1^0 F^\prime S}^R \mathcal{G}_{FF^\prime h_i}^R \mathcal{G}_{\tilde{\chi}_1^0 FS}^L~, \qquad\qquad \zeta_{RRR} = \mathcal{G}_{\tilde{\chi}_1^0 F^\prime S}^R \mathcal{G}_{FF^\prime h_i}^R \mathcal{G}_{\tilde{\chi}_1^0 FS}^R~.
\end{align*}

\underline{{\bf Topology-(\ref{fig:topology1}c):}}
\begin{align}
i\Gamma_{\tilde{\chi}_1^0\tilde{\chi}_1^0h_i}^{(c)} =& \frac{i}{16\pi^2}\Big[\mathbf{P_L}\Bigl\{\Lambda_{LLL}m_{\tilde{\chi}_1^0} m_{F^\prime}\big(2-d\big)\mathbf{C}_2 + \Lambda_{LRL}m_{\tilde{\chi}_1^0} m_F \big(2-d\big)\big(\mathbf{C}_0+\mathbf{C}_2\big) + \Lambda_{LRR}m_{\tilde{\chi}_1^0}^2\big(d-4\big)\big(\mathbf{C}_0\nonumber\\
&+\mathbf{C}_1+\mathbf{C}_2\big) + \Lambda_{RLL}\bigl\{d\mathbf{B}_0 + \big(4m_{\tilde{\chi}_1^0}^2 + m_V^2d - 2q^2\big)\mathbf{C}_0 + \big(4m_{\tilde{\chi}_1^0}^2 + m_{\tilde{\chi}_1^0}^2d - 2q^2\big)\big(\mathbf{C}_1+\mathbf{C}_2\big)\bigr\} \nonumber\\
&+ \Lambda_{RLR}m_{\tilde{\chi}_1^0}m_F\big(2-d\big)\mathbf{C}_1 + \Lambda_{RRL}m_Fm_{F^\prime}d\mathbf{C}_0+\Lambda_{RRR}m_{\tilde{\chi}_1^0}m_{F^\prime}\big(2-d\big)\big(\mathbf{C}_0+\mathbf{C}_1\big)\Bigr\} \nonumber\\
&+ \mathbf{P_R}\Bigl\{\Lambda_{LLL}m_{\tilde{\chi}_1^0}m_{F^\prime}\big(2-d\big)\big(\mathbf{C}_0+\mathbf{C}_1\big)+\Lambda_{LLR}m_F m_{F^\prime}d\mathbf{C}_0 + \Lambda_{LRL}m_{\tilde{\chi}_1^0}m_F \big(2-d\big)\mathbf{C}_1 + \Lambda_{LRR}\nonumber\\
&\times \bigl\{d\mathbf{B}_0 + \big(4m_{\tilde{\chi}_1^0}^2 + m_V^2d -2q^2\big)\mathbf{C}_0 + \big(4m_{\tilde{\chi}_1^0}^2 + m_{\tilde{\chi}_1^0}^2d -2q^2\big)\big(\mathbf{C}_1+\mathbf{C}_2\big)\bigr\} + \Lambda_{RLL}m_{\tilde{\chi}_1^0}^2\big(d-4\big)\big(\mathbf{C}_0\nonumber\\
&+\mathbf{C}_1+\mathbf{C}_2\big) + \Lambda_{RLR}m_{\tilde{\chi}_1^0}m_F\big(2-d\big)\big(\mathbf{C}_0+\mathbf{C}_2\big)+ \Lambda_{RRR}m_{\tilde{\chi}_1^0}m_{F^\prime}\big(2-d\big)\mathbf{C}_2\Bigr\}\Big]~,
\label{eq3:topology2c}
\end{align}

where $\mathbf{B}_0=\mathbf{B}_0\big(q^2, m_F, m_{F^\prime}\big)$, $\mathbf{C}_i = \mathbf{C}_i\big(m_{\tilde{\chi}_1^0}^2, q^2, m_{\tilde{\chi}_1^0}^2 ; m_V, m_F, m_{F^\prime}\big)$ and 
\begin{align*}
\Lambda_{LLL} = \mathcal{G}_{\tilde{\chi}_1^0 F^\prime V}^L \mathcal{G}_{F F^\prime h_i}^L \mathcal{G}_{\tilde{\chi}_1^0 F V}^L~,\qquad\qquad \Lambda_{LLR} = \mathcal{G}_{\tilde{\chi}_1^0 F^\prime V}^L \mathcal{G}_{F F^\prime h_i}^L \mathcal{G}_{\tilde{\chi}_1^0 F V}^R~,\\
\Lambda_{LRL} = \mathcal{G}_{\tilde{\chi}_1^0 F^\prime V}^L \mathcal{G}_{F F^\prime h_i}^R \mathcal{G}_{\tilde{\chi}_1^0 F V}^L~,\qquad\qquad \Lambda_{LRR} = \mathcal{G}_{\tilde{\chi}_1^0 F^\prime V}^L \mathcal{G}_{F F^\prime h_i}^R \mathcal{G}_{\tilde{\chi}_1^0 F V}^R~,\\
\Lambda_{RLL} = \mathcal{G}_{\tilde{\chi}_1^0 F^\prime V}^R \mathcal{G}_{F F^\prime h_i}^L \mathcal{G}_{\tilde{\chi}_1^0 F V}^L~,\qquad\qquad \Lambda_{RLR} = \mathcal{G}_{\tilde{\chi}_1^0 F^\prime V}^R \mathcal{G}_{F F^\prime h_i}^L \mathcal{G}_{\tilde{\chi}_1^0 F V}^R~,\\
\Lambda_{RRL} = \mathcal{G}_{\tilde{\chi}_1^0 F^\prime V}^R \mathcal{G}_{F F^\prime h_i}^R \mathcal{G}_{\tilde{\chi}_1^0 F V}^L~,\qquad\qquad \Lambda_{RRR} = \mathcal{G}_{\tilde{\chi}_1^0 F^\prime V}^R \mathcal{G}_{F F^\prime h_i}^R \mathcal{G}_{\tilde{\chi}_1^0 F V}^R~.
\end{align*}

\underline{{\bf Topology-(\ref{fig:topology1}d):}}
\begin{align}
i\Gamma_{\tilde{\chi}_1^0\tilde{\chi}_1^0h_i}^{(d)} =& -\frac{i}{16\pi^2}\Big[\mathbf{P_L}\Bigl\{\eta_{LL}m_{\tilde{\chi}_1^0}\big(d-2\big)\mathbf{C}_2 + \eta_{RL}m_F d \mathbf{C}_0+\eta_{RR}m_{\tilde{\chi}_1^0}\big(d-2\big)\mathbf{C}_1\Bigr\}\nonumber\\
&+\mathbf{P_R}\Bigl\{\eta_{LL}m_{\tilde{\chi}_1^0}\big(d-2\big)\mathbf{C}_1 + \eta_{LR}m_F d \mathbf{C}_0+\eta_{RR}m_{\tilde{\chi}_1^0}\big(d-2\big)\mathbf{C}_2\Bigr\}\Big]~,
\label{eq4:topology2d}
\end{align}
where $\mathbf{C}_i = \mathbf{C}_i\big(m_{\tilde{\chi}_1^0}^2, q^2, m_{\tilde{\chi}_1^0}^2 ; m_F, m_V, m_{V}\big)$ and 
\begin{align*}
\eta_{LL} = \mathcal{G}_{VVh_i}\mathcal{G}_{\tilde{\chi}_1^0 F V}^L \mathcal{G}_{\tilde{\chi}_1^0 F V}^L~, \qquad\qquad \eta_{LR} = \mathcal{G}_{VVh_i}\mathcal{G}_{\tilde{\chi}_1^0 F V}^L \mathcal{G}_{\tilde{\chi}_1^0 F V}^R~,\\
\eta_{RL} = \mathcal{G}_{VVh_i}\mathcal{G}_{\tilde{\chi}_1^0 F V}^R \mathcal{G}_{\tilde{\chi}_1^0 F V}^L~, \qquad\qquad \eta_{RR} = \mathcal{G}_{VVh_i}\mathcal{G}_{\tilde{\chi}_1^0 F V}^R \mathcal{G}_{\tilde{\chi}_1^0 F V}^R~.
\end{align*}

\underline{{\bf Topology-(\ref{fig:topology1}e):}}
\begin{align}
i\Gamma_{\tilde{\chi}_1^0\tilde{\chi}_1^0h_i}^{(e)} =& \frac{i}{16\pi^2}\Big[\mathbf{P_L}\Bigl\{\psi_{LL}m_{\tilde{\chi}_1^0}m_F\big(\mathbf{C}_2-\mathbf{C}_0\big) + \psi_{LR}m_{\tilde{\chi}_1^0}^2 \big(\mathbf{C}_1+2\mathbf{C}_2\big) + \psi_{RL}\bigl\{-d\mathbf{C}_{00} - m_{\tilde{\chi}_1^0}^2\big(\mathbf{C}_{22}+2\mathbf{C}_{12}\nonumber\\
&+\mathbf{C}_{11}+2\mathbf{C}_1\big)+q^2\mathbf{C}_{12} + \big(2q^2-3m_{\tilde{\chi}_1^0}^2\big)\mathbf{C}_2\bigr\} + \psi_{RR}m_{\tilde{\chi}_1^0}m_F\big(\mathbf{C}_1+2\mathbf{C}_0\big)\Bigr\} + \mathbf{P_R}\Bigl\{\psi_{LL}m_{\tilde{\chi}_1^0}m_F\nonumber\\
&\times\big(\mathbf{C}_1+2\mathbf{C}_0\big)+\psi_{LR}\bigl\{-d\mathbf{C}_{00} - m_{\tilde{\chi}_1^0}^2\big(\mathbf{C}_{22}+2\mathbf{C}_{12}+\mathbf{C}_{11}+2\mathbf{C}_1\big)+q^2\mathbf{C}_{12} + \big(2q^2-3m_{\tilde{\chi}_1^0}^2\big)\mathbf{C}_2\bigr\}\nonumber\\
& + \psi_{RL}m_{\tilde{\chi}_1^0}^2\big(\mathbf{C}_1+2\mathbf{C}_2\big) + \psi_{RR}m_{\tilde{\chi}_1^0}m_F\big(\mathbf{C}_2-\mathbf{C}_0\big)\Bigr\}\Big]~,
\label{eq5:topology2e}
\end{align}
where $\mathbf{C}_i = \mathbf{C}_i\big(m_{\tilde{\chi}_1^0}^2, q^2, m_{\tilde{\chi}_1^0}^2 ; m_F, m_S, m_V\big)$, $\mathbf{C}_{ij} = \mathbf{C}_{ij}\big(m_{\tilde{\chi}_1^0}^2, q^2, m_{\tilde{\chi}_1^0}^2 ; m_F, m_S, m_V\big)$ and 
\begin{align*}
\psi_{LL} = \mathcal{G}_{h_i SV}\mathcal{G}_{\tilde{\chi}_1^0FV}^L \mathcal{G}_{\tilde{\chi}_1^0FS}^L~,\qquad\qquad \psi_{LR} = \mathcal{G}_{h_i SV}\mathcal{G}_{\tilde{\chi}_1^0FV}^L \mathcal{G}_{\tilde{\chi}_1^0FS}^R~, \\
\psi_{RL} = \mathcal{G}_{h_i SV}\mathcal{G}_{\tilde{\chi}_1^0FV}^R \mathcal{G}_{\tilde{\chi}_1^0FS}^L~,\qquad\qquad \psi_{RR} = \mathcal{G}_{h_i SV}\mathcal{G}_{\tilde{\chi}_1^0FV}^R \mathcal{G}_{\tilde{\chi}_1^0FS}^R~.
\end{align*}

\underline{{\bf Topology-(\ref{fig:topology1}f):}}
\begin{align}
i\Gamma_{\tilde{\chi}_1^0\tilde{\chi}_1^0h_i}^{(f)} =& \frac{i}{16\pi^2}\Big[\mathbf{P_L}\Bigl\{\Xi_{LL}\bigl\{d\mathbf{C}_{00} + m_{\tilde{\chi}_1^0}^2\big(\mathbf{C}_{22}+2\mathbf{C}_{12}+\mathbf{C}_{11}+2\mathbf{C}_2+3\mathbf{C}_1\big)-q^2\big(\mathbf{C}_{12}+2\mathbf{C}_1\big)\bigr\}+\Xi_{LR}\nonumber\\
&\times m_{\tilde{\chi}_1^0}m_F\big(\mathbf{C}_0-\mathbf{C}_1\big)-\Xi_{RL}m_{\tilde{\chi}_1^0} m_F\big(\mathbf{C}_2+2\mathbf{C}_0\big)-\Xi_{RR} m_{\tilde{\chi}_1^0}^2\big(\mathbf{C}_2+2\mathbf{C}_1\big)\Bigr\} + \mathbf{P_R}\Bigl\{-\Xi_{LL}m_{\tilde{\chi}_1^0}^2\nonumber\\
&\times\big(\mathbf{C}_2+2\mathbf{C}_1\big)-\Xi_{LR}m_{\tilde{\chi}_1^0}m_F\big(\mathbf{C}_2+2\mathbf{C}_0\big) + \Xi_{RL} m_{\tilde{\chi}_1^0}m_F\big(\mathbf{C}_0-\mathbf{C}_1\big)+ \Xi_{RR}\bigl\{d\mathbf{C}_{00} + m_{\tilde{\chi}_1^0}^2\big(\mathbf{C}_{22}\nonumber\\
&+2\mathbf{C}_{12}+\mathbf{C}_{11}+2\mathbf{C}_2+3\mathbf{C}_1\big)-q^2\big(\mathbf{C}_{12}+2\mathbf{C}_1\big)\bigr\}\Bigr\}\Big]~,
\label{eq6:topology2f}
\end{align}
where $\mathbf{C}_i = \mathbf{C}_i\big(m_{\tilde{\chi}_1^0}^2, q^2, m_{\tilde{\chi}_1^0}^2 ; m_F, m_V, m_S\big)$, $\mathbf{C}_{ij} = \mathbf{C}_{ij}\big(m_{\tilde{\chi}_1^0}^2, q^2, m_{\tilde{\chi}_1^0}^2 ; m_F, m_V, m_S\big)$ and 
\begin{align*}
\Xi_{LL} = \mathcal{G}_{h_i SV}\mathcal{G}_{\tilde{\chi}_1^0FS}^L \mathcal{G}_{\tilde{\chi}_1^0FV}^L~, \qquad\qquad \Xi_{LR} = \mathcal{G}_{h_i SV}\mathcal{G}_{\tilde{\chi}_1^0FS}^L \mathcal{G}_{\tilde{\chi}_1^0FV}^R~,\\
\Xi_{RL} = \mathcal{G}_{h_i SV}\mathcal{G}_{\tilde{\chi}_1^0FS}^R \mathcal{G}_{\tilde{\chi}_1^0FV}^L~, \qquad\qquad \Xi_{RR} = \mathcal{G}_{h_i SV}\mathcal{G}_{\tilde{\chi}_1^0FS}^R \mathcal{G}_{\tilde{\chi}_1^0FV}^R~.
\end{align*}

In the above, $\mathbf{B_0}$, $\mathbf{C}_i$, and $\mathbf{C}_{ij}$ represent the Passarino-Veltman functions and can be evaluated using $\mathtt{LoopTools}$~\cite{Hahn:1998yk} or $\mathtt{Package}$-X~\cite{Patel:2015tea, Patel:2016fam}. Now, the total vertex corrections can be obtained as 
\begin{align}
\Gamma_{\tilde{\chi}_1^0\tilde{\chi}_1^0h_i} =& \Gamma_{\tilde{\chi}_1^0\tilde{\chi}_1^0h_i}^{(a)} + \Gamma_{\tilde{\chi}_1^0\tilde{\chi}_1^0h_i}^{(b)}+ \Gamma_{\tilde{\chi}_1^0\tilde{\chi}_1^0h_i}^{(c)}+
\Gamma_{\tilde{\chi}_1^0\tilde{\chi}_1^0h_i}^{(d)}+
\Gamma_{\tilde{\chi}_1^0\tilde{\chi}_1^0h_i}^{(e)}+
\Gamma_{\tilde{\chi}_1^0\tilde{\chi}_1^0h_i}^{(f)}\nonumber\\
=& C_L^{\rm 1L}\mathbf{P_L} + C_R^{\rm 1L} \mathbf{P_R}~,
\label{totalvertex:corrections}
\end{align}
 where $C^{\rm 1L}_{L, R}$ refer to total one-loop corrections to the coefficients of the left- and right-handed projection operators in the $\tilde{\chi}_1^0\tilde{\chi}_1^0h_i$ vertex.
 
\subsection{Renormalization of the Chargino and Neutralino Sectors: A brief reprisal}
\label{sec:Renormalization}

 In this part, we briefly discuss the various schemes used to renormalize the chargino and neutralino sectors of the MSSM.
The details of the renormalization, which include counterterms and renormalization constants, can be found in 
Refs.~\cite{Eberl:2001eu,Fritzsche:2002bi, 
Oller:2003ge, Oller:2005xg,Drees:2006um, Fowler:2009ay,Heinemeyer:2011gk,
Chatterjee:2012hkk,Bharucha:2012re}. The SUSY parameters that define charged and neutral fermions are
the electroweak gaugino mass parameters $M_1$, $M_2$, and the supersymmetric Higgsino mass parameter $\mu$.
The mass matrices involve the masses of the electroweak gauge bosons with mixing angle $\theta_W$ and $\tan\beta$; all these parameters are renormalized independently from the chargino and neutralino sectors.
The implementation parts have been discussed in Ref.~\cite{Fritzsche:2013fta}, which also covers the Feynman rules of the counterterms for a general Complex MSSM. Although we consider CP-conserving MSSM, we keep our discussion general following Ref.~\cite{Bharucha:2012re}. We start with the Fourier-transformed MSSM Lagrangian, which is bilinear in the chargino and
neutralino fields,
\begin{align}
\mathcal{L}_{\tilde{\chi}^{\pm}\tilde{\chi}^0} &= \bar{\tilde{\chi}}_i^{\pm}\cancel{p}\mathbf{P_L}\tilde{\chi}_i^{\pm} + \bar{\tilde{\chi}}_i^{\pm}\cancel{p}\mathbf{P_R}\tilde{\chi}_i^{\pm} - \bar{\tilde{\chi}}_i^{\pm}\big[\mathbb{V}^{*}\overline{\mathbb{M}}_{\tilde{\chi}^\pm}^{\rm T}\mathbb{U}^{\dagger}\big]_{ij}\mathbf{P_L}\tilde{\chi}_j^{\pm} - \bar{\tilde{\chi}}_i^{\pm}\big[\mathbb{U}\overline{\mathbb{M}}_{\tilde{\chi}^\pm}^{*}\mathbb{V}^{\rm T}\big]_{ij}\mathbf{P_R}\tilde{\chi}_j^{\pm}\nonumber\\
&+\frac{1}{2}\Big(\bar{\tilde{\chi}}_m^{0}\cancel{p}\mathbf{P_L}\tilde{\chi}_m^{0} + \bar{\tilde{\chi}}_m^{0}\cancel{p}\mathbf{P_R}\tilde{\chi}_m^{0} -\bar{\tilde{\chi}}_m^{0}\big[\mathbb{N}^{*}\overline{\mathbb{M}}_{\tilde{\chi}^0}\mathbb{N}^{\dagger}\big]_{mn}\mathbf{P_L}\tilde{\chi}_n^{0}-\bar{\tilde{\chi}}_m^{0}\big[\mathbb{N}\overline{\mathbb{M}}_{\tilde{\chi}^0}^{*}\mathbb{N}^{\rm T}\big]_{mn}\mathbf{P_R}\tilde{\chi}_n^{0}\Big)
\end{align}
where $i,j=1,2$,~$m,n=1,...,4$. We recall that, $\mathbb{U}$, $\mathbb{V}$ and $\mathbb{N}$ diagonalize the chargino and neutralino mass matrices $\overline{\mathbb{M}}_{\tilde{\chi}^\pm}$ and $\overline{\mathbb{M}}_{\tilde{\chi}^0}$ respectively 
(see Sec.~\ref{neutralinocharginosector}).

We note the following replacements of the parameters and fields.
\begin{align}
M_1&\rightarrow M_1 + \delta M_1~,\\
M_2&\rightarrow M_2 + \delta M_2~,\\
\mu &\rightarrow \mu + \delta \mu~,\\
\mathbf{P_L}\tilde{\chi}_i^{\pm} &\rightarrow \Bigg[\mathds{1} + \frac{1}{2}\delta\mathbb{Z}_{\tilde{\chi}^{\pm}}^{L}\Bigg]_{ij}\mathbf{P_L}\tilde{\chi}_{j}^{\pm}~,\\
\mathbf{P_R}\tilde{\chi}_i^{\pm} &\rightarrow \Bigg[\mathds{1} + \frac{1}{2}\delta\mathbb{Z}_{\tilde{\chi}^{\pm}}^{R}\Bigg]_{ij}\mathbf{P_R}\tilde{\chi}_{j}^{\pm}~,\\
\mathbf{P_L}\tilde{\chi}_m^{0} &\rightarrow \Bigg[\mathds{1} + \frac{1}{2}\delta\mathbb{Z}_{\tilde{\chi}^{0}}\Bigg]_{mn}\mathbf{P_L}\tilde{\chi}_{n}^{0}~,\\
\mathbf{P_R}\tilde{\chi}_m^{0} &\rightarrow \Bigg[\mathds{1} + \frac{1}{2}\delta\mathbb{Z}_{\tilde{\chi}^{0}}^{*}\Bigg]_{mn}\mathbf{P_R}\tilde{\chi}_{n}^{0}~,
\end{align} 
where 
$\delta\mathbb{Z}_{\tilde{\chi}^{\pm},\tilde{\chi}^{0}}$ refer to field renormalization constants for the physical states, general $2\times 2$ or $4\times 4$ matrices respectively. The parameter counterterms are generally complex; we need two renormalization conditions to fix those counterterms (one for the real part and another for the complex part).
The transformation matrices are not renormalized; therefore, one can write the matrix in terms of the renormalized one and a counterterm matrix in the following way
\begin{align}
\overline{\mathbb{M}}_{\tilde{\chi}^\pm} \rightarrow&\, \overline{\mathbb{M}}_{\tilde{\chi}^\pm} + \delta\overline{\mathbb{M}}_{\tilde{\chi}^\pm}~,\\
\overline{\mathbb{M}}_{\tilde{\chi}^0} \rightarrow& \,\overline{\mathbb{M}}_{\tilde{\chi}^0} + \delta\overline{\mathbb{M}}_{\tilde{\chi}^0}~,
\end{align}
with
\begin{align}
\delta\overline{\mathbb{M}}_{\tilde{\chi}^\pm}=\left(\begin{array}{c c }
\delta M_2 & \sqrt{2}\,\delta(M_W s_\beta)\\
\sqrt{2}\,\delta(M_W c_\beta) & \delta\mu\\
\end{array}\right)~,
\label{eq:Counter_chargino_mass_matrix}
\end{align}
and 
\begin{align}
\delta\overline{\mathbb{M}}_{\tilde{\chi}^0}=\left(\begin{array}{c c c c}
\delta M_1 & 0 & -\delta(M_Zs_W c_\beta) & \delta(M_Zs_W s_\beta) \\
0 & \delta M_2 & \delta(M_Zc_W c_\beta) & -\delta(M_Zc_W s_\beta)\\
-\delta(M_Zs_W c_\beta) & \delta(M_Zc_W c_\beta) & 0 & -\delta\mu\\
\delta(M_Zs_W s_\beta) & -\delta(M_Zc_W s_\beta) & -\delta\mu & 0\\
\end{array}\right).
\label{eq:Counter_nutralino_mass_matrix}
\end{align}

Also the replacements of the diagonalized matrices $\mathbb{M}_{\tilde{\chi}^{\pm}}$ and $\mathbb{M}_{\tilde{\chi}^{0}}$ can be written as 
\begin{align}
\mathbb{M}_{\tilde{\chi}^{\pm}}&\rightarrow \mathbb{M}_{\tilde{\chi}^{\pm}} + \delta \mathbb{M}_{\tilde{\chi}^{\pm}} = \mathbb{M}_{\tilde{\chi}^{\pm}} + \mathbb{V}^{*}\delta\overline{\mathbb{M}}_{\tilde{\chi}^\pm}^{\rm T} \mathbb{U}^{\dagger}~,\\
\mathbb{M}_{\tilde{\chi}^{0}}&\rightarrow \mathbb{M}_{\tilde{\chi}^{0}} + \delta \mathbb{M}_{\tilde{\chi}^{0}} = \mathbb{M}_{\tilde{\chi}^{0}} + \mathbb{N}^{*}\delta\overline{\mathbb{M}}_{\tilde{\chi}^0}^{\rm T} \mathbb{N}^{\dagger}~.
\end{align}

We can decompose the self energies into left- and right-handed vector and scalar coefficients in the following way
\begin{align}
\big[\Sigma_{\tilde{\chi}}(p^2)\big]_{\ell m} = \cancel{p}\mathbf{P_L} \big[\Sigma_{\tilde{\chi}}^L(p^2)\big]_{\ell m} + \cancel{p}\mathbf{P_R} \big[\Sigma_{\tilde{\chi}}^R(p^2)\big]_{\ell m} + \mathbf{P_L} \big[\Sigma_{\tilde{\chi}}^{SL}(p^2)\big]_{\ell m} + \mathbf{P_R}\big[\Sigma_{\tilde{\chi}}^{SR}(p^2)\big]_{\ell m}.
\label{Self_energy}
\end{align}

The coefficients of the renormalized self-energies can be written as 
\begin{align}
\big[\widehat{\Sigma}_{\tilde{\chi}^{\pm}}^{L}(p^2)\big]_{ij} &= \big[\Sigma_{\tilde{\chi}^{\pm}}^{L}(p^2)\big]_{ij} + \frac{1}{2} \big[\delta\mathbb{Z}_{\tilde{\chi}^{\pm}}^{L} + \delta\mathbb{Z}_{\tilde{\chi}^{\pm}}^{L\dagger}\big]_{ij}~,
\label{RenoSelfEnergy1}\\
\big[\widehat{\Sigma}_{\tilde{\chi}^{\pm}}^{R}(p^2)\big]_{ij} &= \big[\Sigma_{\tilde{\chi}^{\pm}}^{R}(p^2)\big]_{ij} + \frac{1}{2} \big[\delta\mathbb{Z}_{\tilde{\chi}^{\pm}}^{R} + \delta\mathbb{Z}_{\tilde{\chi}^{\pm}}^{R\dagger}\big]_{ij}~,\\
\big[\widehat{\Sigma}_{\tilde{\chi}^{\pm}}^{SL}(p^2)\big]_{ij} &= \big[\Sigma_{\tilde{\chi}^{\pm}}^{SL}(p^2)\big]_{ij} -  \big[\frac{1}{2}\delta\mathbb{Z}_{\tilde{\chi}^{\pm}}^{R\dagger} \mathbb{M}_{\tilde{\chi}^{\pm}} + \frac{1}{2}\mathbb{M}_{\tilde{\chi}^{\pm}}\delta\mathbb{Z}_{\tilde{\chi}^{\pm}}^{L} + \delta\mathbb{M}_{\tilde{\chi}^{\pm}}\big]_{ij}~,\\
\big[\widehat{\Sigma}_{\tilde{\chi}^{\pm}}^{SR}(p^2)\big]_{ij} &= \big[\Sigma_{\tilde{\chi}^{\pm}}^{SR}(p^2)\big]_{ij} -  \big[\frac{1}{2}\delta\mathbb{Z}_{\tilde{\chi}^{\pm}}^{L\dagger} \mathbb{M}_{\tilde{\chi}^{\pm}}^{\dagger} + \frac{1}{2}\mathbb{M}_{\tilde{\chi}^{\pm}}^{\dagger}\delta\mathbb{Z}_{\tilde{\chi}^{\pm}}^{R} + \delta\mathbb{M}_{\tilde{\chi}^{\pm}}^{\dagger}\big]_{ij}~,\\
\big[\widehat{\Sigma}_{\tilde{\chi}^{0}}^{L}(p^2)\big]_{n\ell} &= \big[\Sigma_{\tilde{\chi}^{0}}^{L}(p^2)\big]_{n\ell} + \frac{1}{2} \big[\delta\mathbb{Z}_{\tilde{\chi}^{0}} + \delta\mathbb{Z}_{\tilde{\chi}^{0}}^{\dagger}\big]_{n\ell}~,\\
\big[\widehat{\Sigma}_{\tilde{\chi}^{0}}^{R}(p^2)\big]_{n\ell} &= \big[\Sigma_{\tilde{\chi}^{0}}^{R}(p^2)\big]_{n\ell} + \frac{1}{2} \big[\delta\mathbb{Z}_{\tilde{\chi}^{0}}^{*} + \delta\mathbb{Z}_{\tilde{\chi}^{0}}^{\rm T}\big]_{n\ell}~,\\
\big[\widehat{\Sigma}_{\tilde{\chi}^{0}}^{SL}(p^2)\big]_{n\ell} &= \big[\Sigma_{\tilde{\chi}^{0}}^{SL}(p^2)\big]_{n\ell} -  \big[\frac{1}{2}\delta\mathbb{Z}_{\tilde{\chi}^{0}}^{\rm T} \mathbb{M}_{\tilde{\chi}^{0}} + \frac{1}{2}\mathbb{M}_{\tilde{\chi}^{0}}\delta\mathbb{Z}_{\tilde{\chi}^{0}} + \delta\mathbb{M}_{\tilde{\chi}^{0}}\big]_{n\ell}~,\\
\big[\widehat{\Sigma}_{\tilde{\chi}^{0}}^{SR}(p^2)\big]_{n\ell} &= \big[\Sigma_{\tilde{\chi}^{0}}^{SR}(p^2)\big]_{n\ell} -  \big[\frac{1}{2}\delta\mathbb{Z}_{\tilde{\chi}^{0}}^{\dagger} \mathbb{M}_{\tilde{\chi}^{0}}^{\dagger} + \frac{1}{2}\mathbb{M}_{\tilde{\chi}^{0}}^{\dagger}\delta\mathbb{Z}_{\tilde{\chi}^{0}}^{*} + \delta\mathbb{M}_{\tilde{\chi}^{0}}^{\dagger}\big]_{n\ell}~.
\label{RenoSelfEnergy8}
\end{align}

With the above machinery, in the on-shell renormalization scheme for the charginos and neutralinos, we may evaluate the counterterms $\delta M_1$, $\delta \mu$, and $\delta M_2$ by requiring that the masses of $\tilde{\chi}_{1,2}^\pm$ and one of the neutralino $\tilde{\chi}_n^0$ ($n\in \{1,...,4\}$) are defined as the poles of the corresponding tree-level propagators. 
This scheme is called $\mathtt{CCN}[n]$ where ``$\mathtt{C}$" stands for chargino, ``$\mathtt{N}$" for neutralino, and ``$n$" in the square bracket indicates that $\tilde{\chi}_n^0$ is taken as on-shell. 
One of the choices can be $\mathtt{CCN[1]}$ scheme where the
mass of the dominantly Bino-like lightest neutralino should be chosen on-shell to ensure numerical stability~\cite{Chatterjee_2012}
while a large unphysical contribution may be observed for non-bino-like lightest neutralino~\cite{Baro_2009} if taken as on-shell. 
The scheme fits well even for Bino-dominated mixed LSP scenarios, such as Bino-Higgsino 
 or even for a Bino-Wino-Higgsino neutralino.
 For other hierarchical mass patterns, e.g.,  $\lvert M_2\rvert< \lvert M_1\rvert,|\mu|$ or  $|\mu|< \lvert M_1\rvert, \lvert M_2\rvert$, $\mathtt{CCN}$[1] scheme may fail to yield numerically stable results; thus, different renormalization schemes like $\mathtt{CCN}$[2] or $\mathtt{CCN}$[4] may need to be adopted~\cite{Heinemeyer:2023pcc, Bharucha:2012re}. Wino in the first case and Higgsino for the latter are chosen to be on-shell.
 On the other hand, in the ``$\mathtt{CNN}$" scheme, one of the two charginos and two neutralinos $\tilde{\chi}_\ell^0$ and $\tilde{\chi}_m^0$ are taken to be on-shell~\cite{Drees:2006um, Chatterjee_2012, Heinemeyer:2023pcc}. Since we are interested in the Bino-dominated LSP scenarios, we stick to imposing on-shell conditions for the two charginos and one Bino-like neutralino.
\begin{figure}[H]
	\centering
	\includegraphics[width=0.75\linewidth]{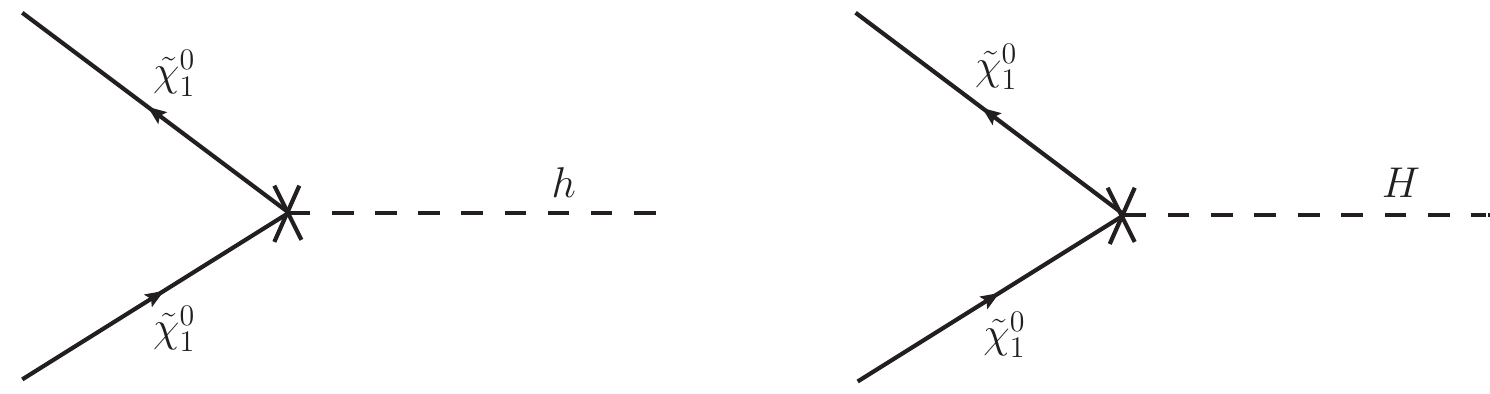}\\
    (a)\qquad\qquad\qquad\qquad\qquad\qquad\qquad\qquad\qquad\qquad\quad(b)
    \caption{Counterterm diagrams for the $\tilde{\chi}^0_1 \tilde{\chi}^0_1 h$ and $\tilde{\chi}^0_1 \tilde{\chi}^0_1 H$ vertices which should be added to the one-loop corrected $\tilde{\chi}^0_1 \tilde{\chi}^0_1 h$ and $\tilde{\chi}^0_1 \tilde{\chi}^0_1 H$ vertices, respectively to get the UV-finite results.}
	\label{fig:chichihicounter}
\end{figure}

The above field renormalization constants can be used in evaluating the vertex counterterms. Finally, we can write the expression for the vertex counterterm as follows (see Fig.~\ref{fig:chichihicounter}a and Fig.~\ref{fig:chichihicounter}b),
\begin{align}
\delta \Gamma_{\tilde{\chi}_1^0\tilde{\chi}_1^0h_i} = \mathbf{P_L} \delta C_{\tilde{\chi}_1^0\tilde{\chi}_1^0h_i}^L + \mathbf{P_R} \delta C_{\tilde{\chi}_1^0\tilde{\chi}_1^0h_i}^R~,
\label{countyerterm:eqngamma}
\end{align}
where, for the lightest CP-even scalar,
\begin{align}
\delta C_{\tilde{\chi}_1^0\tilde{\chi}_1^0h}^{L} =& -\frac{e}{4c_Ws_W^2}\bigg[\frac{4}{c_W^2}\bigl\{\big(c_W^2\delta Z_e+s_W\delta s_W\big)s_W^2\mathcal{N}^{*}_{11} + c_W c_W^2 \big(\delta s_W - s_W\delta Z_e\big)\mathcal{N}^{*}_{12}\bigr\}\big(s_\alpha \mathcal{N}^{*}_{13}+c_\alpha\mathcal{N}^{*}_{14}\big) \nonumber\\
& + s_W\Bigl\{2\big(s_W \mathcal{N}^{*}_{11} - c_W\mathcal{N}^{*}_{12}\big)\bigl\{\big(2\big[\delta\mathbf{Z}^L_{\tilde{\chi}^0}\big]_{11} + \delta Z_{hh}\big)\big(s_\alpha\mathcal{N}^{*}_{13}+c_\alpha\mathcal{N}^{*}_{14}\big) - \delta Z_{hH}\big(c_\alpha \mathcal{N}^{*}_{13}-s_\alpha\mathcal{N}^{*}_{14}\big)\bigr\}\nonumber\\
&+ \big(\big[\delta\mathbf{Z}^L_{\tilde{\chi}^0}\big]_{12} + \big[\delta\mathbf{Z}^L_{\tilde{\chi}^0}\big]_{21}\big)\bigl\{\big(s_\alpha \mathcal{N}^{*}_{13}+c_\alpha\mathcal{N}^{*}_{14}\big)\big(s_W\mathcal{N}^{*}_{21}-c_W\mathcal{N}^{*}_{22}\big)+\big(s_W\mathcal{N}^{*}_{11}-c_W\mathcal{N}^{*}_{12}\big)\big(s_\alpha\mathcal{N}^{*}_{23}\nonumber\\
&+c_\alpha\mathcal{N}^{*}_{24}\big)\bigr\} + \big(\big[\delta\mathbf{Z}^L_{\tilde{\chi}^0}\big]_{13} + \big[\delta\mathbf{Z}^L_{\tilde{\chi}^0}\big]_{31}\big)\bigl\{\big(s_\alpha\mathcal{N}^{*}_{13}+ c_\alpha\mathcal{N}^{*}_{14}\big)\big(s_W\mathcal{N}^{*}_{31}-c_W\mathcal{N}^{*}_{32}\big) + \big(s_W\mathcal{N}^{*}_{11}-c_W\mathcal{N}^{*}_{12}\big)\nonumber\\
&\times\big(s_\alpha\mathcal{N}^{*}_{33}+c_\alpha\mathcal{N}^{*}_{34}\big)\bigr\} + \big(\big[\delta\mathbf{Z}^L_{\tilde{\chi}^0}\big]_{14} + \big[\delta\mathbf{Z}^L_{\tilde{\chi}^0}\big]_{41}\big)\bigl\{\big(s_\alpha\mathcal{N}^{*}_{13}+c_\alpha\mathcal{N}^{*}_{14}\big)\big(s_W\mathcal{N}^{*}_{41}-c_W\mathcal{N}^{*}_{42}\big) + \big(s_W\mathcal{N}^{*}_{11}\nonumber\\
& -c_W\mathcal{N}^{*}_{12}\big)\big(s_\alpha\mathcal{N}^{*}_{43}+c_\alpha\mathcal{N}^{*}_{44}\big)\bigr\}\Bigr\}\bigg]
\end{align}
and 
\begin{align}
\delta C_{\tilde{\chi}_1^0\tilde{\chi}_1^0h}^{R} =& -\frac{e}{4c_W s_W^2}\Bigg[\frac{4}{c_W^2}\bigl\{\big(c_W^2\delta Z_e+s_W\delta s_W\big)s_W^2\mathcal{N}_{11} + c_W c_W^2 \big(\delta s_W - s_W\delta Z_e\big)\mathcal{N}_{12}\bigr\}\big(s_\alpha \mathcal{N}_{13}+c_\alpha\mathcal{N}_{14}\big) \nonumber\\
& +s_W\Bigl\{2\big(s_W\mathcal{N}_{11}-c_W\mathcal{N}_{12}\big)\bigl\{\big(s_\alpha\delta Z_{hh} - c_\alpha\delta Z_{hH}\big)\mathcal{N}_{13} + \big(c_\alpha\delta Z_{hh} + s_\alpha\delta Z_{hH}\big)\mathcal{N}_{14} + \big(2\big[\delta \mathbf{Z}_{\tilde{\chi}^0}^{R}\big]_{11}\big)\nonumber\\
&\times \big(s_\alpha\mathcal{N}_{13}+c_\alpha\mathcal{N}_{14}\big)\bigr\} + \big(\big[\delta\mathbf{Z}_{\tilde{\chi}^0}^{R}\big]_{12} + \big[\delta\mathbf{Z}_{\tilde{\chi}^0}^{R}\big]_{21}\big) \bigl\{\big(s_\alpha\mathcal{N}_{13}+c_\alpha\mathcal{N}_{14}\big)\big(s_W\mathcal{N}_{21}-c_W\mathcal{N}_{22}\big) + \big(s_W\mathcal{N}_{11}\nonumber\\
&-c_W\mathcal{N}_{12}\big)\big(s_\alpha\mathcal{N}_{23}+ c_\alpha \mathcal{N}_{24}\big)\bigr\} + \big(\big[\delta\mathbf{Z}_{\tilde{\chi}^0}^R\big]_{13} + \big[\delta\mathbf{Z}_{\tilde{\chi}^0}^R\big]_{31}\big)\bigl\{\big(s_\alpha\mathcal{N}_{13} + c_\alpha\mathcal{N}_{14}\big)\big(s_W\mathcal{N}_{31} - c_W\mathcal{N}_{32}\big) \nonumber\\
&+ \big(s_W\mathcal{N}_{11}-c_W\mathcal{N}_{12}\big)\big(s_\alpha\mathcal{N}_{33}+c_\alpha\mathcal{N}_{34}\big)\bigr\} + \big(\big[\delta\mathbf{Z}_{\tilde{\chi}^0}^R\big]_{14} + \big[\delta\mathbf{Z}_{\tilde{\chi}^0}^R\big]_{41}\big)\bigl\{\big(s_\alpha\mathcal{N}_{13}+c_\alpha\mathcal{N}_{14}\big)\big(s_W\mathcal{N}_{41}\nonumber\\
&-c_W\mathcal{N}_{42}\big) + \big(s_W\mathcal{N}_{11}-c_W\mathcal{N}_{12}\big)\big(s_\alpha\mathcal{N}_{43} + c_\alpha\mathcal{N}_{44}\big)\bigr\}\Bigr\} \Bigg]~.
\end{align}

Similarly, the counterterm for the heavy Higgs can be obtained by the replacements $s_\alpha\to c_\alpha$, $c_\alpha\to -s_\alpha$, $\delta Z_{hh}\to \delta Z_{HH}$, and $\delta Z_{hH}\to -\delta Z_{hH}$.

In the above, $\delta\mathbf{Z}_{\tilde{\chi}^0}^{L, R}$ involves renormalized self-energies and counterterms of the mass matrices of the physical states~\cite{Heinemeyer:2011gk, Bharucha:2012re}. Similarly,
$\delta {Z}_{hH}$, $\delta {Z}_{hh}$ and 
$\delta {Z}_{HH}$ come from the renormalization of the neutral Higgs sector.
  \begin{eqnarray}
M_{h_i}&\to & M_{h_i} + \delta M_{h_i}\, ,
\\
\left(\begin{array}{c} h \\ H \end{array}\right) &\to &
\left (\begin{array}{cc}
1 + \frac{1}{2}\delta Z_{hh} & \frac{1}{2}\delta Z_{hH}\\
\frac{1}{2}\delta Z_{Hh}& 1 + \frac{1}{2}\delta Z_{HH}
\end{array}\right)
\left(\begin{array}{c} h \\ H \end{array}\right)~.
\end{eqnarray}

The other terms in the counterterm vertices are already present in
the renormalization of the SM. Here, we refer to~\cite{Fritzsche:2002bi, Fritzsche_2014} for the relevant expressions. For instance, the renormalization constants, e.g., $\delta Z_e$, $\delta s_W$, are fixed by the on-shell conditions. Thus, as a default option in $\mathtt{FormCalc}$, we use the fine-structure constant $\alpha=\alpha(0)=1/137.0359996$ defined at the Thomson limit.\footnote{We may recall that the renormalization of electric charge can be written as $e^{\rm bare}\to e(0)(1+\delta Z_e^{(0)})=e(M_Z^2)(1+\delta Z_e^{e(M_Z^2)})$+ higher orders, with $e(M_Z^2)=e(0)/(1-\frac{1}{2}\Delta\alpha)$ and $\delta Z_e^{e(M_Z^2)}=\delta Z_e^{(0)} -\frac{1}{2}\Delta\alpha$ where $\Delta\alpha$ is a finite quantity involving the contributions from the $e,\mu,\tau$ leptons and the light quarks (i.e., all except $t$)~\cite{Chatterjee:2012hkk, Denner:1991kt, Hagiwara:2011af, Steinhauser:1998rq}. On the contrary, if one uses the ``running on-shell" value of $\alpha$, i.e., $\alpha(M_Z^2)=1/128.93$ or the running $\overline{\rm MS}$ value $\hat{\alpha}(M_Z)= 1/127.932$ (which usually spectrum-generator like $\mathtt{SPheno}$ considers), then the definition of $\delta Z_e^{e(M_Z^2)}$ has to be adopted. To this end, $\mathtt{FormCalc}$ calculates the charge renormalization constant at the Thomson limit, i.e., $\delta Z_e^{(0)}$~\cite{Fritzsche:2013fta}, so we always use $\alpha(0)$ or $e(0)$.}
Similarly, the on-shell definition of $s_W$ has been fixed as, $s_W^2 =1 - \frac{M^2_W}{M^2_Z}$. Though $M_W$ is normally computed using the fine-structure constant in the Thomson limit $\alpha(0)$, the Fermi constant $G_\mu$ and mass of the $Z$ boson, here we stick to $M_W=80.3484$ in the
analysis. This is within the $\sim 1 \sigma$ variation if $W$ boson mass measurements are performed by the ATLAS, LHCb, and D0 experiments, excluding the recent CDF results: $M_W=80.3692 \pm 0.0133$ GeV~\cite{Amoroso:2023pey}.
The relatively large theoretical uncertainty arises due to parton distribution functions.
\footnote{In the former calculation, higher-order corrections involving the standard model and the MSSM are needed. Thus, if we compute the $M_W$ instead, the resultant change in the
	SI-DD cross-section is $\le 1\%$.}

Finally, we club the vertex corrections and counterterms as $\Gamma_{\tilde{\chi}_1^0\tilde{\chi}_1^0h_i} + \delta \Gamma_{\tilde{\chi}_1^0\tilde{\chi}_1^0h_i}$ to
obtain the UV-finite amplitude where $\Gamma_{\tilde{\chi}_1^0\tilde{\chi}_1^0 h_i}$ and $\delta \Gamma_{\tilde{\chi}_1^0\tilde{\chi}_1^0h_i}$ are defined in  Eq.~\eqref{totalvertex:corrections} and Eq.~\eqref{countyerterm:eqngamma}, respectively. As we will see, to get the UV-finite result, we have to use tree-level masses for the physical states inside the loop.

\section{Precision measurements and constraints from direct searches}
\label{sec:pre_dir}
Here, we summarize different avenues of precision and collider phenomenology, which can be marked along with the SI-DD of the neutralino DM in different parts of the MSSM parameter space. We consider
3 GeV
theoretical uncertainty in calculating SUSY Higgs mass leads to the following range~\cite{Bahl:2019hmm} for the SM-like Higgs mass in the MSSM.
\begin{equation}
122<m_h<128 ~\rm GeV.
\end{equation}
Otherwise, we respect the constraints originating from the $B$-physics measurements\footnote{$B$-physics constraints are satisfied using $\mathtt{SPheno}$-4.0.4 that uses the model file for the MSSM from $\mathtt{SARAH}$ where the input mass parameters for the SUSY models are defined in $\overline{\rm DR}$-scheme.} at $2\sigma$ variations, e.g., $ 3.02 \times 10^{-4} < BR(b \to s\gamma) < 3.62 \times 10^{-4}$~\cite{HFLAV:2019otj}, 
$2.23 \times 10^{-9} < BR(B_s\to \mu^+\mu^-) <  3.63 \times 10^{-9}$~\cite{Altmannshofer:2021qrr}. We recall that our primary interest is to observe 
the role of the renormalized $\tilde \chi_1^0 \tilde \chi_1^0 h_i$ vertex in the SI-DD 
where $\delta a_\mu$ can be satisfied using SUSY contributions. Assuming the $\tilde{\chi}_1^0$ to be the only source for DM, 
we note the acceptable value of the relic abundance data~\cite{WMAP:2012nax, Planck:2018vyg}
 \begin{equation}
   \Omega_{\rm DM}h^2=0.1198\pm 0.0012.
   \label{eq:relic}
 \end{equation}
\noindent
However, as noted in Sec.~\ref{sec:intro}, the constraint Eq.~\eqref{eq:relic} is not always endorsed as a necessary condition,
especially for understanding the parametric dependence to highlight the region of higher
NLO corrections.
It is well known that lighter EW spectra with masses not far away from a few hundred GeV are preferred for compliance with $\delta a_\mu$. The direct search constraints from LHC or LEP
can be potentially important for consideration. 
As mentioned, we set squarks and gluino masses at $\geq$~4 TeV to cope with the LHC constraints~\cite{ATLAS:2020syg, CMS:2019zmd}.
Thus, it is instructive to lay down a brief discussion of the recent results on the anomalous magnetic moment of the muon and the status of the LHC searches on the MSSM parameter space. It may be added here that the LHC constraints on SUSY searches are finally verified using $\mathtt{SModelS-2.3.0}$~\cite{Kraml:2013mwa,Dutta:2018ioj,Khosa:2020zar,Alguero:2021dig}.

\subsection{Anomalus magnetic moment of muon ($\delta a_\mu$) in the MSSM}
\label{sec:amu}
The recent $a_\mu$ measurement by FNAL~\cite{Muong-2:2021vma, Muong-2:2021ojo} has confirmed the earlier result by the E821 experiment at Brookhaven, yielding the experimental average $a_\mu^{\rm EXP}= 116592061(41)\times10^{-11}$ which leads to 
a 4.2$\sigma$ discrepancy~\cite{Muong-2:2021ojo} compared to the SM value $a_\mu^{\rm SM} = (116591810 \pm 43)\times 10^{-11}$~\cite{Aoyama:2020ynm}, which is mainly based on the Refs.~\cite{Davier:2017zfy, Keshavarzi:2018mgv, Colangelo:2018mtw, Davier:2019can, Keshavarzi:2019abf, Kurz:2014wya, Hoferichter:2019mqg, Melnikov:2003xd, Masjuan:2017tvw, Colangelo:2017fiz, Hoferichter:2018kwz, Gerardin:2019vio, Bijnens:2019ghy, Colangelo:2019uex, Colangelo:2014qya, Blum:2019ugy, Aoyama:2012wk, atoms7010028, Czarnecki:2002nt, Gnendiger:2013pva}.
\begin{align}
\delta a_\mu = a_\mu^{\rm EXP}-a_\mu^{\rm SM} = 251\pm 59\times10^{-11}.
\end{align}

The E989 experiment at Fermilab recently released an update regarding the measurement of $a_\mu$ from Run-2 and Run-3. 
The new combined value yields a deviation of\footnote{The value of $(g-2)_\mu$ from Run-2 and Run-3 is $a_\mu^{\rm Run-2,3} = (116592055 \pm 24)\times 10^{-11}$. Therefore, the new experimental average becomes $a_\mu^{\rm Exp(New)} = (116592059 \pm 22)\times 10^{-11}$~\cite{Muong-2:2023cdq}.}
\begin{align}
   \delta a_\mu^{\rm New} = (249 \pm 48)\times 10^{-11} 
   \label{eq:g2aoyama}
\end{align}
which leads to a $5.1\sigma$ discrepancy. However,
 $\delta a_\mu^{\rm New}$ quoted in Eq.\eqref{eq:g2aoyama} is subject to 
 SM theory prediction, mainly the leading-order hadronic vacuum polarisation (HVP) contributions. Here, we stick to ~\cite{Aoyama:2020ynm} where dispersive techniques
are used to extract the leading-order HVP contribution from the
$e^+e^-\to$ hadrons data. Instead, if the lattice-QCD result for HVP by BMW collaboration is used, $\delta a_\mu^{\rm New}$
     reduces to 1.6$\sigma$, leading to 2.1$\sigma$ tension with the ${e^+e^-}$ determination of the HVP contribution. In this regard, Ref. \cite{Colangelo:2022vok} discusses how windows in Euclidean time can help to reduce the potential conflicts between evaluations of the HVP contribution to the $(g-2)_\mu$ in lattice-QCD \footnote{Refs.~\cite{FermilabLatticeHPQCD:2023jof, ExtendedTwistedMass:2022jpw, Ce:2022kxy} also studied recently the window observable for the HVP contribution to $(g-2)_\mu$ from lattice QCD calculations.} and from $e^+e^-\to$ hadrons cross-section data.
Along the same line, Ref. \cite{Wittig:2023pcl} also manifested the tension between the lattice QCD approach and the traditional data-driven approach, while for the latter, the recent CMD-3 result was not used. 
    Recently, Ref.\cite{CMD-3:2023rfe} calculated the $(g-2)_\mu$ using the data-driven approach. They measured the cross-section of the dominant channel $e^{+}e^{-}\to \pi^{+}\pi^{-}$ using the CMD-3 detector at a center of mass energy below 1 GeV, 
    though the result seems to be incompatible with previous determinations~\cite{CMD-3:2023alj, CMD-2:2003gqi, KLOE-2:2017fda, BESIII:2015equ, BaBar:2012bdw}

\begin{figure}[H]
	\centering
	\includegraphics[width=0.90\linewidth]{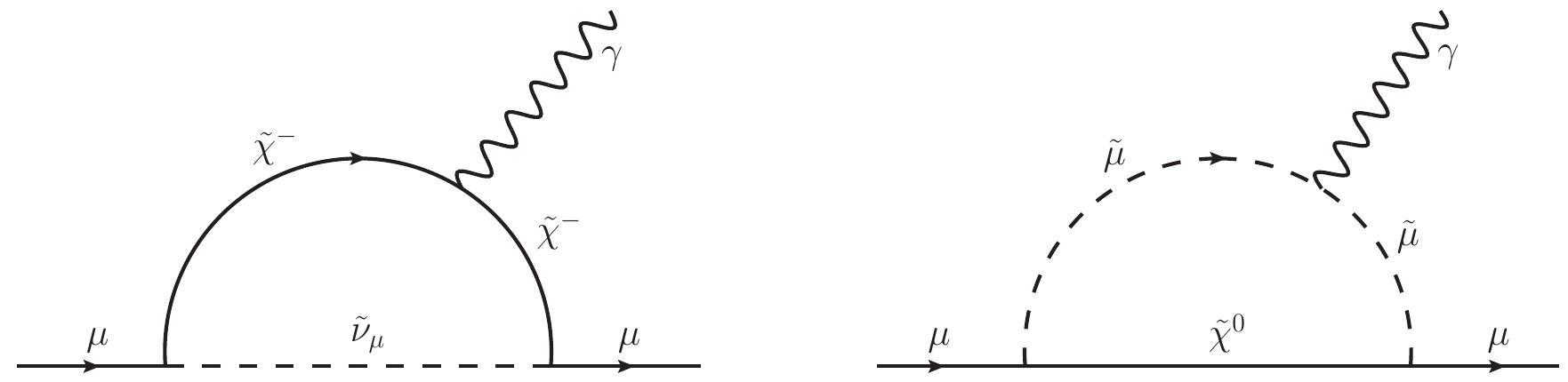}\\
	(a)\qquad\qquad\qquad\qquad\qquad\qquad\qquad\qquad\qquad\qquad\qquad\quad\,\,(b)
	\caption{One-loop contributions to $a_\mu$ in MSSM}
	\label{fig:g2mu}
\end{figure}

In the MSSM, the one-loop contributions to the anomalous magnetic moment of muon or $a_\mu$, as shown in Fig.~\ref{fig:g2mu}, are mainly mediated by $\tilde{\chi}^{-}-\tilde{\nu}_\mu$ and $\tilde{\mu}-\tilde{\chi}^{0}$~\cite{Lopez:1993vi, Chattopadhyay:1995ae, Moroi:1995yh, Chattopadhyay:2000ws, Martin:2001st, Heinemeyer:2003dq, Stockinger:2006zn, Cho:2011rk, Endo:2013lva, Endo:2013bba, Chakraborti:2014gea, Chakraborti:2015mra, Chowdhury:2015rja,  Kowalska:2015zja, Gomez:2018efz, Chakraborti:2021mbr, Chakraborti:2021kkr, Ali:2021kxa, Lindner:2016bgg}  

The $(g-2)_\mu$ prior to the Fermilab Run-1 result are studied in the Refs.~\cite{Hagiwara:2017lse, Endo:2020mqz, Chakraborti:2020vjp}.
  In the aftermath of Fermilab Run-1, the $(g-2)_\mu$ was studied in the Refs.~\cite{Aboubrahim:2021xfi, Athron:2021iuf, VanBeekveld:2021tgn, Endo:2021zal, Chakraborti:2021dli, Chakraborti:2021bmv}.

The contributions can be written as~\cite{Martin:2001st, Stockinger:2006zn},
\begin{align}
\delta a_\mu^{\tilde{\chi}^0} =& \frac{m_\mu}{16\pi^2} \sum_{\ell=1}^{4} \sum_{m=1}^{2}\Bigg[-\frac{m_\mu}{12m^2_{\tilde{\mu}_m}}\big(\lvert n_{\ell m}^{L}\rvert^2 + \lvert n_{\ell m}^{R}\rvert^2 \big)\mathcal{F}_{1}^{\mathcal{N}}\big(x_{\ell m}\big) + \frac{m_{\tilde{\chi}_\ell^0}}{3m^2_{\tilde{\mu}_m}}{\rm Re}\big[n_{\ell m}^L n_{\ell m}^R\big]\mathcal{F}_2^\mathcal{N}\big(x_{\ell m}\big)\Bigg]~,\\
\delta a_\mu^{\tilde{\chi}^\pm} =& \frac{m_\mu}{16\pi^2} \sum_{k=1}^{2} \Bigg[\frac{m_\mu}{12m^2_{\tilde{\nu}_\mu}}\big(\lvert c_{k}^{L}\rvert^2 + \lvert c_{k}^{R}\rvert^2 \big)\mathcal{F}_{1}^{C}\big(x_{k}\big) + \frac{2 m_{\tilde{\chi}_k^\pm}}{3m^2_{\tilde{\nu}_\mu}}{\rm Re}\big[c_{k}^L c_{k}^R\big]\mathcal{F}_2^C\big(x_{k}\big)\Bigg]~,
\end{align}
where the summations label the neutralino, smuon, and chargino mass eigenstates, respectively, and 
\begin{align}
n_{\ell m}^L =& \frac{1}{\sqrt{2}}\big(g_1\mathcal{N}_{\ell 1} + g_2\mathcal{N}_{\ell 2}\big)X_{m1}^{*} - y_\mu \mathcal{N}_{\ell 3}X_{m2}^{*}~,\\
n_{\ell m}^R =& \sqrt{2}g_1\mathcal{N}_{\ell 1} X_{m2} + y_\mu \mathcal{N}_{\ell 3}X_{m1}~,\\
c_{k}^L =& -g_2 \mathcal{V}_{k1}~,\\
c_{k}^{R} =& y_\mu \mathcal{U}_{k2}~,
\end{align}
with $y_\mu $ is the muon Yukawa coupling. The loop functions are given by 
\begin{align}
\mathcal{F}_1^\mathcal{N}\big(x\big) =& \frac{2}{\big(1-x\big)^4}\Big[1-6x + 3x^2 +2x^3 -6x^2\ln\big(x\big)\Big]~,\\
\mathcal{F}_2^\mathcal{N}\big(x\big) =& \frac{3}{\big(1-x\big)^3}\Big[1-x^2 + 2x\ln\big(x\big)\Big]~,\\
\mathcal{F}_1^C\big(x\big) =& \frac{2}{\big(1-x\big)^4}\Big[2+3x-6x^2+x^3 +6x\ln\big(x\big)\Big]~,\\
\mathcal{F}_1^C\big(x\big) =& -\frac{3}{2\big(1-x\big)^3}\Big[3-4x+x^2+2\ln\big(x\big)\Big]~, 
\end{align}
where the definition of the variables $x_{\ell m}= m^2_{\tilde{\chi}_\ell^0}/m^2_{\tilde{\mu}_m}$ and $x_{k}= m^2_{\tilde{\chi}_k^\pm}/m^2_{\tilde{\nu}_\mu}$ have been used. Since $\delta a_\mu^{\rm SUSY}>0$ for $\mu>0$ and $\delta a_\mu^{\rm SUSY}<0$ for $\mu<0$~\cite{Chattopadhyay:2002zq, Chattopadhyay:2002jx}, here, we restrict ourselves to the case where $\mu$ is real and positive, i.e., $\mu>0$ in order to have the positive SUSY contributions to $(g-2)_\mu$.
For a light Bino-like neutralino, i.e., for the scenario $M_1<<M_2,\mu$, the loops contain only a light Bino and the smuons. In that case, one can write~\cite{Martin:2001st},
\begin{align}
\delta a_\mu^{\rm Bino-like} = \frac{g_1^2}{48\pi^2}\frac{m_\mu^2M_1{\rm Re}\big[\mu\tan\beta-A_\mu^{*}\big]}{m^2_{\tilde{\mu}_2}-m^2_{\tilde{\mu}_1}}\Bigg[\frac{\mathcal{F}_2^\mathcal{N}\big(x_{11}\big)}{m^2_{\tilde{\mu}_1}} - \frac{\mathcal{F}_2^\mathcal{N}\big(x_{12}\big)}{m^2_{\tilde{\mu}_2}}\Bigg]~,
\label{Eq:g2binolike}
\end{align}
where $x_{1m}= M_1^2/m^2_{\tilde{\mu}_m}$. For numerical evaluations for
flavor constraints including $(g-2)_\mu$, we
use $\mathtt{SPheno}$-4.0.4~\cite{Porod:2003um, Porod:2011nf} that uses $\mathtt{SARAH}$-4.14.5~\cite{Staub:2013tta, Staub:2015kfa} for generating the MSSM model files.

It is instructive to note that $\mathtt{SPheno}$ calculates all the one-loop SUSY contributions to $(g-2)_\mu$. But $(g-2)_\mu$ also may receive 
contributions from the two-loop (mainly Barr-Zee type) diagrams involving fermion/sfermion in the loop~\cite{Fargnoli:2013zda, Fargnoli:2013zia} which 
in the present case may not offer any significant changes. \footnote{ For the two-loop contributions for
$(g-2)_\mu$ we refer to $\mathtt{GM2Calc}$~\cite{Athron:2015rva, Athron:2021evk, vonWeitershausen:2010zr, Fargnoli:2013zia, Bach:2015doa, Cherchiglia:2016eui}. 
 $\mathtt{GM2Calc}$ uses the on-shell masses for the following parameters
	\begin{align}
	M_1~,\qquad M_2~,\qquad \mu~,\qquad m_{\tilde{\mu}_L}~,\qquad m_{\tilde{\mu}_R}~,
	\label{eq:onshellpara}
	\end{align}
	where  $m_{\tilde{\mu}_{L}}$ and $m_{\tilde{\mu}_{R}}$ are the smuon mass parameters. On the other hand, $\mathtt{SPheno}$, which uses the model file for the MSSM from $\mathtt{SARAH}$, defines the input mass parameters in $\overline{\rm DR}$-scheme.
   One may find that $\mathtt{GM2Calc}$~\cite{Athron:2015rva, Athron:2021evk, vonWeitershausen:2010zr, Fargnoli:2013zia, Bach:2015doa, Cherchiglia:2016eui} is a more reliable tool for the scenarios where two-loop results can be important.}

\subsection {LHC and LEP bounds on Electroweakinos and Sleptons} 
\label{sec:collider}
For the searches of charginos/neutralinos and sleptons at $\sqrt{s}=$13~TeV by ATLAS and CMS we refer the reader~\cite{ATLAS:2018ojr, CMS:2017moi, ATLAS:2018eui,ATLAS:2019lff,ATLAS:2019lng,ATLAS:2020pgy,ATLAS:2019wgx,ATLAS:2021moa} and~\cite{CMS:2018kag,CMS:2018szt,CMS:2018eqb,CMS:2020bfa,CMS:2021edw}. Also, the direct production of charginos, neutralinos, and sleptons in the final states with two leptons have been searched by ATLAS at $\sqrt{s}=$8~TeV~\cite{ATLAS:2014zve}.
Following the ATLAS searches~\cite{ATLAS:2019lng, ATLAS:2017vat}, Higgsino-like neutralinos or charginos above the LEP limit can be constrained for a mass difference $\Delta m (\tilde \chi_2^0/\tilde{\chi}^\pm_1, \tilde \chi_1^0)\geq 2.4$ GeV. Similarly, a lower limit  $m_{\tilde{\chi}^\pm_1} \simeq m_{\tilde{\chi}^0_2} \geq 193$ GeV can be set for a mass splitting 
of 9.3 GeV. Using CMS results $m_{\tilde{\chi}^\pm_1}\simeq 150 \rm~ GeV$ for a mass difference
$\sim 3$ GeV~\cite{CMS:2021edw} can be placed. For a recent review of searches for Electroweakinos at the LHC, see~\cite{Adam:2021rrw}.

Usually,
in the Higgsino-like LSP models, a few combinations of electroweak states may be important: $\tilde{\chi}^0_1\tilde{\chi}^0_2,
\tilde{\chi}^0_1\tilde{\chi}^\pm_1,\tilde{\chi}^0_2\tilde{\chi}^\pm_1,\tilde{\chi}^+_1\tilde{\chi}^-_1$. The possibility of having a lighter Electroweakino in the MSSM without confronting the LHC searches requires a compressed mass spectra; thus it relies on the soft leptons or jets arising in the decays of charginos and neutralinos via off-shell EW gauge bosons $\tilde{\chi}^\pm_1 \to W^{(*)}+\tilde{\chi}^0_1$ and $\tilde{\chi}^0_i \to Z^{(*)}/h_{\rm SM} + \tilde{\chi}^0_1$ 
($h_{\rm SM}$ refers to an SM-like Higgs scalar in any BSM model). 
In the present context,  $\tilde{\chi}^0_1$ can be $\tilde B$ dominated, whereas relatively heavier neutralinos $\tilde{\chi}^0_2,\tilde{\chi}^0_3$
and $\tilde{\chi}^\pm_1$ may become Higgsino-like. 
Even a better-compressed scenario can be conceived when a Wino-like
$ \tilde \chi_2^0$
is lighter than Higgsino-like $\tilde{\chi}^0_3,\tilde{\chi}^0_4$.
 However, Higgsino-like states cannot be too light since moderate/large gaugino-Higgsino mixings have been excluded via the SI-DD results. For instance, the direct detection of Bino-like $\tilde \chi_1^0$-nucleon cross-section set a limit of $\mu \geq 600$ GeV for an LSP mass of 100 GeV (see, e.g., Fig.~\ref{fig:sigmabwh11}). With this in mind, the presence of lighter sleptons and sneutrinos becomes necessary to satisfy $\delta a_\mu$.
Then Higgsino-like heavier charginos/neutralinos may decay through $\tilde \ell(\ell) \nu(\tilde \nu)$ or $\tilde \ell \ell,\tilde \nu \nu$ or even via Wino-like neutralino states.

We summarise here the potentially important final states comprised of $l^+l^-$ ($l \in e,\mu,\tau$)
pair, jets, and missing transverse momentum through pair production of charginos, neutralinos, and sleptons, searched at the ATLAS and CMS collaborations. 
\begin{itemize}
    \item 
    $PP\to \tilde \chi_1^\pm \tilde \chi_2^0 \to Z \chi_1^0 W^\pm \chi_1^0$ ($W$ and $Z$ bosons can be off-shell) are considered in Ref.~\cite{ATLAS:2022zwa,ATLAS:2021moa,ATLAS:2019wgx,CMS:2018szt,ATLAS:2018eui}. For on-shell vector bosons, $Z$ and $W$ decay to leptonic and leptonic (hadronic) final states, respectively. 
   An ISR jet may lead the required handle to detect the soft leptons above the SM background~\cite{ATLAS:2019lng, ATLAS:2018eui}.    
   The lower limits for equal-mass $\tilde \chi_1^\pm \tilde \chi_2^0$ are $\sim$ 800 GeV for a massless $\tilde \chi_1^0$~\cite{ATLAS:2022zwa, CMS:2022sfi}, for decaying with 100\% BR in the gauge boson final states.
    The second lightest neutralino may decay through $h_{\rm SM}$, $\tilde \chi_2^0 \to h_{\rm SM} + \tilde \chi_1^0$ with 100\% BR, is considered
    in Ref.~\cite{ATLAS:2021moa,ATLAS:2020pgy,CMS:2018szt}. For the $Wh_{\rm SM}$ mediated signals, and, 
$\Delta m (\tilde \chi_2^0,\tilde \chi_1^0) \geq m_{h_{\rm SM}}$ minimum  $\tilde \chi_1^\pm \tilde \chi_2^0$ mass set at 190 GeV~\cite{ATLAS:2021moa}. Additionally, pair production of 
charginos followed by its decay to $\tilde \ell(\ell) \nu(\tilde \nu)$ was considered in 
Ref.~\cite{ATLAS:2019lff}.
    \item
In the parameter space of our concern, all the Electroweakinos may be below the TeV scale. 
Thus, pair production of heavier Electroweakinos may be important, especially if each of them decays into a lighter Electroweakino and an on-shell $W$, $Z$ and SM-like Higgs boson~\cite{ATLAS:2021yqv}.

 \item 
$PP\to \tilde \ell \tilde \ell \to \ell  \tilde \chi_1^0 \ell \tilde \chi_1^0 : $ Direct pair-production of sleptons with $\tilde l$ refers mainly $\tilde e,\tilde \mu$
with each decaying into a charged lepton and $\tilde \chi_1^0$,  have been searched at~\cite{ATLAS:2022hbt,ATLAS:2019lff,ATLAS:2019lng,ATLAS:2018ojr,CMS:2018eqb}. 
Usually, for a lighter $\tilde e$ and $\tilde \mu$ with masses $\leq 150$ GeV, the mass splitting $\Delta m ({\tilde \ell}, {\tilde \chi_1^0}) \leq$  
50 GeV is desired to have an acceptable parameter space point.
We always keep track of $\Delta m ({\tilde \ell}, m_{\tilde \chi_1^0})$ and consider  $m_{\tilde{\ell}}\geq 100$~GeV.
For relatively heavier sleptons, $m_{\tilde{\ell}}-m_{\tilde \chi_1^0}$ plane is depicted
in Ref.~\cite{CMS:2018eqb,ATLAS:2022hbt}.

\item Charginos/neutralinos can potentially decay into sleptons, which then decay into leptons, see, e.g., \cite{ATLAS:2018ojr}. The mass limits for $\tilde{\chi}_2^0, \tilde{\chi}_1^\pm$ can be excluded up to 1.1 TeV for neutralino masses less than 550 GeV. These channels are important if the respective BRs are $100\%$. We always keep track of these constraints and find them to be not very important in most of the parameter space.
\end{itemize}
The aforesaid potentially important searches related to sleptons and  EW particles are already included in the recent $\mathtt{SModelS-2.3.0}$. This includes the new ATLAS and CMS results relevant in the present context,~\cite{ATLAS:2022hbt, ATLAS:2019lff, ATLAS:2022zwa, ATLAS:2021yqv, CMS:2022sfi, CMS:2022vpy}.
\section{Methodology to implement one-loop $\tilde \chi_1^0\tilde \chi_1^0 h_i$ vertex to $\mathtt{MicrOMEGAs}$}
\label{sec:methodology}
We list here the necessary steps
followed to evaluate the one-loop renormalized $\tilde \chi_1^0\tilde \chi_1^0 h_i$ vertex numerically, hence the SI-DD
of the LSP. 
We use $\mathtt{FeynArts}$-3.11~\cite{Hahn:2000kx, KUBLBECK1990165, Hahn:2001rv, Fritzsche:2013fta}, $\mathtt{FormCalc}$-9.9~\cite{Hahn:1998yk, Fritzsche:2013fta}, and $\mathtt{LoopTools}$-2.16~\cite{Hahn:1998yk} at different stages as discussed below.
\begin{itemize}
	\item 
 $\mathtt{FeynArts}$ contains the model files $\mathtt{MSSM}$.mod and $\mathtt{MSSMCT}$.mod in which all the Feynman rules are implemented. 
 We generate all the relevant one-loop diagrams for the $\tilde{\chi}_1^0\tilde{\chi}_1^0h_i$ (where $h_i=h, H$) vertex using $\mathtt{FeynArts}$. 
	\item We choose the Feynman gauge for computing the loops, which is also the default choice of $\mathtt{FeynArts}$. We include all the diagrams, including  Goldstone bosons, to get the gauge invariant result. There are 234 diagrams for the $h$-mediated or the $H$-mediated processes for consideration.
	\item 
 The total amplitude for all the one-loop diagrams, 
as evaluated by $\mathtt{FeynArts}$, leaves the momentum integrals unevaluated.
 $\mathtt{FormCalc}$ evaluates all the momentum integrals and writes the amplitude in a simplified form through its internal abbreviation functions.
		\item The vertex correction parts 
 can be cast as $\Gamma_{\tilde{\chi}_1^0\tilde{\chi}_1^0h_i} = {C}^{\rm 1L}_L \mathbf{P_L} + {C}^{\rm 1L}_R \mathbf{P_R}$, where $\mathbf{P_L}$ and $\mathbf{P_R}$ are left- and right-handed projection operators. We extract the ${C}^{\rm 1L}_{L, R}$-parts and 
 convert it to $\mathtt{Fortran}$ codes using routines in $\mathtt{FormCalc}$.
  \item
We use the spectrum calculator $\mathtt{SPheno}$-4.0.4~\cite{Porod:2003um, Porod:2011nf} for numerical evaluations. The model 
 files for the MSSM are generated by $\mathtt{SARAH}$-4.14.5~\cite{Staub:2013tta, Staub:2015kfa}. Then SM and MSSM inputs and MSSM outputs are fed to our code that calculates  ${C}^{\rm 1L}_L$ and ${C}^{\rm 1L}_R$. At this stage, $\mathtt{LoopTools}$ has been used to
 evaluate the Passarino-Veltman scalar integrals.

	\item The loop-corrected $\tilde{\chi}_1^0\tilde{\chi}_1^0h_i$ vertex contain UV-divergencies, unless
 $\tilde{\chi}_1^0$ is a pure state.
  The renormalization of $\tilde{\chi}_1^0\tilde{\chi}_1^0h_i$ vertex is done by using $\mathtt{FeynArts}$, $\mathtt{FormCalc}$, and $\mathtt{LoopTools}$.
 We generate the counterterm diagrams for the $\tilde{\chi}_1^0\tilde{\chi}_1^0h_i$ vertex and create the amplitudes using $\mathtt{FeynArts}$. 
 As said before, the relevant expressions may be found in Refs.~\cite{Eberl:2001eu, Fritzsche:2002bi, 
Oller:2003ge, Oller:2005xg,Drees:2006um, Fowler:2009ay,Heinemeyer:2011gk,
Chatterjee:2012hkk,Bharucha:2012re}.
	 Thereafter, we choose an appropriate renormalization scheme for our scenarios. Since we focus on the Bino-Higgsino-like and Bino-Wino-Higgsino-like mixed neutralino scenarios, dominated by $\tilde B$ component, the suitable scheme is $\mathtt{CCN}[1]$ which is also the default choice in $\mathtt{FormCalc}$.
 Using $\mathtt{FormCalc}$ and adopting $\mathtt{CCN}[1]$ scheme, we evaluate all the relevant renormalization constants that are contained in the amplitude of the vertex counterterms.
 The latter has the structure $\delta C_L \mathbf{P_L} + \delta C_R \mathbf{P_R}$ 
 which is the same as that of the vertex with $\delta C_L$ 
 and $\delta C_R$ refer to the amplitudes of the vertex counterterms. This leads to the final
 structure of the corrected 
 $\tilde{\chi}_1^0\tilde{\chi}_1^0h_i$ 
 vertex as $(C^{\rm 1L}_L+\delta C_L)\mathbf{P_L} + (C^{\rm 1L}_R +\delta C_R) \mathbf{P_R}$. 

	\item 
 The corresponding $\mathtt{Fortran}$ code is used to check the UV-finiteness of our loop-corrected vertices. There is a parameter ``$\Delta$" which is equivalent to $\frac{2}{\epsilon}-\gamma+\log(4\pi)$ and $\mathtt{LoopTools}$ takes its default value to be zero. The UV-finiteness requires the final result not to depend on the parameter ``$\Delta$" up to a certain numerical precision. So we vary the parameter ``$\Delta$" up to $10^{7}$, and we find that the total corrections ( i.e., $C^{\rm 1L}_L+\delta C_L$ and $C^{\rm 1L}_R+\delta C_R$) do not change. 
 This manifests that $(C^{\rm 1L}_{L,R}+\delta C_{L,R})$ is UV-finite. 
  Note that we use tree-level masses for all the particles appearing in the loop to get the UV-finite result. 
	\item  
 We endow the renormalized vertices $\tilde{\chi}_1^0\tilde{\chi}_1^0h$ and $\tilde{\chi}_1^0\tilde{\chi}_1^0H$
 to 
 $\mathtt{MicrOMEGAs}$-5.0.4~\cite{Belanger:2001fz,Belanger:2006is,Belanger:2008sj,Belanger:2013oya} 
 to calculate the DM-related observables, e.g., SI-DD cross-section and the relic density. Both off-shell Higgs states $(h, H)$ assume loop corrected masses.
 A necessary cross-check at this point
 is to verify 
 the $B_{\tilde H},B_{\tilde W\tilde H}$ scenarios under the recent LHC constraints. 
 With the latest $\mathtt{SModelS}$-2.3.0~\cite{Kraml:2013mwa,Dutta:2018ioj,Khosa:2020zar,Alguero:2021dig}
we delineate the parameter space, which is still allowed under the collider searches.
\end{itemize}

\section{Numerical Results}
\label{sec:numeical}
Within this section, we present the numerical outcomes, demonstrating the impact of the dark matter direct detection cross-section on the MSSM parameter space induced by the one-loop corrections to the $\tilde{\chi}_1^0 \tilde{\chi}_1^0 h$ and $\tilde{\chi}_1^0 \tilde{\chi}_1^0 H$ vertices. Specifically, we are interested in assessing
numerically (I) the relative rise in the one-loop renormalized $\tilde \chi_1^0 \tilde \chi_1^0 h_i$ vertex to its LO value, (II) the updated SI-DD cross-section $\sigma^{\rm NLO}_{\rm SI}$ for LSP mass $m_{\tilde \chi_1^0}$ and (III) the resultant and revised contours, depicting the lowest band of $\mu$ with varying $M_1$ based upon the recent \textbf{LUX-ZEPLIN (LZ)} experimental limits on $\tilde \chi_1^0$-nucleon cross-section.

We begin with highlighting the parts of the parameter space in the electroweak MSSM that satisfy $\delta a_\mu$ or other $B$ physics observables.
We choose $\tan\beta=30,10$ for both $\tilde{B}_{\tilde{H}}$ and $\tilde{B}_{\tilde{W}\tilde{H}}$ cases, and additionally, the large $\tan\beta$(=50) limit in the $\tilde{B}_{\tilde{W}\tilde{H}}$ scenario for numerical presentations. 
The DM constraints for $\tilde \chi_1^0$ or a critical checking of the validity of each parameter space point under SUSY searches is partially endorsed as a necessary parameter space criterion. 
While studying the parametric dependence to delineate the effects of one-loop
calculations in (I) and (II), relic abundance or limits from SUSY searches can be observed to be relaxed. However,
in predicting $M_1 - \mu$ plane in (III) or the bench-mark points (BMPs) in Tab.~\ref{tab:my_label}, relic constraint (vide Eq.~\eqref{eq:relic}) and the limits from 
$\mathtt{SModelS}$-2.3.0 are always respected.

\subsection {$\tilde B_{\tilde H}$ DM and $\sigma^{\rm NLO}_{\rm SI}$~:}
Here, we perform a scan over the relevant parameters (all masses are in GeV):
 \begin{equation}
 \mathbf{50\le M_1\le 300,\,\,\, 400\le \mu\le 1000,\,\,\, 100\le m_{\tilde{\mu}_L, \tilde{\mu}_R}\le 350,\,\,\, 100\le m_{\tilde{e}_L, \tilde{e}_R}\le 350}~.
 \label{BHRP} \end{equation}
We set $\mu \geq 400$ GeV for an efficient parametric scan. For our
choice of $M_1$, the lower values of $\mu$ are disfavored, even from the LO SI-DD results. It will be further discussed when we elaborate on our results in Fig.~\ref{fig:sigmabwh11}. 
The Wino is almost decoupled with $M_2=1.5$ TeV; consequently, the neutralino is composed of a dominantly Bino-like state and Higgsino. Also, lighter sleptons ($\tilde{e}$ and  $\tilde{\mu}$) are preferred to comply with the anomalous magnetic moment of muon $\delta a_\mu$. Relatively heavier 
staus are considered so as to satisfy the LHC constraints easily. Additionally, all the points satisfy the constraints from B-physics mentioned earlier. 
The LO $\tilde{\chi}_1^0$-nucleon cross-section 
is related to gaugino-Higgsino mixings induced by the tree-level 
$\tilde \chi_1^0 \tilde \chi_1^0 h_i $ vertex,
noted it as $C^{\rm LO}_{L, R}$ 
in Eq.~\eqref{Eq:lochichihi}. 
Similarly, $C^{\rm NLO}_{L,R}$, the NLO vertex includes $C^{\rm LO}_{L, R}$, one-loop vertex corrections $C^{1\rm L}_{L, R}$, and contributions from the counterterms $\delta C_{L, R}$ as 
\begin{align}
    C^{\rm NLO}_{L,R}= C^{\rm LO}_{L, R} + C^{\rm 1L}_{L, R} + \delta C_{L, R}. 
    \label{eqn:nlo}
\end{align}
With
the set of parameters, stated in Eq.~\eqref{BHRP}, we calculate $C^{\rm LO}_{L,R}$ and 
$C^{\rm NLO}_{L,R}$. The latter is subsequently fed to  $\mathtt{MicrOMEGAs}$-5.0.4 for numerical evaluations of 
$\sigma^{\rm NLO}_{\rm SI}$.

\begin{figure}[H]
	\centering
	\includegraphics[width=0.35\linewidth]{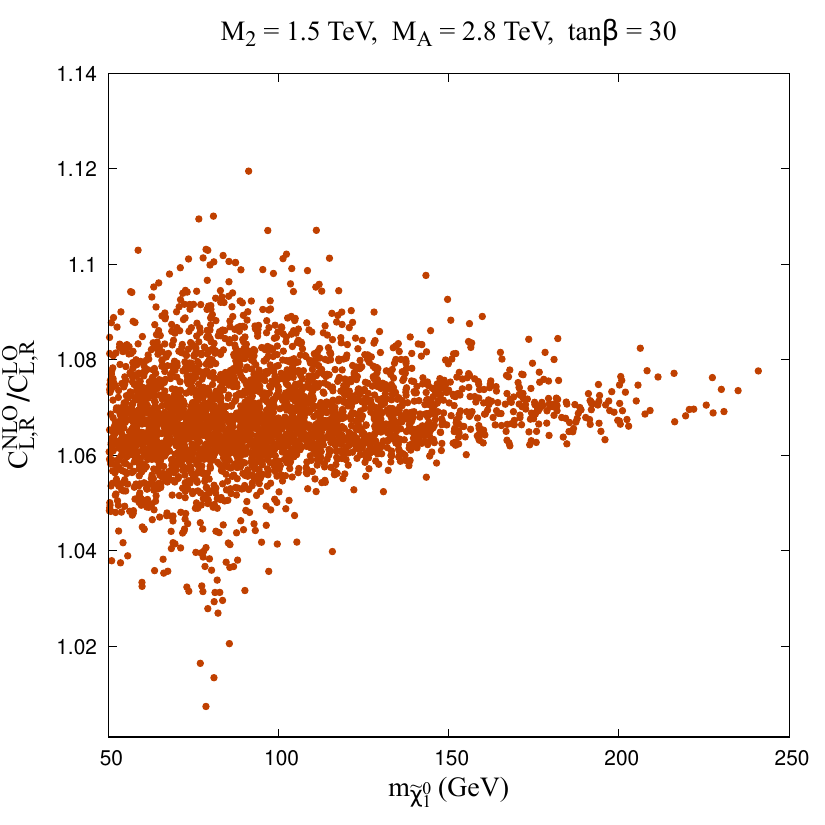}
   \includegraphics[width=0.35\linewidth]{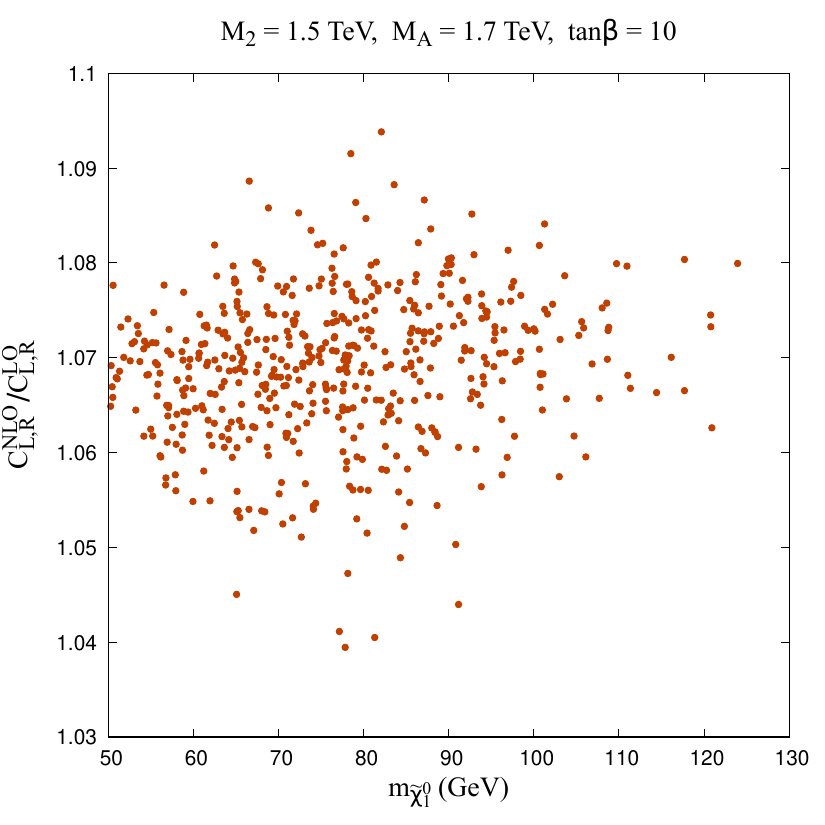}\\
   \qquad\,(a)\qquad\quad\qquad\qquad\qquad\qquad\qquad\qquad\qquad\,(b)\\
	\includegraphics[width=0.35\linewidth]{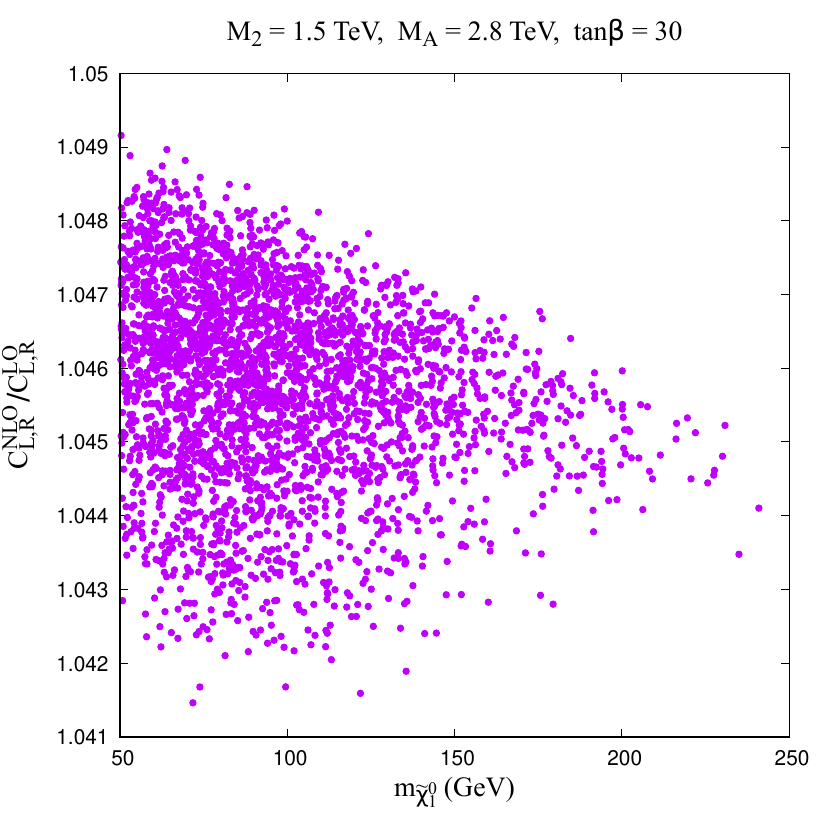}
	\includegraphics[width=0.35\linewidth]{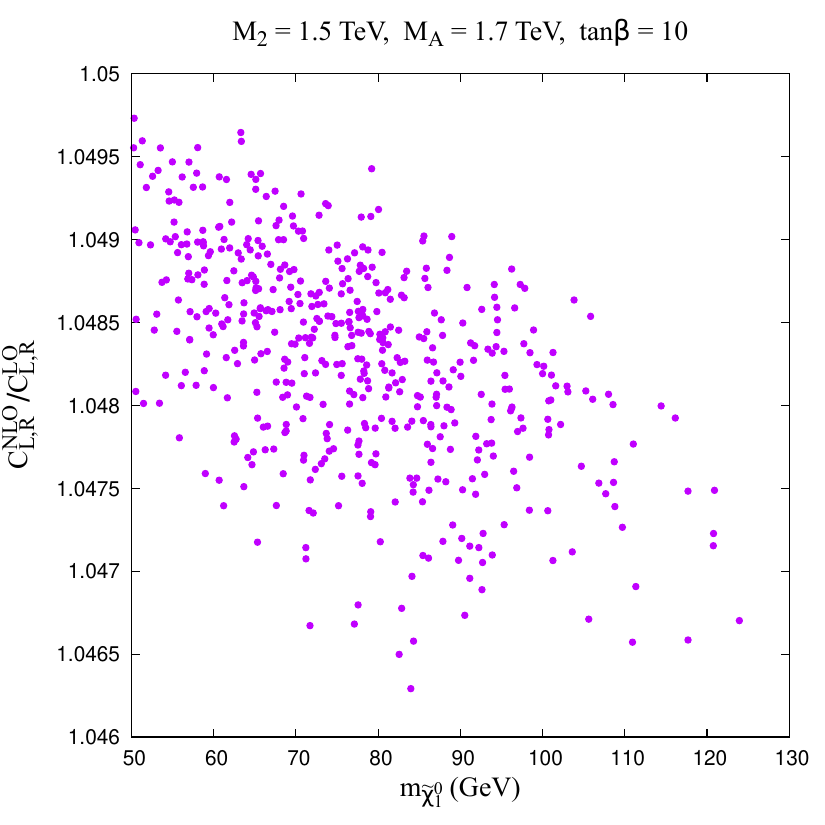}\\
	\qquad\,\,\,\,(c)\qquad\quad\qquad\qquad\qquad\qquad\qquad\qquad\qquad\,\,(d)\\
     \caption{(a) and (b) show the variations of $\frac{C_{L, R}^{\rm NLO}}{C_{L, R}^{\rm LO}}$ ($\equiv \mathcal{R}$) with the mass of LSP for the SM-like Higgs scalar whereas (c) and (d) show the same for the heavier Higgs state. 
     The choice of parameters is discussed in the text. While $(g-2)_\mu$ and the $B$-physics constraints are always satisfied, the cosmological relic abundance data (see Eq.~\eqref{eq:relic}) and the SI-DD bounds are not strictly endorsed. It is apparent that as large as $\sim$ $12\%$ rise in $\mathcal{R}_{h}$ and that of $\sim 5\%$ in $\mathcal{R}_{H}$ can be observed after including the one-loop radiative corrections along with the counterterm results with the LO results.}
		\label{fig:chi1treeloophcrr11}
\end{figure}

Since we are mainly interested in the relative rise of $C^{\rm NLO}_{L,R}$ over its LO value, a quantity of our interest could be the ratio of their numerical values, $\frac{C_{L, R}^{\rm NLO}}{C_{L, R}^{\rm LO}}$. With this in mind,
we plot $\frac{C_{L, R}^{\rm NLO}}{C_{L, R}^{\rm LO}}$, defined henceforth as $\mathcal{R}$, with the mass of LSP for the SM-like Higgs scalar (Fig.~\ref{fig:chi1treeloophcrr11}a and \ref{fig:chi1treeloophcrr11}b) and the same for the heavier Higgs (\ref{fig:chi1treeloophcrr11}c and \ref{fig:chi1treeloophcrr11}d). 
We recall here that both Higgs scalars interfere in the evaluation of
$\sigma^{\rm NLO}_{\rm SI}$.
The variations are shown for two sets of $(\tan\beta, M_A) \equiv
(30,2.8),(10,1.7)$ where the masses of CP-odd Higgs are in TeV. Fewer points are obtained for $\tan\beta =10$ satisfying the deviation in $(g-2)_\mu$ and other phenomenological constraints. 
The reason is that $(g-2)_\mu$ depends on the muon Yukawa coupling $y_\mu$, which is inversely proportional to $\cos\beta$, 
$y_\mu \propto \frac{1}{\cos\beta}\sim \tan\beta$ (for $\tan\beta \geq 5$). It is also evident from Eq.~\eqref{Eq:g2binolike}. In the case of heavier LSP mass with relatively heavier $\tilde \chi_2^0, \tilde \chi_3^0$, and $\tilde\chi_1^\pm$, the larger value of $\tan\beta$ is favoured to satisfy
   $\delta a_\mu$. We may verify it numerically from Fig.~\ref{fig:chi1treeloophcrr11}. For a moderate $\tan\beta$ (=30), one finds a relatively heavier LSP region ($\sim 250$ GeV) can be reached that can accommodate the $(g-2)_\mu$ compared to the lower value of $\tan\beta$(=10) (which can reach up to the LSP mass of $\sim 130$~GeV). It is also apparent from Fig.~\ref{fig:chi1treeloophcrr11} that an enhancement in $\mathcal{R}$ up to $\mathcal{R}_h=12\%$ and
$\mathcal{R}_H=5\%$ can be obtained to $\tilde{\chi}_1^0 \tilde{\chi}_1^0 h$ and $\tilde{\chi}_1^0 \tilde{\chi}_1^0 H$ couplings respectively after considering the NLO results via Eq.~\eqref{eqn:nlo}.
The maximum value of $\mathcal{R}_{h,H}$ refers to the scenarios where leading order $C_{L, R}^{\rm LO}$ or the Bino-Higgsino mixing hits the minimum value.
\begin{figure}[H]
	\centering
	\includegraphics[width=0.41\linewidth]{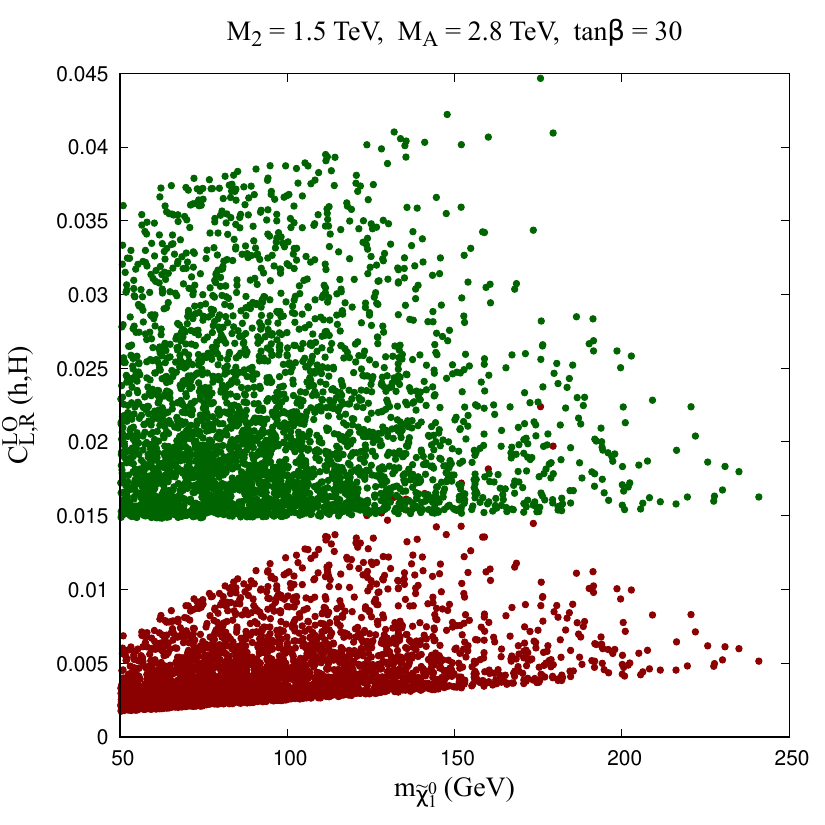}\\
	\caption{Variations of the LO or tree-level couplings of $\tilde{\chi}_1^0\tilde{\chi}_1^0h$ and $\tilde{\chi}_1^0\tilde{\chi}_1^0H$ vertices with $m_{\tilde{\chi}_1^0}$. The points with larger values of the coupling (green) correspond to the $\tilde{\chi}_1^0\tilde{\chi}_1^0H$ vertex, and the points with lower values of the coupling (red) correspond to the $\tilde{\chi}_1^0\tilde{\chi}_1^0h$ vertex. As in Fig.~\ref{fig:chi1treeloophcrr11}, $(g-2)_\mu$ and the $B$-physics constraints are always satisfied, the cosmological relic abundance data and the SI-DD bounds are relaxed.}
	\label{fig:CLRTreehH}
\end{figure}

The relatively subdued effect in  $\mathcal{R}_H$, related to
heavier CP-even Higgs scalar, is due to its higher LO value. It can be noted from 
Eq.~\eqref{Eq:LO_h} and Eq.~\eqref{Eq:LO_H}. 
First,
$\mu$ assumes higher values than $M_1$. And, then the Higgs mixing angle is 
$<<1$. So, for the tree-level couplings associated with the light Higgs boson,
the Bino mass term dominates in Eq.~\eqref{Eq:LO_h} while for $H$ boson, the first
term within the bracket of Eq.~\eqref{Eq:LO_H} contributes mainly. Thus, the LO couplings are higher for $H$ boson.
We show
its numerical values,  
in Fig.~\ref{fig:CLRTreehH} for a representative choice of input parameters, $M_2=1.5~$TeV, $M_A=2.8~$TeV. Similarly, we assume $\tan\beta=30$ for the plot.
The green points show  
$C_{L, R}^{\rm LO}$ for $H$ scalar while red regions present the same 
for $h$ boson for the same range of parameters, as stated earlier.

\begin{figure}[H]
	\centering
	\includegraphics[width=0.36\linewidth]{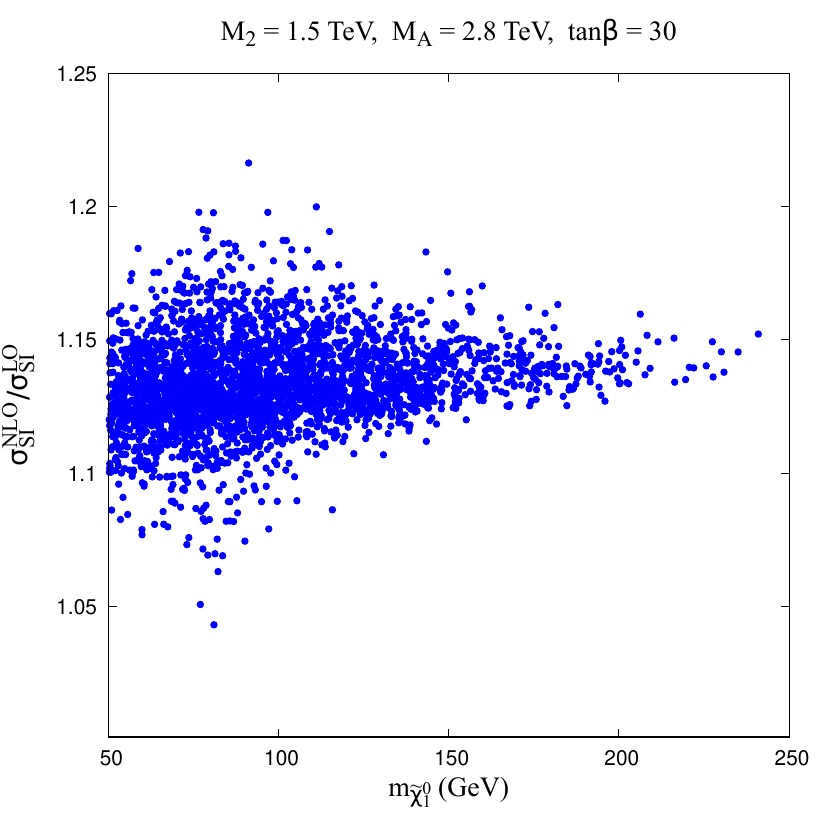}
	\includegraphics[width=0.36\linewidth]{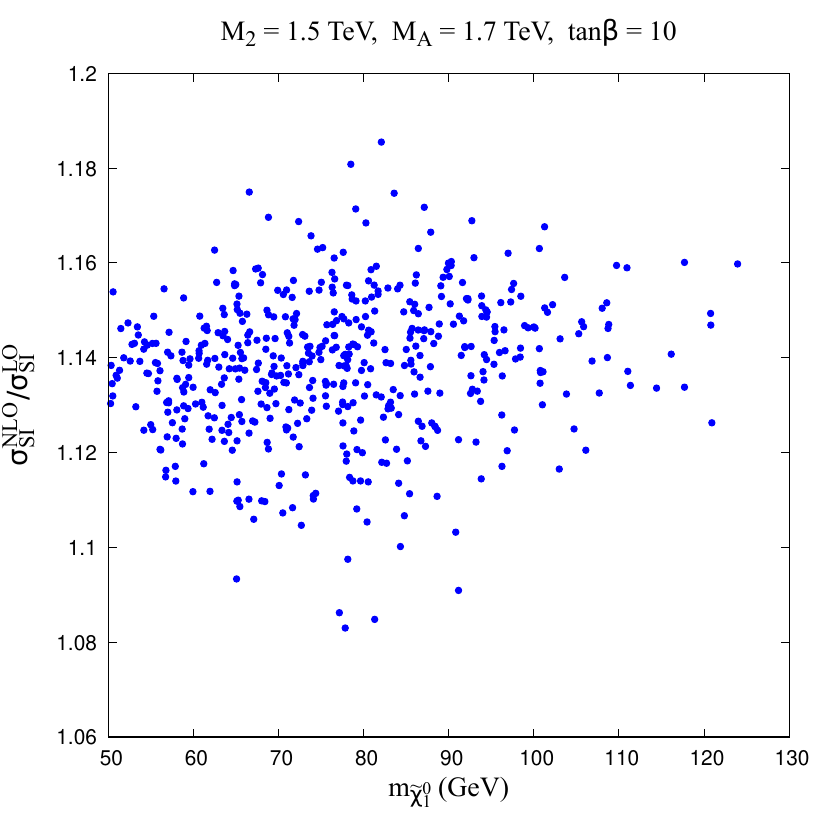}\\
	\,\,\qquad(a)\qquad\qquad\qquad\qquad\qquad\qquad\qquad\qquad\qquad\,\,(b)\\
	\includegraphics[width=0.36\linewidth]{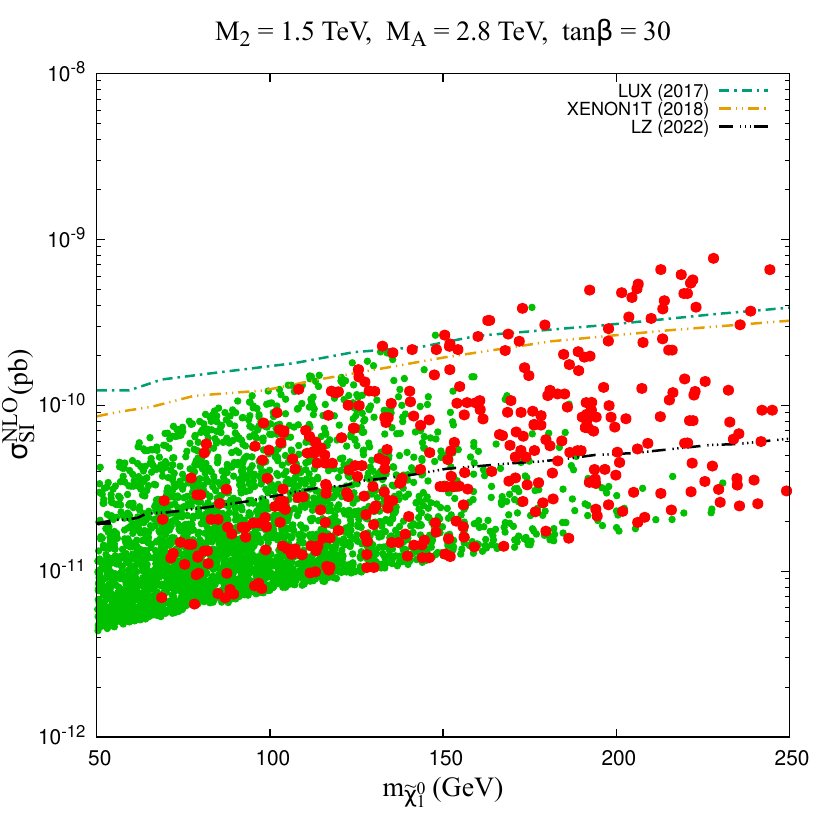}\\
	\qquad(c)
	\caption{(a) and (b) represent the variations of $\frac{\sigma_{\rm SI}^{\rm NLO}}{\sigma_{\rm SI}^{\rm LO}}$ or $\mathcal{R}^{\sigma}$ with the mass of LSP, while the variation of $\sigma_{\rm SI}^{\rm NLO}$ with $m_{\tilde \chi_1^0}$ is shown in (c) for $\tan\beta=30$. The red points in (c)
		are allowed by $\mathtt{SModelS}$-2.3.0. Note that in this case, the Bino fraction in $\tilde{\chi}_1^0$ is $\mathcal{N}_{11}^2\geq 97\%$.
		It is evident from (a) and (b) that we get an enhancement up to $\sim 20\%$ in $\mathcal{R}^{\sigma}$ after including the one-loop renormalized vertices.}
	\label{fig:chi1ddcross-secproton10treeneg11}
\end{figure}

Following the relative dominance of $h$ scalar in the 
$C_{L, R}^{\rm NLO}$, we now turn our attention to quantifying the relative increase of $\mathcal{R}_{h,H}$
in the $\chi_1^0$-nucleon($p$) cross-section, through $\frac{\sigma_{\rm SI}^{\rm NLO}}{\sigma_{\rm SI}^{\rm LO}}$ (see Fig.~\ref{fig:chi1ddcross-secproton10treeneg11}a and \ref{fig:chi1ddcross-secproton10treeneg11}b). We define the ratio as $\mathcal{R}^{\sigma}$ for simplicity. The parameters are the same as in Eq.~\eqref{BHRP}. It is evident from Fig.~\ref{fig:chi1ddcross-secproton10treeneg11}a and \ref{fig:chi1ddcross-secproton10treeneg11}b that we get an enhancement in $\mathcal{R}^{\sigma}$ up to $\sim 20\%$ in the SI-DD cross-sections after including the one-loop renormalized vertices.
In a simple scenario,
with either $h$ or $H$ presents in the spectra, the rise in $\sigma_{\rm SI}^{\rm NLO}$ can directly be correlated to the variations in 
$\tilde{\chi}_1^0 \tilde{\chi}_1^0 h$ or $\tilde{\chi}_1^0 \tilde{\chi}_1^0 H$ vertex 
    $\sigma_{\rm SI}^{\rm NLO} = \sigma_{\rm SI}^{\rm LO} \frac{(C_{L, R}^{\rm NLO})^2}{(C_{L, R}^{\rm LO})^2}
   =\sigma_{\rm SI}^{\rm LO} \Bigg[1+\frac{2C^{\rm 1L}_{L,R}}{C_{L,R}^{\rm LO}} +\frac{2\delta C_{L,R}}{C_{L,R}^{\rm LO}}\Bigg]$.
When more than one scalar is present, both of the CP-even scalar bosons 
will interfere and
the overall rise in the 
$\mathcal{R}^{\sigma}$
can not be apprehended easily.

Finally, we present the NLO cross-section  $\sigma_{\rm SI}^{\rm NLO}$ with the mass of LSP (Fig.~\ref{fig:chi1ddcross-secproton10treeneg11}c) for $\tan\beta=30$ only. At this stage, it is customary to check the validity of the parameter space under SUSY searches. While the green regions depict part of the MSSM parameter space that otherwise satisfies $B$-physics constraints and $(g-2)_\mu$,
the red points in Fig.~\ref{fig:chi1ddcross-secproton10treeneg11}c are additionally consistent with SUSY searches. As noted, the SUSY searches are validated with $\mathtt{SModelS}$-2.3.0. Even with the one-loop corrected
SI-DD, $\sigma_{\rm SI}^{\rm NLO}$, some parts of the parameter space are
still not excluded by the latest \textbf{LZ} data. It may be 
noted here that the exclusion limits from \textbf{LUX}, \textbf{XENON-1T}
or \textbf{LZ} are shown assuming the central values only.

\subsection{$\tilde B_{\tilde W\tilde H}$ DM and $\sigma^{\rm NLO}_{\rm SI}$~:}
\begin{figure}[H]
	\centering
	\includegraphics[width=0.42\linewidth]{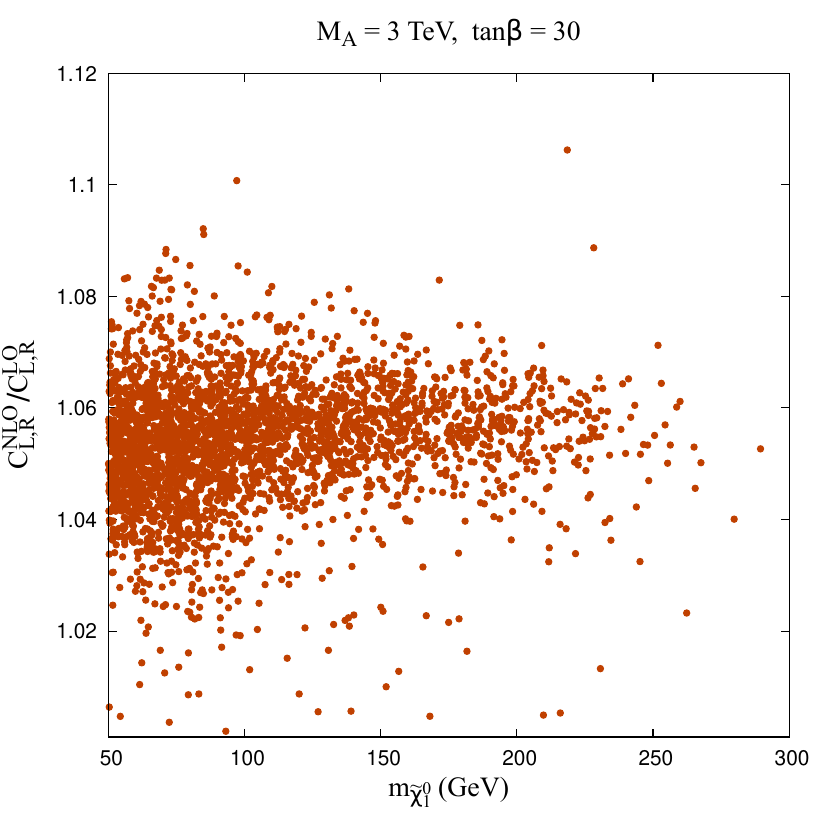}
	\includegraphics[width=0.42\linewidth]{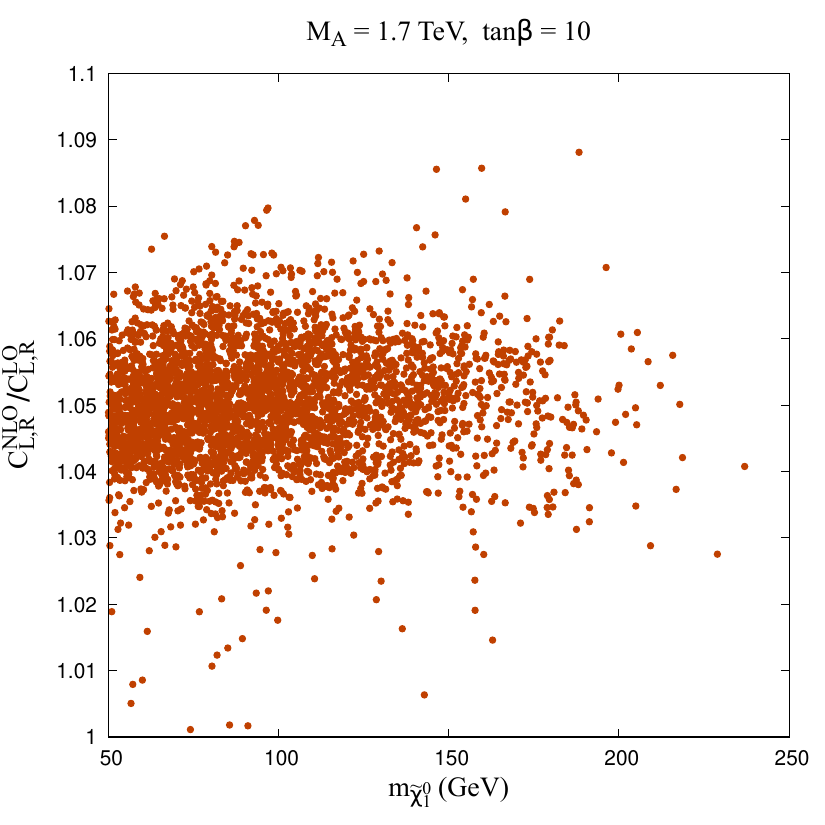}\\
	\quad\quad\,(a)\,\,\qquad\qquad\qquad\qquad\qquad\qquad\qquad\qquad\qquad\quad\,\,\,\,\,(b)\\
	\includegraphics[width=0.42\linewidth]{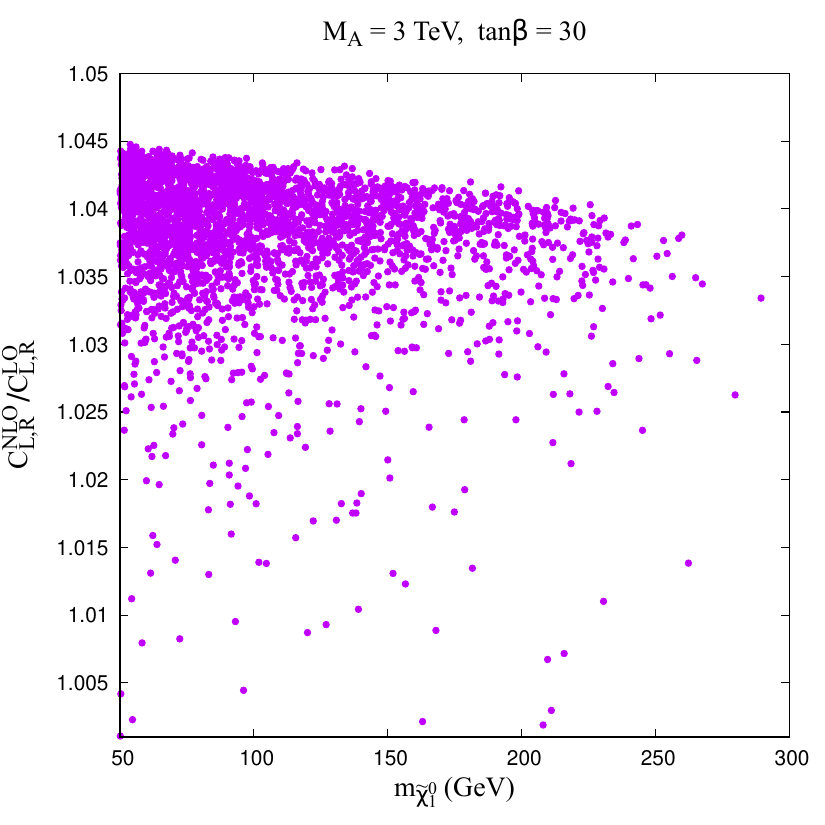}
	\includegraphics[width=0.42\linewidth]{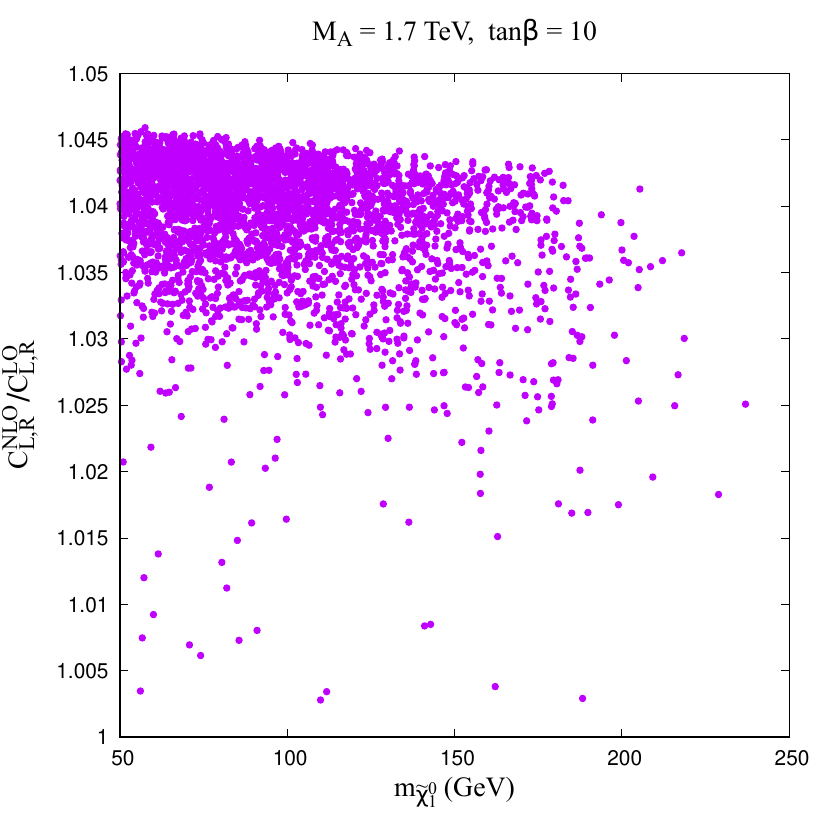}\\
	\quad\quad\,\,(c)\,\,\qquad\qquad\qquad\qquad\qquad\qquad\qquad\qquad\qquad\quad\,\,\,\,\,(d)
	\caption{(a) and (b) show the variations of $\mathcal{R}=\frac{C_{L, R}^{\rm NLO}}{C_{L, R}^{\rm LO}}$ with the mass of LSP for the SM-like Higgs scalar whereas (c) and (d) show the same for the heavier Higgs. These are the same as the $\tilde B_{\tilde H}$ case except that, here, we have taken the lighter Wino with $M_2\in [150, 600]$ GeV (see Eq.~\eqref{BWHinputs}). As before, $(g-2)_\mu$ and the $B$-physics constraints are always satisfied; the cosmological relic abundance data and SI-DD constraints are not strictly endorsed.}
	\label{fig:muddn132n142nrg212}
\end{figure}

Having analyzed the $\tilde B_{\tilde H}$ DM, we now present the numerical results when the neutralino is composed of a dominantly Bino, Higgsinos, and a relatively larger component of Wino, which already 
referred to as $\tilde B_{\tilde W\tilde H}$ scenario.
Specifically, we consider $M_1< M_2\lesssim\mu$ among the EW inputs with a numerical scan over the following ranges of the parameters, 

 \begin{equation}
\mathbf{50\le M_1\le 300,\,\,\, 150\le M_2\le 600,\,\,\, 400\le \mu\le 1000,\,\,\, 100\le m_{\tilde{\mu}_L, \tilde{\mu}_R}\le 350,\,\,\, 100\le m_{\tilde{e}_L, \tilde{e}_R}\le 350}~,
\label{BWHinputs}
\end{equation}
In the above, all the masses are in GeV.
The ranges of the parameters are the same as in $\tilde B_{\tilde H}$ case except that, here, we have taken the lighter Wino with $M_2\in [150, 600]$ GeV to raise the Wino composition in $\tilde \chi_1^0$ 
(see Eq.~\eqref{BWHinputs}) compared to $\tilde{B}_{\tilde{H}}$ scenario. As for the $\tilde B_{\tilde H}$ DM, the relative rise in 
$C_{L, R}^{\rm NLO}$ compared to $C_{L, R}^{\rm LO}$, quantified as
$\frac{C_{L, R}^{\rm NLO}}{C_{L, R}^{\rm LO}}$ (or $\mathcal{R}$, for the sake of brevity) is shown with $m_{\tilde{\chi}_1^0}$ for the SM-like Higgs scalar (Fig.~\ref{fig:muddn132n142nrg212}a and \ref{fig:muddn132n142nrg212}b) and the heavier Higgs (Fig.~\ref{fig:muddn132n142nrg212}c and \ref{fig:muddn132n142nrg212}d). 
As before, we primarily consider
$\tan\beta=30,10$
with two different values of 
$M_A=3, 1.7$ TeV. Additionally, we consider $\tan\beta=50$ with $M_A=3$~TeV to reach the larger mass region of the LSP. 
In Fig.~\ref{fig:muddn132n142nrg212}(a,b) ($\tan\beta=30,10$), we can achieve a maximum value for $\mathcal{R}_h$, $\sim 9\%$.
The relative increase $\mathcal{R}_H$ is much smaller $\sim 4.5\%$ 
irrespective of the value of $\tan\beta$ which follows from the fact that 
$\tilde{\chi}_1^0 \tilde{\chi}_1^0 H$ takes higher value at the LO. So the ratio $\mathcal{R}_H$
resides on the lower side. 

As in the $\tilde B_{\tilde H}$ case, we may again observe that $\tan\beta=30$ helps to reach larger values of LSP masses (up to 300 GeV). Since the rise in $m_{\tilde{\chi}_1^0}$, one has to raise the masses of $\tilde{\chi}_1^\pm$, a larger $\tan\beta$ would be necessary (see e.g., Eq.~\eqref{Eq:g2binolike}). Moreover, unlike the previous scenario, a light Wino-like chargino also helps to enhance the BSM contributions to $(g-2)_\mu$. So, in contrast to Fig.~\ref{fig:chi1treeloophcrr11}, one can even go to larger masses for the LSP and Higgsino-like states.
Along the same line, it may be interesting to see how far we can reach in the LSP mass at a large $\tan\beta$ value (e.g., $\tan\beta=50$).

Thus, we contemplate a scenario for large $\tan\beta$ (=50). Here, heavier chargino and neutralino masses ($\sim 600$ GeV) can be reached in compliance with $(g-2)_\mu$\footnote{The higher LSP region of masses up to $\sim600$ GeV can be reached with the large value of $\tan\beta$. The region of the allowed parameter space is mentioned in \textbf{https://atlas.web.cern.ch/Atlas/GROUPS/PHYSICS/PUBNOTES/ATL-PHYS-PUB-2023-025/fig$_16$.png}.}. As before, we compute the EW corrections to the $\tilde{\chi}_1^0\tilde{\chi}_1^0h_i$ coupling and the corresponding SI-DD cross-section.. The ranges of the parameters we consider in this scenario are the following.
\begin{equation}
\mathbf{50\le M_1\le 600,\,\,\, 150\le M_2\le 1000,\,\,\, 400\le \mu\le 1500,\,\,\, 100\le m_{\tilde{\mu}_L, \tilde{\mu}_R}\le 650,\,\,\, 100\le m_{\tilde{e}_L, \tilde{e}_R}\le 650}~,
\label{BWHinputs50}
\end{equation}

\begin{figure}[H]
	\centering
	\includegraphics[width=0.45\linewidth]{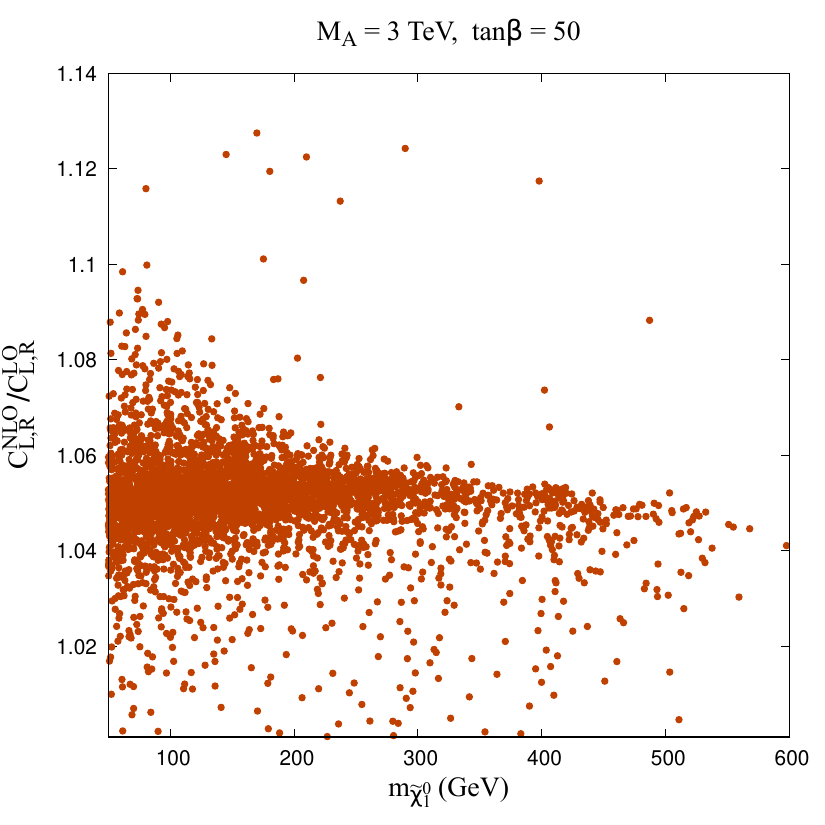}
	\includegraphics[width=0.45\linewidth]{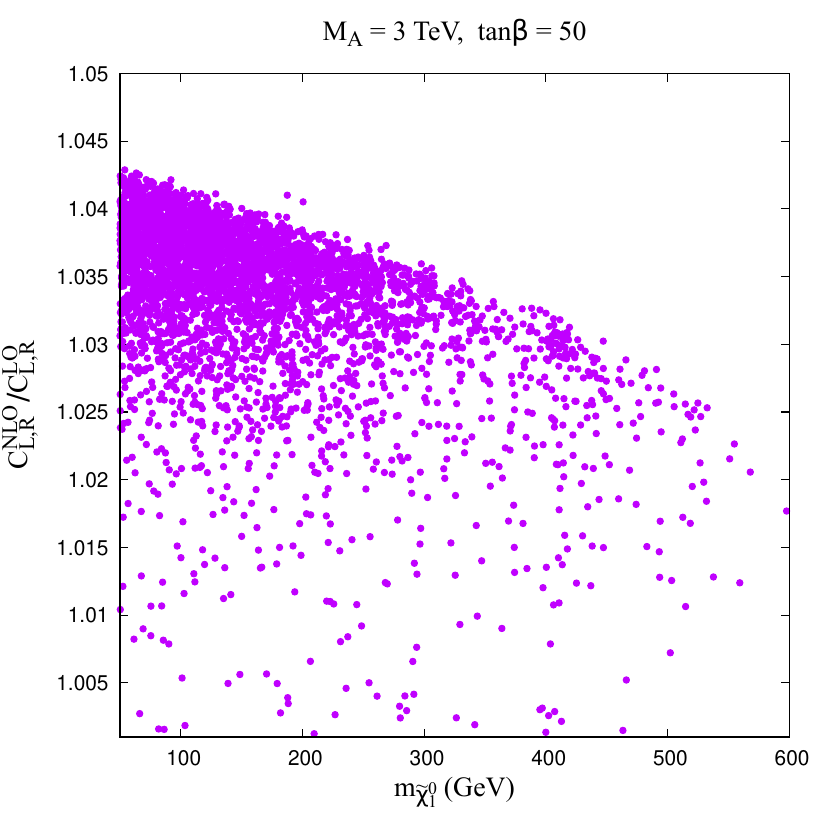}\\
	\quad\quad\,\,\,\,(a)\qquad\qquad\qquad\qquad\qquad\qquad\qquad\qquad\qquad\qquad\qquad\,(b)
	\caption{(a) and (b) represent the variations of  $\frac{C_{L,R}^{\rm NLO}}{C_{L,R}^{\rm LO}}$ with the mass of LSP for $\tan\beta=50$. In this scenario, the LSP has a Bino fraction of $\mathcal{N}_{11}^2\geq85\%$. Here, we obtain $\sim 13\%$ corrections in the $\tilde{\chi}_1^0\tilde{\chi}_1^0h$ coupling, and $\sim 4.3\%$ corrections in the  $\tilde{\chi}_1^0\tilde{\chi}_1^0H$ coupling. As before, $(g-2)_\mu$ and the $B$-physics constraints are always respected; the cosmological relic abundance data and SI-DD bounds are not considered.}
	\label{fig:sigmabwh50}
\end{figure}
where all the masses are in GeV. From Fig.~\ref{fig:sigmabwh50}, we obtain up to $\sim 13\%$ NLO corrections to the $\tilde{\chi}_1^0\tilde{\chi}_1^0h$ coupling and that of $\sim 4.3\%$ corrections to the $\tilde{\chi}_1^0\tilde{\chi}_1^0H$ coupling.

\begin{figure}[H]
	\centering
	\includegraphics[width=0.42\linewidth]{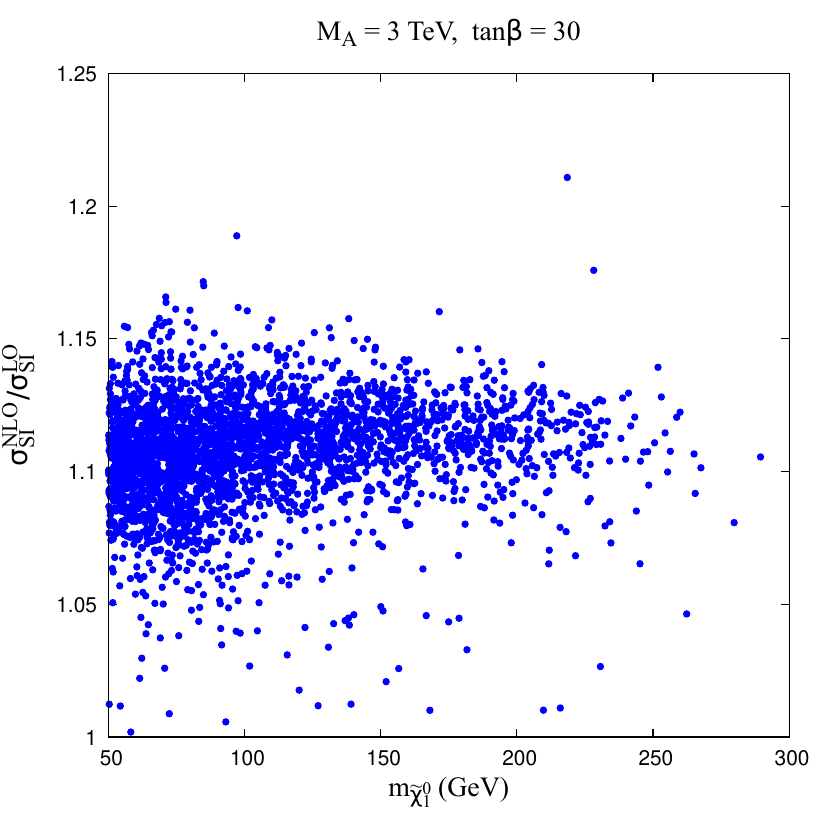}
	\includegraphics[width=0.42\linewidth]{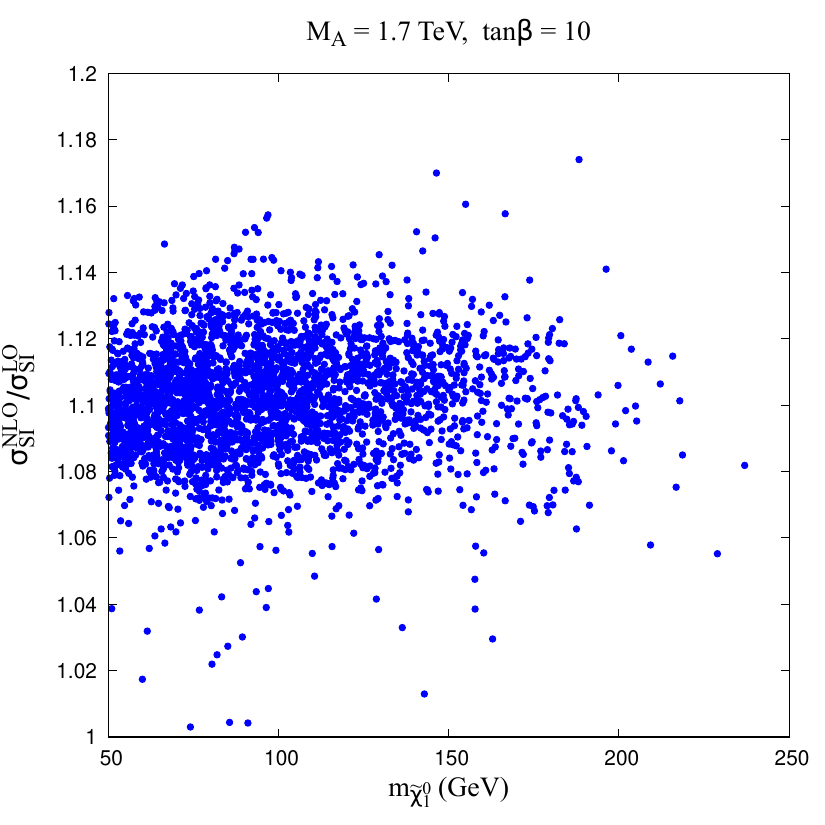}\\       
	\quad\quad(a)\,\,\quad\qquad\qquad\qquad\qquad\qquad\qquad\qquad\qquad\qquad\,\,\,\,\,\,(b)\\
	\includegraphics[width=0.42\linewidth]{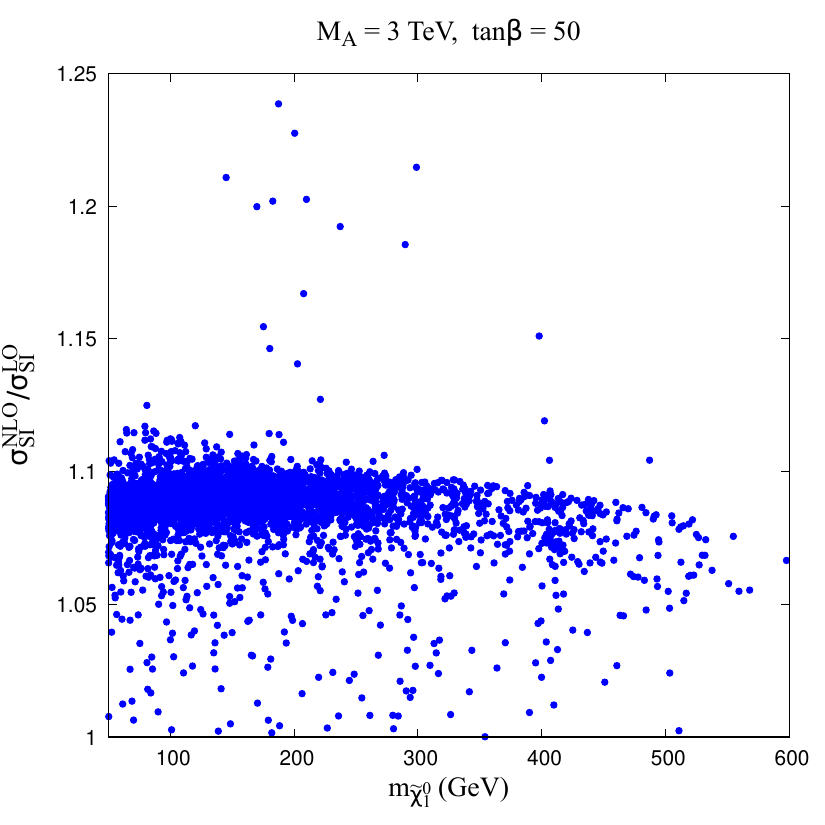}
	\includegraphics[width=0.42\linewidth]{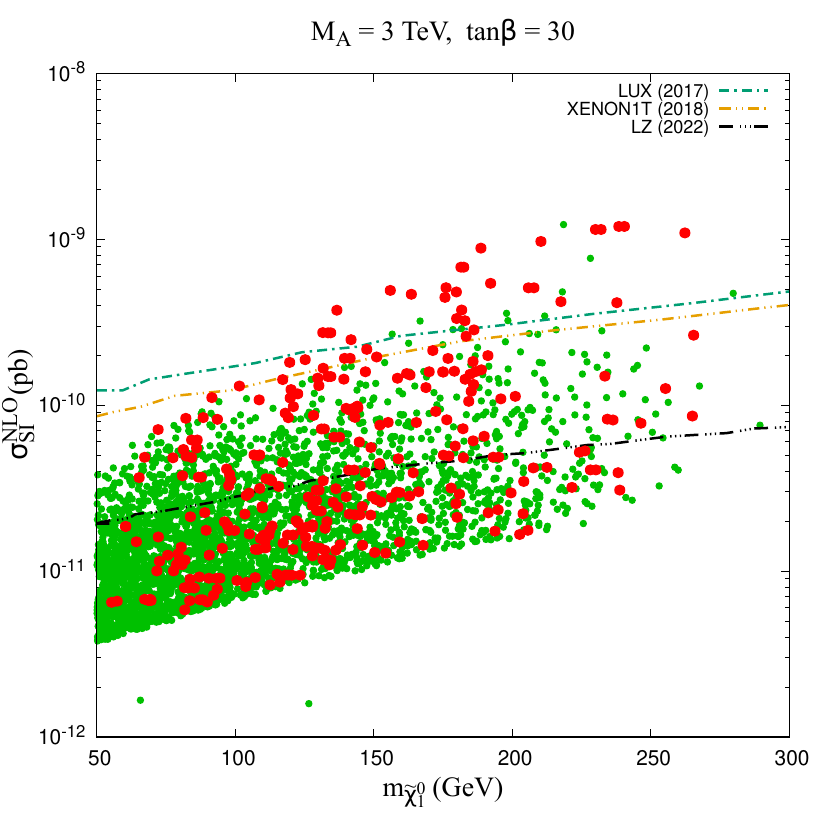}\\ \quad\quad(c)\,\quad\qquad\qquad\qquad\qquad\qquad\qquad\qquad\qquad\qquad\,\,\,\,\,\,(d)\\

	\caption{(a), (b), and (c) represent the variations of  $\frac{\sigma_{\rm SI}^{\rm NLO}}{\sigma_{\rm SI}^{\rm LO}}$ 
		with the mass of LSP, and (d) $\sigma_{\rm SI}^{\rm NLO}$ with the same for $\tan\beta=30$. In (d), the red points satisfy the present SUSY search constraints, verified by $\mathtt{SModelS-2.3.0}$. Here, the LSP has a Bino fraction of $\mathcal{N}_{11}^2\geq97\%$.
		Note that we get $\sim 20\%$ corrections in the cross-sections for $\tan\beta=30,50$ and $\sim 18\%$ for $\tan\beta=10$.}
	\label{fig:sigmabwh}
\end{figure}

The resultant change in the SI-DD cross-section $\mathcal{R}^\sigma=\frac{\sigma_{\rm SI}^{\rm NLO}}{\sigma_{\rm SI}^{\rm LO}}$ with the mass of LSP for $\tan\beta=30,10$, and the large $\tan\beta=50$, and the variations of $\sigma_{\rm SI}^{\rm NLO}$ with the same for $\tan\beta=30$ are shown in Fig.~\ref{fig:sigmabwh}. As can be seen, 
$\mathcal{R}^\sigma$ reads $\sim 20\%$ corrections 
for our choices of the $\tan\beta$. Additionally, a large part
of the parameter space still satisfies the stringent \textbf{LZ} limits.
As before, the red points in Fig.~\ref{fig:sigmabwh}c satisfy the SUSY search limits verified by $\mathtt{SModelS-2.3.0}$.

\begin{figure}[H]
		\centering
        \includegraphics[width=0.329\linewidth]{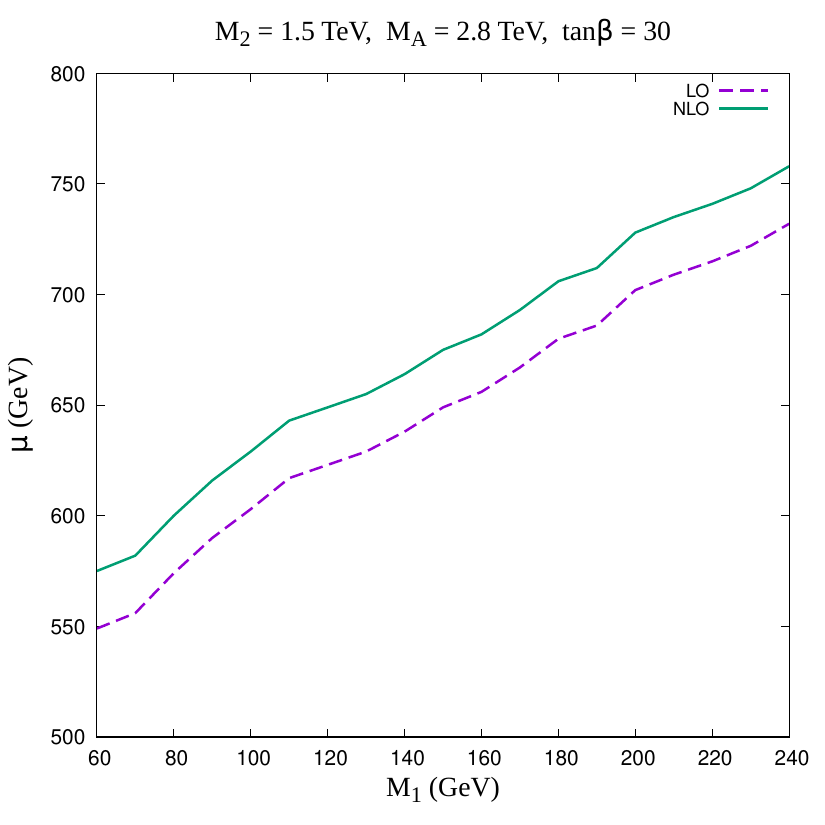}
        \includegraphics[width=0.329\linewidth]{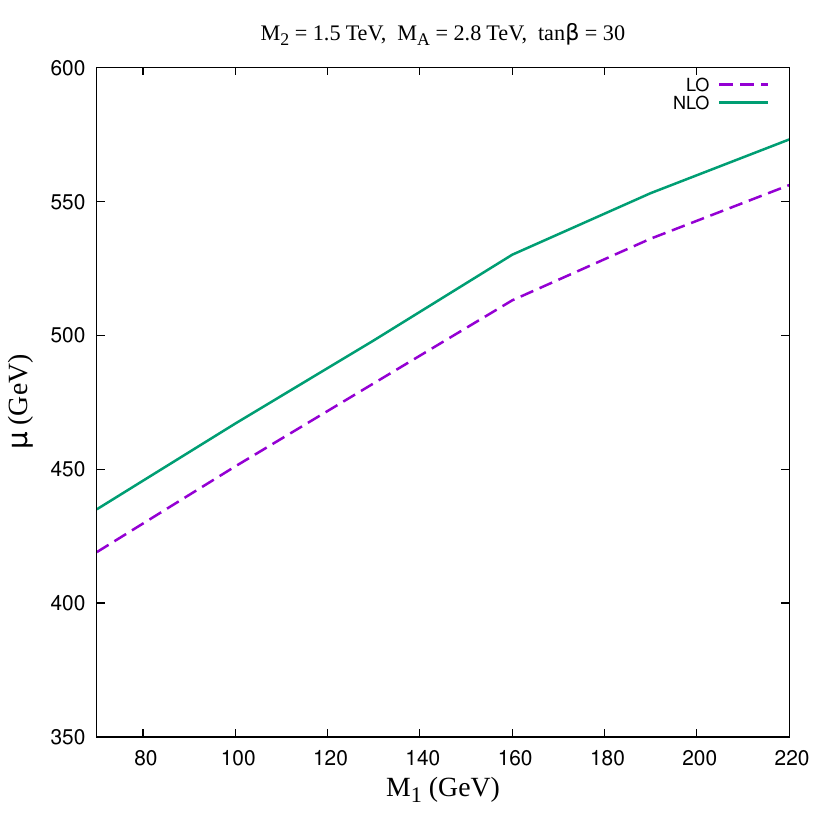}
        \includegraphics[width=0.329\linewidth]{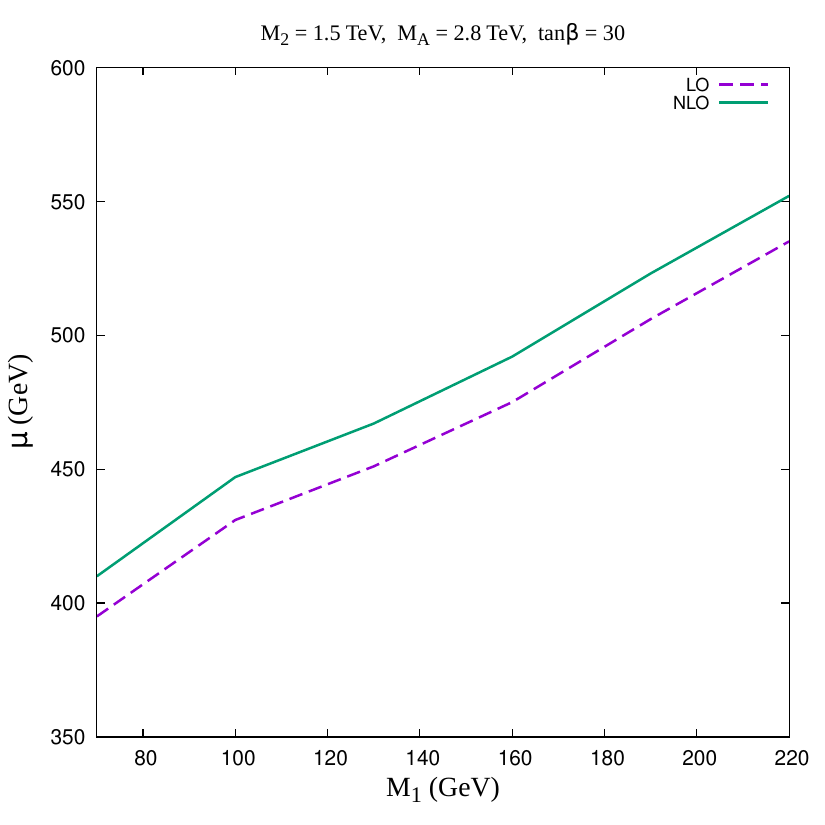}\\
        \,\,\quad(a)\hskip5.5cm \,(b)\hskip5.5cm \,\,(c)
    	\caption{
     Contours depicting the lowest values of $\mu$ for a given $M_1$, computed from 
$C_{L, R}^{\rm LO}$ and $C_{L, R}^{\rm NLO}$ for the $\chi_1^0\chi_1^0h_i$
coupling using (a) $90\%$ confidence limit of \textbf{LZ}, (b) $2\sigma$ upper bound ($+2\sigma$) of $\textbf{LZ}$, and (c) $90\%$ confidence limit of $\textbf{XENON-1T(2018)}$. 
Overall, the lower bound in $\mu$ shifts upward, leading to more stringent limits in the $M_1 - \mu$ plane if the NLO corrections are included.
The $B$-physics constraints, $\delta a_\mu$, the cosmological relic abundance data, and the constraints following $\mathtt{SModelS-2.3.0}$ are always respected.}

\label{fig:sigmabwh11}
\end{figure}

With the knowledge gathered from the previous exercise, we may anticipate that for higher values of $\sigma_{\rm SI}^{\rm NLO}$, the MSSM parameter space will be curbed further. As a result, a more stringent bound on the gaugino-Higgsino mixing parameter may be derived. Following the fact that their mass parameters drive the mixing, we observe a rise in the lowest band of the Higgsino mass $\mu$ for a given value of $M_1$, as depicted in Figs.~\ref{fig:sigmabwh11}a-\ref{fig:sigmabwh11}c. For the experimental input, we use the $90\%$ confidence limit of \textbf{LZ} in Fig.~\ref{fig:sigmabwh11}a, $2\sigma$ upper bound ($+2\sigma$) of \textbf{LZ} in Fig.~\ref{fig:sigmabwh11}b, and $90\%$ confidence limit of \textbf{XENON-1T} in Fig.~\ref{fig:sigmabwh11}c on the $\sigma_{\tilde \chi_1^0-p}$ cross-section. We stick to a fixed value of $\tan\beta=30$ for simplicity. Similarly, $M_2=1.5$~TeV and $M_A=2.8$~TeV are assumed. Thus, it leads to $\tilde B_{\tilde H}$ DM scenario. In Figs.~\ref{fig:sigmabwh11}a-\ref{fig:sigmabwh11}c, we observe two contours in the $M_1 - \mu$ plane, showing the lowest value of the parameters allowed from the SI-DD cross-section if one uses (I) 
$C_{L, R}^{\rm LO}$ and (II) $C_{L, R}^{\rm NLO}$ in the calculations.
As before, the $B$-physics constraints, $(g-2)_\mu$, and the LHC searches 
on the SUSY parameters are always respected. Moreover, the observed relic density, vide Eq.~\eqref{eq:relic} is strictly adhered to.

In Fig.~~\ref{fig:sigmabwh11}a, with the LO of $\tilde{\chi}_1^0\tilde{\chi}_1^0h_i$ vertex, we find $\mu \geq 550$ GeV for $M_1 = 60$ GeV. After including the NLO corrections (vide Eq.~\eqref{eqn:nlo}), the lower limit becomes $\mu \geq 575$ for the same value of $M_1$. Similarly, for $M_1=240$ GeV, the $\mu$ value shifts from 
732 GeV to 758 GeV upon including the NLO corrections. Therefore, $\mu$ shifts upward by $\sim 25$~GeV in this case. Similarly, in Figs.~\ref{fig:sigmabwh11}b and ~\ref{fig:sigmabwh11}c, $\mu$ shifts upward by $\sim 20$~GeV.
The underlying reason for the relatively lower shift in Figs.~\ref{fig:sigmabwh11}b and ~\ref{fig:sigmabwh11}c depends on the fact that the maximum value for the Bino-Higgsino mixing at the LO in the first case (i.e., Fig.~~\ref{fig:sigmabwh11}a) is relatively smaller than the latter cases, which suppresses the LO value of the $\tilde{\chi}_1^0\tilde{\chi}_1^0h(H)$. Therefore, the NLO corrections in the first case are larger, and a relatively higher shift in the $\mu$ value is obtained.   
Overall, the lower bound in $\mu$ shifts upward, leading to more stringent limits in the $M_1 - \mu$ plane when the NLO corrections are included. Typically, one may extend the contours
for lower or higher values of $M_1$. But, then 
 $(g-2)_\mu$ and $\mathtt{SModelS-2.3.0}$ are somewhat restrictive on the MSSM parameter space.

\begin{table}[H]
	\centering
	\begin{tabular}{|c|c|c|c|c|c|c|c|c|c|c|c|}
		\hline
		\multicolumn{11}{|c|}{ $\tilde B_{\tilde H}$ LSP} \\ 		
		\hline
		
		BMPs&$\tan\beta$& $\mu$ &  $M_1$ & $M_{2}$ &  $M_A$  &$M_H$ & $m_{\tilde{\mu}_{L}}$ &  $m_{\tilde{\mu}_{R}}$ & $m_{\tilde{e}_{L}}$ & $m_{\tilde{e}_{R}}$ \\	
		\hline
		I& 30 & 603 & 100  & 1500 &  2800 & 2268 & 178 & 135 & 177 & 131 \\
		\hline
		
		BMPs& $m_{\tilde{\chi}_1^0}$ & $m_{\tilde{\chi}_1^\pm}, m_{\tilde{\chi}_2^0}$& $\delta a_\mu$ & $\Omega h^2$ &$C_{L,R}^{\rm LO}(h)$ & $C_{L,R}^{\rm NLO}(h)$& $C_{L,R}^{\rm LO}(H)$& $C_{L,R}^{\rm NLO}(H)$ & $\sigma_{\rm SI}^{\rm LO}$& $\sigma_{\rm SI}^{\rm NLO}$\\
		\hline
		I& 99 & 624 &$2.12\times 10^{-9}$ &0.118 &0.00583& 0.00622 & 0.02515 & 0.02625 & $2.760\times 10^{-11}$& $3.130\times 10^{-11}$ \\
		\hline
		\hline
		\multicolumn{11}{|c|}{ $\tilde B_{\tilde W\tilde H}$ LSP} \\ 
		\hline
		BMPs&$\tan\beta$& $\mu$&  $M_1$ & $M_{2}$&  $M_A$ &$M_H$ & $m_{\tilde{\mu}_{L}}$ &  $m_{\tilde{\mu}_{R}}$ & $m_{\tilde{e}_{L}}$ & $m_{\tilde{e}_{R}}$ \\	
		\hline
		II& 30 & 710  & 190 & 265 & 3000 & 2392  & 344 & 248 & 254 & 204 \\
		\hline
		
		BMPs& $m_{\tilde{\chi}_1^0}$ & $m_{\tilde{\chi}_1^\pm}, m_{\tilde{\chi}_2^0}$& $\delta a_\mu$ & $\Omega h^2$&$C_{L,R}^{\rm LO}(h)$ & $C_{L,R}^{\rm NLO}(h)$& $C_{L,R}^{\rm LO}(H)$& $C_{L,R}^{\rm NLO}(H)$ & $\sigma_{\rm SI}^{\rm LO}$& $\sigma_{\rm SI}^{\rm NLO}$ \\	
		\hline
		II& 189 & 282 &$3.54\times 10^{-9}$& 0.119 & 0.00812 & 0.00858 & 0.02433 & 0.02519 & $4.709\times 10^{-11}$ & $5.241\times 10^{-11}$ \\
		\hline
	\end{tabular}
	\caption{ A few exemplary points presented where $\sigma^{ \rm NLO}_{ \rm SI}$
		excludes a parameter space point, which otherwise is allowed when one
		considers $\sigma^{\rm LO}_{\rm SI}$. Here, all the mass parameters are in GeV, and the cross-sections are in pb. For BMP-I, the Bino, Wino, and Higgsino compositions are $\mathcal{N}_{11}=0.9973$, $\mathcal{N}_{12}=-9.9266\times 10^{-4}$, $\mathcal{N}_{13}=7.2147\times 10^{-2}$, and $\mathcal{N}_{14}=-1.3909\times 10^{-2}$. Similarly, for BMP-II, $\mathcal{N}_{11}=-0.9975$, $\mathcal{N}_{12}=1.7352\times 10^{-2}$, $\mathcal{N}_{13}=-6.6141\times 10^{-2}$, and $\mathcal{N}_{14}=1.9271\times 10^{-2}$.}
	\label{tab:my_label}
\end{table}

 We now examine the allowed parameter space through a few BMPs that, as before,  satisfy all the necessary $B$-physics constraints, DD bounds on DM, $\delta a_\mu$,
 and the present SUSY search\footnote{The first BMP is allowed by the condition $\Delta m(\tilde{\ell}, \tilde{\chi}_1^0)\leq50$ GeV for $\tilde{e}$ and $\tilde{\mu}$ masses $\leq$150~GeV (see Sec.~\ref{sec:collider} for details) whereas the second BMP is allowed as mentioned in https://atlas.web.cern.ch/Atlas/GROUPS /PHYSICS/PUBNOTES/ATL-PHYS-PUB-2023-025/fig$_16$.png.}.  
 We mainly present a few BMPs, which are allowed by the SI-DD cross-section based on $C_{L, R}^{\rm LO}$, but become excluded when NLO corrected 
 $\chi_1^0\chi_1^0h_i$ vertex is considered instead.
For BMP-I in Tab.~\ref{tab:my_label}, where the LSP is $\tilde B_{\tilde H}$, having $m_{\tilde{\chi}_1^0} =99$ GeV, we get $6.7\%$ rise to $\tilde{\chi}_1^0\tilde{\chi}_1^0h$ vertex and $4.4\%$ rise to $\tilde{\chi}_1^0\tilde{\chi}_1^0H$ vertex
following the inclusion of the NLO corrections. Finally, we obtain an overall $13.4\%$ enhancement to the SI-DD cross-section. The LO cross-section for this BMP is $2.760\times 10^{-11}$ pb, which resides below the central line of \textbf{LZ} (\textbf{LZ} limit is $2.9\times10^{-11}$ pb for this point), thus, allowed by the DD bound. After including the NLO corrections, the DD cross-section becomes $3.130\times 10^{-11}$ pb, which is ruled out by the DD limit of \textbf{LZ}.

In another example, we consider the BMP-II\footnote{Here, charginos/neutralinos can decay into selectrons with a non-negligible BR. However, the BRs of the decays $\tilde{\chi}_1^\pm,\tilde{\chi}_2^0$ to $\tilde{\nu}_e e$, 
		$\tilde{\ell}_{eL} e_L$ can be made insignificant by slightly pushing
		the selectron masses, which does not cause any change in the rest of the analysis.} where $m_{\tilde{\chi}_1^0} = 189$ GeV, we obtain $5.7\%$ and $3.5\%$ rise to $\mathcal {R}_h$ and $\mathcal {R}_H$
respectively and $11.3\%$ corrections to the DD cross-section. In this case, the LO cross-section is $4.709\times 10^{-11}$ pb, again below the central line of \textbf{LZ} (\textbf{LZ} limit is $5.0\times10^{-11}$ pb for this point), hence allowed by the SI-DD searches. After incorporating the NLO corrections, we obtain the SI-DD cross-section $5.241\times 10^{-11}$ pb, which is now above
the \textbf{LZ}-line and hence excluded by the SI-DD search of \textbf{LZ}.

\section{Conclusions}
\label{sec:conclusion}
A dominantly Bino-like $\tilde{\chi}_1^0$, but having (i) a minimal Higgsino component and (ii) a minimal Wino-Higgsino component in the MSSM can accommodate $(g-2)_\mu$, the recent SUSY search constraints, and the LO DM-nucleon scattering cross-section for the DM searches.
Defining them as $\tilde B_{\tilde H}$ and $\tilde B_{\tilde W\tilde H}$, we have computed the NLO corrections to the LSP-Higgs interaction vertices, mainly focusing on the electroweak particles only. There are a total of 234 diagrams for the $\tilde{\chi}_1^0\tilde{\chi}_1^0h$ vertex corrections and 234 for the $\tilde{\chi}_1^0\tilde{\chi}_1^0H$ vertex corrections at the particle level. We have assembled all the diagrams by six topologies and presented the analytical expressions for each topology. To get the UV-finite result, we have included the vertex counterterms.
For the $\tilde B_{\tilde H}$ LSP, including NLO corrections, we have obtained up to $12\%$ and $5\%$ enhancement to the $\tilde{\chi}_1^0\tilde{\chi}_1^0h$ and $\tilde{\chi}_1^0\tilde{\chi}_1^0H$ couplings respectively which in turn leads to an enhancement up to $20\%$ to the SI-DD cross-section. Similarly, for the $\tilde B_{\tilde W\tilde H}$ LSP, we have obtained up to $20\%$ enhancement to the SI-DD cross-section. Through the detailed numerical studies, we have shown that the relative enhancement has only a mild dependence over $\tan\beta$. With the improved corrections, the MSSM parameter space is further squeezed, though somewhat moderately. 
Finally, we reanalyze the exclusion limits in the $M_1-\mu$ plane computed from the SI-DD cross-section for the $\tilde B_{\tilde H}$ LSP, using leading-order and NLO corrected couplings. Here, $\tan\beta=30$ is assumed for the presentation. Overall, a rise of about $\sim 25$ GeV in the Higgsino mass parameter can be observed for $M_1\in [60,240]$ GeV after incorporating the NLO corrections if one uses the $90\%$ confidence limit of the \textbf{LZ} results on the SI-DD cross-section. For other values of $\tan\beta$, a similar shift in the $\mu$ parameter can easily be anticipated. Moreover, a higher value of $\tan\beta(=50)$ is favoured to satisfy the $(g-2)_\mu$ for a heavier LSP (with mass $\sim 600$~GeV). However, even here, the relative rise in the NLO coupling or cross-section is of the same size, which leads to a similar rise in the $\mu$ parameter in the $M_1-\mu$ plane.

\section{Acknowledgements}
Our computations were supported in part by SAMKHYA: the High-Performance Computing Facility provided by the Institute of Physics (IoP), Bhubaneswar, India. The authors thank B. De for his contributions in the early stages of his work. We acknowledge C. Schappacher, A. Pukhov, and S. Heinemeyer for their valuable discussions. A. C. also acknowledges the local hospitality at IoP, Bhubaneswar, during the meeting IMHEP-19 and  IMHEP-22, where some parts of the work were done. AC and SAP also acknowledge the hospitality at IoP, Bhubaneswar, during a visit. SB acknowledges the local hospitality at SNIoE, Greater Noida, during the meeting at WPAC-2023, where this work was finalized. The authors also acknowledge the support received from the SERB project.

\appendix
\section{Couplings}
\label{eq:couplings;}
For the sake of completeness, here we present all the vertex factors following Ref.~\cite{Drees:2004jm} that appeared in the triangular topologies presented in Sec.~\ref{subsec:ver}.

\underline{{\bf Topology-(\ref{fig:topology1}a):}}
\vskip0.2cm
(1) $h_i=h/H$, $F=\tilde{\chi}_{\ell}^0$, and $S=S^\prime=h$.
\begin{align}
\xi_{LL}= \lambda_{h_ihh}\mathcal{G}^{L}_{\tilde{\chi}_1^0\tilde{\chi}_{\ell}^0h}\mathcal{G}^{R*}_{\tilde{\chi}_1^0\tilde{\chi}_{\ell}^0h}~,\qquad\qquad\xi_{LR}= \lambda_{h_ihh}\mathcal{G}^{L}_{\tilde{\chi}_1^0\tilde{\chi}_{\ell}^0h}\mathcal{G}^{L*}_{\tilde{\chi}_1^0\tilde{\chi}_{\ell}^0h}~,\\
\xi_{RL}= \lambda_{h_ihh}\mathcal{G}^{R}_{\tilde{\chi}_1^0\tilde{\chi}_{\ell}^0h}\mathcal{G}^{R*}_{\tilde{\chi}_1^0\tilde{\chi}_{\ell}^0h}~,\qquad\qquad \xi_{RR}= \lambda_{h_ihh}\mathcal{G}^{R}_{\tilde{\chi}_1^0\tilde{\chi}_{\ell}^0h}\mathcal{G}^{L*}_{\tilde{\chi}_1^0\tilde{\chi}_{\ell}^0h}~,
\end{align}

where $\ell=1,...,4$;\qquad $\lambda_{h_ih_ih_i} = -3\dfrac{g_2M_Z}{2c_W}B_{h_i},\,\, {\rm with}\,\, B_{h_i}= \left\{
\begin{array}{ll}
c_{2\alpha} s_{\beta+\alpha}; & h_i=h \\
c_{2\alpha} c_{\beta+\alpha}; & h_i=H \\
\end{array} 
\right.$,
\vskip0.3cm

 $\mathcal{G}^{L}_{\tilde{\chi}_1^0\tilde{\chi}_{\ell}^0h_{i}} = \left\{
\begin{array}{ll}
g_2\big(Q_{\ell 1}^{\prime\prime*}s_\alpha+S_{\ell 1}^{\prime\prime*}c_\alpha\big); & h_i=h \\
g_2\big(-Q_{\ell 1}^{\prime\prime*}c_\alpha+S_{\ell 1}^{\prime\prime*}s_\alpha\big); & h_i=H \\
\end{array} 
\right.~,$\,\,
$\mathcal{G}^{R}_{\tilde{\chi}_1^0\tilde{\chi}_{\ell}^0h_{i}} = \left\{
\begin{array}{ll}
g_2\big(Q_{1\ell}^{\prime\prime}s_\alpha+S_{1\ell}^{\prime\prime}c_\alpha\big); & h_i=h \\
g_2\big(-Q_{1\ell}^{\prime\prime}c_\alpha+S_{1\ell}^{\prime\prime}s_\alpha\big); & h_i=H \\
\end{array} 
\right..$
\vskip0.8cm

(2) $h_i=h/H$, $F=\tilde{\chi}_{\ell}^0$, and $S=h$, $S^\prime=H$ or $S=H$, $S^\prime=h$. 
\begin{align}
\xi_{LL}= \lambda_{h_ihH}\mathcal{G}^{L}_{\tilde{\chi}_1^0\tilde{\chi}_{\ell}^0H}\mathcal{G}^{R*}_{\tilde{\chi}_1^0\tilde{\chi}_{\ell}^0h}~,\qquad\qquad\xi_{LR}= \lambda_{h_ihH}\mathcal{G}^{L}_{\tilde{\chi}_1^0\tilde{\chi}_{\ell}^0H}\mathcal{G}^{L*}_{\tilde{\chi}_1^0\tilde{\chi}_{\ell}^0h}~,\\
\xi_{RL}= \lambda_{h_ihH}\mathcal{G}^{R}_{\tilde{\chi}_1^0\tilde{\chi}_{\ell}^0H}\mathcal{G}^{R*}_{\tilde{\chi}_1^0\tilde{\chi}_{\ell}^0h}~,\qquad\qquad \xi_{RR}= \lambda_{h_ihH}\mathcal{G}^{R}_{\tilde{\chi}_1^0\tilde{\chi}_{\ell}^0H}\mathcal{G}^{L*}_{\tilde{\chi}_1^0\tilde{\chi}_{\ell}^0h}~,
\end{align}

where $\lambda_{h_ihH}=\dfrac{g_2M_Z}{2c_W}C_{h_i}$, with $C_{h_i} = \left\{
\begin{array}{ll}
-2s_{2\alpha} s_{\beta+\alpha}+c_{\beta+\alpha}c_{2\alpha}; & h_i=h \\
\,\,\,\,\,2s_{2\alpha}c_{\beta+\alpha}+s_{\beta+\alpha}c_{2\alpha}; & h_i=H \\
\end{array} 
\right..$
\vskip0.3cm

(3) $h_i=h/H$, $F=\tilde{\chi}_{\ell}^0$, and $S=S^\prime=H$.
\begin{align}
\xi_{LL}= \lambda_{h_iHH}\mathcal{G}^{L}_{\tilde{\chi}_1^0\tilde{\chi}_{\ell}^0H}\mathcal{G}^{R*}_{\tilde{\chi}_1^0\tilde{\chi}_{\ell}^0H}~,\qquad\qquad\xi_{LR}= \lambda_{h_iHH}\mathcal{G}^{L}_{\tilde{\chi}_1^0\tilde{\chi}_{\ell}^0H}\mathcal{G}^{L*}_{\tilde{\chi}_1^0\tilde{\chi}_{\ell}^0H}~,\\
\xi_{RL}= \lambda_{h_iHH}\mathcal{G}^{R}_{\tilde{\chi}_1^0\tilde{\chi}_{\ell}^0H}\mathcal{G}^{R*}_{\tilde{\chi}_1^0\tilde{\chi}_{\ell}^0H}~,\qquad\qquad \xi_{RR}= \lambda_{h_iHH}\mathcal{G}^{R}_{\tilde{\chi}_1^0\tilde{\chi}_{\ell}^0H}\mathcal{G}^{L*}_{\tilde{\chi}_1^0\tilde{\chi}_{\ell}^0H}~,
\end{align}
\vskip0.8cm

(4) $h_i=h/H$, $F=\tilde{\chi}_{\ell}^0$, and $S=S^\prime=A$.
\begin{align}
\xi_{LL}= \lambda_{h_iAA}\mathcal{G}^{L}_{\tilde{\chi}_1^0\tilde{\chi}_{\ell}^0A}\mathcal{G}^{R*}_{\tilde{\chi}_1^0\tilde{\chi}_{\ell}^0A}~,\qquad\qquad\xi_{LR}= \lambda_{h_iAA}\mathcal{G}^{L}_{\tilde{\chi}_1^0\tilde{\chi}_{\ell}^0A}\mathcal{G}^{L*}_{\tilde{\chi}_1^0\tilde{\chi}_{\ell}^0A}~,\\
\xi_{RL}= \lambda_{h_iAA}\mathcal{G}^{R}_{\tilde{\chi}_1^0\tilde{\chi}_{\ell}^0A}\mathcal{G}^{R*}_{\tilde{\chi}_1^0\tilde{\chi}_{\ell}^0A}~,\qquad\qquad \xi_{RR}= \lambda_{h_iAA}\mathcal{G}^{R}_{\tilde{\chi}_1^0\tilde{\chi}_{\ell}^0A}\mathcal{G}^{L*}_{\tilde{\chi}_1^0\tilde{\chi}_{\ell}^0A}~,
\end{align}

where $\lambda_{h_iAA}=-\dfrac{g_2M_Z}{2c_W} c_{2\beta}\,D_{h_i}$, with $D_{h_i} = \left\{
\begin{array}{ll}
\,\,\,\,\,s_{\beta+\alpha}; & h_i=h \\
-c_{\beta+\alpha}; & h_i=H \\
\end{array} 
\right.,$
\vskip0.3cm

$\mathcal{G}^{L}_{\tilde{\chi}_1^0\tilde{\chi}_{\ell}^0A}=i\big(Q_{\ell 1}^{\prime\prime*}s_\beta-S_{\ell 1}^{\prime\prime*}c_\beta\big)$, and \,\,$\mathcal{G}^{R}_{\tilde{\chi}_1^0\tilde{\chi}_{\ell}^0A}=i\big(-Q_{1\ell}^{\prime\prime}s_\beta+S_{1\ell}^{\prime\prime}c_\beta\big)$.
\vskip0.8cm

(5) $h_i=h/H$, $F=\tilde{\chi}_{\ell}^0$, and $S=A$, $S^\prime=G$ or $S=G$, $S^\prime=A$.
\begin{align}
\xi_{LL}= \lambda_{h_iAG}\mathcal{G}^{L}_{\tilde{\chi}_1^0\tilde{\chi}_{\ell}^0G}\mathcal{G}^{R*}_{\tilde{\chi}_1^0\tilde{\chi}_{\ell}^0A}~,\qquad\qquad\xi_{LR}= \lambda_{h_iAG}\mathcal{G}^{L}_{\tilde{\chi}_1^0\tilde{\chi}_{\ell}^0G}\mathcal{G}^{L*}_{\tilde{\chi}_1^0\tilde{\chi}_{\ell}^0A}~,\\
\xi_{RL}= \lambda_{h_iAG}\mathcal{G}^{R}_{\tilde{\chi}_1^0\tilde{\chi}_{\ell}^0G}\mathcal{G}^{R*}_{\tilde{\chi}_1^0\tilde{\chi}_{\ell}^0A}~,\qquad\qquad \xi_{RR}= \lambda_{h_iAG}\mathcal{G}^{R}_{\tilde{\chi}_1^0\tilde{\chi}_{\ell}^0G}\mathcal{G}^{L*}_{\tilde{\chi}_1^0\tilde{\chi}_{\ell}^0A}~,
\end{align}

where $\lambda_{h_iAG} = -\dfrac{g_2M_Z}{2c_W}s_{2\beta}\, D_{h_i}$,\,\, 
$\mathcal{G}^{L}_{\tilde{\chi}_1^0\tilde{\chi}_{\ell}^0G} = ig_2\big(-Q_{\ell 1}^{\prime\prime*}c_\beta - S_{\ell 1}^{\prime\prime*}s_\beta\big)$, and $\mathcal{G}^{R}_{\tilde{\chi}_1^0\tilde{\chi}_{\ell}^0G} = ig_2\big(Q_{1\ell}^{\prime\prime}c_\beta + S_{1\ell}^{\prime\prime}s_\beta\big)$.
\vskip0.8cm

(6) $h_i=h/H$, $F=\tilde{\chi}_{\ell}^0$, and $S=S^\prime=G$.
\begin{align}
\xi_{LL}= \lambda_{h_iGG}\mathcal{G}^{L}_{\tilde{\chi}_1^0\tilde{\chi}_{\ell}^0G}\mathcal{G}^{R*}_{\tilde{\chi}_1^0\tilde{\chi}_{\ell}^0G}~,\qquad\qquad\xi_{LR}= \lambda_{h_iGG}\mathcal{G}^{L}_{\tilde{\chi}_1^0\tilde{\chi}_{\ell}^0G}\mathcal{G}^{L*}_{\tilde{\chi}_1^0\tilde{\chi}_{\ell}^0G}~,\\
\xi_{RL}= \lambda_{h_iGG}\mathcal{G}^{R}_{\tilde{\chi}_1^0\tilde{\chi}_{\ell}^0G}\mathcal{G}^{R*}_{\tilde{\chi}_1^0\tilde{\chi}_{\ell}^0G}~,\qquad\qquad \xi_{RR}= \lambda_{h_iGG}\mathcal{G}^{R}_{\tilde{\chi}_1^0\tilde{\chi}_{\ell}^0G}\mathcal{G}^{L*}_{\tilde{\chi}_1^0\tilde{\chi}_{\ell}^0G}~,
\end{align}

where $\lambda_{h_iGG}=-\dfrac{g_2M_Z}{2c_W}c_{2\beta}\,D^\prime_{h_i}$,\qquad with $D^\prime_{h_i} = \left\{
\begin{array}{ll}
-s_{\beta+\alpha}; & h_i=h \\
\,\,\,\,\,c_{\beta+\alpha}; & h_i=H \\
\end{array} 
\right.,$
\vskip0.8cm

(7) $h_i=h/H$, $F=\tilde{\chi}_{\ell}^{\pm}$, and $S=S^\prime=H^{\pm}$.
\begin{align}
\xi_{LL}= \lambda_{h_iH^\pm H^\pm}\mathcal{G}^{L}_{\tilde{\chi}_1^0\tilde{\chi}_{\ell}^\pm H^\pm}\mathcal{G}^{R*}_{\tilde{\chi}_1^0\tilde{\chi}_{\ell}^\pm H^\pm}~,\qquad\qquad\xi_{LR}= \lambda_{h_iH^\pm H^\pm}\mathcal{G}^{L}_{\tilde{\chi}_1^0\tilde{\chi}_{\ell}^\pm H^\pm}\mathcal{G}^{L*}_{\tilde{\chi}_1^0\tilde{\chi}_{\ell}^\pm H^\pm}~,\\
\xi_{RL}= \lambda_{h_iH^\pm H^\pm}\mathcal{G}^{R}_{\tilde{\chi}_1^0\tilde{\chi}_{\ell}^\pm H^\pm}\mathcal{G}^{R*}_{\tilde{\chi}_1^0\tilde{\chi}_{\ell}^\pm H^\pm}~,\qquad\qquad \xi_{RR}= \lambda_{h_iH^\pm H^\pm}\mathcal{G}^{R}_{\tilde{\chi}_1^0\tilde{\chi}_{\ell}^\pm H^\pm}\mathcal{G}^{L*}_{\tilde{\chi}_1^0\tilde{\chi}_{\ell}^\pm H^\pm}~,
\end{align}

where, $\lambda_{h_iH^\pm H^\pm} = -g_2 A_{h_i}$, with $A_{h_i} = \left\{
\begin{array}{ll}
M_Ws_{\beta-\alpha}+\dfrac{M_Z}{2c_W}c_{2\beta} s_{\beta+\alpha}; & h_i=h \\
M_W c_{\beta-\alpha}-\dfrac{M_Z}{2c_W}c_{2\beta}c_{\beta+\alpha}; & h_i=H \\
\end{array} 
\right.,$ 
\vskip0.3cm

 $\mathcal{G}^{L}_{\tilde{\chi}_1^0\tilde{\chi}_{\ell}^\pm H^\pm}=-g_2 Q^{\prime L}_{1\ell}$, and $\mathcal{G}^{R}_{\tilde{\chi}_1^0\tilde{\chi}_{\ell}^\pm H^\pm} = -g_2 Q^{\prime R}_{1\ell}~.$
\vskip0.8cm

(8) $h_i=h/H$, $F=\tilde{\chi}_{\ell}^{\pm}$, and $S=H^\pm$, $S^\prime=G^{\pm}$ or $S=G^\pm$, $S^\prime=H^{\pm}$.
\begin{align}
\xi_{LL}= \lambda_{h_iH^\pm G^\pm}\mathcal{G}^{L}_{\tilde{\chi}_1^0\tilde{\chi}_{\ell}^\pm G^\pm}\mathcal{G}^{R*}_{\tilde{\chi}_1^0\tilde{\chi}_{\ell}^\pm H^\pm}~,\qquad\qquad\xi_{LR}= \lambda_{h_iH^\pm G^\pm}\mathcal{G}^{L}_{\tilde{\chi}_1^0\tilde{\chi}_{\ell}^\pm G^\pm}\mathcal{G}^{L*}_{\tilde{\chi}_1^0\tilde{\chi}_{\ell}^\pm H^\pm}~,\\
\xi_{RL}= \lambda_{h_iH^\pm G^\pm}\mathcal{G}^{R}_{\tilde{\chi}_1^0\tilde{\chi}_{\ell}^\pm G^\pm}\mathcal{G}^{R*}_{\tilde{\chi}_1^0\tilde{\chi}_{\ell}^\pm H^\pm}~,\qquad\qquad \xi_{RR}= \lambda_{h_iH^\pm G^\pm}\mathcal{G}^{R}_{\tilde{\chi}_1^0\tilde{\chi}_{\ell}^\pm G^\pm}\mathcal{G}^{L*}_{\tilde{\chi}_1^0\tilde{\chi}_{\ell}^\pm H^\pm}~,
\end{align}

where $\lambda_{h_iH^\pm G^\pm} = -\frac{g_2M_W}{2}A^{\prime}_{h_i}$, with $A^\prime_{h_i} = \left\{
\begin{array}{ll}
\,\,\,\,\,\dfrac{s_{2\beta} s_{\beta+\alpha}}{c_W^2}-c_{\beta-\alpha}; & h_i=h \\
-\dfrac{s_{2\beta} c_{\beta+\alpha}}{c_W^2}-s_{\beta-\alpha}; & h_i=H \\
\end{array} 
\right.,$
\vskip0.3cm

 $\mathcal{G}^{L}_{\tilde{\chi}_1^0\tilde{\chi}_{\ell}^\pm G^\pm} = -g_2 t_\beta Q_{1\ell}^{\prime L}$, and $\mathcal{G}^{R}_{\tilde{\chi}_1^0\tilde{\chi}_{\ell}^\pm G^\pm} = \dfrac{g_2}{t_\beta} Q_{1\ell}^{\prime R}$.
\vskip0.8cm 

(9) $h_i=h/H$, $F=\tilde{\chi}_{\ell}^{\pm}$, and $S=S^\prime=G^{\pm}$.
\begin{align}
\xi_{LL}= \lambda_{h_iG^\pm G^\pm}\mathcal{G}^{L}_{\tilde{\chi}_1^0\tilde{\chi}_{\ell}^\pm G^\pm}\mathcal{G}^{R*}_{\tilde{\chi}_1^0\tilde{\chi}_{\ell}^\pm G^\pm}~,\qquad\qquad\xi_{LR}= \lambda_{h_iG^\pm G^\pm}\mathcal{G}^{L}_{\tilde{\chi}_1^0\tilde{\chi}_{\ell}^\pm G^\pm}\mathcal{G}^{L*}_{\tilde{\chi}_1^0\tilde{\chi}_{\ell}^\pm G^\pm}~,\\
\xi_{RL}= \lambda_{h_iG^\pm G^\pm}\mathcal{G}^{R}_{\tilde{\chi}_1^0\tilde{\chi}_{\ell}^\pm G^\pm}\mathcal{G}^{R*}_{\tilde{\chi}_1^0\tilde{\chi}_{\ell}^\pm G^\pm}~,\qquad\qquad \xi_{RR}= \lambda_{h_iG^\pm G^\pm}\mathcal{G}^{R}_{\tilde{\chi}_1^0\tilde{\chi}_{\ell}^\pm G^\pm}\mathcal{G}^{L*}_{\tilde{\chi}_1^0\tilde{\chi}_{\ell}^\pm G^\pm}~,
\end{align} 

where $\lambda_{h_iG^\pm G^\pm} = -\dfrac{g_2M_Z}{2c_W} c_{2\beta} \,D^\prime_{h_i}$.
\vskip0.8cm

(10) $h_i = h/H$, $F=\nu_n$, $S=\tilde{\nu}_{\ell}$, $S^\prime=\tilde{\nu}_{m}$.
\begin{align}
\xi_{LL}=0~,\,\,\qquad\qquad\qquad\qquad\qquad\qquad\qquad\xi_{LR}= \lambda_{h_i\tilde{\nu}_\ell \tilde{\nu}_m}\mathcal{G}^{L}_{\tilde{\chi}_1^0\nu_n\tilde{\nu}_{\ell}}\mathcal{G}^{L*}_{\tilde{\chi}_1^0\nu_n\tilde{\nu}_{m}}~,\\
\xi_{RL}=0~,\qquad\qquad\qquad\qquad\qquad\qquad\qquad\qquad\qquad\qquad\qquad\quad\,\, \xi_{RR}= 0~,
\end{align} 

where $\ell,m,n=1,2,3$; \qquad$\mathcal{G}^{L}_{\tilde{\chi}_1^0\nu_n\tilde{\nu}_{m}} = G^{\nu}_{nm1}$, \,\,$\lambda_{h_i\tilde{\nu}_\ell \tilde{\nu}_m} = \left\{
\begin{array}{ll}
\,\,\,\,\,c_g[\tilde{\nu}]s_{\alpha+\beta}\delta_{\ell m}; & h_i=h \\
-c_g[\tilde{\nu}]c_{\alpha+\beta}\delta_{\ell m}; & h_i=H \\
\end{array} 
\right.,\,\,\,\,$

 with $c_g[\tilde{\nu}]\equiv \dfrac{g_2 M_W}{2}\big(1+t_W^2\big)$.
\vskip1.0cm

(11) $h_i = h/H$, $F=e_n$, $S=\tilde{e}_{\ell}$, $S^\prime=\tilde{e}_{m}$.
\begin{align}
\xi_{LL}= \lambda_{h_i\tilde{e}_\ell \tilde{e}_m}\mathcal{G}^{L}_{e_n\tilde{e}_{m}\tilde{\chi}_1^0}\mathcal{G}^{R*}_{e_n\tilde{e}_{\ell}\tilde{\chi}_1^0}~,\qquad\qquad\xi_{LR}= \lambda_{h_i\tilde{e}_\ell \tilde{e}_m}\mathcal{G}^{L}_{e_n\tilde{e}_{m}\tilde{\chi}_1^0}\mathcal{G}^{L*}_{e_n\tilde{e}_{\ell}\tilde{\chi}_1^0}~,\\
\xi_{RL}= \lambda_{h_i\tilde{e}_\ell \tilde{e}_m}\mathcal{G}^{R}_{e_n\tilde{e}_{m}\tilde{\chi}_1^0}\mathcal{G}^{R*}_{e_n\tilde{e}_{\ell}\tilde{\chi}_1^0}~,\qquad\qquad \xi_{RR}= \lambda_{h_i\tilde{e}_\ell \tilde{e}_m}\mathcal{G}^{R}_{e_n\tilde{e}_{m}\tilde{\chi}_1^0}\mathcal{G}^{L*}_{e_n\tilde{e}_{\ell}\tilde{\chi}_1^0}~,
\end{align} 

where $n=1,2,3$;\qquad $\ell,m=1,...,6$;\qquad $\mathcal{G}^{L}_{e_n\tilde{e}_{m}\tilde{\chi}_1^0} = G_{nm1}^{e_L}$,\,\,  $\mathcal{G}^{R}_{e_n\tilde{e}_{m}\tilde{\chi}_1^0} = G_{nm1}^{e_R}$, and
\vskip0.3cm
  $\lambda_{h_i\tilde{e}_\ell \tilde{e}_m} = \left\{
\begin{array}{ll}
-c_A[\tilde{e}_\ell, \tilde{e}_m]s_{\alpha}+ c_\mu [\tilde{e}_\ell, \tilde{e}_m]c_{\alpha} + c_g[\tilde{e}_\ell, \tilde{e}_m]s_{\alpha+\beta} ; & h_i=h \\
\,\,\,\,\,c_A[\tilde{e}_\ell, \tilde{e}_m]c_{\alpha}+ c_\mu [\tilde{e}_\ell, \tilde{e}_m]s_{\alpha} - c_g[\tilde{e}_\ell, \tilde{e}_m]c_{\alpha+\beta}; & h_i=H \\
\end{array} 
\right.,\,\,\,\,$

where 
\begin{align*}
c_A[\tilde{e}_\ell, \tilde{e}_m] \equiv& \frac{g_2}{M_Wc_\beta}\Bigl\{-\sum_{i=1}^{3}m^2_{e_i}\Big[W_{i\ell}^{\tilde{e}*}W_{im}^{\tilde{e}} + W_{i+3\,\ell}^{\tilde{e}*}W_{i+3\,m}^{\tilde{e}}\Big] + \frac{1}{2}\sum_{i,j=1}^{3}\Big[\big(\mathbf{m_eA^{e\dagger}}\big)_{ij}W_{i\ell}^{\tilde{e}*}W_{j+3\,m}^{\tilde{e}}\nonumber\\
&+\big(\mathbf{m_eA^{e}}\big)_{ij}W_{j+3\,\ell}^{\tilde{e}*}W_{im}^{\tilde{e}}\Big]\Bigr\}~,\\
c_\mu[\tilde{e}_\ell, \tilde{e}_m] \equiv& \frac{g_2}{2M_Wc_\beta}\sum_{i=1}^{3}m_{e_i}\Big[\mu W_{i\ell}^{\tilde{e*}} W_{i+3\,m}^{\tilde{e}} + \mu^{*} W_{i+3\,\ell}^{\tilde{e*}} W_{im}^{\tilde{e}}\Big]~,\\
c_\mu[\tilde{e}_\ell, \tilde{e}_m] \equiv& \frac{g_2M_W}{2}\sum_{i=1}^{3}\Big[W_{i\ell}^{\tilde{e*}}W_{im}^{\tilde{e}}\big(t_W^2-1\big) - 2t_W^2 W_{i+3\,\ell}^{\tilde{e*}}W_{i+3\,m}^{\tilde{e}}\Big]
\end{align*}

\underline{{\bf Topology-(\ref{fig:topology1}b):}}
\vskip0.2cm
(1) $h_i=h/H$, $S = h/H$, $F=\tilde{\chi}_\ell^0$, $F^\prime=\tilde{\chi}_n^0$.
\begin{align*}
\zeta_{LLL} = \mathcal{G}_{\tilde{\chi}_1^0 \tilde{\chi}_n^0 h_i}^L \mathcal{G}_{\tilde{\chi}_\ell^0\tilde{\chi}_n^0 h_i}^L \mathcal{G}_{\tilde{\chi}_1^0 \tilde{\chi}_\ell^0 h_i}^{R*}~, \qquad\qquad \zeta_{LLR} = \mathcal{G}_{\tilde{\chi}_1^0 \tilde{\chi}_n^0 h_i}^L \mathcal{G}_{\tilde{\chi}_\ell^0\tilde{\chi}_n^0 h_i}^L \mathcal{G}_{\tilde{\chi}_1^0 \tilde{\chi}_\ell^0 h_i}^{L*}~,\\
\zeta_{LRL} = \mathcal{G}_{\tilde{\chi}_1^0 \tilde{\chi}_n^0 h_i}^L \mathcal{G}_{\tilde{\chi}_\ell^0\tilde{\chi}_n^0 h_i}^R \mathcal{G}_{\tilde{\chi}_1^0 \tilde{\chi}_\ell^0 h_i}^{R*}~, \qquad\qquad \zeta_{LRR} = \mathcal{G}_{\tilde{\chi}_1^0 \tilde{\chi}_n^0 h_i}^L \mathcal{G}_{\tilde{\chi}_\ell^0\tilde{\chi}_n^0 h_i}^R \mathcal{G}_{\tilde{\chi}_1^0 \tilde{\chi}_\ell^0 h_i}^{L*}~,\\
\zeta_{RLL} = \mathcal{G}_{\tilde{\chi}_1^0 \tilde{\chi}_n^0 h_i}^R \mathcal{G}_{\tilde{\chi}_\ell^0\tilde{\chi}_n^0 h_i}^L \mathcal{G}_{\tilde{\chi}_1^0 \tilde{\chi}_\ell^0 h_i}^{R*}~, \qquad\qquad \zeta_{RLR} = \mathcal{G}_{\tilde{\chi}_1^0 \tilde{\chi}_n^0 h_i}^R \mathcal{G}_{\tilde{\chi}_\ell^0\tilde{\chi}_n^0 h_i}^L \mathcal{G}_{\tilde{\chi}_1^0 \tilde{\chi}_\ell^0 h_i}^{L*}~,\\
\zeta_{RRL} = \mathcal{G}_{\tilde{\chi}_1^0 \tilde{\chi}_n^0 h_i}^R \mathcal{G}_{\tilde{\chi}_\ell^0\tilde{\chi}_n^0 h_i}^R \mathcal{G}_{\tilde{\chi}_1^0 \tilde{\chi}_\ell^0 h_i}^{R*}~, \qquad\qquad \zeta_{RRR} = \mathcal{G}_{\tilde{\chi}_1^0 \tilde{\chi}_n^0 h_i}^R \mathcal{G}_{\tilde{\chi}_\ell^0\tilde{\chi}_n^0 h_i}^R \mathcal{G}_{\tilde{\chi}_1^0 \tilde{\chi}_\ell^0 h_i}^{L*}~,
\end{align*}

where $\mathcal{G}_{\tilde{\chi}_\ell^0\tilde{\chi}_n^0 h_i}^L = \left\{
\begin{array}{ll}
\,\,\,\,\, g_2\big(Q_{\ell n}^{\prime\prime*}\,s_\alpha+S_{\ell n}^{\prime\prime*}\,c_\alpha\big); & h_i=h \\
-g_2\big(Q_{\ell n}^{\prime\prime*}\,c_\alpha-S_{\ell n}^{\prime\prime*}\,s_\alpha\big); & h_i=H \\
\end{array} 
\right.$\, and\, $\mathcal{G}_{\tilde{\chi}_\ell^0\tilde{\chi}_n^0 h_i}^R = \left\{
\begin{array}{ll}
\,\,\,\,\, g_2\big(Q_{n\ell}^{\prime\prime}\,s_\alpha+S_{n\ell}^{\prime\prime}\,c_\alpha\big); & h_i=h \\
-g_2\big(Q_{n\ell}^{\prime\prime}\,c_\alpha-S_{n\ell}^{\prime\prime}\,s_\alpha\big); & h_i=H \\
\end{array} 
\right..$
\vskip1.0cm

(2) $h_i=h/H$, $S = A$, $F=\tilde{\chi}_\ell^0$, $F^\prime=\tilde{\chi}_n^0$.
\begin{align*}
\zeta_{LLL} = \mathcal{G}_{\tilde{\chi}_1^0 \tilde{\chi}_n^0 A}^L \mathcal{G}_{\tilde{\chi}_\ell^0\tilde{\chi}_n^0 h_i}^L \mathcal{G}_{\tilde{\chi}_1^0 \tilde{\chi}_\ell^0 A}^{R*}~, \qquad\qquad \zeta_{LLR} = \mathcal{G}_{\tilde{\chi}_1^0 \tilde{\chi}_n^0 A}^L \mathcal{G}_{\tilde{\chi}_\ell^0\tilde{\chi}_n^0 h_i}^L \mathcal{G}_{\tilde{\chi}_1^0 \tilde{\chi}_\ell^0 A}^{L*}~,\\
\zeta_{LRL} = \mathcal{G}_{\tilde{\chi}_1^0 \tilde{\chi}_n^0 A}^L \mathcal{G}_{\tilde{\chi}_\ell^0\tilde{\chi}_n^0 h_i}^R \mathcal{G}_{\tilde{\chi}_1^0 \tilde{\chi}_\ell^0 A}^{R*}~, \qquad\qquad \zeta_{LRR} = \mathcal{G}_{\tilde{\chi}_1^0 \tilde{\chi}_n^0 A}^L \mathcal{G}_{\tilde{\chi}_\ell^0\tilde{\chi}_n^0 h_i}^R \mathcal{G}_{\tilde{\chi}_1^0 \tilde{\chi}_\ell^0 A}^{L*}~,\\
\zeta_{RLL} = \mathcal{G}_{\tilde{\chi}_1^0 \tilde{\chi}_n^0 A}^R \mathcal{G}_{\tilde{\chi}_\ell^0\tilde{\chi}_n^0 h_i}^L \mathcal{G}_{\tilde{\chi}_1^0 \tilde{\chi}_\ell^0 A}^{R*}~, \qquad\qquad \zeta_{RLR} = \mathcal{G}_{\tilde{\chi}_1^0 \tilde{\chi}_n^0 A}^R \mathcal{G}_{\tilde{\chi}_\ell^0\tilde{\chi}_n^0 h_i}^L \mathcal{G}_{\tilde{\chi}_1^0 \tilde{\chi}_\ell^0 A}^{L*}~,\\
\zeta_{RRL} = \mathcal{G}_{\tilde{\chi}_1^0 \tilde{\chi}_n^0 A}^R \mathcal{G}_{\tilde{\chi}_\ell^0\tilde{\chi}_n^0 h_i}^R \mathcal{G}_{\tilde{\chi}_1^0 \tilde{\chi}_\ell^0 A}^{R*}~, \qquad\qquad \zeta_{RRR} = \mathcal{G}_{\tilde{\chi}_1^0 \tilde{\chi}_n^0 A}^R \mathcal{G}_{\tilde{\chi}_\ell^0\tilde{\chi}_n^0 h_i}^R \mathcal{G}_{\tilde{\chi}_1^0 \tilde{\chi}_\ell^0 A}^{L*}~,
\end{align*}

where $\mathcal{G}_{\tilde{\chi}_1^0 \tilde{\chi}_n^0 A}^L = i\big(Q_{n1}^{\prime\prime*}s_\beta - S_{n1}^{\prime\prime*}c_\beta\big)$\, and \, $\mathcal{G}_{\tilde{\chi}_1^0 \tilde{\chi}_n^0 A}^R = i\big(-Q_{1n}^{\prime\prime}s_\beta + S_{1n}^{\prime\prime}c_\beta\big)$.
\vskip1.0cm

(3) $h_i=h/H$, $S = G$, $F=\tilde{\chi}_\ell^0$, $F^\prime=\tilde{\chi}_n^0$.
\begin{align*}
\zeta_{LLL} = \mathcal{G}_{\tilde{\chi}_1^0 \tilde{\chi}_n^0 G}^L \mathcal{G}_{\tilde{\chi}_\ell^0\tilde{\chi}_n^0 h_i}^L \mathcal{G}_{\tilde{\chi}_1^0 \tilde{\chi}_\ell^0 G}^{R*}~, \qquad\qquad \zeta_{LLR} = \mathcal{G}_{\tilde{\chi}_1^0 \tilde{\chi}_n^0 G}^L \mathcal{G}_{\tilde{\chi}_\ell^0\tilde{\chi}_n^0 h_i}^L \mathcal{G}_{\tilde{\chi}_1^0 \tilde{\chi}_\ell^0 G}^{L*}~,\\
\zeta_{LRL} = \mathcal{G}_{\tilde{\chi}_1^0 \tilde{\chi}_n^0 G}^L \mathcal{G}_{\tilde{\chi}_\ell^0\tilde{\chi}_n^0 h_i}^R \mathcal{G}_{\tilde{\chi}_1^0 \tilde{\chi}_\ell^0 G}^{R*}~, \qquad\qquad \zeta_{LRR} = \mathcal{G}_{\tilde{\chi}_1^0 \tilde{\chi}_n^0 G}^L \mathcal{G}_{\tilde{\chi}_\ell^0\tilde{\chi}_n^0 h_i}^R \mathcal{G}_{\tilde{\chi}_1^0 \tilde{\chi}_\ell^0 G}^{L*}~,\\
\zeta_{RLL} = \mathcal{G}_{\tilde{\chi}_1^0 \tilde{\chi}_n^0 G}^R \mathcal{G}_{\tilde{\chi}_\ell^0\tilde{\chi}_n^0 h_i}^L \mathcal{G}_{\tilde{\chi}_1^0 \tilde{\chi}_\ell^0 G}^{R*}~, \qquad\qquad \zeta_{RLR} = \mathcal{G}_{\tilde{\chi}_1^0 \tilde{\chi}_n^0 G}^R \mathcal{G}_{\tilde{\chi}_\ell^0\tilde{\chi}_n^0 h_i}^L \mathcal{G}_{\tilde{\chi}_1^0 \tilde{\chi}_\ell^0 G}^{L*}~,\\
\zeta_{RRL} = \mathcal{G}_{\tilde{\chi}_1^0 \tilde{\chi}_n^0 G}^R \mathcal{G}_{\tilde{\chi}_\ell^0\tilde{\chi}_n^0 h_i}^R \mathcal{G}_{\tilde{\chi}_1^0 \tilde{\chi}_\ell^0 G}^{R*}~, \qquad\qquad \zeta_{RRR} = \mathcal{G}_{\tilde{\chi}_1^0 \tilde{\chi}_n^0 G}^R \mathcal{G}_{\tilde{\chi}_\ell^0\tilde{\chi}_n^0 h_i}^R \mathcal{G}_{\tilde{\chi}_1^0 \tilde{\chi}_\ell^0 G}^{L*}~.
\end{align*}
\vskip0.8cm

(4) $h_i=h/H$, $S = H^\pm$, $F=\tilde{\chi}_\ell^\pm$, $F^\prime=\tilde{\chi}_n^\pm$.
\begin{align*}
\zeta_{LLL} = \mathcal{G}_{\tilde{\chi}_1^0 \tilde{\chi}_n^\pm H^\pm}^L \mathcal{G}_{\tilde{\chi}_\ell^\pm\tilde{\chi}_n^\pm h_i}^L \mathcal{G}_{\tilde{\chi}_1^0 \tilde{\chi}_\ell^\pm H^\pm}^{R*}~, \qquad\qquad \zeta_{LLR} = \mathcal{G}_{\tilde{\chi}_1^0 \tilde{\chi}_n^\pm H^\pm}^L \mathcal{G}_{\tilde{\chi}_\ell^\pm\tilde{\chi}_n^\pm h_i}^L \mathcal{G}_{\tilde{\chi}_1^0 \tilde{\chi}_\ell^\pm H^\pm}^{L*}~,\\
\zeta_{LRL} = \mathcal{G}_{\tilde{\chi}_1^0 \tilde{\chi}_n^\pm H^\pm}^L \mathcal{G}_{\tilde{\chi}_\ell^\pm\tilde{\chi}_n^\pm h_i}^R \mathcal{G}_{\tilde{\chi}_1^0 \tilde{\chi}_\ell^\pm H^\pm}^{R*}~, \qquad\qquad \zeta_{LRR} = \mathcal{G}_{\tilde{\chi}_1^0 \tilde{\chi}_n^\pm H^\pm}^L \mathcal{G}_{\tilde{\chi}_\ell^\pm\tilde{\chi}_n^\pm h_i}^R \mathcal{G}_{\tilde{\chi}_1^0 \tilde{\chi}_\ell^\pm H^\pm}^{L*}~,\\
\zeta_{RLL} = \mathcal{G}_{\tilde{\chi}_1^0 \tilde{\chi}_n^\pm H^\pm}^R \mathcal{G}_{\tilde{\chi}_\ell^\pm\tilde{\chi}_n^\pm h_i}^L \mathcal{G}_{\tilde{\chi}_1^0 \tilde{\chi}_\ell^\pm H^\pm}^{R*}~, \qquad\qquad \zeta_{RLR} = \mathcal{G}_{\tilde{\chi}_1^0 \tilde{\chi}_n^\pm H^\pm}^R \mathcal{G}_{\tilde{\chi}_\ell^\pm\tilde{\chi}_n^\pm h_i}^L \mathcal{G}_{\tilde{\chi}_1^0 \tilde{\chi}_\ell^\pm H^\pm}^{L*}~,\\
\zeta_{RRL} = \mathcal{G}_{\tilde{\chi}_1^0 \tilde{\chi}_n^\pm H^\pm}^R \mathcal{G}_{\tilde{\chi}_\ell^\pm\tilde{\chi}_n^\pm h_i}^R \mathcal{G}_{\tilde{\chi}_1^0 \tilde{\chi}_\ell^\pm H^\pm}^{R*}~, \qquad\qquad \zeta_{RRR} = \mathcal{G}_{\tilde{\chi}_1^0 \tilde{\chi}_n^\pm H^\pm}^R \mathcal{G}_{\tilde{\chi}_\ell^\pm\tilde{\chi}_n^\pm h_i}^R \mathcal{G}_{\tilde{\chi}_1^0 \tilde{\chi}_\ell^\pm H^\pm}^{L*}~.
\end{align*}
\vskip0.8cm

(5) $h_i=h/H$, $S = G^\pm$, $F=\tilde{\chi}_\ell^\pm$, $F^\prime=\tilde{\chi}_n^\pm$.
\begin{align*}
\zeta_{LLL} = \mathcal{G}_{\tilde{\chi}_1^0 \tilde{\chi}_n^\pm G^\pm}^L \mathcal{G}_{\tilde{\chi}_\ell^\pm\tilde{\chi}_n^\pm h_i}^L \mathcal{G}_{\tilde{\chi}_1^0 \tilde{\chi}_\ell^\pm G^\pm}^{R*}~, \qquad\qquad \zeta_{LLR} = \mathcal{G}_{\tilde{\chi}_1^0 \tilde{\chi}_n^\pm G^\pm}^L \mathcal{G}_{\tilde{\chi}_\ell^\pm\tilde{\chi}_n^\pm h_i}^L \mathcal{G}_{\tilde{\chi}_1^0 \tilde{\chi}_\ell^\pm G^\pm}^{L*}~,\\
\zeta_{LRL} = \mathcal{G}_{\tilde{\chi}_1^0 \tilde{\chi}_n^\pm G^\pm}^L \mathcal{G}_{\tilde{\chi}_\ell^\pm\tilde{\chi}_n^\pm h_i}^R \mathcal{G}_{\tilde{\chi}_1^0 \tilde{\chi}_\ell^\pm G^\pm}^{R*}~, \qquad\qquad \zeta_{LRR} = \mathcal{G}_{\tilde{\chi}_1^0 \tilde{\chi}_n^\pm G^\pm}^L \mathcal{G}_{\tilde{\chi}_\ell^\pm\tilde{\chi}_n^\pm h_i}^R \mathcal{G}_{\tilde{\chi}_1^0 \tilde{\chi}_\ell^\pm G^\pm}^{L*}~,\\
\zeta_{RLL} = \mathcal{G}_{\tilde{\chi}_1^0 \tilde{\chi}_n^\pm G^\pm}^R \mathcal{G}_{\tilde{\chi}_\ell^\pm\tilde{\chi}_n^\pm h_i}^L \mathcal{G}_{\tilde{\chi}_1^0 \tilde{\chi}_\ell^\pm G^\pm}^{R*}~, \qquad\qquad \zeta_{RLR} = \mathcal{G}_{\tilde{\chi}_1^0 \tilde{\chi}_n^\pm G^\pm}^R \mathcal{G}_{\tilde{\chi}_\ell^\pm\tilde{\chi}_n^\pm h_i}^L \mathcal{G}_{\tilde{\chi}_1^0 \tilde{\chi}_\ell^\pm G^\pm}^{L*}~,\\
\zeta_{RRL} = \mathcal{G}_{\tilde{\chi}_1^0 \tilde{\chi}_n^\pm G^\pm}^R \mathcal{G}_{\tilde{\chi}_\ell^\pm\tilde{\chi}_n^\pm h_i}^R \mathcal{G}_{\tilde{\chi}_1^0 \tilde{\chi}_\ell^\pm G^\pm}^{R*}~, \qquad\qquad \zeta_{RRR} = \mathcal{G}_{\tilde{\chi}_1^0 \tilde{\chi}_n^\pm G^\pm}^R \mathcal{G}_{\tilde{\chi}_\ell^\pm\tilde{\chi}_n^\pm h_i}^R \mathcal{G}_{\tilde{\chi}_1^0 \tilde{\chi}_\ell^\pm G^\pm}^{L*}~.
\end{align*}
\vskip0.8cm

(6) $h_i=h/H$, $S = \tilde{e}_m$, $F=e_\ell$, $F^\prime=e_n$.
\begin{align*}
\zeta_{LLL} = \mathcal{G}_{\tilde{\chi}_1^0 \tilde{e}_m e_n}^L \mathcal{G}_{e_\ell e_n h_i}^L \mathcal{G}_{\tilde{\chi}_1^0 \tilde{e}_m e_\ell}^{R*}~, \qquad\qquad \zeta_{LLR} = \mathcal{G}_{\tilde{\chi}_1^0 \tilde{e}_m e_n}^L \mathcal{G}_{e_\ell e_n h_i}^L \mathcal{G}_{\tilde{\chi}_1^0 \tilde{e}_m e_\ell}^{L*}~,\\
\zeta_{LRL} = \mathcal{G}_{\tilde{\chi}_1^0 \tilde{e}_m e_n}^L \mathcal{G}_{e_\ell e_n h_i}^R \mathcal{G}_{\tilde{\chi}_1^0 \tilde{e}_m e_\ell}^{R*}~, \qquad\qquad \zeta_{LRR} = \mathcal{G}_{\tilde{\chi}_1^0 \tilde{e}_m e_n}^L \mathcal{G}_{e_\ell e_n h_i}^R \mathcal{G}_{\tilde{\chi}_1^0 \tilde{e}_m e_\ell}^{L*}~,\\
\zeta_{RLL} = \mathcal{G}_{\tilde{\chi}_1^0 \tilde{e}_m e_n}^R \mathcal{G}_{e_\ell e_n h_i}^L \mathcal{G}_{\tilde{\chi}_1^0 \tilde{e}_m e_\ell}^{R*}~, \qquad\qquad \zeta_{RLR} = \mathcal{G}_{\tilde{\chi}_1^0 \tilde{e}_m e_n}^R \mathcal{G}_{e_\ell e_n h_i}^L \mathcal{G}_{\tilde{\chi}_1^0 \tilde{e}_m e_\ell}^{L*}~,\\
\zeta_{RRL} = \mathcal{G}_{\tilde{\chi}_1^0 \tilde{e}_m e_n}^R \mathcal{G}_{e_\ell e_n h_i}^R \mathcal{G}_{\tilde{\chi}_1^0 \tilde{e}_m e_\ell}^{R*}~, \qquad\qquad \zeta_{RRR} = \mathcal{G}_{\tilde{\chi}_1^0 \tilde{e}_m e_n}^R \mathcal{G}_{e_\ell e_n h_i}^R \mathcal{G}_{\tilde{\chi}_1^0 \tilde{e}_m e_\ell}^{L*}~.
\end{align*}

where $\ell,n=1,2,3$;\qquad $m=1,...,6$;\qquad $\mathcal{G}^{L}_{\tilde{\chi}_1^0\tilde{e}_m e_{n}} = G_{nm1}^{e_L}$,\,\, $\mathcal{G}^{R}_{\tilde{\chi}_1^0\tilde{e}_m e_{n}} = G_{nm1}^{e_R}$.
\vskip0.8cm

\underline{{\bf Topology-(\ref{fig:topology1}c):}}
\vskip0.2cm
(1) $h_i=h/H$, $F=\tilde{\chi}_\ell^0$, $F^\prime=\tilde{\chi}_n^0$, $V=Z$
\begin{align*}
\Lambda_{LLL} = \mathcal{G}_{\tilde{\chi}_1^0 \tilde{\chi}_n^0 Z}^L \mathcal{G}_{\tilde{\chi}_\ell^0 \tilde{\chi}_n^0 h_i}^L \mathcal{G}_{\tilde{\chi}_1^0 \tilde{\chi}_\ell^0 Z}^{L*}~,\qquad\qquad \Lambda_{LLR} = -\mathcal{G}_{\tilde{\chi}_1^0 \tilde{\chi}_n^0 Z}^L \mathcal{G}_{\tilde{\chi}_\ell^0 \tilde{\chi}_n^0 h_i}^L \mathcal{G}_{\tilde{\chi}_1^0 \tilde{\chi}_\ell^0 Z}^L~,\\
\Lambda_{LRL} = \mathcal{G}_{\tilde{\chi}_1^0 \tilde{\chi}_n^0 Z}^L \mathcal{G}_{\tilde{\chi}_\ell^0 \tilde{\chi}_n^0 h_i}^R \mathcal{G}_{\tilde{\chi}_1^0 \tilde{\chi}_\ell^0 Z}^{L*}~,\qquad\qquad \Lambda_{LRR} = -\mathcal{G}_{\tilde{\chi}_1^0 \tilde{\chi}_n^0 Z}^L \mathcal{G}_{\tilde{\chi}_\ell^0 \tilde{\chi}_n^0 h_i}^R \mathcal{G}_{\tilde{\chi}_1^0 \tilde{\chi}_\ell^0 Z}^L~,\\
\Lambda_{RLL} = \mathcal{G}_{\tilde{\chi}_1^0 \tilde{\chi}_n^0 Z}^R \mathcal{G}_{\tilde{\chi}_\ell^0 \tilde{\chi}_n^0 h_i}^L \mathcal{G}_{\tilde{\chi}_1^0 \tilde{\chi}_\ell^0 Z}^{L*}~,\qquad\qquad \Lambda_{RLR} = -\mathcal{G}_{\tilde{\chi}_1^0 \tilde{\chi}_n^0 Z}^R \mathcal{G}_{\tilde{\chi}_\ell^0 \tilde{\chi}_n^0 h_i}^L \mathcal{G}_{\tilde{\chi}_1^0 \tilde{\chi}_\ell^0 Z}^L~,\\
\Lambda_{RRL} = \mathcal{G}_{\tilde{\chi}_1^0 \tilde{\chi}_n^0 Z}^R \mathcal{G}_{\tilde{\chi}_\ell^0 \tilde{\chi}_n^0 h_i}^R \mathcal{G}_{\tilde{\chi}_1^0 \tilde{\chi}_\ell^0 Z}^{L*}~,\qquad\qquad \Lambda_{RRR} = -\mathcal{G}_{\tilde{\chi}_1^0 \tilde{\chi}_n^0 Z}^R \mathcal{G}_{\tilde{\chi}_\ell^0 \tilde{\chi}_n^0 h_i}^R \mathcal{G}_{\tilde{\chi}_1^0 \tilde{\chi}_\ell^0 Z}^L~,
\end{align*}

where $\mathcal{G}_{\tilde{\chi}_\ell^0 \tilde{\chi}_n^0 Z}^L = \dfrac{g_2}{c_W}N_{\ell n}^L$\, and \, $\mathcal{G}_{\tilde{\chi}_\ell^0 \tilde{\chi}_n^0 Z}^R = \dfrac{g_2}{c_W}N_{\ell n}^R$.
\vskip0.8cm

(2) $h_i=h/H$, $F=\tilde{\chi}_\ell^\pm$, $F^\prime=\tilde{\chi}_n^\pm$, $V=W^\pm$
 \begin{align*}
\Lambda_{LLL} = \mathcal{G}_{\tilde{\chi}_1^0 \tilde{\chi}_n^\pm W^\pm}^L \mathcal{G}_{\tilde{\chi}_\ell^\pm \tilde{\chi}_n^\pm h_i}^L \mathcal{G}_{\tilde{\chi}_1^0 \tilde{\chi}_\ell^\pm W^\pm}^{R*}~,\qquad\qquad \Lambda_{LLR} = \mathcal{G}_{\tilde{\chi}_1^0 \tilde{\chi}_n^\pm W^\pm}^L \mathcal{G}_{\tilde{\chi}_\ell^\pm \tilde{\chi}_n^\pm h_i}^L \mathcal{G}_{\tilde{\chi}_1^0 \tilde{\chi}_\ell^\pm W^\pm}^{L*}~,\\
\Lambda_{LRL} = \mathcal{G}_{\tilde{\chi}_1^0 \tilde{\chi}_n^\pm W^\pm}^L \mathcal{G}_{\tilde{\chi}_\ell^\pm \tilde{\chi}_n^\pm h_i}^R \mathcal{G}_{\tilde{\chi}_1^0 \tilde{\chi}_\ell^\pm W^\pm}^{R*}~,\qquad\qquad \Lambda_{LRR} = \mathcal{G}_{\tilde{\chi}_1^0 \tilde{\chi}_n^\pm W^\pm}^L \mathcal{G}_{\tilde{\chi}_\ell^\pm \tilde{\chi}_n^\pm h_i}^R \mathcal{G}_{\tilde{\chi}_1^0 \tilde{\chi}_\ell^\pm W^\pm}^{L*}~,\\
\Lambda_{RLL} = \mathcal{G}_{\tilde{\chi}_1^0 \tilde{\chi}_n^\pm W^\pm}^R \mathcal{G}_{\tilde{\chi}_\ell^\pm \tilde{\chi}_n^\pm h_i}^L \mathcal{G}_{\tilde{\chi}_1^0 \tilde{\chi}_\ell^\pm W^\pm}^{R*}~,\qquad\qquad \Lambda_{RLR} = \mathcal{G}_{\tilde{\chi}_1^0 \tilde{\chi}_n^\pm W^\pm}^R \mathcal{G}_{\tilde{\chi}_\ell^\pm \tilde{\chi}_n^\pm h_i}^L \mathcal{G}_{\tilde{\chi}_1^0 \tilde{\chi}_\ell^\pm W^\pm}^{L*}~,\\
\Lambda_{RRL} = \mathcal{G}_{\tilde{\chi}_1^0 \tilde{\chi}_n^\pm W^\pm}^R \mathcal{G}_{\tilde{\chi}_\ell^\pm \tilde{\chi}_n^\pm h_i}^R \mathcal{G}_{\tilde{\chi}_1^0 \tilde{\chi}_\ell^\pm W^\pm}^{R*}~,\qquad\qquad \Lambda_{RRR} = \mathcal{G}_{\tilde{\chi}_1^0 \tilde{\chi}_n^\pm W^\pm}^R \mathcal{G}_{\tilde{\chi}_\ell^\pm \tilde{\chi}_n^\pm h_i}^R \mathcal{G}_{\tilde{\chi}_1^0 \tilde{\chi}_\ell^\pm W^\pm}^{L*}~,
\end{align*}

where $\mathcal{G}_{\tilde{\chi}_\ell^0 \tilde{\chi}_n^\pm W^\pm}^L = g_2 C_{\ell n}^L$\, and \, $\mathcal{G}_{\tilde{\chi}_\ell^0 \tilde{\chi}_n^\pm W^\pm}^R = g_2 C_{\ell n}^R$.
\vskip0.8cm

\underline{{\bf Topology-(\ref{fig:topology1}d):}}
\vskip0.2cm
(1) $h_i=h/H$, $F=\tilde{\chi}_\ell^0$, $V=Z$.
\begin{align*}
\eta_{LL} = \mathcal{G}_{ZZh_i}\mathcal{G}_{\tilde{\chi}_1^0 \tilde{\chi}_\ell^0 Z}^L \mathcal{G}_{\tilde{\chi}_1^0 \tilde{\chi}_\ell^0 Z}^{L*}~, \qquad\qquad \eta_{LR} = -\mathcal{G}_{ZZh_i}\mathcal{G}_{\tilde{\chi}_1^0 \tilde{\chi}_\ell^0 Z}^L \mathcal{G}_{\tilde{\chi}_1^0 \tilde{\chi}_\ell^0 Z}^L~,\\
\eta_{RL} = \mathcal{G}_{ZZh_i}\mathcal{G}_{\tilde{\chi}_1^0 \tilde{\chi}_\ell^0 Z}^R \mathcal{G}_{\tilde{\chi}_1^0 \tilde{\chi}_\ell^0 Z}^{L*}~, \qquad\qquad \eta_{RR} = -\mathcal{G}_{ZZh_i}\mathcal{G}_{\tilde{\chi}_1^0 \tilde{\chi}_\ell^0 Z}^R \mathcal{G}_{\tilde{\chi}_1^0 \tilde{\chi}_\ell^0 Z}^L~,
\end{align*}

where $\mathcal{G}_{ZZh_i} = g_2M_Z g^{\mu\nu}Y_{h_i}$,\, with\, $Y_{h_i} = \left\{
\begin{array}{ll}
\dfrac{s_{\beta-\alpha}}{c_W} ; & h_i=h \\
\dfrac{c_{\beta-\alpha}}{c_W}; & h_i=H \\
\end{array} 
\right..$ 
\vskip0.8cm

(2) $h_i=h/H$, $F=\tilde{\chi}_\ell^\pm$, $V=W^\pm$.
\begin{align*}
\eta_{LL} = \mathcal{G}_{W^\pm W^\pm h_i}\mathcal{G}_{\tilde{\chi}_1^0 \tilde{\chi}_\ell^\pm W^\pm}^L \mathcal{G}_{\tilde{\chi}_1^0 \tilde{\chi}_\ell^\pm W^\pm}^{R*}~, \qquad\qquad \eta_{LR} = \mathcal{G}_{W^\pm W^\pm h_i}\mathcal{G}_{\tilde{\chi}_1^0 \tilde{\chi}_\ell^\pm W^\pm}^L \mathcal{G}_{\tilde{\chi}_1^0 \tilde{\chi}_\ell^\pm W^\pm}^{L*}~,\\
\eta_{RL} = \mathcal{G}_{W^\pm W^\pm h_i}\mathcal{G}_{\tilde{\chi}_1^0 \tilde{\chi}_\ell^\pm W^\pm}^R \mathcal{G}_{\tilde{\chi}_1^0 \tilde{\chi}_\ell^\pm W^\pm}^{R*}~, \qquad\qquad \eta_{RR} = \mathcal{G}_{W^\pm W^\pm h_i}\mathcal{G}_{\tilde{\chi}_1^0 \tilde{\chi}_\ell^\pm W^\pm}^R \mathcal{G}_{\tilde{\chi}_1^0 \tilde{\chi}_\ell^\pm W^\pm}^{L*}~,
\end{align*}

where $\mathcal{G}_{W^\pm W^\pm h_i} = g_2M_W g^{\mu\nu}Y^\prime_{h_i}$,\, with\, $Y^\prime_{h_i} = \left\{
\begin{array}{ll}
s_{\beta-\alpha} ; & h_i=h \\
c_{\beta-\alpha}; & h_i=H \\
\end{array} 
\right..$ 
\vskip0.8cm

\underline{{\bf Topology-(\ref{fig:topology1}e):}}
\vskip0.2cm
(1) $h_i=h/H$, $F=\tilde{\chi}_\ell^0$, $S=A$, $V=Z$.
\begin{align*}
\psi_{LL} = \mathcal{G}_{h_i AZ}\mathcal{G}_{\tilde{\chi}_1^0\tilde{\chi}_\ell^0 Z}^L \mathcal{G}_{\tilde{\chi}_1^0\tilde{\chi}_\ell^0 A}^{R*}~,\qquad\qquad \psi_{LR} = \mathcal{G}_{h_i AZ}\mathcal{G}_{\tilde{\chi}_1^0\tilde{\chi}_\ell^0 Z}^L \mathcal{G}_{\tilde{\chi}_1^0\tilde{\chi}_\ell^0 A}^{L*}~, \\
\psi_{RL} = \mathcal{G}_{h_i AZ}\mathcal{G}_{\tilde{\chi}_1^0\tilde{\chi}_\ell^0 Z}^R \mathcal{G}_{\tilde{\chi}_1^0\tilde{\chi}_\ell^0 A}^{R*}~,\qquad\qquad \psi_{RR} = \mathcal{G}_{h_i AZ}\mathcal{G}_{\tilde{\chi}_1^0\tilde{\chi}_\ell^0 Z}^R \mathcal{G}_{\tilde{\chi}_1^0\tilde{\chi}_\ell^0 A}^{L*}~,
\end{align*}

where $\mathcal{G}_{h_i AZ} = \dfrac{g_2}{2c_W}Y^{\prime\prime}_{h_i}$,\, with\, $Y^{\prime\prime}_{h_i} = \left\{
\begin{array}{ll}
\,\,\,\,c_{\beta-\alpha} ; & h_i=h \\
-s_{\beta-\alpha}; & h_i=H \\
\end{array} 
\right..$
\vskip0.8cm

(2) $h_i=h/H$, $F=\tilde{\chi}_\ell^0$, $S=G$, $V=Z$.
\begin{align*}
\psi_{LL} = \mathcal{G}_{h_i GZ}\mathcal{G}_{\tilde{\chi}_1^0\tilde{\chi}_\ell^0 Z}^L \mathcal{G}_{\tilde{\chi}_1^0\tilde{\chi}_\ell^0 G}^{R*}~,\qquad\qquad \psi_{LR} = \mathcal{G}_{h_i GZ}\mathcal{G}_{\tilde{\chi}_1^0\tilde{\chi}_\ell^0 Z}^L \mathcal{G}_{\tilde{\chi}_1^0\tilde{\chi}_\ell^0 G}^{L*}~, \\
\psi_{RL} = \mathcal{G}_{h_i GZ}\mathcal{G}_{\tilde{\chi}_1^0\tilde{\chi}_\ell^0 Z}^R \mathcal{G}_{\tilde{\chi}_1^0\tilde{\chi}_\ell^0 G}^{R*}~,\qquad\qquad \psi_{RR} = \mathcal{G}_{h_i GZ}\mathcal{G}_{\tilde{\chi}_1^0\tilde{\chi}_\ell^0 Z}^R \mathcal{G}_{\tilde{\chi}_1^0\tilde{\chi}_\ell^0 G}^{L*}~,
\end{align*}

where $\mathcal{G}_{h_i GZ} = \left\{
\begin{array}{ll}
\dfrac{g_2}{2c_W}s_{\beta-\alpha} ; & h_i=h \\
\dfrac{g_2}{2c_W}c_{\beta-\alpha}; & h_i=H \\
\end{array} 
\right..$
\vskip0.8cm

(3) $h_i=h/H$, $F=\tilde{\chi}_\ell^\pm$, $S=H^\pm$, $V=W^\pm$.
\begin{align*}
\psi_{LL} = \mathcal{G}_{h_i H^\pm W^\pm}\mathcal{G}_{\tilde{\chi}_1^0\tilde{\chi}_\ell^\pm W^\pm}^L \mathcal{G}_{\tilde{\chi}_1^0\tilde{\chi}_\ell^\pm H^\pm
}^{R*}~,\qquad\qquad \psi_{LR} = \mathcal{G}_{h_i H^\pm W^\pm}\mathcal{G}_{\tilde{\chi}_1^0\tilde{\chi}_\ell^\pm W^\pm}^L \mathcal{G}_{\tilde{\chi}_1^0\tilde{\chi}_\ell^\pm H^\pm}^{L*}~, \\
\psi_{RL} = \mathcal{G}_{h_i H^\pm W^\pm}\mathcal{G}_{\tilde{\chi}_1^0\tilde{\chi}_\ell^\pm W^\pm}^R \mathcal{G}_{\tilde{\chi}_1^0\tilde{\chi}_\ell^\pm H^\pm}^{R*}~,\qquad\qquad \psi_{RR} = \mathcal{G}_{h_i H^\pm W^\pm}\mathcal{G}_{\tilde{\chi}_1^0\tilde{\chi}_\ell^\pm W^\pm}^R \mathcal{G}_{\tilde{\chi}_1^0\tilde{\chi}_\ell^\pm H^\pm}^{L*}~,
\end{align*}

where $\mathcal{G}_{h_i H^\pm W^\pm} = \dfrac{g_2}{2}Y^{\prime\prime}_{h_i}$.
\vskip0.8cm

(4) $h_i=h/H$, $F=\tilde{\chi}_\ell^\pm$, $S=G^\pm$, $V=W^\pm$.
\begin{align*}
\psi_{LL} = \mathcal{G}_{h_i G^\pm W^\pm}\mathcal{G}_{\tilde{\chi}_1^0\tilde{\chi}_\ell^\pm W^\pm}^L \mathcal{G}_{\tilde{\chi}_1^0\tilde{\chi}_\ell^\pm G^\pm
}^{R*}~,\qquad\qquad \psi_{LR} = \mathcal{G}_{h_i G^\pm W^\pm}\mathcal{G}_{\tilde{\chi}_1^0\tilde{\chi}_\ell^\pm W^\pm}^L \mathcal{G}_{\tilde{\chi}_1^0\tilde{\chi}_\ell^\pm G^\pm}^{L*}~, \\
\psi_{RL} = \mathcal{G}_{h_i G^\pm W^\pm}\mathcal{G}_{\tilde{\chi}_1^0\tilde{\chi}_\ell^\pm W^\pm}^R \mathcal{G}_{\tilde{\chi}_1^0\tilde{\chi}_\ell^\pm G^\pm}^{R*}~,\qquad\qquad \psi_{RR} = \mathcal{G}_{h_i G^\pm W^\pm}\mathcal{G}_{\tilde{\chi}_1^0\tilde{\chi}_\ell^\pm W^\pm}^R \mathcal{G}_{\tilde{\chi}_1^0\tilde{\chi}_\ell^\pm G^\pm}^{L*}~,
\end{align*}

where $\mathcal{G}_{h_i G^\pm W^\pm} =  \left\{
\begin{array}{ll}
-\dfrac{g_2}{2}s_{\beta-\alpha} ; & h_i=h \\
-\dfrac{g_2}{2}c_{\beta-\alpha}; & h_i=H \\
\end{array} 
\right..$
\vskip0.8cm

\underline{{\bf Topology-(\ref{fig:topology1}f):}}
\vskip0.2cm
(1) $h_i=h/H$, $F=\tilde{\chi}_\ell^0$, $S=A$, $V=Z$.
\begin{align*}
\Xi_{LL} = \mathcal{G}_{h_i AZ}\mathcal{G}_{\tilde{\chi}_1^0 \tilde{\chi}_\ell^0 A}^L \mathcal{G}_{\tilde{\chi}_1^0 \tilde{\chi}_\ell^0 Z}^{L*}~, \qquad\qquad \Xi_{LR} = -\mathcal{G}_{h_i AZ}\mathcal{G}_{\tilde{\chi}_1^0 \tilde{\chi}_\ell^0 A}^L \mathcal{G}_{\tilde{\chi}_1^0 \tilde{\chi}_\ell^0 Z}^{L}~,\\
\Xi_{RL} = \mathcal{G}_{h_i AZ}\mathcal{G}_{\tilde{\chi}_1^0 \tilde{\chi}_\ell^0 A}^R \mathcal{G}_{\tilde{\chi}_1^0 \tilde{\chi}_\ell^0 Z}^{L*}~, \qquad\qquad \Xi_{RR} = -\mathcal{G}_{h_i AZ}\mathcal{G}_{\tilde{\chi}_1^0 \tilde{\chi}_\ell^0 A}^R \mathcal{G}_{\tilde{\chi}_1^0 \tilde{\chi}_\ell^0 Z}^{L}~.
\end{align*}
\vskip0.8cm
(2) $h_i=h/H$, $F=\tilde{\chi}_\ell^0$, $S=G$, $V=Z$.
\begin{align*}
\Xi_{LL} = \mathcal{G}_{h_i GZ}\mathcal{G}_{\tilde{\chi}_1^0 \tilde{\chi}_\ell^0 G}^L \mathcal{G}_{\tilde{\chi}_1^0 \tilde{\chi}_\ell^0 Z}^{L*}~, \qquad\qquad \Xi_{LR} = -\mathcal{G}_{h_i GZ}\mathcal{G}_{\tilde{\chi}_1^0 \tilde{\chi}_\ell^0 G}^L \mathcal{G}_{\tilde{\chi}_1^0 \tilde{\chi}_\ell^0 Z}^{L}~,\\
\Xi_{RL} = \mathcal{G}_{h_i GZ}\mathcal{G}_{\tilde{\chi}_1^0 \tilde{\chi}_\ell^0 G}^R \mathcal{G}_{\tilde{\chi}_1^0 \tilde{\chi}_\ell^0 Z}^{L*}~, \qquad\qquad \Xi_{RR} = -\mathcal{G}_{h_i GZ}\mathcal{G}_{\tilde{\chi}_1^0 \tilde{\chi}_\ell^0 G}^R \mathcal{G}_{\tilde{\chi}_1^0 \tilde{\chi}_\ell^0 Z}^{L}~.
\end{align*}
\vskip0.8cm
(3) $h_i=h/H$, $F=\tilde{\chi}_\ell^\pm$, $S=H^\pm$, $V=W^\pm$.
\begin{align*}
\Xi_{LL} = \mathcal{G}_{h_i H^\pm W^\pm}\mathcal{G}_{\tilde{\chi}_1^0\tilde{\chi}_\ell^\pm H^\pm}^L \mathcal{G}_{\tilde{\chi}_1^0\tilde{\chi}_\ell^\pm W^\pm}^{R*}~, \qquad\qquad \Xi_{LR} = \mathcal{G}_{h_i H^\pm W^\pm}\mathcal{G}_{\tilde{\chi}_1^0\tilde{\chi}_\ell^\pm H^\pm}^L \mathcal{G}_{\tilde{\chi}_1^0\tilde{\chi}_\ell^\pm W^\pm}^{L*}~,\\
\Xi_{RL} = \mathcal{G}_{h_i H^\pm W^\pm}\mathcal{G}_{\tilde{\chi}_1^0\tilde{\chi}_\ell^\pm H^\pm}^R \mathcal{G}_{\tilde{\chi}_1^0\tilde{\chi}_\ell^\pm W^\pm}^{R*}~, \qquad\qquad \Xi_{RR} = \mathcal{G}_{h_i H^\pm W^\pm}\mathcal{G}_{\tilde{\chi}_1^0\tilde{\chi}_\ell^\pm H^\pm}^R \mathcal{G}_{\tilde{\chi}_1^0\tilde{\chi}_\ell^\pm W^\pm}^{L*}~.
\end{align*}
\vskip0.8cm
(3) $h_i=h/H$, $F=\tilde{\chi}_\ell^\pm$, $S=G^\pm$, $V=W^\pm$.
\begin{align*}
\Xi_{LL} = \mathcal{G}_{h_i G^\pm W^\pm}\mathcal{G}_{\tilde{\chi}_1^0\tilde{\chi}_\ell^\pm G^\pm}^L \mathcal{G}_{\tilde{\chi}_1^0\tilde{\chi}_\ell^\pm W^\pm}^{R*}~, \qquad\qquad \Xi_{LR} = \mathcal{G}_{h_i G^\pm W^\pm}\mathcal{G}_{\tilde{\chi}_1^0\tilde{\chi}_\ell^\pm G^\pm}^L \mathcal{G}_{\tilde{\chi}_1^0\tilde{\chi}_\ell^\pm W^\pm}^{L*}~,\\
\Xi_{RL} = \mathcal{G}_{h_i G^\pm W^\pm}\mathcal{G}_{\tilde{\chi}_1^0\tilde{\chi}_\ell^\pm G^\pm}^R \mathcal{G}_{\tilde{\chi}_1^0\tilde{\chi}_\ell^\pm W^\pm}^{R*}~, \qquad\qquad \Xi_{RR} = \mathcal{G}_{h_i G^\pm W^\pm}\mathcal{G}_{\tilde{\chi}_1^0\tilde{\chi}_\ell^\pm G^\pm}^R \mathcal{G}_{\tilde{\chi}_1^0\tilde{\chi}_\ell^\pm W^\pm}^{L*}~.
\end{align*}
\vskip0.5cm
In the above, we have used the following.

\begin{align*}
    C_{\ell k}^{L} &= \mathcal{N}_{\ell 2} \mathcal{V}_{k1}^{*} -\frac{1}{\sqrt{2}} \mathcal{N}_{\ell 4} \mathcal{V}_{k2}^{*}~,\\
    C_{\ell k}^{R} &= \mathcal{N}_{\ell 2}^{*} \mathcal{U}_{k1} +\frac{1}{\sqrt{2}} \mathcal{N}_{\ell 3}^{*} \mathcal{U}_{k2}~,\\
    N_{\ell n}^{L} &= \frac{1}{2}\big(-\mathcal{N}_{\ell 3} \mathcal{N}_{n 3}^{*} + \mathcal{N}_{\ell 4} \mathcal{N}_{n 4}^{*} \big)~,\\
    N_{\ell n}^{R} &= - \big(N_{\ell n}^{L}\big)^{*}~,\\
    Q_{k\ell} &= \frac{1}{2} \mathcal{V}_{k1} \mathcal{U}_{\ell 2}~,\\
    S_{k\ell} &= \frac{1}{2} \mathcal{V}_{k2} \mathcal{U}_{\ell 1}~,\\
    Q_{\ell k}^{\prime L} &= c_\beta\Big[\mathcal{N}_{\ell 4}^{*}\mathcal{V}^{*}_{k1}+\frac{1}{\sqrt{2}} \mathcal{V}^{*}_{k2}\big(\mathcal{N}^{*}_{\ell 2} + t_{W} \mathcal{N}^{*}_{\ell 1}\big)\Big]~,
        \end{align*} 
    \begin{align*}
    Q_{\ell k}^{\prime R} &= s_\beta\Big[\mathcal{N}_{\ell 3}\mathcal{U}_{k1}-\frac{1}{\sqrt{2}} \mathcal{U}_{k2}\big(\mathcal{N}_{\ell 2} + t_{W} \mathcal{N}_{\ell 1}\big)\Big]~,\\
    Q_{n\ell}^{\prime\prime} &= \frac{1}{2}\big[\mathcal{N}_{n3}\big(\mathcal{N}_{\ell 2}-t_{W}\mathcal{N}_{\ell 1}\big) + \mathcal{N}_{\ell 3}\big(\mathcal{N}_{n 2}-t_{W}\mathcal{N}_{n 1}\big)\big]~,\\ 
    S_{n\ell}^{\prime\prime} &= \frac{1}{2}\big[\mathcal{N}_{n4}\big(\mathcal{N}_{\ell 2}-t_{W}\mathcal{N}_{\ell 1}\big) + \mathcal{N}_{\ell 4}\big(\mathcal{N}_{n 2}-t_{W}\mathcal{N}_{n 1}\big)\big]~,\\
    G_{nm1}^{\nu}&= -\frac{1}{\sqrt{2}}g_2\big(\mathcal{N}_{12}^{*}-t_W\mathcal{N}_{11}^{*}\big)U_{nm}^{\tilde{\nu}*}~,\\ 
    G_{nm1}^{e_L} &= \frac{1}{\sqrt{2}}g_2 \big(\mathcal{N}_{12}^{*}+ t_W \mathcal{N}_{11}^{*}\big)W_{nm}^{\tilde{e}*} - \frac{g_2}{\sqrt{2}M_W c_\beta}m_{e_n}\mathcal{N}_{13}^{*}W_{n+3\,\, m}^{\tilde{e}*}~,\\
    G_{nm1}^{e_R} &= -\sqrt{2}g_2 t_W \mathcal{N}_{11} W_{n+3\,\, m}^{\tilde{e}*} - \frac{g_2}{\sqrt{2}M_W c_\beta} m_{e_i}\mathcal{N}_{13}W_{nm}^{\tilde{e}*}~.
\end{align*}

\section{Diagonalization of the Neutralino Mass Matrix}
\label{eq:appB}
We present the approximate analytical solutions for the eigenvalues of $\overline{\mathbb{M}}_{\tilde{\chi}^0}$ defined in Eq.~\eqref{eq:nu_mass} (see, e.g.,~\cite{Gunion:1987yh}) and the composition of the lightest neutralino, which will be relevant for the discussion. The $4\times 4$ neutralino mass matrix in the basis $(\tilde{B}, \tilde{W}^0, \tilde{H}^0_d, \tilde{H}^0_u)$ can be read as
\begin{align}
\overline{\mathbb{M}}_{\tilde{\chi}^0}=\left(\begin{array}{c c c c}
M_1 & 0 & -M_Z s_W c_\beta & M_Z s_W s_\beta \\
0 & M_2 & M_Z c_W c_\beta & -M_Z c_W s_\beta\\
-M_Z s_W c_\beta & M_Z c_W c_\beta & 0 & -\mu\\
M_Z s_W s_\beta & -M_Z c_W s_\beta & -\mu & 0\\
\end{array}\right).
\label{eq:nutralino_mass_matrix:1}
\end{align}
The lightest neutralino $\tilde{\chi}_1^0$, in the above basis can be written as 
\begin{align}
    \tilde{\chi}_1^0 = \mathcal{N}_{11} \tilde{B} + \mathcal{N}_{12}\tilde{W}^0 + \mathcal{N}_{13}\tilde{H}_d^0 + \mathcal{N}_{14} \tilde{H}_u^0 .
    \label{neutralino:compo}
\end{align}
In order to calculate the mass eigenvalues of Eq.~\eqref{eq:nutralino_mass_matrix:1} and the compositions of $\tilde{\chi}_1^0$, $N_{1j}$ (see Eq.~\eqref{neutralino:compo}), we rotate the neutralino mass matrix to a basis ($\tilde{B}, \tilde{W}^0, \tilde{H}_1^0, \tilde{H}_2^0$), where $\tilde{H}^0_1=\frac{\tilde{H}_u^0 - \tilde{H}_d^0}{\sqrt{2}}$ and $\tilde{H}^0_2=\frac{\tilde{H}_u^0 + \tilde{H}_d^0}{\sqrt{2}}$. Then, we can write by orthogonal transformation,
\begin{align}
    \mathbf{M} = &\mathbf{U}\,\overline{\mathbb{M}}_{\tilde{\chi}^0}\,\mathbf{U}^T\\
    =&\left(\begin{array}{c c c c}
M_1 & 0 & -\frac{1}{\sqrt{2}}M_Zs_W (s_\beta + c_\beta) & \frac{1}{\sqrt{2}}M_Zs_W (s_\beta - c_\beta) \\
0 & M_2 & \frac{1}{\sqrt{2}}M_Zc_W (s_\beta + c_\beta) & -\frac{1}{\sqrt{2}}M_Zc_W (s_\beta - c_\beta)\\
-\frac{1}{\sqrt{2}}M_Zs_W (s_\beta + c_\beta) & \frac{1}{\sqrt{2}}M_Zc_W (s_\beta + c_\beta) & \mu & 0\\
\frac{1}{\sqrt{2}}M_Zs_W (s_\beta - c_\beta) & -\frac{1}{\sqrt{2}}M_Zc_W (s_\beta - c_\beta) & 0 & -\mu\\
\end{array}\right),
\label{eq:nutralino_mass_matrix:rotated}
\end{align}
where the matrix $\mathbf{U}$ is given by, 
\begin{align}
    \mathbf{U} = \left(\begin{array}{c c c c}
1 & 0 & 0 & 0 \\
0 & 1 & 0 & 0\\
0 & 0 & \frac{1}{\sqrt{2}} & -\frac{1}{\sqrt{2}}\\
0 & 0 & \frac{1}{\sqrt{2}} & \frac{1}{\sqrt{2}}\\
\end{array}\right).
\label{rotating:u}
\end{align}

We can write the matrix in Eq.~\eqref{eq:nutralino_mass_matrix:rotated} in the following way
\begin{align}
    \mathbf{M} = \left(\begin{array}{c c c c}
M_1 & 0 & 0 & 0 \\
0 & M_2 & 0 & 0\\
0 & 0 & \mu & 0\\
0 & 0 & 0 & -\mu\\
\end{array}\right) + \left(\begin{array}{c c c c}
0 & 0 & a_1 & a_2 \\
0 & 0 & a_3 & a_4\\
a_1 & a_3 & 0 & 0\\
a_2 & a_4 & 0 & 0\\
\end{array}\right) = \mathbf{M}_D + \mathbf{M}_P~.
\label{eq:perturbation}
\end{align}
The off-diagonal matrix $\mathbf{M}_P$ will be treated as perturbations as the diagonal eigenvalues in $\mathbf{M}_D$ are typically larger than $M_Z$. Now, we use time-independent perturbation theory to calculate the eigenvalues of $\mathbf{M}$. The eigenvectors of the unperturbed matrix $\mathbf{M}_D$ can be written as, 

\begin{align}
    \lvert \phi_1\rangle = \left(\begin{array}{c}
1 \\
0 \\
0 \\
0 \\
\end{array}\right),\,\, \lvert \phi_2\rangle = \left(\begin{array}{c}
0 \\
1 \\
0 \\
0 \\
\end{array}\right),\,\, \lvert \phi_3\rangle = \left(\begin{array}{c}
0 \\
0 \\
1 \\
0 \\
\end{array}\right),\,\, {\rm and}\,\, \lvert \phi_4\rangle = \left(\begin{array}{c}
0 \\
0 \\
0 \\
1 \\
\end{array}\right), 
\end{align}
with the mass eigenvalues at the $0^{\rm th}$ order (unperturbed eigenvalues) are  given by,  
\begin{align}
    m_{\tilde{\chi}_1^0}^{(0)} = M_1,\,\,\,\, m_{\tilde{\chi}_2^0}^{(0)} = M_2,\,\,\,\, m_{\tilde{\chi}_3^0}^{(0)} = \mu,\,\,\,\, m_{\tilde{\chi}_4^0}^{(0)} = -\mu.
\end{align}

Now, the first-order corrections to the mass eigenvalues in the non-degenerate perturbation theory can be written as,
\begin{align}
     m_{\tilde{\chi}_1^0}^{(1)} &= \langle\phi_1 \lvert \mathbf{M}_P\rvert \phi_1\rangle = 0~,\nonumber\\
     m_{\tilde{\chi}_2^0}^{(1)} &= \langle\phi_2 \lvert \mathbf{M}_P\rvert \phi_2\rangle = 0~,\nonumber\\
     m_{\tilde{\chi}_3^0}^{(1)} &= \langle\phi_3 \lvert \mathbf{M}_P\rvert \phi_3\rangle = 0~,\nonumber\\
     m_{\tilde{\chi}_4^0}^{(1)} &= \langle\phi_4 \lvert \mathbf{M}_P\rvert \phi_4\rangle = 0~.
\end{align}
Therefore, we see that the first-order corrections to the mass eigenvalues vanish. Let us now consider the second-order corrections, which for the non-degenerate perturbation theory read as,
\begin{align}
     m_{\tilde{\chi}_n^0}^{(2)} = \sum_{\ell\ne n}^{} \frac{\Big\lvert\langle \phi_\ell \lvert \mathbf{M}_P\rvert\phi_n\rangle\Big\rvert^2}{m_{\tilde{\chi}_n^0}^{(0)} - m_{\tilde{\chi}_\ell^0}^{(0)}}.
\end{align}
Consequently, the second-order corrections are computed as follows. 
\begin{align}
     m_{\tilde{\chi}_1^0}^{(2)} =& \sum_{\ell = 2,3,4}^{} \frac{\Big\lvert\langle \phi_\ell \lvert \mathbf{M}_P\rvert\phi_1\rangle\Big\rvert^2}{m_{\tilde{\chi}_1^0}^{(0)} - m_{\tilde{\chi}_\ell^0}^{(0)}}= \frac{\lvert a_1\rvert^2}{m_{\tilde{\chi}_1^0}^{(0)} - m_{\tilde{\chi}_3^0}^{(0)}} + \frac{\lvert a_2\rvert^2}{m_{\tilde{\chi}_1^0}^{(0)} - m_{\tilde{\chi}_4^0}^{(0)}}
     = \frac{M_Z^2 s_W^2 (M_1+\mu s_{2\beta})}{M_1^2-\mu^2}.\\
     m_{\tilde{\chi}_2^0}^{(2)} =& \sum_{\ell = 1,3,4}^{} \frac{\Big\lvert\langle \phi_\ell \lvert \mathbf{M}_P\rvert\phi_2\rangle\Big\rvert^2}{m_{\tilde{\chi}_2^0}^{(0)} - m_{\tilde{\chi}_\ell^0}^{(0)}}= \frac{\lvert a_3\rvert^2}{m_{\tilde{\chi}_2^0}^{(0)} - m_{\tilde{\chi}_3^0}^{(0)}} + \frac{\lvert a_4\rvert^2}{m_{\tilde{\chi}_2^0}^{(0)} - m_{\tilde{\chi}_4^0}^{(0)}}
     = \frac{M_Z^2 c^2_W (M_2+\mu s_{2\beta})}{M_2^2-\mu^2}.\\
     m_{\tilde{\chi}_3^0}^{(2)} =& \sum_{\ell = 1,2,4}^{} \frac{\Big\lvert\langle \phi_\ell \lvert \mathbf{M}_P\rvert\phi_3\rangle\Big\rvert^2}{m_{\tilde{\chi}_3^0}^{(0)} - m_{\tilde{\chi}_\ell^0}^{(0)}}= \frac{\lvert a_1\rvert^2}{m_{\tilde{\chi}_3^0}^{(0)} - m_{\tilde{\chi}_1^0}^{(0)}} + \frac{\lvert a_3\rvert^2}{m_{\tilde{\chi}_3^0}^{(0)} - m_{\tilde{\chi}_2^0}^{(0)}}
     = \frac{M_Z^2 (1+s_{2\beta}) (\mu - M_1c^2_W - M_2s^2_W)}{2(\mu -M_1)(\mu-M_2)}.\\
     m_{\tilde{\chi}_4^0}^{(2)} =& \sum_{\ell = 1,2,3}^{} \frac{\Big\lvert\langle \phi_\ell \lvert \mathbf{M}_P\rvert\phi_4\rangle\Big\rvert^2}{m_{\tilde{\chi}_4^0}^{(0)} - m_{\tilde{\chi}_\ell^0}^{(0)}}= \frac{\lvert a_2\rvert^2}{m_{\tilde{\chi}_4^0}^{(0)} - m_{\tilde{\chi}_1^0}^{(0)}} + \frac{\lvert a_4\rvert^2}{m_{\tilde{\chi}_4^0}^{(0)} - m_{\tilde{\chi}_2^0}^{(0)}}
     = \frac{M_Z^2 (1-s_{2\beta}) (\mu + M_1c^2_W + M_2s^2_W)}{2(\mu +M_1)(\mu+M_2)}.
\end{align}

Therefore, the masses of the neutralinos in the order $m_{\tilde{\chi}_1^0}<m_{\tilde{\chi}_2^0}<m_{\tilde{\chi}_3^0}<m_{\tilde{\chi}_4^0}$, can be expressed as 
\begin{align}
    m_{\tilde{\chi}_1^0} =& M_1 + \frac{M_Z^2 s_W^2 (M_1+\mu s_{2\beta})}{M_1^2-\mu^2} + ...\\
    m_{\tilde{\chi}_2^0} =& M_2 + \frac{M_Z^2 c^2_W (M_2+\mu s_{2\beta})}{M_2^2-\mu^2} + ...\\
    m_{\tilde{\chi}_3^0} =& \lvert\mu\rvert + \frac{M_Z^2 (1-s_{2\beta}) (\mu + M_1c^2_W + M_2s^2_W){sgn}(\mu)}{2(\mu +M_1)(\mu+M_2)} + ...\\
    m_{\tilde{\chi}_4^0} =& \lvert\mu\rvert + \frac{M_Z^2 (1+s_{2\beta}) (\mu - M_1c^2_W - M_2s^2_W){ sgn}(\mu)}{2(\mu -M_1)(\mu-M_2)} + ...
\end{align}
The matrix $\mathbf{M}$ can be diagonalized by an orthogonal transformation 
$\mathbf{M_{diag}}=\mathbf{V} \mathbf{M} \mathbf{V}^T$. Thus 
 the elements $\mathcal{N}_{1\ell}$ (in the original basis) is expressed as,
\begin{align}
    \mathcal{N}_{k\ell} = \mathbf{V}_{kn}\mathbf{U}_{n\ell} .
\end{align}
At the first-order of non-degenerate perturbation theory, we can write
\begin{align}
    \mathbf{V}_{kn}^{(1)} = \sum_{k\ne n}^{} \frac{\langle \phi_n \lvert \mathbf{M}_P\rvert\phi_k\rangle}{m_{\tilde{\chi}_k^0}^{(0)} - m_{\tilde{\chi}_n^0}^{(0)}}.
\end{align}
Now we can calculate the first-order corrections as, 
\begin{align}
    \mathbf{V}_{12}^{(1)} = & \frac{\langle \phi_2 \lvert \mathbf{M}_P\rvert\phi_1\rangle}{m_{\tilde{\chi}_1^0}^{(0)} - m_{\tilde{\chi}_2^0}^{(0)}} = 0.\\
    \mathbf{V}_{13}^{(1)} = & \frac{\langle \phi_3 \lvert \mathbf{M}_P\rvert\phi_1\rangle}{m_{\tilde{\chi}_1^0}^{(0)} - m_{\tilde{\chi}_3^0}^{(0)}} = \frac{a_1}{m_{\tilde{\chi}_1^0}^{(0)} - m_{\tilde{\chi}_3^0}^{(0)}} = \frac{M_Z s_W (s_\beta + c_\beta)}{\sqrt{2}(\mu-M_1)}.\\
    \mathbf{V}_{14}^{(1)} = & \frac{\langle \phi_4 \lvert \mathbf{M}_P\rvert\phi_1\rangle}{m_{\tilde{\chi}_1^0}^{(0)} - m_{\tilde{\chi}_4^0}^{(0)}} = \frac{a_2}{m_{\tilde{\chi}_1^0}^{(0)} - m_{\tilde{\chi}_4^0}^{(0)}} = \frac{M_Z s_W (s_\beta - c_\beta)}{\sqrt{2}(\mu+M_1)}.
\end{align}
Since the first-order correction $\mathbf{V}_{12}^{(1)}=0$, we need to calculate the second-order correction to $\mathbf{V}_{12}$. The second-order corrections can be written as,
\begin{align}
    \mathbf{V}_{kn}^{(2)} = \sum_{m\ne k}^{} \Bigg[\sum_{n\ne k}^{} \frac{\langle \phi_n\lvert \mathbf{M}_P\rvert\phi_m\rangle \langle \phi_m\lvert\mathbf{M}_P\rvert\phi_k\rangle}{\big(m_{\tilde{\chi}_k^0}^{(0)}-m_{\tilde{\chi}_m^0}^{(0)}\big)\big(m_{\tilde{\chi}_k^0}^{(0)}-m_{\tilde{\chi}_n^0}^{(0)}\big)}-\frac{\langle \phi_n\lvert \mathbf{M}_P\rvert\phi_k\rangle \langle \phi_k\lvert\mathbf{M}_P\rvert\phi_k\rangle}{\big(m_{\tilde{\chi}_k^0}^{(0)}-m_{\tilde{\chi}_n^0}^{(0)}\big)^2}\Bigg].
\end{align}
Therefore, we have 
\begin{align}
    \mathbf{V}_{12}^{(2)} = &\sum_{m=2,3,4}^{} \frac{\langle \phi_2\lvert \mathbf{M}_P\rvert\phi_m\rangle \langle \phi_m\lvert\mathbf{M}_P\rvert\phi_1\rangle}{\big(m_{\tilde{\chi}_1^0}^{(0)}-m_{\tilde{\chi}_m^0}^{(0)}\big)\big(m_{\tilde{\chi}_1^0}^{(0)}-m_{\tilde{\chi}_2^0}^{(0)}\big)}-\frac{\langle \phi_2\lvert \mathbf{M}_P\rvert\phi_1\rangle \langle \phi_1\lvert\mathbf{M}_P\rvert\phi_1\rangle}{\big(m_{\tilde{\chi}_1^0}^{(0)}-m_{\tilde{\chi}_2^0}^{(0)}\big)^2},\\
    =&\frac{\langle \phi_2\lvert \mathbf{M}_P\rvert\phi_3\rangle \langle \phi_3\lvert\mathbf{M}_P\rvert\phi_1\rangle}{\big(m_{\tilde{\chi}_1^0}^{(0)}-m_{\tilde{\chi}_3^0}^{(0)}\big)\big(m_{\tilde{\chi}_1^0}^{(0)}-m_{\tilde{\chi}_2^0}^{(0)}\big)} + \frac{\langle \phi_2\lvert \mathbf{M}_P\rvert\phi_4\rangle \langle \phi_4\lvert\mathbf{M}_P\rvert\phi_1\rangle}{\big(m_{\tilde{\chi}_1^0}^{(0)}-m_{\tilde{\chi}_4^0}^{(0)}\big)\big(m_{\tilde{\chi}_1^0}^{(0)}-m_{\tilde{\chi}_2^0}^{(0)}\big)},\\
    =&\frac{a_1 a_3}{(M_1-\mu)(M_1-M_2)} + \frac{a_2 a_4}{(M_1+\mu)(M_1-M_2)},\\
    =&-\frac{M_Z^2 s_{2W}(s_\beta+c_\beta)^2}{4(\mu-M_1)(M_2-M_1)} + \frac{M_Z^2 s_{2W} (s_\beta-c_\beta)^2}{4(\mu+M_1)(M_2-M_1)}.
\end{align}
We derive the components of $\mathcal{N}_{1\ell}$.
\begin{align}
    \mathcal{N}_{11}\simeq& 1.\\
    \mathcal{N}_{12}\simeq& \mathbf{V}_{12}^{(2)} = -\frac{M_Z^2 s_{2W}(s_\beta+c_\beta)^2}{4(\mu-M_1)(M_2-M_1)} + \frac{M_Z^2 s_{2W} (s_\beta-c_\beta)^2}{4(\mu+M_1)(M_2-M_1)}.\\
    \mathcal{N}_{13}\simeq& \frac{1}{\sqrt{2}}\mathbf{V}_{13}+ \frac{1}{\sqrt{2}}\mathbf{V}_{14} = \frac{M_Z s_W (\mu s_\beta + M_1 c_\beta)}{\mu^2-M_1^2}.\\
    \mathcal{N}_{14}\simeq& -\frac{1}{\sqrt{2}}\mathbf{V}_{13}+ \frac{1}{\sqrt{2}}\mathbf{V}_{14} = -\frac{M_Z s_W (\mu c_\beta + M_1 s_\beta)}{\mu^2-M_1^2}.
\end{align}

Finally, the masses and mixings of
a Bino-like, Wino-like, and Higgsino-like neutralinos $\tilde \chi_i^0$ in the limit $\tan\beta \geq 10$ (in particular, $s_\beta\to 1$, $c_\beta\to 0$) take a simple form (assuming $sgn(\mu)=+1$) in terms of the fundamental model parameters. 
\begin{align}
m_{\tilde \chi_1^0} \simeq &M_1 + \frac{M_Z^2 s_W^2 M_1}{M_1^2-\mu^2}~,~~~~~~
m_{\tilde \chi_2^0}\simeq M_2 +\frac{M_Z^2 c^2_W M_2}{M_2^2-\mu^2}.\\
m_{\tilde \chi_3^0}\simeq &|\mu| +\frac{M_Z^2  (\mu - M_1c^2_W - M_2s^2_W)}{2(\mu -M_1)(\mu-M_2)}~,~
m_{\tilde \chi_4^0} \simeq |\mu| + \frac{M_Z^2  (\mu + M_1c^2_W + M_2s^2_W)}{2(\mu +M_1)(\mu+M_2)}.\\
\mathcal{N}_{11}\simeq& 1.\\
    \mathcal{N}_{12} \simeq & -\frac{M_Z^2 s_{2W}}{4(\mu-M_1)(M_2-M_1)} + \frac{M_Z^2 s_{2W} }{4(\mu+M_1)(M_2-M_1)}.\\
    \mathcal{N}_{13} \simeq &\frac{M_Z s_W \mu }{\mu^2-M_1^2}.\\
    \mathcal{N}_{14} \simeq &-\frac{M_Z s_W M_1}{\mu^2-M_1^2}.
    \end{align}
 $\mathbf{{\tilde B}_{\tilde H}}$ LSP~: The physical state $\tilde\chi_1^0$ becomes Bino-Higgsino-like when $M_1 <\lvert\mu\rvert \ll M_2$ approximately holds. Since $\tilde\chi_2^0$ decoupled, masses for the Higgsino-like states can be approximated to 
\begin{align}
m_{\tilde \chi_{3,4}^0}\simeq &|\mu| +\frac{M_Z^2 s_W^2}{2(\mu \mp M_1)}.
\label{eq:higgapp}
\end{align}
Following SI direct detection limits, $\tilde \chi_1^0$ can only have moderate or minimal Higgsino components; thus, $m_{\tilde \chi_{1,3,4}^0}$ can 
be further simplified neglecting $M_1$ which results to
\begin{equation}
\Delta m (\tilde\chi_{3,4}^0, \tilde{\chi}_{1}^0) = m_{\tilde\chi_{3,4}^0}-m_{\tilde\chi_{1}^0} \simeq |\mu|-M_1 + \frac{M_Z^2 s_W^2}{2 \mu} + 
\frac{M_Z^2 s_W^2 M_1}{\mu^2}. 
\end{equation}
\noindent
The other important mass splitting for our study is
$\Delta m (\tilde\chi^\pm_{1},\tilde{\chi}_{1}^0)={m_{\tilde\chi^\pm_{1}}}-m_{{\tilde\chi_{1}^0}}$. Apart from the $\tilde{\chi}_{1}^0-\tilde\chi^\pm_{1}$ coannihilations, the LHC limits on lighter charginos depend critically on the
$\Delta m (\tilde\chi^\pm_{1},\tilde{\chi}_{1}^0)$. Adapting 
the same route as before one obtains (assuming $|M_2 \mu| > M_W^2 s_{2\beta}$)
 \begin{align}
m_{\tilde \chi_1^\pm} \simeq &M_2 + 
M_W^2\left[\frac{M_2+\mu s_{2\beta}}{M_2^2-\mu^2}\right].\\ \nonumber
m_{\tilde \chi_2^\pm} \simeq & |\mu| - 
M_W^2 sgn(\mu)\left[\frac{\mu+M_2 s_{2\beta}}{M_2^2-\mu^2}\right].\end{align}
\noindent
In the limit of large $\tan\beta$ and heavy Wino, the
mass splitting between Higgsino-like chargino and $\tilde \chi_1^0$ can be approximated to
\begin{equation}
\Delta m (\tilde\chi^\pm_{1},\tilde{\chi}_{1}^0) \equiv |\mu| -M_1 - sgn(\mu)\frac{M_W^2\mu}{M_2^2} + \frac{M_Z^2 s_W^2 M_1}{\mu^2}.
\end{equation}
\noindent
$\mathbf{{\tilde B}_{\tilde W \tilde H}}$ LSP~:
It refers to a limit $M_1 \leq M_2 < |\mu|$. 
Even after the latest LHC Run-2 data, the muon $g-2$ anomaly
can still be accommodated with Winos lighter than Higgsinos. The relevant mass splittings can be calculated from the above equations.

\bigskip
\bibliographystyle{JHEPCust.bst}
\bibliography{Bino_like_DM}

\providecommand{\href}[2]{#2}\begingroup\raggedright\begin{thebibliography}{100}

\bibitem{Jungman:1995df}
G.~Jungman, M.~Kamionkowski and K.~Griest, \emph{{Supersymmetric dark matter}},
  \href{http://dx.doi.org/10.1016/0370-1573(95)00058-5}{\emph{Phys. Rept.} {\bf
  267} (1996) 195--373}, [\href{http://arxiv.org/abs/hep-ph/9506380}{{\tt
  hep-ph/9506380}}].

\bibitem{Bertone:2004pz}
G.~Bertone, D.~Hooper and J.~Silk, \emph{{Particle dark matter: Evidence,
  candidates and constraints}},
  \href{http://dx.doi.org/10.1016/j.physrep.2004.08.031}{\emph{Phys. Rept.}
  {\bf 405} (2005) 279--390}, [\href{http://arxiv.org/abs/hep-ph/0404175}{{\tt
  hep-ph/0404175}}].

\bibitem{Arkani_Hamed_2006}
N.~Arkani-Hamed, A.~Delgado and G.~Giudice, \emph{The well-tempered
  neutralino},
  \href{http://dx.doi.org/10.1016/j.nuclphysb.2006.02.010}{\emph{Nuclear
  Physics B} {\bf 741} (may, 2006) 108--130}.

\bibitem{Chan:1997bi}
K.~L. Chan, U.~Chattopadhyay and P.~Nath, \emph{{Naturalness, weak scale
  supersymmetry and the prospect for the observation of supersymmetry at the
  Tevatron and at the CERN LHC}},
  \href{http://dx.doi.org/10.1103/PhysRevD.58.096004}{\emph{Phys. Rev. D} {\bf
  58} (1998) 096004}, [\href{http://arxiv.org/abs/hep-ph/9710473}{{\tt
  hep-ph/9710473}}].

\bibitem{Chattopadhyay:2003xi}
U.~Chattopadhyay, A.~Corsetti and P.~Nath, \emph{{WMAP constraints, SUSY dark
  matter and implications for the direct detection of SUSY}},
  \href{http://dx.doi.org/10.1103/PhysRevD.68.035005}{\emph{Phys. Rev. D} {\bf
  68} (2003) 035005}, [\href{http://arxiv.org/abs/hep-ph/0303201}{{\tt
  hep-ph/0303201}}].

\bibitem{Chattopadhyay:2005mv}
U.~Chattopadhyay, D.~Choudhury, M.~Drees, P.~Konar and D.~P. Roy,
  \emph{{Looking for a heavy Higgsino LSP in collider and dark matter
  experiments}},
  \href{http://dx.doi.org/10.1016/j.physletb.2005.09.088}{\emph{Phys. Lett. B}
  {\bf 632} (2006) 114--126}, [\href{http://arxiv.org/abs/hep-ph/0508098}{{\tt
  hep-ph/0508098}}].

\bibitem{Akula:2011jx}
S.~Akula, M.~Liu, P.~Nath and G.~Peim, \emph{{Naturalness, Supersymmetry and
  Implications for LHC and Dark Matter}},
  \href{http://dx.doi.org/10.1016/j.physletb.2012.01.077}{\emph{Phys. Lett. B}
  {\bf 709} (2012) 192--199}, [\href{http://arxiv.org/abs/1111.4589}{{\tt
  1111.4589}}].

\bibitem{Baer:2011ab}
H.~Baer, V.~Barger and A.~Mustafayev, \emph{{Implications of a 125 GeV Higgs
  scalar for LHC SUSY and neutralino dark matter searches}},
  \href{http://dx.doi.org/10.1103/PhysRevD.85.075010}{\emph{Phys. Rev. D} {\bf
  85} (2012) 075010}, [\href{http://arxiv.org/abs/1112.3017}{{\tt 1112.3017}}].

\bibitem{Ellis:2012aa}
J.~Ellis and K.~A. Olive, \emph{{Revisiting the Higgs Mass and Dark Matter in
  the CMSSM}},
  \href{http://dx.doi.org/10.1140/epjc/s10052-012-2005-2}{\emph{Eur. Phys. J.
  C} {\bf 72} (2012) 2005}, [\href{http://arxiv.org/abs/1202.3262}{{\tt
  1202.3262}}].

\bibitem{Buchmueller:2013rsa}
O.~Buchmueller et~al., \emph{{The CMSSM and NUHM1 after LHC Run 1}},
  \href{http://dx.doi.org/10.1140/epjc/s10052-014-2922-3}{\emph{Eur. Phys. J.
  C} {\bf 74} (2014) 2922}, [\href{http://arxiv.org/abs/1312.5250}{{\tt
  1312.5250}}].

\bibitem{Baer:2015tva}
H.~Baer, V.~Barger, P.~Huang, D.~Mickelson, M.~Padeffke-Kirkland and X.~Tata,
  \emph{{Natural SUSY with a bino- or wino-like LSP}},
  \href{http://dx.doi.org/10.1103/PhysRevD.91.075005}{\emph{Phys. Rev. D} {\bf
  91} (2015) 075005}, [\href{http://arxiv.org/abs/1501.06357}{{\tt
  1501.06357}}].

\bibitem{He:2023lgi}
Y.~He, L.~Meng, Y.~Yue and D.~Zhang, \emph{{Impact of recent measurement of
  $(g-2)_\mu$, LHC search for supersymmetry, and LZ experiment on Minimal
  Supersymmetric Standard Model}},  \href{http://arxiv.org/abs/2303.02360}{{\tt
  2303.02360}}.

\bibitem{Chattopadhyay:2006xb}
U.~Chattopadhyay, D.~Das, P.~Konar and D.~P. Roy, \emph{{Looking for a heavy
  wino LSP in collider and dark matter experiments}},
  \href{http://dx.doi.org/10.1103/PhysRevD.75.073014}{\emph{Phys. Rev. D} {\bf
  75} (2007) 073014}, [\href{http://arxiv.org/abs/hep-ph/0610077}{{\tt
  hep-ph/0610077}}].

\bibitem{Hisano:2006nn}
J.~Hisano, S.~Matsumoto, M.~Nagai, O.~Saito and M.~Senami,
  \emph{{Non-perturbative effect on thermal relic abundance of dark matter}},
  \href{http://dx.doi.org/10.1016/j.physletb.2007.01.012}{\emph{Phys. Lett. B}
  {\bf 646} (2007) 34--38}, [\href{http://arxiv.org/abs/hep-ph/0610249}{{\tt
  hep-ph/0610249}}].

\bibitem{Masiero_2005}
A.~Masiero, S.~Profumo and P.~Ullio, \emph{Neutralino dark matter detection in
  split supersymmetry scenarios},
  \href{http://dx.doi.org/10.1016/j.nuclphysb.2005.01.028}{\emph{Nuclear
  Physics B} {\bf 712} (apr, 2005) 86--114}.

\bibitem{Cohen:2013ama}
T.~Cohen, M.~Lisanti, A.~Pierce and T.~R. Slatyer, \emph{{Wino Dark Matter
  Under Siege}},
  \href{http://dx.doi.org/10.1088/1475-7516/2013/10/061}{\emph{JCAP} {\bf 10}
  (2013) 061}, [\href{http://arxiv.org/abs/1307.4082}{{\tt 1307.4082}}].

\bibitem{Bhattacherjee:2014dya}
B.~Bhattacherjee, M.~Ibe, K.~Ichikawa, S.~Matsumoto and K.~Nishiyama,
  \emph{{Wino Dark Matter and Future dSph Observations}},
  \href{http://dx.doi.org/10.1007/JHEP07(2014)080}{\emph{JHEP} {\bf 07} (2014)
  080}, [\href{http://arxiv.org/abs/1405.4914}{{\tt 1405.4914}}].

\bibitem{Baumgart:2014saa}
M.~Baumgart, I.~Z. Rothstein and V.~Vaidya, \emph{{Constraints on Galactic Wino
  Densities from Gamma Ray Lines}},
  \href{http://dx.doi.org/10.1007/JHEP04(2015)106}{\emph{JHEP} {\bf 04} (2015)
  106}, [\href{http://arxiv.org/abs/1412.8698}{{\tt 1412.8698}}].

\bibitem{Hryczuk:2014hpa}
A.~Hryczuk, I.~Cholis, R.~Iengo, M.~Tavakoli and P.~Ullio, \emph{{Indirect
  Detection Analysis: Wino Dark Matter Case Study}},
  \href{http://dx.doi.org/10.1088/1475-7516/2014/07/031}{\emph{JCAP} {\bf 07}
  (2014) 031}, [\href{http://arxiv.org/abs/1401.6212}{{\tt 1401.6212}}].

\bibitem{Ibe:2015tma}
M.~Ibe, S.~Matsumoto, S.~Shirai and T.~T. Yanagida, \emph{{Wino Dark Matter in
  light of the AMS-02 2015 Data}},
  \href{http://dx.doi.org/10.1103/PhysRevD.91.111701}{\emph{Phys. Rev. D} {\bf
  91} (2015) 111701}, [\href{http://arxiv.org/abs/1504.05554}{{\tt
  1504.05554}}].

\bibitem{Beneke:2016ync}
M.~Beneke, A.~Bharucha, F.~Dighera, C.~Hellmann, A.~Hryczuk, S.~Recksiegel
  et~al., \emph{{Relic density of wino-like dark matter in the MSSM}},
  \href{http://dx.doi.org/10.1007/JHEP03(2016)119}{\emph{JHEP} {\bf 03} (2016)
  119}, [\href{http://arxiv.org/abs/1601.04718}{{\tt 1601.04718}}].

\bibitem{Beneke:2016jpw}
M.~Beneke, A.~Bharucha, A.~Hryczuk, S.~Recksiegel and P.~Ruiz-Femenia,
  \emph{{The last refuge of mixed wino-Higgsino dark matter}},
  \href{http://dx.doi.org/10.1007/JHEP01(2017)002}{\emph{JHEP} {\bf 01} (2017)
  002}, [\href{http://arxiv.org/abs/1611.00804}{{\tt 1611.00804}}].

\bibitem{Beneke:2020vff}
M.~Beneke, R.~Szafron and K.~Urban, \emph{{Sommerfeld-corrected relic abundance
  of wino dark matter with NLO electroweak potentials}},
  \href{http://dx.doi.org/10.1007/JHEP02(2021)020}{\emph{JHEP} {\bf 02} (2021)
  020}, [\href{http://arxiv.org/abs/2009.00640}{{\tt 2009.00640}}].

\bibitem{Beneke:2019qaa}
M.~Beneke, R.~Szafron and K.~Urban, \emph{{Wino potential and Sommerfeld effect
  at NLO}}, \href{http://dx.doi.org/10.1016/j.physletb.2019.135112}{\emph{Phys.
  Lett. B} {\bf 800} (2020) 135112},
  [\href{http://arxiv.org/abs/1909.04584}{{\tt 1909.04584}}].

\bibitem{WMAP:2012nax}
{\scshape WMAP} collaboration, G.~Hinshaw et~al., \emph{{Nine-Year Wilkinson
  Microwave Anisotropy Probe (WMAP) Observations: Cosmological Parameter
  Results}},
  \href{http://dx.doi.org/10.1088/0067-0049/208/2/19}{\emph{Astrophys. J.
  Suppl.} {\bf 208} (2013) 19}, [\href{http://arxiv.org/abs/1212.5226}{{\tt
  1212.5226}}].

\bibitem{Planck:2018vyg}
{\scshape Planck} collaboration, N.~Aghanim et~al., \emph{{Planck 2018 results.
  VI. Cosmological parameters}},
  \href{http://dx.doi.org/10.1051/0004-6361/201833910}{\emph{Astron.
  Astrophys.} {\bf 641} (2020) A6},
  [\href{http://arxiv.org/abs/1807.06209}{{\tt 1807.06209}}]. [Erratum:
  Astron.Astrophys. 652, C4 (2021)].

\bibitem{Chakraborti:2017dpu}
M.~Chakraborti, U.~Chattopadhyay and S.~Poddar, \emph{{How light a higgsino or
  a wino dark matter can become in a compressed scenario of MSSM}},
  \href{http://dx.doi.org/10.1007/JHEP09(2017)064}{\emph{JHEP} {\bf 09} (2017)
  064}, [\href{http://arxiv.org/abs/1702.03954}{{\tt 1702.03954}}].

\bibitem{Feng:1999zg}
J.~L. Feng, K.~T. Matchev and T.~Moroi, \emph{{Focus points and naturalness in
  supersymmetry}},
  \href{http://dx.doi.org/10.1103/PhysRevD.61.075005}{\emph{Phys. Rev. D} {\bf
  61} (2000) 075005}, [\href{http://arxiv.org/abs/hep-ph/9909334}{{\tt
  hep-ph/9909334}}].

\bibitem{Feng:1999mn}
J.~L. Feng, K.~T. Matchev and T.~Moroi, \emph{{Multi - TeV scalars are natural
  in minimal supergravity}},
  \href{http://dx.doi.org/10.1103/PhysRevLett.84.2322}{\emph{Phys. Rev. Lett.}
  {\bf 84} (2000) 2322--2325}, [\href{http://arxiv.org/abs/hep-ph/9908309}{{\tt
  hep-ph/9908309}}].

\bibitem{Feng:2000gh}
J.~L. Feng, K.~T. Matchev and F.~Wilczek, \emph{{Neutralino dark matter in
  focus point supersymmetry}},
  \href{http://dx.doi.org/10.1016/S0370-2693(00)00512-8}{\emph{Phys. Lett. B}
  {\bf 482} (2000) 388--399}, [\href{http://arxiv.org/abs/hep-ph/0004043}{{\tt
  hep-ph/0004043}}].

\bibitem{Chattopadhyay:2000fj}
U.~Chattopadhyay, T.~Ibrahim and D.~P. Roy, \emph{{Electron and neutron
  electric dipole moments in the focus point scenario of SUGRA model}},
  \href{http://dx.doi.org/10.1103/PhysRevD.64.013004}{\emph{Phys. Rev. D} {\bf
  64} (2001) 013004}, [\href{http://arxiv.org/abs/hep-ph/0012337}{{\tt
  hep-ph/0012337}}].

\bibitem{Das:2007jn}
S.~P. Das, A.~Datta, M.~Guchait, M.~Maity and S.~Mukherjee, \emph{{Focus Point
  SUSY at the LHC Revisited}},
  \href{http://dx.doi.org/10.1140/epjc/s10052-008-0561-2}{\emph{Eur. Phys. J.
  C} {\bf 54} (2008) 645--653}, [\href{http://arxiv.org/abs/0708.2048}{{\tt
  0708.2048}}].

\bibitem{Chattopadhyay:2000qa}
U.~Chattopadhyay, A.~Datta, A.~Datta, A.~Datta and D.~P. Roy, \emph{{LHC
  signature of the minimal SUGRA model with a large soft scalar mass}},
  \href{http://dx.doi.org/10.1016/S0370-2693(00)01120-5}{\emph{Phys. Lett. B}
  {\bf 493} (2000) 127--134}, [\href{http://arxiv.org/abs/hep-ph/0008228}{{\tt
  hep-ph/0008228}}].

\bibitem{Baer_200522}
H.~Baer, T.~Krupovnickas, A.~Mustafayev, E.-K. Park, S.~Profumo and X.~Tata,
  \emph{Exploring the {BWCA} (bino-wino co-annihilation) scenario for
  neutralino dark matter},
  \href{http://dx.doi.org/10.1088/1126-6708/2005/12/011}{\emph{Journal of High
  Energy Physics} {\bf 2005} (dec, 2005) 011--011}.

\bibitem{Chattopadhyay:2007di}
U.~Chattopadhyay, D.~Das, A.~Datta and S.~Poddar, \emph{{Non-zero trilinear
  parameter in the mSUGRA model: Dark matter and collider signals at Tevatron
  and LHC}}, \href{http://dx.doi.org/10.1103/PhysRevD.76.055008}{\emph{Phys.
  Rev. D} {\bf 76} (2007) 055008}, [\href{http://arxiv.org/abs/0705.0921}{{\tt
  0705.0921}}].

\bibitem{Chattopadhyay:2008hk}
U.~Chattopadhyay and D.~Das, \emph{{Higgs funnel region of SUSY dark matter for
  small tan beta, RG effects on pseudoscalar Higgs boson with scalar mass
  non-universality}},
  \href{http://dx.doi.org/10.1103/PhysRevD.79.035007}{\emph{Phys. Rev. D} {\bf
  79} (2009) 035007}, [\href{http://arxiv.org/abs/0809.4065}{{\tt 0809.4065}}].

\bibitem{Chattopadhyay:2009fr}
U.~Chattopadhyay, D.~Das and D.~P. Roy, \emph{{Mixed Neutralino Dark Matter in
  Nonuniversal Gaugino Mass Models}},
  \href{http://dx.doi.org/10.1103/PhysRevD.79.095013}{\emph{Phys. Rev. D} {\bf
  79} (2009) 095013}, [\href{http://arxiv.org/abs/0902.4568}{{\tt 0902.4568}}].

\bibitem{Chattopadhyay:2010vp}
U.~Chattopadhyay, D.~Das, D.~K. Ghosh and M.~Maity, \emph{{Probing the light
  Higgs pole resonance annihilation of dark matter in the light of XENON100 and
  CDMS-II observations}},
  \href{http://dx.doi.org/10.1103/PhysRevD.82.075013}{\emph{Phys. Rev. D} {\bf
  82} (2010) 075013}, [\href{http://arxiv.org/abs/1006.3045}{{\tt 1006.3045}}].

\bibitem{Chakraborti:2014fha}
M.~Chakraborti, U.~Chattopadhyay, S.~Rao and D.~P. Roy, \emph{{Higgsino Dark
  Matter in Nonuniversal Gaugino Mass Models}},
  \href{http://dx.doi.org/10.1103/PhysRevD.91.035022}{\emph{Phys. Rev. D} {\bf
  91} (2015) 035022}, [\href{http://arxiv.org/abs/1411.4517}{{\tt 1411.4517}}].

\bibitem{Feldman:2007fq}
D.~Feldman, Z.~Liu and P.~Nath, \emph{{Light Higgses at the Tevatron and at the
  LHC and Observable Dark Matter in SUGRA and D Branes}},
  \href{http://dx.doi.org/10.1016/j.physletb.2008.02.063}{\emph{Phys. Lett. B}
  {\bf 662} (2008) 190--198}, [\href{http://arxiv.org/abs/0711.4591}{{\tt
  0711.4591}}].

\bibitem{Feldman:2008jy}
D.~Feldman, Z.~Liu and P.~Nath, \emph{{Decoding the Mechanism for the Origin of
  Dark Matter in the Early Universe Using LHC Data}},
  \href{http://dx.doi.org/10.1103/PhysRevD.78.083523}{\emph{Phys. Rev. D} {\bf
  78} (2008) 083523}, [\href{http://arxiv.org/abs/0808.1595}{{\tt 0808.1595}}].

\bibitem{Drees:2015aeo}
M.~Drees and J.~S. Kim, \emph{{Minimal natural supersymmetry after the LHC8}},
  \href{http://dx.doi.org/10.1103/PhysRevD.93.095005}{\emph{Phys. Rev. D} {\bf
  93} (2016) 095005}, [\href{http://arxiv.org/abs/1511.04461}{{\tt
  1511.04461}}].

\bibitem{Barducci:2015ffa}
D.~Barducci, A.~Belyaev, A.~K.~M. Bharucha, W.~Porod and V.~Sanz,
  \emph{{Uncovering Natural Supersymmetry via the interplay between the LHC and
  Direct Dark Matter Detection}},
  \href{http://dx.doi.org/10.1007/JHEP07(2015)066}{\emph{JHEP} {\bf 07} (2015)
  066}, [\href{http://arxiv.org/abs/1504.02472}{{\tt 1504.02472}}].

\bibitem{Baer:2016usl}
H.~Baer, V.~Barger, M.~Savoy and X.~Tata, \emph{{Multichannel assault on
  natural supersymmetry at the high luminosity LHC}},
  \href{http://dx.doi.org/10.1103/PhysRevD.94.035025}{\emph{Phys. Rev. D} {\bf
  94} (2016) 035025}, [\href{http://arxiv.org/abs/1604.07438}{{\tt
  1604.07438}}].

\bibitem{Chatterjee:2017nyx}
A.~Chatterjee, J.~Dutta and S.~K. Rai, \emph{{Natural SUSY at LHC with
  Right-Sneutrino LSP}},
  \href{http://dx.doi.org/10.1007/JHEP06(2018)042}{\emph{JHEP} {\bf 06} (2018)
  042}, [\href{http://arxiv.org/abs/1710.10617}{{\tt 1710.10617}}].

\bibitem{Randall:1998uk}
L.~Randall and R.~Sundrum, \emph{{Out of this world supersymmetry breaking}},
  \href{http://dx.doi.org/10.1016/S0550-3213(99)00359-4}{\emph{Nucl. Phys. B}
  {\bf 557} (1999) 79--118}, [\href{http://arxiv.org/abs/hep-th/9810155}{{\tt
  hep-th/9810155}}].

\bibitem{Giudice:1998xp}
G.~F. Giudice, M.~A. Luty, H.~Murayama and R.~Rattazzi, \emph{{Gaugino mass
  without singlets}},
  \href{http://dx.doi.org/10.1088/1126-6708/1998/12/027}{\emph{JHEP} {\bf 12}
  (1998) 027}, [\href{http://arxiv.org/abs/hep-ph/9810442}{{\tt
  hep-ph/9810442}}].

\bibitem{Gherghetta:1999sw}
T.~Gherghetta, G.~F. Giudice and J.~D. Wells, \emph{{Phenomenological
  consequences of supersymmetry with anomaly induced masses}},
  \href{http://dx.doi.org/10.1016/S0550-3213(99)00429-0}{\emph{Nucl. Phys. B}
  {\bf 559} (1999) 27--47}, [\href{http://arxiv.org/abs/hep-ph/9904378}{{\tt
  hep-ph/9904378}}].

\bibitem{ATLAS:2020syg}
{\scshape ATLAS} collaboration, G.~Aad et~al., \emph{{Search for squarks and
  gluinos in final states with jets and missing transverse momentum using 139
  fb$^{-1}$ of $\sqrt{s}$ =13 TeV $pp$ collision data with the ATLAS
  detector}}, \href{http://dx.doi.org/10.1007/JHEP02(2021)143}{\emph{JHEP} {\bf
  02} (2021) 143}, [\href{http://arxiv.org/abs/2010.14293}{{\tt 2010.14293}}].

\bibitem{CMS:2019zmd}
{\scshape CMS} collaboration, T.~C. Collaboration et~al., \emph{{Search for
  supersymmetry in proton-proton collisions at 13 TeV in final states with jets
  and missing transverse momentum}},
  \href{http://dx.doi.org/10.1007/JHEP10(2019)244}{\emph{JHEP} {\bf 10} (2019)
  244}, [\href{http://arxiv.org/abs/1908.04722}{{\tt 1908.04722}}].

\bibitem{Canepa:2019hph}
A.~Canepa, \emph{{Searches for Supersymmetry at the Large Hadron Collider}},
  \href{http://dx.doi.org/10.1016/j.revip.2019.100033}{\emph{Rev. Phys.} {\bf
  4} (2019) 100033}.

\bibitem{Chakraborti:2015mra}
M.~Chakraborti, U.~Chattopadhyay, A.~Choudhury, A.~Datta and S.~Poddar,
  \emph{{Reduced LHC constraints for higgsino-like heavier electroweakinos}},
  \href{http://dx.doi.org/10.1007/JHEP11(2015)050}{\emph{JHEP} {\bf 11} (2015)
  050}, [\href{http://arxiv.org/abs/1507.01395}{{\tt 1507.01395}}].

\bibitem{Muong-2:2006rrc}
{\scshape Muon g-2} collaboration, G.~W. Bennett et~al., \emph{{Final Report of
  the Muon E821 Anomalous Magnetic Moment Measurement at BNL}},
  \href{http://dx.doi.org/10.1103/PhysRevD.73.072003}{\emph{Phys. Rev. D} {\bf
  73} (2006) 072003}, [\href{http://arxiv.org/abs/hep-ex/0602035}{{\tt
  hep-ex/0602035}}].

\bibitem{Muong-2:2021ojo}
{\scshape Muon g-2} collaboration, B.~Abi et~al., \emph{{Measurement of the
  Positive Muon Anomalous Magnetic Moment to 0.46 ppm}},
  \href{http://dx.doi.org/10.1103/PhysRevLett.126.141801}{\emph{Phys. Rev.
  Lett.} {\bf 126} (2021) 141801}, [\href{http://arxiv.org/abs/2104.03281}{{\tt
  2104.03281}}].

\bibitem{Aoyama:2020ynm}
T.~Aoyama et~al., \emph{{The anomalous magnetic moment of the muon in the
  Standard Model}},
  \href{http://dx.doi.org/10.1016/j.physrep.2020.07.006}{\emph{Phys. Rept.}
  {\bf 887} (2020) 1--166}, [\href{http://arxiv.org/abs/2006.04822}{{\tt
  2006.04822}}].

\bibitem{Davier:2017zfy}
M.~Davier, A.~Hoecker, B.~Malaescu and Z.~Zhang, \emph{{Reevaluation of the
  hadronic vacuum polarisation contributions to the Standard Model predictions
  of the muon $g-2$ and ${\alpha (m_Z^2)}$ using newest hadronic cross-section
  data}}, \href{http://dx.doi.org/10.1140/epjc/s10052-017-5161-6}{\emph{Eur.
  Phys. J. C} {\bf 77} (2017) 827},
  [\href{http://arxiv.org/abs/1706.09436}{{\tt 1706.09436}}].

\bibitem{Keshavarzi:2018mgv}
A.~Keshavarzi, D.~Nomura and T.~Teubner, \emph{{Muon $g-2$ and $\alpha(M_Z^2)$:
  a new data-based analysis}},
  \href{http://dx.doi.org/10.1103/PhysRevD.97.114025}{\emph{Phys. Rev. D} {\bf
  97} (2018) 114025}, [\href{http://arxiv.org/abs/1802.02995}{{\tt
  1802.02995}}].

\bibitem{Colangelo:2018mtw}
G.~Colangelo, M.~Hoferichter and P.~Stoffer, \emph{{Two-pion contribution to
  hadronic vacuum polarization}},
  \href{http://dx.doi.org/10.1007/JHEP02(2019)006}{\emph{JHEP} {\bf 02} (2019)
  006}, [\href{http://arxiv.org/abs/1810.00007}{{\tt 1810.00007}}].

\bibitem{Davier:2019can}
M.~Davier, A.~Hoecker, B.~Malaescu and Z.~Zhang, \emph{{A new evaluation of the
  hadronic vacuum polarisation contributions to the muon anomalous magnetic
  moment and to $\mathbf{\boldsymbol\alpha(m_Z^2)}$}},
  \href{http://dx.doi.org/10.1140/epjc/s10052-020-7792-2}{\emph{Eur. Phys. J.
  C} {\bf 80} (2020) 241}, [\href{http://arxiv.org/abs/1908.00921}{{\tt
  1908.00921}}]. [Erratum: Eur.Phys.J.C 80, 410 (2020)].

\bibitem{Keshavarzi:2019abf}
A.~Keshavarzi, D.~Nomura and T.~Teubner, \emph{{$g-2$ of charged leptons,
  $\alpha (M^2_Z)$ , and the hyperfine splitting of muonium}},
  \href{http://dx.doi.org/10.1103/PhysRevD.101.014029}{\emph{Phys. Rev. D} {\bf
  101} (2020) 014029}, [\href{http://arxiv.org/abs/1911.00367}{{\tt
  1911.00367}}].

\bibitem{Kurz:2014wya}
A.~Kurz, T.~Liu, P.~Marquard and M.~Steinhauser, \emph{{Hadronic contribution
  to the muon anomalous magnetic moment to next-to-next-to-leading order}},
  \href{http://dx.doi.org/10.1016/j.physletb.2014.05.043}{\emph{Phys. Lett. B}
  {\bf 734} (2014) 144--147}, [\href{http://arxiv.org/abs/1403.6400}{{\tt
  1403.6400}}].

\bibitem{Hoferichter:2019mqg}
M.~Hoferichter, B.-L. Hoid and B.~Kubis, \emph{{Three-pion contribution to
  hadronic vacuum polarization}},
  \href{http://dx.doi.org/10.1007/JHEP08(2019)137}{\emph{JHEP} {\bf 08} (2019)
  137}, [\href{http://arxiv.org/abs/1907.01556}{{\tt 1907.01556}}].

\bibitem{Melnikov:2003xd}
K.~Melnikov and A.~Vainshtein, \emph{{Hadronic light-by-light scattering
  contribution to the muon anomalous magnetic moment revisited}},
  \href{http://dx.doi.org/10.1103/PhysRevD.70.113006}{\emph{Phys. Rev. D} {\bf
  70} (2004) 113006}, [\href{http://arxiv.org/abs/hep-ph/0312226}{{\tt
  hep-ph/0312226}}].

\bibitem{Masjuan:2017tvw}
P.~Masjuan and P.~Sanchez-Puertas, \emph{{Pseudoscalar-pole contribution to the
  $(g_{\mu}-2)$: a rational approach}},
  \href{http://dx.doi.org/10.1103/PhysRevD.95.054026}{\emph{Phys. Rev. D} {\bf
  95} (2017) 054026}, [\href{http://arxiv.org/abs/1701.05829}{{\tt
  1701.05829}}].

\bibitem{Colangelo:2017fiz}
G.~Colangelo, M.~Hoferichter, M.~Procura and P.~Stoffer, \emph{{Dispersion
  relation for hadronic light-by-light scattering: two-pion contributions}},
  \href{http://dx.doi.org/10.1007/JHEP04(2017)161}{\emph{JHEP} {\bf 04} (2017)
  161}, [\href{http://arxiv.org/abs/1702.07347}{{\tt 1702.07347}}].

\bibitem{Hoferichter:2018kwz}
M.~Hoferichter, B.-L. Hoid, B.~Kubis, S.~Leupold and S.~P. Schneider,
  \emph{{Dispersion relation for hadronic light-by-light scattering: pion
  pole}}, \href{http://dx.doi.org/10.1007/JHEP10(2018)141}{\emph{JHEP} {\bf 10}
  (2018) 141}, [\href{http://arxiv.org/abs/1808.04823}{{\tt 1808.04823}}].

\bibitem{Gerardin:2019vio}
A.~G\'erardin, H.~B. Meyer and A.~Nyffeler, \emph{{Lattice calculation of the
  pion transition form factor with $N_f=2+1$ Wilson quarks}},
  \href{http://dx.doi.org/10.1103/PhysRevD.100.034520}{\emph{Phys. Rev. D} {\bf
  100} (2019) 034520}, [\href{http://arxiv.org/abs/1903.09471}{{\tt
  1903.09471}}].

\bibitem{Bijnens:2019ghy}
J.~Bijnens, N.~Hermansson-Truedsson and A.~Rodr\'\i{}guez-S\'anchez,
  \emph{{Short-distance constraints for the HLbL contribution to the muon
  anomalous magnetic moment}},
  \href{http://dx.doi.org/10.1016/j.physletb.2019.134994}{\emph{Phys. Lett. B}
  {\bf 798} (2019) 134994}, [\href{http://arxiv.org/abs/1908.03331}{{\tt
  1908.03331}}].

\bibitem{Colangelo:2019uex}
G.~Colangelo, F.~Hagelstein, M.~Hoferichter, L.~Laub and P.~Stoffer,
  \emph{{Longitudinal short-distance constraints for the hadronic
  light-by-light contribution to $(g-2)_\mu$ with large-$N_c$ Regge models}},
  \href{http://dx.doi.org/10.1007/JHEP03(2020)101}{\emph{JHEP} {\bf 03} (2020)
  101}, [\href{http://arxiv.org/abs/1910.13432}{{\tt 1910.13432}}].

\bibitem{Colangelo:2014qya}
G.~Colangelo, M.~Hoferichter, A.~Nyffeler, M.~Passera and P.~Stoffer,
  \emph{{Remarks on higher-order hadronic corrections to the muon
  g\ensuremath{-}2}},
  \href{http://dx.doi.org/10.1016/j.physletb.2014.06.012}{\emph{Phys. Lett. B}
  {\bf 735} (2014) 90--91}, [\href{http://arxiv.org/abs/1403.7512}{{\tt
  1403.7512}}].

\bibitem{Blum:2019ugy}
T.~Blum, N.~Christ, M.~Hayakawa, T.~Izubuchi, L.~Jin, C.~Jung et~al.,
  \emph{{Hadronic Light-by-Light Scattering Contribution to the Muon Anomalous
  Magnetic Moment from Lattice QCD}},
  \href{http://dx.doi.org/10.1103/PhysRevLett.124.132002}{\emph{Phys. Rev.
  Lett.} {\bf 124} (2020) 132002}, [\href{http://arxiv.org/abs/1911.08123}{{\tt
  1911.08123}}].

\bibitem{Aoyama:2012wk}
T.~Aoyama, M.~Hayakawa, T.~Kinoshita and M.~Nio, \emph{{Complete Tenth-Order
  QED Contribution to the Muon g-2}},
  \href{http://dx.doi.org/10.1103/PhysRevLett.109.111808}{\emph{Phys. Rev.
  Lett.} {\bf 109} (2012) 111808}, [\href{http://arxiv.org/abs/1205.5370}{{\tt
  1205.5370}}].

\bibitem{atoms7010028}
T.~Aoyama, T.~Kinoshita and M.~Nio, \emph{Theory of the anomalous magnetic
  moment of the electron},
  \href{http://dx.doi.org/10.3390/atoms7010028}{\emph{Atoms} {\bf 7} (2019) }.

\bibitem{Czarnecki:2002nt}
A.~Czarnecki, W.~J. Marciano and A.~Vainshtein, \emph{{Refinements in
  electroweak contributions to the muon anomalous magnetic moment}},
  \href{http://dx.doi.org/10.1103/PhysRevD.67.073006}{\emph{Phys. Rev. D} {\bf
  67} (2003) 073006}, [\href{http://arxiv.org/abs/hep-ph/0212229}{{\tt
  hep-ph/0212229}}]. [Erratum: Phys.Rev.D 73, 119901 (2006)].

\bibitem{Gnendiger:2013pva}
C.~Gnendiger, D.~St\"ockinger and H.~St\"ockinger-Kim, \emph{{The electroweak
  contributions to $(g-2)_\mu$ after the Higgs boson mass measurement}},
  \href{http://dx.doi.org/10.1103/PhysRevD.88.053005}{\emph{Phys. Rev. D} {\bf
  88} (2013) 053005}, [\href{http://arxiv.org/abs/1306.5546}{{\tt 1306.5546}}].

\bibitem{Muong-2:2023cdq}
{\scshape Muon g-2} collaboration, D.~P. Aguillard et~al., \emph{{Measurement
  of the Positive Muon Anomalous Magnetic Moment to 0.20 ppm}},
  \href{http://arxiv.org/abs/2308.06230}{{\tt 2308.06230}}.

\bibitem{PandaX-II:2020oim}
{\scshape PandaX-II} collaboration, Q.~Wang et~al., \emph{{Results of dark
  matter search using the full PandaX-II exposure}},
  \href{http://dx.doi.org/10.1088/1674-1137/abb658}{\emph{Chin. Phys. C} {\bf
  44} (2020) 125001}, [\href{http://arxiv.org/abs/2007.15469}{{\tt
  2007.15469}}].

\bibitem{XENON:2023cxc}
{\scshape XENON} collaboration, E.~Aprile et~al., \emph{{First Dark Matter
  Search with Nuclear Recoils from the XENONnT Experiment}},
  \href{http://dx.doi.org/10.1103/PhysRevLett.131.041003}{\emph{Phys. Rev.
  Lett.} {\bf 131} (2023) 041003}, [\href{http://arxiv.org/abs/2303.14729}{{\tt
  2303.14729}}].

\bibitem{XENON:2018voc}
{\scshape XENON} collaboration, E.~Aprile et~al., \emph{{Dark Matter Search
  Results from a One Ton-Year Exposure of XENON1T}},
  \href{http://dx.doi.org/10.1103/PhysRevLett.121.111302}{\emph{Phys. Rev.
  Lett.} {\bf 121} (2018) 111302}, [\href{http://arxiv.org/abs/1805.12562}{{\tt
  1805.12562}}].

\bibitem{LUX:2017ree}
{\scshape LUX} collaboration, D.~S. Akerib et~al., \emph{{Limits on
  spin-dependent WIMP-nucleon cross section obtained from the complete LUX
  exposure}},
  \href{http://dx.doi.org/10.1103/PhysRevLett.118.251302}{\emph{Phys. Rev.
  Lett.} {\bf 118} (2017) 251302}, [\href{http://arxiv.org/abs/1705.03380}{{\tt
  1705.03380}}].

\bibitem{PandaX-II:2017hlx}
{\scshape PandaX-II} collaboration, X.~Cui et~al., \emph{{Dark Matter Results
  From 54-Ton-Day Exposure of PandaX-II Experiment}},
  \href{http://dx.doi.org/10.1103/PhysRevLett.119.181302}{\emph{Phys. Rev.
  Lett.} {\bf 119} (2017) 181302}, [\href{http://arxiv.org/abs/1708.06917}{{\tt
  1708.06917}}].

\bibitem{XENON:2020kmp}
{\scshape XENON} collaboration, E.~Aprile et~al., \emph{{Projected WIMP
  sensitivity of the XENONnT dark matter experiment}},
  \href{http://dx.doi.org/10.1088/1475-7516/2020/11/031}{\emph{JCAP} {\bf 11}
  (2020) 031}, [\href{http://arxiv.org/abs/2007.08796}{{\tt 2007.08796}}].

\bibitem{PICO:2016kso}
{\scshape PICO} collaboration, C.~Amole et~al., \emph{{Improved dark matter
  search results from PICO-2L Run 2}},
  \href{http://dx.doi.org/10.1103/PhysRevD.93.061101}{\emph{Phys. Rev. D} {\bf
  93} (2016) 061101}, [\href{http://arxiv.org/abs/1601.03729}{{\tt
  1601.03729}}].

\bibitem{LUX:2016sci}
{\scshape LUX} collaboration, D.~S. Akerib et~al., \emph{{Results on the
  Spin-Dependent Scattering of Weakly Interacting Massive Particles on Nucleons
  from the Run 3 Data of the LUX Experiment}},
  \href{http://dx.doi.org/10.1103/PhysRevLett.116.161302}{\emph{Phys. Rev.
  Lett.} {\bf 116} (2016) 161302}, [\href{http://arxiv.org/abs/1602.03489}{{\tt
  1602.03489}}].

\bibitem{PandaX-II:2016wea}
{\scshape PandaX-II} collaboration, C.~Fu et~al., \emph{{Spin-Dependent
  Weakly-Interacting-Massive-Particle\textendash{}Nucleon Cross Section Limits
  from First Data of PandaX-II Experiment}},
  \href{http://dx.doi.org/10.1103/PhysRevLett.118.071301}{\emph{Phys. Rev.
  Lett.} {\bf 118} (2017) 071301}, [\href{http://arxiv.org/abs/1611.06553}{{\tt
  1611.06553}}]. [Erratum: Phys.Rev.Lett. 120, 049902 (2018)].

\bibitem{LUX-ZEPLIN:2022xrq}
{\scshape LUX-ZEPLIN} collaboration, J.~Aalbers et~al., \emph{{First Dark
  Matter Search Results from the LUX-ZEPLIN (LZ) Experiment}},
  \href{http://dx.doi.org/10.1103/PhysRevLett.131.041002}{\emph{Phys. Rev.
  Lett.} {\bf 131} (2023) 041002}, [\href{http://arxiv.org/abs/2207.03764}{{\tt
  2207.03764}}].

\bibitem{Baer_2016}
H.~Baer, V.~Barger and H.~Serce, \emph{{SUSY} under siege from direct and
  indirect {WIMP} detection experiments},
  \href{http://dx.doi.org/10.1103/physrevd.94.115019}{\emph{Physical Review D}
  {\bf 94} (dec, 2016) }.

\bibitem{Badziak:2017the}
M.~Badziak, M.~Olechowski and P.~Szczerbiak, \emph{{Is well-tempered neutralino
  in MSSM still alive after 2016 LUX results?}},
  \href{http://dx.doi.org/10.1016/j.physletb.2017.04.059}{\emph{Phys. Lett. B}
  {\bf 770} (2017) 226--235}, [\href{http://arxiv.org/abs/1701.05869}{{\tt
  1701.05869}}].

\bibitem{Profumo:2017ntc}
S.~Profumo, T.~Stefaniak and L.~Stephenson~Haskins, \emph{{The Not-So-Well
  Tempered Neutralino}},
  \href{http://dx.doi.org/10.1103/PhysRevD.96.055018}{\emph{Phys. Rev. D} {\bf
  96} (2017) 055018}, [\href{http://arxiv.org/abs/1706.08537}{{\tt
  1706.08537}}].

\bibitem{Abdughani:2019wai}
M.~Abdughani, K.-I. Hikasa, L.~Wu, J.~M. Yang and J.~Zhao, \emph{{Testing
  electroweak SUSY for muon $g$ \ensuremath{-} 2 and dark matter at the LHC and
  beyond}}, \href{http://dx.doi.org/10.1007/JHEP11(2019)095}{\emph{JHEP} {\bf
  11} (2019) 095}, [\href{http://arxiv.org/abs/1909.07792}{{\tt 1909.07792}}].

\bibitem{Cheung_2013}
C.~Cheung, L.~J. Hall, D.~Pinner and J.~T. Ruderman, \emph{Prospects and blind
  spots for neutralino dark matter},
  \href{http://dx.doi.org/10.1007/jhep05(2013)100}{\emph{Journal of High Energy
  Physics} {\bf 2013} (may, 2013) }.

\bibitem{Cheung_2014}
C.~Cheung and D.~Sanford, \emph{Simplified models of mixed dark matter},
  \href{http://dx.doi.org/10.1088/1475-7516/2014/02/011}{\emph{Journal of
  Cosmology and Astroparticle Physics} {\bf 2014} (feb, 2014) 011--011}.

\bibitem{Huang:2014xua}
P.~Huang and C.~E.~M. Wagner, \emph{{Blind Spots for neutralino Dark Matter in
  the MSSM with an intermediate $m_A$}},
  \href{http://dx.doi.org/10.1103/PhysRevD.90.015018}{\emph{Phys. Rev. D} {\bf
  90} (2014) 015018}, [\href{http://arxiv.org/abs/1404.0392}{{\tt 1404.0392}}].

\bibitem{Das:2020ozo}
D.~Das, B.~De and S.~Mitra, \emph{{Cancellation in Dark Matter-Nucleon
  Interactions: the Role of Non-Standard-Model-like Yukawa Couplings}},
  \href{http://dx.doi.org/10.1016/j.physletb.2021.136159}{\emph{Phys. Lett. B}
  {\bf 815} (2021) 136159}, [\href{http://arxiv.org/abs/2011.13225}{{\tt
  2011.13225}}].

\bibitem{Duan:2018rls}
G.~H. Duan, K.-I. Hikasa, J.~Ren, L.~Wu and J.~M. Yang, \emph{{Probing
  bino-wino coannihilation dark matter below the neutrino floor at the LHC}},
  \href{http://dx.doi.org/10.1103/PhysRevD.98.015010}{\emph{Phys. Rev. D} {\bf
  98} (2018) 015010}, [\href{http://arxiv.org/abs/1804.05238}{{\tt
  1804.05238}}].

\bibitem{Giudice:1995np}
G.~F. Giudice and A.~Pomarol, \emph{{Mass degeneracy of the Higgsinos}},
  \href{http://dx.doi.org/10.1016/0370-2693(96)00060-3}{\emph{Phys. Lett. B}
  {\bf 372} (1996) 253--258}, [\href{http://arxiv.org/abs/hep-ph/9512337}{{\tt
  hep-ph/9512337}}].

\bibitem{Drees:1996pk}
M.~Drees, M.~M. Nojiri, D.~P. Roy and Y.~Yamada, \emph{{Light Higgsino dark
  matter}}, \href{http://dx.doi.org/10.1103/PhysRevD.64.039901}{\emph{Phys.
  Rev. D} {\bf 56} (1997) 276--290},
  [\href{http://arxiv.org/abs/hep-ph/9701219}{{\tt hep-ph/9701219}}]. [Erratum:
  Phys.Rev.D 64, 039901 (2001)].

\bibitem{Drees:1993bu}
M.~Drees and M.~Nojiri, \emph{{Neutralino - nucleon scattering revisited}},
  \href{http://dx.doi.org/10.1103/PhysRevD.48.3483}{\emph{Phys. Rev. D} {\bf
  48} (1993) 3483--3501}, [\href{http://arxiv.org/abs/hep-ph/9307208}{{\tt
  hep-ph/9307208}}].

\bibitem{Bisal:2023fgb}
S.~Bisal, A.~Chatterjee, D.~Das and S.~A. Pasha, \emph{{Radiative Corrections
  to Aid the Direct Detection of the Higgsino-like Neutralino Dark Matter:
  Spin-Independent Interactions}},  \href{http://arxiv.org/abs/2311.09937}{{\tt
  2311.09937}}.

\bibitem{Hisano:2004pv}
J.~Hisano, S.~Matsumoto, M.~M. Nojiri and O.~Saito, \emph{{Direct detection of
  the Wino and Higgsino-like neutralino dark matters at one-loop level}},
  \href{http://dx.doi.org/10.1103/PhysRevD.71.015007}{\emph{Phys. Rev. D} {\bf
  71} (2005) 015007}, [\href{http://arxiv.org/abs/hep-ph/0407168}{{\tt
  hep-ph/0407168}}].

\bibitem{Harz:2023llw}
J.~Harz, B.~Herrmann, M.~Klasen, K.~Kova\v{r}\'\i{}k and L.~P. Wiggering,
  \emph{{Precision predictions for dark matter with DM@NLO in the MSSM}},
  \href{http://arxiv.org/abs/2312.17206}{{\tt 2312.17206}}.

\bibitem{Klasen:2016qyz}
M.~Klasen, K.~Kovarik and P.~Steppeler, \emph{{SUSY-QCD corrections for direct
  detection of neutralino dark matter and correlations with relic density}},
  \href{http://dx.doi.org/10.1103/PhysRevD.94.095002}{\emph{Phys. Rev. D} {\bf
  94} (2016) 095002}, [\href{http://arxiv.org/abs/1607.06396}{{\tt
  1607.06396}}].

\bibitem{Cirelli:2005uq}
M.~Cirelli, N.~Fornengo and A.~Strumia, \emph{{Minimal dark matter}},
  \href{http://dx.doi.org/10.1016/j.nuclphysb.2006.07.012}{\emph{Nucl. Phys. B}
  {\bf 753} (2006) 178--194}, [\href{http://arxiv.org/abs/hep-ph/0512090}{{\tt
  hep-ph/0512090}}].

\bibitem{Hisano:2011cs}
J.~Hisano, K.~Ishiwata, N.~Nagata and T.~Takesako, \emph{{Direct Detection of
  Electroweak-Interacting Dark Matter}},
  \href{http://dx.doi.org/10.1007/JHEP07(2011)005}{\emph{JHEP} {\bf 07} (2011)
  005}, [\href{http://arxiv.org/abs/1104.0228}{{\tt 1104.0228}}].

\bibitem{Hisano:2010fy}
J.~Hisano, K.~Ishiwata and N.~Nagata, \emph{{A complete calculation for direct
  detection of Wino dark matter}},
  \href{http://dx.doi.org/10.1016/j.physletb.2010.05.047}{\emph{Phys. Lett. B}
  {\bf 690} (2010) 311--315}, [\href{http://arxiv.org/abs/1004.4090}{{\tt
  1004.4090}}].

\bibitem{Hisano:2010ct}
J.~Hisano, K.~Ishiwata and N.~Nagata, \emph{{Gluon contribution to the dark
  matter direct detection}},
  \href{http://dx.doi.org/10.1103/PhysRevD.82.115007}{\emph{Phys. Rev. D} {\bf
  82} (2010) 115007}, [\href{http://arxiv.org/abs/1007.2601}{{\tt 1007.2601}}].

\bibitem{Drees:2004jm}
M.~Drees, R.~Godbole and P.~Roy, \emph{{Theory and phenomenology of sparticles:
  An account of four-dimensional N=1 supersymmetry in high energy physics}}.
\newblock 2004.

\bibitem{Gunion:1987yh}
J.~F. Gunion and H.~E. Haber, \emph{{Two-body Decays of Neutralinos and
  Charginos}}, \href{http://dx.doi.org/10.1103/PhysRevD.37.2515}{\emph{Phys.
  Rev. D} {\bf 37} (1988) 2515}.

\bibitem{ElKheishen:1992yv}
M.~M. El~Kheishen, A.~A. Aboshousha and A.~A. Shafik, \emph{{Analytic formulas
  for the neutralino masses and the neutralino mixing matrix}},
  \href{http://dx.doi.org/10.1103/PhysRevD.45.4345}{\emph{Phys. Rev. D} {\bf
  45} (1992) 4345--4348}.

\bibitem{Barger:1993gh}
V.~D. Barger, M.~S. Berger and P.~Ohmann, \emph{{The Supersymmetric particle
  spectrum}}, \href{http://dx.doi.org/10.1103/PhysRevD.49.4908}{\emph{Phys.
  Rev. D} {\bf 49} (1994) 4908--4930},
  [\href{http://arxiv.org/abs/hep-ph/9311269}{{\tt hep-ph/9311269}}].

\bibitem{Goodman:1984dc}
M.~W. Goodman and E.~Witten, \emph{{Detectability of Certain Dark Matter
  Candidates}}, \href{http://dx.doi.org/10.1103/PhysRevD.31.3059}{\emph{Phys.
  Rev. D} {\bf 31} (1985) 3059}.

\bibitem{Griest:1988ma}
K.~Griest, \emph{{Cross-Sections, Relic Abundance and Detection Rates for
  Neutralino Dark Matter}},
  \href{http://dx.doi.org/10.1103/PhysRevD.38.2357}{\emph{Phys. Rev. D} {\bf
  38} (1988) 2357}. [Erratum: Phys.Rev.D 39, 3802 (1989)].

\bibitem{Ellis:1987sh}
J.~R. Ellis and R.~A. Flores, \emph{{Realistic Predictions for the Detection of
  Supersymmetric Dark Matter}},
  \href{http://dx.doi.org/10.1016/0550-3213(88)90111-3}{\emph{Nucl. Phys. B}
  {\bf 307} (1988) 883--908}.

\bibitem{Barbieri:1988zs}
R.~Barbieri, M.~Frigeni and G.~F. Giudice, \emph{{Dark Matter Neutralinos in
  Supergravity Theories}},
  \href{http://dx.doi.org/10.1016/0550-3213(89)90404-5}{\emph{Nucl. Phys. B}
  {\bf 313} (1989) 725--735}.

\bibitem{Ellis:2000ds}
J.~R. Ellis, A.~Ferstl and K.~A. Olive, \emph{{Reevaluation of the elastic
  scattering of supersymmetric dark matter}},
  \href{http://dx.doi.org/10.1016/S0370-2693(00)00459-7}{\emph{Phys. Lett. B}
  {\bf 481} (2000) 304--314}, [\href{http://arxiv.org/abs/hep-ph/0001005}{{\tt
  hep-ph/0001005}}].

\bibitem{Vergados:2006sy}
J.~D. Vergados, \emph{{On the direct detection of dark matter- exploring all
  the signatures of the neutralino-nucleus interaction}},
  \href{http://dx.doi.org/10.1007/978-3-540-71013-4_3}{\emph{Lect. Notes Phys.}
  {\bf 720} (2007) 69--100}, [\href{http://arxiv.org/abs/hep-ph/0601064}{{\tt
  hep-ph/0601064}}].

\bibitem{Oikonomou:2006mh}
V.~K. Oikonomou, J.~D. Vergados and C.~C. Moustakidis, \emph{{Direct Detection
  of Dark Matter-Rates for Various Wimps}},
  \href{http://dx.doi.org/10.1016/j.nuclphysb.2007.03.014}{\emph{Nucl. Phys. B}
  {\bf 773} (2007) 19--42}, [\href{http://arxiv.org/abs/hep-ph/0612293}{{\tt
  hep-ph/0612293}}].

\bibitem{Ellis:2008hf}
J.~R. Ellis, K.~A. Olive and C.~Savage, \emph{{Hadronic Uncertainties in the
  Elastic Scattering of Supersymmetric Dark Matter}},
  \href{http://dx.doi.org/10.1103/PhysRevD.77.065026}{\emph{Phys. Rev. D} {\bf
  77} (2008) 065026}, [\href{http://arxiv.org/abs/0801.3656}{{\tt 0801.3656}}].

\bibitem{Nath:1994ci}
P.~Nath and R.~L. Arnowitt, \emph{{Event rates in dark matter detectors for
  neutralinos including constraints from the b ---\ensuremath{>} s gamma
  decay}}, \href{http://dx.doi.org/10.1103/PhysRevLett.74.4592}{\emph{Phys.
  Rev. Lett.} {\bf 74} (1995) 4592--4595},
  [\href{http://arxiv.org/abs/hep-ph/9409301}{{\tt hep-ph/9409301}}].

\bibitem{Hisano:2017jmz}
J.~Hisano, \emph{{Effective theory approach to direct detection of dark
  matter}},  \href{http://arxiv.org/abs/1712.02947}{{\tt 1712.02947}}.

\bibitem{PhysRevD.59.055009}
T.~Falk, A.~Ferstl and K.~A. Olive, \emph{New contributions to neutralino
  elastic cross sections from cp violating phases in the minimal supersymmetric
  standard model},
  \href{http://dx.doi.org/10.1103/PhysRevD.59.055009}{\emph{Phys. Rev. D} {\bf
  59} (Feb, 1999) 055009}.

\bibitem{SHIFMAN1978443}
M.~Shifman, A.~Vainshtein and V.~Zakharov, \emph{Remarks on higgs-boson
  interactions with nucleons},
  \href{http://dx.doi.org/https://doi.org/10.1016/0370-2693(78)90481-1}{\emph{Physics
  Letters B} {\bf 78} (1978) 443--446}.

\bibitem{Thomas:2012tg}
A.~Thomas, P.~Shanahan and R.~Young, \emph{{Strangeness in the nucleon: what
  have we learned?}},
  \href{http://dx.doi.org/10.1393/ncc/i2012-11292-7}{\emph{Nuovo Cim. C} {\bf
  035N04} (2012) 3--10}, [\href{http://arxiv.org/abs/1202.6407}{{\tt
  1202.6407}}].

\bibitem{Belanger:2013oya}
G.~Belanger, F.~Boudjema, A.~Pukhov and A.~Semenov, \emph{{micrOMEGAs$_3$: A
  program for calculating dark matter observables}},
  \href{http://dx.doi.org/10.1016/j.cpc.2013.10.016}{\emph{Comput. Phys.
  Commun.} {\bf 185} (2014) 960--985},
  [\href{http://arxiv.org/abs/1305.0237}{{\tt 1305.0237}}].

\bibitem{Alarcon:2011zs}
J.~Alarcon, J.~Martin~Camalich and J.~Oller, \emph{{The chiral representation
  of the $\pi N$ scattering amplitude and the pion-nucleon sigma term}},
  \href{http://dx.doi.org/10.1103/PhysRevD.85.051503}{\emph{Phys. Rev. D} {\bf
  85} (2012) 051503}, [\href{http://arxiv.org/abs/1110.3797}{{\tt 1110.3797}}].

\bibitem{Crivellin:2013ipa}
A.~Crivellin, M.~Hoferichter and M.~Procura, \emph{{Accurate evaluation of
  hadronic uncertainties in spin-independent WIMP-nucleon scattering:
  Disentangling two- and three-flavor effects}},
  \href{http://dx.doi.org/10.1103/PhysRevD.89.054021}{\emph{Phys. Rev. D} {\bf
  89} (2014) 054021}, [\href{http://arxiv.org/abs/1312.4951}{{\tt 1312.4951}}].

\bibitem{Hoferichter:2015dsa}
M.~Hoferichter, J.~Ruiz~de Elvira, B.~Kubis and U.-G. Mei\ss{}ner,
  \emph{{High-Precision Determination of the Pion-Nucleon \ensuremath{\sigma}
  Term from Roy-Steiner Equations}},
  \href{http://dx.doi.org/10.1103/PhysRevLett.115.092301}{\emph{Phys. Rev.
  Lett.} {\bf 115} (2015) 092301}, [\href{http://arxiv.org/abs/1506.04142}{{\tt
  1506.04142}}].

\bibitem{Hooper:2009zm}
D.~Hooper, \emph{{Particle Dark Matter}},  in \emph{{Proceedings of Theoretical
  Advanced Study Institute in Elementary Particle Physics on The dawn of the
  LHC era (TASI 2008)}: {Boulder, USA, June 2-27, 2008}}, pp.~709--764, 2010.
\newblock \href{http://arxiv.org/abs/0901.4090}{{\tt 0901.4090}}.
\newblock \href{http://dx.doi.org/10.1142/9789812838360\_0014}{DOI}.

\bibitem{Patel:2015tea}
H.~H. Patel, \emph{{Package-X: A Mathematica package for the analytic
  calculation of one-loop integrals}},
  \href{http://dx.doi.org/10.1016/j.cpc.2015.08.017}{\emph{Comput. Phys.
  Commun.} {\bf 197} (2015) 276--290},
  [\href{http://arxiv.org/abs/1503.01469}{{\tt 1503.01469}}].

\bibitem{Patel:2016fam}
H.~H. Patel, \emph{{Package-X 2.0: A Mathematica package for the analytic
  calculation of one-loop integrals}},
  \href{http://dx.doi.org/10.1016/j.cpc.2017.04.015}{\emph{Comput. Phys.
  Commun.} {\bf 218} (2017) 66--70},
  [\href{http://arxiv.org/abs/1612.00009}{{\tt 1612.00009}}].

\bibitem{Hahn:1998yk}
T.~Hahn and M.~Perez-Victoria, \emph{{Automatized one loop calculations in
  four-dimensions and D-dimensions}},
  \href{http://dx.doi.org/10.1016/S0010-4655(98)00173-8}{\emph{Comput. Phys.
  Commun.} {\bf 118} (1999) 153--165},
  [\href{http://arxiv.org/abs/hep-ph/9807565}{{\tt hep-ph/9807565}}].

\bibitem{Eberl:2001eu}
H.~Eberl, M.~Kincel, W.~Majerotto and Y.~Yamada, \emph{{One loop corrections to
  the chargino and neutralino mass matrices in the on-shell scheme}},
  \href{http://dx.doi.org/10.1103/PhysRevD.64.115013}{\emph{Phys. Rev. D} {\bf
  64} (2001) 115013}, [\href{http://arxiv.org/abs/hep-ph/0104109}{{\tt
  hep-ph/0104109}}].

\bibitem{Fritzsche:2002bi}
T.~Fritzsche and W.~Hollik, \emph{{Complete one loop corrections to the mass
  spectrum of charginos and neutralinos in the MSSM}},
  \href{http://dx.doi.org/10.1007/s10052-002-0992-0}{\emph{Eur. Phys. J. C}
  {\bf 24} (2002) 619--629}, [\href{http://arxiv.org/abs/hep-ph/0203159}{{\tt
  hep-ph/0203159}}].

\bibitem{Oller:2003ge}
W.~Oller, H.~Eberl, W.~Majerotto and C.~Weber, \emph{{Analysis of the chargino
  and neutralino mass parameters at one loop level}},
  \href{http://dx.doi.org/10.1140/epjc/s2003-01246-9}{\emph{Eur. Phys. J. C}
  {\bf 29} (2003) 563--572}, [\href{http://arxiv.org/abs/hep-ph/0304006}{{\tt
  hep-ph/0304006}}].

\bibitem{Oller:2005xg}
W.~Oller, H.~Eberl and W.~Majerotto, \emph{{Precise predictions for chargino
  and neutralino pair production in e+ e- annihilation}},
  \href{http://dx.doi.org/10.1103/PhysRevD.71.115002}{\emph{Phys. Rev. D} {\bf
  71} (2005) 115002}, [\href{http://arxiv.org/abs/hep-ph/0504109}{{\tt
  hep-ph/0504109}}].

\bibitem{Drees:2006um}
M.~Drees, W.~Hollik and Q.~Xu, \emph{{One-loop calculations of the decay of the
  next-to-lightest neutralino in the MSSM}},
  \href{http://dx.doi.org/10.1088/1126-6708/2007/02/032}{\emph{JHEP} {\bf 02}
  (2007) 032}, [\href{http://arxiv.org/abs/hep-ph/0610267}{{\tt
  hep-ph/0610267}}].

\bibitem{Fowler:2009ay}
A.~C. Fowler and G.~Weiglein, \emph{{Precise Predictions for Higgs Production
  in Neutralino Decays in the Complex MSSM}},
  \href{http://dx.doi.org/10.1007/JHEP01(2010)108}{\emph{JHEP} {\bf 01} (2010)
  108}, [\href{http://arxiv.org/abs/0909.5165}{{\tt 0909.5165}}].

\bibitem{Heinemeyer:2011gk}
S.~Heinemeyer, F.~von~der Pahlen and C.~Schappacher, \emph{{Chargino Decays in
  the Complex MSSM: A Full One-Loop Analysis}},
  \href{http://dx.doi.org/10.1140/epjc/s10052-012-1892-6}{\emph{Eur. Phys. J.
  C} {\bf 72} (2012) 1892}, [\href{http://arxiv.org/abs/1112.0760}{{\tt
  1112.0760}}].

\bibitem{Chatterjee:2012hkk}
A.~Chatterjee, M.~Drees and S.~Kulkarni, \emph{{Radiative Corrections to the
  Neutralino Dark Matter Relic Density - an Effective Coupling Approach}},
  \href{http://dx.doi.org/10.1103/PhysRevD.86.105025}{\emph{Phys. Rev. D} {\bf
  86} (2012) 105025}, [\href{http://arxiv.org/abs/1209.2328}{{\tt 1209.2328}}].

\bibitem{Bharucha:2012re}
A.~Bharucha, S.~Heinemeyer, F.~von~der Pahlen and C.~Schappacher,
  \emph{{Neutralino Decays in the Complex MSSM at One-Loop: a Comparison of
  On-Shell Renormalization Schemes}},
  \href{http://dx.doi.org/10.1103/PhysRevD.86.075023}{\emph{Phys. Rev. D} {\bf
  86} (2012) 075023}, [\href{http://arxiv.org/abs/1208.4106}{{\tt 1208.4106}}].

\bibitem{Fritzsche:2013fta}
T.~Fritzsche, T.~Hahn, S.~Heinemeyer, F.~von~der Pahlen, H.~Rzehak and
  C.~Schappacher, \emph{{The Implementation of the Renormalized Complex MSSM in
  FeynArts and FormCalc}},
  \href{http://dx.doi.org/10.1016/j.cpc.2014.02.005}{\emph{Comput. Phys.
  Commun.} {\bf 185} (2014) 1529--1545},
  [\href{http://arxiv.org/abs/1309.1692}{{\tt 1309.1692}}].

\bibitem{Chatterjee_2012}
A.~Chatterjee, M.~Drees, S.~Kulkarni and Q.~Xu, \emph{On-shell renormalization
  of the chargino and neutralino masses in the {MSSM}},
  \href{http://dx.doi.org/10.1103/physrevd.85.075013}{\emph{Physical Review D}
  {\bf 85} (apr, 2012) }.

\bibitem{Baro_2009}
N.~Baro and F.~Boudjema, \emph{Automatized full one-loop renormalization of the
  {MSSM}. {II}. the chargino-neutralino sector, the sfermion sector, and some
  applications},
  \href{http://dx.doi.org/10.1103/physrevd.80.076010}{\emph{Physical Review D}
  {\bf 80} (oct, 2009) }.

\bibitem{Heinemeyer:2023pcc}
S.~Heinemeyer and F.~von~der Pahlen, \emph{{Automated Choice for the Best
  Renormalization Scheme in BSM Models}},
  \href{http://arxiv.org/abs/2302.12187}{{\tt 2302.12187}}.

\bibitem{Fritzsche_2014}
T.~Fritzsche, T.~Hahn, S.~Heinemeyer, F.~von~der Pahlen, H.~Rzehak and
  C.~Schappacher, \emph{The implementation of the renormalized complex {MSSM}
  in {FeynArts} and {FormCalc}},
  \href{http://dx.doi.org/10.1016/j.cpc.2014.02.005}{\emph{Computer Physics
  Communications} {\bf 185} (jun, 2014) 1529--1545}.

\bibitem{Denner:1991kt}
A.~Denner, \emph{{Techniques for calculation of electroweak radiative
  corrections at the one loop level and results for W physics at LEP-200}},
  \href{http://dx.doi.org/10.1002/prop.2190410402}{\emph{Fortsch. Phys.} {\bf
  41} (1993) 307--420}, [\href{http://arxiv.org/abs/0709.1075}{{\tt
  0709.1075}}].

\bibitem{Hagiwara:2011af}
K.~Hagiwara, R.~Liao, A.~D. Martin, D.~Nomura and T.~Teubner,
  \emph{{$(g-2)_\mu$ and $\alpha(M^2_Z)$ re-evaluated using new precise data}},
  \href{http://dx.doi.org/10.1088/0954-3899/38/8/085003}{\emph{J. Phys. G} {\bf
  38} (2011) 085003}, [\href{http://arxiv.org/abs/1105.3149}{{\tt 1105.3149}}].

\bibitem{Steinhauser:1998rq}
M.~Steinhauser, \emph{{Leptonic contribution to the effective electromagnetic
  coupling constant up to three loops}},
  \href{http://dx.doi.org/10.1016/S0370-2693(98)00503-6}{\emph{Phys. Lett. B}
  {\bf 429} (1998) 158--161}, [\href{http://arxiv.org/abs/hep-ph/9803313}{{\tt
  hep-ph/9803313}}].

\bibitem{Amoroso:2023pey}
S.~Amoroso et~al., \emph{{Compatibility and combination of world W-boson mass
  measurements}},  \href{http://arxiv.org/abs/2308.09417}{{\tt 2308.09417}}.

\bibitem{Bahl:2019hmm}
H.~Bahl, S.~Heinemeyer, W.~Hollik and G.~Weiglein, \emph{{Theoretical
  uncertainties in the MSSM Higgs boson mass calculation}},
  \href{http://dx.doi.org/10.1140/epjc/s10052-020-8079-3}{\emph{Eur. Phys. J.
  C} {\bf 80} (2020) 497}, [\href{http://arxiv.org/abs/1912.04199}{{\tt
  1912.04199}}].

\bibitem{HFLAV:2019otj}
{\scshape HFLAV} collaboration, Y.~S. Amhis et~al., \emph{{Averages of
  b-hadron, c-hadron, and $\tau $-lepton properties as of 2018}},
  \href{http://dx.doi.org/10.1140/epjc/s10052-020-8156-7}{\emph{Eur. Phys. J.
  C} {\bf 81} (2021) 226}, [\href{http://arxiv.org/abs/1909.12524}{{\tt
  1909.12524}}].

\bibitem{Altmannshofer:2021qrr}
W.~Altmannshofer and P.~Stangl, \emph{{New physics in rare B decays after
  Moriond 2021}},
  \href{http://dx.doi.org/10.1140/epjc/s10052-021-09725-1}{\emph{Eur. Phys. J.
  C} {\bf 81} (2021) 952}, [\href{http://arxiv.org/abs/2103.13370}{{\tt
  2103.13370}}].

\bibitem{Kraml:2013mwa}
S.~Kraml, S.~Kulkarni, U.~Laa, A.~Lessa, W.~Magerl, D.~Proschofsky-Spindler
  et~al., \emph{{SModelS: a tool for interpreting simplified-model results from
  the LHC and its application to supersymmetry}},
  \href{http://dx.doi.org/10.1140/epjc/s10052-014-2868-5}{\emph{Eur. Phys. J.
  C} {\bf 74} (2014) 2868}, [\href{http://arxiv.org/abs/1312.4175}{{\tt
  1312.4175}}].

\bibitem{Dutta:2018ioj}
J.~Dutta, S.~Kraml, A.~Lessa and W.~Waltenberger, \emph{{SModelS extension with
  the CMS supersymmetry search results from Run 2}},
  \href{http://dx.doi.org/10.31526/LHEP.1.2018.02}{\emph{LHEP} {\bf 1} (2018)
  5--12}, [\href{http://arxiv.org/abs/1803.02204}{{\tt 1803.02204}}].

\bibitem{Khosa:2020zar}
C.~K. Khosa, S.~Kraml, A.~Lessa, P.~Neuhuber and W.~Waltenberger,
  \emph{{SModelS Database Update v1.2.3}},
  \href{http://dx.doi.org/10.31526/lhep.2020.158}{\emph{LHEP} {\bf 2020} (2020)
  158}, [\href{http://arxiv.org/abs/2005.00555}{{\tt 2005.00555}}].

\bibitem{Alguero:2021dig}
G.~Alguero, J.~Heisig, C.~K. Khosa, S.~Kraml, S.~Kulkarni, A.~Lessa et~al.,
  \emph{{Constraining new physics with SModelS version 2}},
  \href{http://dx.doi.org/10.1007/JHEP08(2022)068}{\emph{JHEP} {\bf 08} (2022)
  068}, [\href{http://arxiv.org/abs/2112.00769}{{\tt 2112.00769}}].

\bibitem{Muong-2:2021vma}
{\scshape Muon g-2} collaboration, T.~Albahri et~al., \emph{{Measurement of the
  anomalous precession frequency of the muon in the Fermilab Muon $g-2$
  Experiment}},
  \href{http://dx.doi.org/10.1103/PhysRevD.103.072002}{\emph{Phys. Rev. D} {\bf
  103} (2021) 072002}, [\href{http://arxiv.org/abs/2104.03247}{{\tt
  2104.03247}}].

\bibitem{Colangelo:2022vok}
G.~Colangelo, A.~X. El-Khadra, M.~Hoferichter, A.~Keshavarzi, C.~Lehner,
  P.~Stoffer et~al., \emph{{Data-driven evaluations of Euclidean windows to
  scrutinize hadronic vacuum polarization}},
  \href{http://dx.doi.org/10.1016/j.physletb.2022.137313}{\emph{Phys. Lett. B}
  {\bf 833} (2022) 137313}, [\href{http://arxiv.org/abs/2205.12963}{{\tt
  2205.12963}}].

\bibitem{FermilabLatticeHPQCD:2023jof}
{\scshape Fermilab Lattice, HPQCD,, MILC} collaboration, A.~Bazavov et~al.,
  \emph{{Light-quark connected intermediate-window contributions to the muon
  g-2 hadronic vacuum polarization from lattice QCD}},
  \href{http://dx.doi.org/10.1103/PhysRevD.107.114514}{\emph{Phys. Rev. D} {\bf
  107} (2023) 114514}, [\href{http://arxiv.org/abs/2301.08274}{{\tt
  2301.08274}}].

\bibitem{ExtendedTwistedMass:2022jpw}
{\scshape Extended Twisted Mass} collaboration, C.~Alexandrou et~al.,
  \emph{{Lattice calculation of the short and intermediate time-distance
  hadronic vacuum polarization contributions to the muon magnetic moment using
  twisted-mass fermions}},
  \href{http://dx.doi.org/10.1103/PhysRevD.107.074506}{\emph{Phys. Rev. D} {\bf
  107} (2023) 074506}, [\href{http://arxiv.org/abs/2206.15084}{{\tt
  2206.15084}}].

\bibitem{Ce:2022kxy}
M.~C\`e et~al., \emph{{Window observable for the hadronic vacuum polarization
  contribution to the muon g-2 from lattice QCD}},
  \href{http://dx.doi.org/10.1103/PhysRevD.106.114502}{\emph{Phys. Rev. D} {\bf
  106} (2022) 114502}, [\href{http://arxiv.org/abs/2206.06582}{{\tt
  2206.06582}}].

\bibitem{Wittig:2023pcl}
H.~Wittig, \emph{{Progress on $(g-2)_\mu$ from Lattice QCD}},  in \emph{{57th
  Rencontres de Moriond on Electroweak Interactions and Unified Theories}}, 6,
  2023.
\newblock \href{http://arxiv.org/abs/2306.04165}{{\tt 2306.04165}}.

\bibitem{CMD-3:2023rfe}
{\scshape CMD-3} collaboration, F.~V. Ignatov et~al., \emph{{Measurement of the
  pion formfactor with CMD-3 detector and its implication to the hadronic
  contribution to muon (g-2)}},  \href{http://arxiv.org/abs/2309.12910}{{\tt
  2309.12910}}.

\bibitem{CMD-3:2023alj}
{\scshape CMD-3} collaboration, F.~V. Ignatov et~al., \emph{{Measurement of the
  $e^+e^-\to\pi^+\pi^-$ cross section from threshold to 1.2 GeV with the CMD-3
  detector}},  \href{http://arxiv.org/abs/2302.08834}{{\tt 2302.08834}}.

\bibitem{CMD-2:2003gqi}
{\scshape CMD-2} collaboration, R.~R. Akhmetshin et~al., \emph{{Reanalysis of
  hadronic cross-section measurements at CMD-2}},
  \href{http://dx.doi.org/10.1016/j.physletb.2003.10.108}{\emph{Phys. Lett. B}
  {\bf 578} (2004) 285--289}, [\href{http://arxiv.org/abs/hep-ex/0308008}{{\tt
  hep-ex/0308008}}].

\bibitem{KLOE-2:2017fda}
{\scshape KLOE-2} collaboration, A.~Anastasi et~al., \emph{{Combination of KLOE
  $\sigma\big(e^+e^-\rightarrow\pi^+\pi^-\gamma(\gamma)\big)$ measurements and
  determination of $a_{\mu}^{\pi^+\pi^-}$ in the energy range $0.10 < s < 0.95$
  GeV$^2$}}, \href{http://dx.doi.org/10.1007/JHEP03(2018)173}{\emph{JHEP} {\bf
  03} (2018) 173}, [\href{http://arxiv.org/abs/1711.03085}{{\tt 1711.03085}}].

\bibitem{BESIII:2015equ}
{\scshape BESIII} collaboration, M.~Ablikim et~al., \emph{{Measurement of the
  $e^+ e^- \to \pi^+ \pi^-$ cross section between 600 and 900 MeV using initial
  state radiation}},
  \href{http://dx.doi.org/10.1016/j.physletb.2015.11.043}{\emph{Phys. Lett. B}
  {\bf 753} (2016) 629--638}, [\href{http://arxiv.org/abs/1507.08188}{{\tt
  1507.08188}}]. [Erratum: Phys.Lett.B 812, 135982 (2021)].

\bibitem{BaBar:2012bdw}
{\scshape BaBar} collaboration, J.~P. Lees et~al., \emph{{Precise Measurement
  of the $e^+ e^- \to \pi^+\pi^- (\gamma)$ Cross Section with the Initial-State
  Radiation Method at BABAR}},
  \href{http://dx.doi.org/10.1103/PhysRevD.86.032013}{\emph{Phys. Rev. D} {\bf
  86} (2012) 032013}, [\href{http://arxiv.org/abs/1205.2228}{{\tt 1205.2228}}].

\bibitem{Lopez:1993vi}
J.~L. Lopez, D.~V. Nanopoulos and X.~Wang, \emph{{Large (g-2)-mu in SU(5) x
  U(1) supergravity models}},
  \href{http://dx.doi.org/10.1103/PhysRevD.49.366}{\emph{Phys. Rev. D} {\bf 49}
  (1994) 366--372}, [\href{http://arxiv.org/abs/hep-ph/9308336}{{\tt
  hep-ph/9308336}}].

\bibitem{Chattopadhyay:1995ae}
U.~Chattopadhyay and P.~Nath, \emph{{Probing supergravity grand unification in
  the Brookhaven g-2 experiment}},
  \href{http://dx.doi.org/10.1103/PhysRevD.53.1648}{\emph{Phys. Rev. D} {\bf
  53} (1996) 1648--1657}, [\href{http://arxiv.org/abs/hep-ph/9507386}{{\tt
  hep-ph/9507386}}].

\bibitem{Moroi:1995yh}
T.~Moroi, \emph{{The Muon anomalous magnetic dipole moment in the minimal
  supersymmetric standard model}},
  \href{http://dx.doi.org/10.1103/PhysRevD.53.6565}{\emph{Phys. Rev. D} {\bf
  53} (1996) 6565--6575}, [\href{http://arxiv.org/abs/hep-ph/9512396}{{\tt
  hep-ph/9512396}}]. [Erratum: Phys.Rev.D 56, 4424 (1997)].

\bibitem{Chattopadhyay:2000ws}
U.~Chattopadhyay, D.~K. Ghosh and S.~Roy, \emph{{Constraining anomaly mediated
  supersymmetry breaking framework via on going muon g-2 experiment at
  Brookhaven}}, \href{http://dx.doi.org/10.1103/PhysRevD.62.115001}{\emph{Phys.
  Rev. D} {\bf 62} (2000) 115001},
  [\href{http://arxiv.org/abs/hep-ph/0006049}{{\tt hep-ph/0006049}}].

\bibitem{Martin:2001st}
S.~P. Martin and J.~D. Wells, \emph{{Muon Anomalous Magnetic Dipole Moment in
  Supersymmetric Theories}},
  \href{http://dx.doi.org/10.1103/PhysRevD.64.035003}{\emph{Phys. Rev. D} {\bf
  64} (2001) 035003}, [\href{http://arxiv.org/abs/hep-ph/0103067}{{\tt
  hep-ph/0103067}}].

\bibitem{Heinemeyer:2003dq}
S.~Heinemeyer, D.~Stockinger and G.~Weiglein, \emph{{Two loop SUSY corrections
  to the anomalous magnetic moment of the muon}},
  \href{http://dx.doi.org/10.1016/j.nuclphysb.2004.04.017}{\emph{Nucl. Phys. B}
  {\bf 690} (2004) 62--80}, [\href{http://arxiv.org/abs/hep-ph/0312264}{{\tt
  hep-ph/0312264}}].

\bibitem{Stockinger:2006zn}
D.~Stockinger, \emph{{The Muon Magnetic Moment and Supersymmetry}},
  \href{http://dx.doi.org/10.1088/0954-3899/34/2/R01}{\emph{J. Phys. G} {\bf
  34} (2007) R45--R92}, [\href{http://arxiv.org/abs/hep-ph/0609168}{{\tt
  hep-ph/0609168}}].

\bibitem{Cho:2011rk}
G.-C. Cho, K.~Hagiwara, Y.~Matsumoto and D.~Nomura, \emph{{The MSSM confronts
  the precision electroweak data and the muon g-2}},
  \href{http://dx.doi.org/10.1007/JHEP11(2011)068}{\emph{JHEP} {\bf 11} (2011)
  068}, [\href{http://arxiv.org/abs/1104.1769}{{\tt 1104.1769}}].

\bibitem{Endo:2013lva}
M.~Endo, K.~Hamaguchi, T.~Kitahara and T.~Yoshinaga, \emph{{Probing Bino
  contribution to muon $g - 2$}},
  \href{http://dx.doi.org/10.1007/JHEP11(2013)013}{\emph{JHEP} {\bf 11} (2013)
  013}, [\href{http://arxiv.org/abs/1309.3065}{{\tt 1309.3065}}].

\bibitem{Endo:2013bba}
M.~Endo, K.~Hamaguchi, S.~Iwamoto and T.~Yoshinaga, \emph{{Muon g-2 vs LHC in
  Supersymmetric Models}},
  \href{http://dx.doi.org/10.1007/JHEP01(2014)123}{\emph{JHEP} {\bf 01} (2014)
  123}, [\href{http://arxiv.org/abs/1303.4256}{{\tt 1303.4256}}].

\bibitem{Chakraborti:2014gea}
M.~Chakraborti, U.~Chattopadhyay, A.~Choudhury, A.~Datta and S.~Poddar,
  \emph{{The Electroweak Sector of the pMSSM in the Light of LHC - 8 TeV and
  Other Data}}, \href{http://dx.doi.org/10.1007/JHEP07(2014)019}{\emph{JHEP}
  {\bf 07} (2014) 019}, [\href{http://arxiv.org/abs/1404.4841}{{\tt
  1404.4841}}].

\bibitem{Chowdhury:2015rja}
D.~Chowdhury and N.~Yokozaki, \emph{{Muon g \ensuremath{-} 2 in anomaly
  mediated SUSY breaking}},
  \href{http://dx.doi.org/10.1007/JHEP08(2015)111}{\emph{JHEP} {\bf 08} (2015)
  111}, [\href{http://arxiv.org/abs/1505.05153}{{\tt 1505.05153}}].

\bibitem{Kowalska:2015zja}
K.~Kowalska, L.~Roszkowski, E.~M. Sessolo and A.~J. Williams,
  \emph{{GUT-inspired SUSY and the muon g \ensuremath{-} 2 anomaly: prospects
  for LHC 14 TeV}},
  \href{http://dx.doi.org/10.1007/JHEP06(2015)020}{\emph{JHEP} {\bf 06} (2015)
  020}, [\href{http://arxiv.org/abs/1503.08219}{{\tt 1503.08219}}].

\bibitem{Gomez:2018efz}
M.~E. Gomez, S.~Lola, R.~Ruiz~de Austri and Q.~Shafi, \emph{{Confronting SUSY
  GUT with Dark Matter, Sparticle Spectroscopy and Muon $(g-2)$}},
  \href{http://dx.doi.org/10.3389/fphy.2018.00127}{\emph{Front. in Phys.} {\bf
  6} (2018) 127}, [\href{http://arxiv.org/abs/1806.11152}{{\tt 1806.11152}}].

\bibitem{Chakraborti:2021mbr}
M.~Chakraborti, S.~Heinemeyer, I.~Saha and C.~Schappacher, \emph{{$(g-2)_\mu $
  and SUSY dark matter: direct detection and collider search complementarity}},
  \href{http://dx.doi.org/10.1140/epjc/s10052-022-10414-w}{\emph{Eur. Phys. J.
  C} {\bf 82} (2022) 483}, [\href{http://arxiv.org/abs/2112.01389}{{\tt
  2112.01389}}].

\bibitem{Chakraborti:2021kkr}
M.~Chakraborti, S.~Heinemeyer and I.~Saha, \emph{{Improved ${(g-2)_\mu }$
  measurements and wino/higgsino dark matter}},
  \href{http://dx.doi.org/10.1140/epjc/s10052-021-09814-1}{\emph{Eur. Phys. J.
  C} {\bf 81} (2021) 1069}, [\href{http://arxiv.org/abs/2103.13403}{{\tt
  2103.13403}}].

\bibitem{Ali:2021kxa}
M.~I. Ali, M.~Chakraborti, U.~Chattopadhyay and S.~Mukherjee, \emph{{Muon and
  electron $(g-2)$ anomalies with non-holomorphic interactions in MSSM}},
  \href{http://dx.doi.org/10.1140/epjc/s10052-023-11216-4}{\emph{Eur. Phys. J.
  C} {\bf 83} (2023) 60}, [\href{http://arxiv.org/abs/2112.09867}{{\tt
  2112.09867}}].

\bibitem{Lindner:2016bgg}
M.~Lindner, M.~Platscher and F.~S. Queiroz, \emph{{A Call for New Physics : The
  Muon Anomalous Magnetic Moment and Lepton Flavor Violation}},
  \href{http://dx.doi.org/10.1016/j.physrep.2017.12.001}{\emph{Phys. Rept.}
  {\bf 731} (2018) 1--82}, [\href{http://arxiv.org/abs/1610.06587}{{\tt
  1610.06587}}].

\bibitem{Hagiwara:2017lse}
K.~Hagiwara, K.~Ma and S.~Mukhopadhyay, \emph{{Closing in on the chargino
  contribution to the muon g-2 in the MSSM: current LHC constraints}},
  \href{http://dx.doi.org/10.1103/PhysRevD.97.055035}{\emph{Phys. Rev. D} {\bf
  97} (2018) 055035}, [\href{http://arxiv.org/abs/1706.09313}{{\tt
  1706.09313}}].

\bibitem{Endo:2020mqz}
M.~Endo, K.~Hamaguchi, S.~Iwamoto and T.~Kitahara, \emph{{Muon $g-2$ vs LHC Run
  2 in supersymmetric models}},
  \href{http://dx.doi.org/10.1007/JHEP04(2020)165}{\emph{JHEP} {\bf 04} (2020)
  165}, [\href{http://arxiv.org/abs/2001.11025}{{\tt 2001.11025}}].

\bibitem{Chakraborti:2020vjp}
M.~Chakraborti, S.~Heinemeyer and I.~Saha, \emph{{Improved $(g-2)_\mu$
  Measurements and Supersymmetry}},
  \href{http://dx.doi.org/10.1140/epjc/s10052-020-08504-8}{\emph{Eur. Phys. J.
  C} {\bf 80} (2020) 984}, [\href{http://arxiv.org/abs/2006.15157}{{\tt
  2006.15157}}].

\bibitem{Aboubrahim:2021xfi}
A.~Aboubrahim, M.~Klasen and P.~Nath, \emph{{What the Fermilab muon $g-$2
  experiment tells us about discovering supersymmetry at high luminosity and
  high energy upgrades to the LHC}},
  \href{http://dx.doi.org/10.1103/PhysRevD.104.035039}{\emph{Phys. Rev. D} {\bf
  104} (2021) 035039}, [\href{http://arxiv.org/abs/2104.03839}{{\tt
  2104.03839}}].

\bibitem{Athron:2021iuf}
P.~Athron, C.~Bal\'azs, D.~H.~J. Jacob, W.~Kotlarski, D.~St\"ockinger and
  H.~St\"ockinger-Kim, \emph{{New physics explanations of a$_{\mu}$ in light of
  the FNAL muon g-2 measurement}},
  \href{http://dx.doi.org/10.1007/JHEP09(2021)080}{\emph{JHEP} {\bf 09} (2021)
  080}, [\href{http://arxiv.org/abs/2104.03691}{{\tt 2104.03691}}].

\bibitem{VanBeekveld:2021tgn}
M.~Van~Beekveld, W.~Beenakker, M.~Schutten and J.~De~Wit, \emph{{Dark matter,
  fine-tuning and $(g-2)_{\mu}$ in the pMSSM}},
  \href{http://dx.doi.org/10.21468/SciPostPhys.11.3.049}{\emph{SciPost Phys.}
  {\bf 11} (2021) 049}, [\href{http://arxiv.org/abs/2104.03245}{{\tt
  2104.03245}}].

\bibitem{Endo:2021zal}
M.~Endo, K.~Hamaguchi, S.~Iwamoto and T.~Kitahara, \emph{{Supersymmetric
  interpretation of the muon g-2 anomaly}},
  \href{http://dx.doi.org/10.1007/JHEP07(2021)075}{\emph{JHEP} {\bf 07} (2021)
  075}, [\href{http://arxiv.org/abs/2104.03217}{{\tt 2104.03217}}].

\bibitem{Chakraborti:2021dli}
M.~Chakraborti, S.~Heinemeyer and I.~Saha, \emph{{The new MUON G-2 result and
  supersymmetry}},
  \href{http://dx.doi.org/10.1140/epjc/s10052-021-09900-4}{\emph{Eur. Phys. J.
  C} {\bf 81} (2021) 1114}, [\href{http://arxiv.org/abs/2104.03287}{{\tt
  2104.03287}}].

\bibitem{Chakraborti:2021bmv}
M.~Chakraborti, L.~Roszkowski and S.~Trojanowski, \emph{{GUT-constrained
  supersymmetry and dark matter in light of the new $(g-2)_\mu$
  determination}}, \href{http://dx.doi.org/10.1007/JHEP05(2021)252}{\emph{JHEP}
  {\bf 05} (2021) 252}, [\href{http://arxiv.org/abs/2104.04458}{{\tt
  2104.04458}}].

\bibitem{Chattopadhyay:2002zq}
U.~Chattopadhyay, A.~Corsetti and P.~Nath, \emph{{Theoretical status of muon
  (g-2)}}, \href{http://dx.doi.org/10.1063/1.1492172}{\emph{AIP Conf. Proc.}
  {\bf 624} (2002) 230--238}, [\href{http://arxiv.org/abs/hep-ph/0202275}{{\tt
  hep-ph/0202275}}].

\bibitem{Chattopadhyay:2002jx}
U.~Chattopadhyay and P.~Nath, \emph{{Interpreting the new Brookhaven muon (g-2)
  result}}, \href{http://dx.doi.org/10.1103/PhysRevD.66.093001}{\emph{Phys.
  Rev. D} {\bf 66} (2002) 093001},
  [\href{http://arxiv.org/abs/hep-ph/0208012}{{\tt hep-ph/0208012}}].

\bibitem{Porod:2003um}
W.~Porod, \emph{{SPheno, a program for calculating supersymmetric spectra, SUSY
  particle decays and SUSY particle production at e+ e- colliders}},
  \href{http://dx.doi.org/10.1016/S0010-4655(03)00222-4}{\emph{Comput. Phys.
  Commun.} {\bf 153} (2003) 275--315},
  [\href{http://arxiv.org/abs/hep-ph/0301101}{{\tt hep-ph/0301101}}].

\bibitem{Porod:2011nf}
W.~Porod and F.~Staub, \emph{{SPheno 3.1: Extensions including flavour,
  CP-phases and models beyond the MSSM}},
  \href{http://dx.doi.org/10.1016/j.cpc.2012.05.021}{\emph{Comput. Phys.
  Commun.} {\bf 183} (2012) 2458--2469},
  [\href{http://arxiv.org/abs/1104.1573}{{\tt 1104.1573}}].

\bibitem{Staub:2013tta}
F.~Staub, \emph{{SARAH 4 : A tool for (not only SUSY) model builders}},
  \href{http://dx.doi.org/10.1016/j.cpc.2014.02.018}{\emph{Comput. Phys.
  Commun.} {\bf 185} (2014) 1773--1790},
  [\href{http://arxiv.org/abs/1309.7223}{{\tt 1309.7223}}].

\bibitem{Staub:2015kfa}
F.~Staub, \emph{{Exploring new models in all detail with SARAH}},
  \href{http://dx.doi.org/10.1155/2015/840780}{\emph{Adv. High Energy Phys.}
  {\bf 2015} (2015) 840780}, [\href{http://arxiv.org/abs/1503.04200}{{\tt
  1503.04200}}].

\bibitem{Fargnoli:2013zda}
H.~G. Fargnoli, C.~Gnendiger, S.~Pa\ss{}ehr, D.~St\"ockinger and
  H.~St\"ockinger-Kim, \emph{{Non-decoupling two-loop corrections to
  $(g-2)_\mu$ from fermion/sfermion loops in the MSSM}},
  \href{http://dx.doi.org/10.1016/j.physletb.2013.09.034}{\emph{Phys. Lett. B}
  {\bf 726} (2013) 717--724}, [\href{http://arxiv.org/abs/1309.0980}{{\tt
  1309.0980}}].

\bibitem{Fargnoli:2013zia}
H.~Fargnoli, C.~Gnendiger, S.~Pa\ss{}ehr, D.~St\"ockinger and
  H.~St\"ockinger-Kim, \emph{{Two-loop corrections to the muon magnetic moment
  from fermion/sfermion loops in the MSSM: detailed results}},
  \href{http://dx.doi.org/10.1007/JHEP02(2014)070}{\emph{JHEP} {\bf 02} (2014)
  070}, [\href{http://arxiv.org/abs/1311.1775}{{\tt 1311.1775}}].

\bibitem{Athron:2015rva}
P.~Athron, M.~Bach, H.~G. Fargnoli, C.~Gnendiger, R.~Greifenhagen, J.-h. Park
  et~al., \emph{{GM2Calc: Precise MSSM prediction for $(g - 2)$ of the muon}},
  \href{http://dx.doi.org/10.1140/epjc/s10052-015-3870-2}{\emph{Eur. Phys. J.
  C} {\bf 76} (2016) 62}, [\href{http://arxiv.org/abs/1510.08071}{{\tt
  1510.08071}}].

\bibitem{Athron:2021evk}
P.~Athron, C.~Balazs, A.~Cherchiglia, D.~H.~J. Jacob, D.~St\"ockinger,
  H.~St\"ockinger-Kim et~al., \emph{{Two-loop prediction of the anomalous
  magnetic moment of the muon in the Two-Higgs Doublet Model with GM2Calc 2}},
  \href{http://dx.doi.org/10.1140/epjc/s10052-022-10148-9}{\emph{Eur. Phys. J.
  C} {\bf 82} (2022) 229}, [\href{http://arxiv.org/abs/2110.13238}{{\tt
  2110.13238}}].

\bibitem{vonWeitershausen:2010zr}
P.~von Weitershausen, M.~Schafer, H.~Stockinger-Kim and D.~Stockinger,
  \emph{{Photonic SUSY Two-Loop Corrections to the Muon Magnetic Moment}},
  \href{http://dx.doi.org/10.1103/PhysRevD.81.093004}{\emph{Phys. Rev. D} {\bf
  81} (2010) 093004}, [\href{http://arxiv.org/abs/1003.5820}{{\tt 1003.5820}}].

\bibitem{Bach:2015doa}
M.~Bach, J.-h. Park, D.~St\"ockinger and H.~St\"ockinger-Kim, \emph{{Large muon
  $(g-2)$ with TeV-scale SUSY masses for $\tan\beta\to\infty$}},
  \href{http://dx.doi.org/10.1007/JHEP10(2015)026}{\emph{JHEP} {\bf 10} (2015)
  026}, [\href{http://arxiv.org/abs/1504.05500}{{\tt 1504.05500}}].

\bibitem{Cherchiglia:2016eui}
A.~Cherchiglia, P.~Kneschke, D.~St\"ockinger and H.~St\"ockinger-Kim,
  \emph{{The muon magnetic moment in the 2HDM: complete two-loop result}},
  \href{http://dx.doi.org/10.1007/JHEP10(2021)242}{\emph{JHEP} {\bf 01} (2017)
  007}, [\href{http://arxiv.org/abs/1607.06292}{{\tt 1607.06292}}]. [Erratum:
  JHEP 10, 242 (2021)].

\bibitem{ATLAS:2018ojr}
{\scshape ATLAS} collaboration, M.~Aaboud et~al., \emph{{Search for electroweak
  production of supersymmetric particles in final states with two or three
  leptons at $\sqrt{s}=13\,$TeV with the ATLAS detector}},
  \href{http://dx.doi.org/10.1140/epjc/s10052-018-6423-7}{\emph{Eur. Phys. J.
  C} {\bf 78} (2018) 995}, [\href{http://arxiv.org/abs/1803.02762}{{\tt
  1803.02762}}].

\bibitem{CMS:2017moi}
{\scshape CMS} collaboration, A.~M. Sirunyan et~al., \emph{{Search for
  electroweak production of charginos and neutralinos in multilepton final
  states in proton-proton collisions at $\sqrt{s}=$ 13 TeV}},
  \href{http://dx.doi.org/10.1007/JHEP03(2018)166}{\emph{JHEP} {\bf 03} (2018)
  166}, [\href{http://arxiv.org/abs/1709.05406}{{\tt 1709.05406}}].

\bibitem{ATLAS:2018eui}
{\scshape ATLAS} collaboration, M.~Aaboud et~al., \emph{{Search for
  chargino-neutralino production using recursive jigsaw reconstruction in final
  states with two or three charged leptons in proton-proton collisions at
  $\sqrt{s}=13$ TeV with the ATLAS detector}},
  \href{http://dx.doi.org/10.1103/PhysRevD.98.092012}{\emph{Phys. Rev. D} {\bf
  98} (2018) 092012}, [\href{http://arxiv.org/abs/1806.02293}{{\tt
  1806.02293}}].

\bibitem{ATLAS:2019lff}
{\scshape ATLAS} collaboration, G.~Aad et~al., \emph{{Search for electroweak
  production of charginos and sleptons decaying into final states with two
  leptons and missing transverse momentum in $\sqrt{s}=13$ TeV $pp$ collisions
  using the ATLAS detector}},
  \href{http://dx.doi.org/10.1140/epjc/s10052-019-7594-6}{\emph{Eur. Phys. J.
  C} {\bf 80} (2020) 123}, [\href{http://arxiv.org/abs/1908.08215}{{\tt
  1908.08215}}].

\bibitem{ATLAS:2019lng}
{\scshape ATLAS} collaboration, G.~Aad et~al., \emph{{Searches for electroweak
  production of supersymmetric particles with compressed mass spectra in
  $\sqrt{s}=$ 13 TeV $pp$ collisions with the ATLAS detector}},
  \href{http://dx.doi.org/10.1103/PhysRevD.101.052005}{\emph{Phys. Rev. D} {\bf
  101} (2020) 052005}, [\href{http://arxiv.org/abs/1911.12606}{{\tt
  1911.12606}}].

\bibitem{ATLAS:2020pgy}
{\scshape ATLAS} collaboration, G.~Aad et~al., \emph{{Search for direct
  production of electroweakinos in final states with one lepton, missing
  transverse momentum and a Higgs boson decaying into two $b$-jets in $pp$
  collisions at $\sqrt{s}=13$ TeV with the ATLAS detector}},
  \href{http://dx.doi.org/10.1140/epjc/s10052-020-8050-3}{\emph{Eur. Phys. J.
  C} {\bf 80} (2020) 691}, [\href{http://arxiv.org/abs/1909.09226}{{\tt
  1909.09226}}].

\bibitem{ATLAS:2019wgx}
{\scshape ATLAS} collaboration, G.~Aad et~al., \emph{{Search for
  chargino-neutralino production with mass splittings near the electroweak
  scale in three-lepton final states in $\sqrt {s}$=13 TeV $pp$ collisions with
  the ATLAS detector}},
  \href{http://dx.doi.org/10.1103/PhysRevD.101.072001}{\emph{Phys. Rev. D} {\bf
  101} (2020) 072001}, [\href{http://arxiv.org/abs/1912.08479}{{\tt
  1912.08479}}].

\bibitem{ATLAS:2021moa}
{\scshape ATLAS} collaboration, G.~Aad et~al., \emph{{Search for
  chargino\textendash{}neutralino pair production in final states with three
  leptons and missing transverse momentum in $\sqrt{s} = 13$~TeV pp collisions
  with the ATLAS detector}},
  \href{http://dx.doi.org/10.1140/epjc/s10052-021-09749-7}{\emph{Eur. Phys. J.
  C} {\bf 81} (2021) 1118}, [\href{http://arxiv.org/abs/2106.01676}{{\tt
  2106.01676}}].

\bibitem{CMS:2018kag}
{\scshape CMS} collaboration, A.~M. Sirunyan et~al., \emph{{Search for new
  physics in events with two soft oppositely charged leptons and missing
  transverse momentum in proton-proton collisions at $\sqrt{s}=$ 13 TeV}},
  \href{http://dx.doi.org/10.1016/j.physletb.2018.05.062}{\emph{Phys. Lett. B}
  {\bf 782} (2018) 440--467}, [\href{http://arxiv.org/abs/1801.01846}{{\tt
  1801.01846}}].

\bibitem{CMS:2018szt}
{\scshape CMS} collaboration, A.~M. Sirunyan et~al., \emph{{Combined search for
  electroweak production of charginos and neutralinos in proton-proton
  collisions at $\sqrt{s} =$ 13 TeV}},
  \href{http://dx.doi.org/10.1007/JHEP03(2018)160}{\emph{JHEP} {\bf 03} (2018)
  160}, [\href{http://arxiv.org/abs/1801.03957}{{\tt 1801.03957}}].

\bibitem{CMS:2018eqb}
{\scshape CMS} collaboration, A.~M. Sirunyan et~al., \emph{{Search for
  supersymmetric partners of electrons and muons in proton-proton collisions at
  $\sqrt{s}=$ 13 TeV}},
  \href{http://dx.doi.org/10.1016/j.physletb.2019.01.005}{\emph{Phys. Lett. B}
  {\bf 790} (2019) 140--166}, [\href{http://arxiv.org/abs/1806.05264}{{\tt
  1806.05264}}].

\bibitem{CMS:2020bfa}
{\scshape CMS} collaboration, A.~M. Sirunyan et~al., \emph{{Search for
  supersymmetry in final states with two oppositely charged same-flavor leptons
  and missing transverse momentum in proton-proton collisions at $\sqrt{s} =$
  13 TeV}}, \href{http://dx.doi.org/10.1007/JHEP04(2021)123}{\emph{JHEP} {\bf
  04} (2021) 123}, [\href{http://arxiv.org/abs/2012.08600}{{\tt 2012.08600}}].

\bibitem{CMS:2021edw}
{\scshape CMS} collaboration, A.~Tumasyan et~al., \emph{{Search for
  supersymmetry in final states with two or three soft leptons and missing
  transverse momentum in proton-proton collisions at $ \sqrt{s} $ = 13 TeV}},
  \href{http://dx.doi.org/10.1007/JHEP04(2022)091}{\emph{JHEP} {\bf 04} (2022)
  091}, [\href{http://arxiv.org/abs/2111.06296}{{\tt 2111.06296}}].

\bibitem{ATLAS:2014zve}
{\scshape ATLAS} collaboration, G.~Aad et~al., \emph{{Search for direct
  production of charginos, neutralinos and sleptons in final states with two
  leptons and missing transverse momentum in $pp$ collisions at $\sqrt{s} =$ 8
  TeV with the ATLAS detector}},
  \href{http://dx.doi.org/10.1007/JHEP05(2014)071}{\emph{JHEP} {\bf 05} (2014)
  071}, [\href{http://arxiv.org/abs/1403.5294}{{\tt 1403.5294}}].

\bibitem{ATLAS:2017vat}
{\scshape ATLAS} collaboration, M.~Aaboud et~al., \emph{{Search for electroweak
  production of supersymmetric states in scenarios with compressed mass spectra
  at $\sqrt{s}=13$ TeV with the ATLAS detector}},
  \href{http://dx.doi.org/10.1103/PhysRevD.97.052010}{\emph{Phys. Rev. D} {\bf
  97} (2018) 052010}, [\href{http://arxiv.org/abs/1712.08119}{{\tt
  1712.08119}}].

\bibitem{Adam:2021rrw}
W.~Adam and I.~Vivarelli, \emph{{Status of searches for electroweak-scale
  supersymmetry after LHC Run 2}},
  \href{http://dx.doi.org/10.1142/S0217751X21300222}{\emph{Int. J. Mod. Phys.
  A} {\bf 37} (2022) 2130022}, [\href{http://arxiv.org/abs/2111.10180}{{\tt
  2111.10180}}].

\bibitem{ATLAS:2022zwa}
{\scshape ATLAS} collaboration, G.~Aad et~al., \emph{{Searches for new
  phenomena in events with two leptons, jets, and missing transverse momentum
  in 139~fb$^{-1}$ of $\sqrt{s}=13$~TeV $pp$ collisions with the ATLAS
  detector}},
  \href{http://dx.doi.org/10.1140/epjc/s10052-023-11434-w}{\emph{Eur. Phys. J.
  C} {\bf 83} (2023) 515}, [\href{http://arxiv.org/abs/2204.13072}{{\tt
  2204.13072}}].

\bibitem{CMS:2022sfi}
{\scshape CMS} collaboration, A.~Tumasyan et~al., \emph{{Search for electroweak
  production of charginos and neutralinos at s=13TeV in final states containing
  hadronic decays of WW, WZ, or WH and missing transverse momentum}},
  \href{http://dx.doi.org/10.1016/j.physletb.2022.137460}{\emph{Phys. Lett. B}
  {\bf 842} (2023) 137460}, [\href{http://arxiv.org/abs/2205.09597}{{\tt
  2205.09597}}].

\bibitem{ATLAS:2021yqv}
{\scshape ATLAS} collaboration, G.~Aad et~al., \emph{{Search for charginos and
  neutralinos in final states with two boosted hadronically decaying bosons and
  missing transverse momentum in $pp$ collisions at $\sqrt {s}$ = 13\,\,TeV
  with the ATLAS detector}},
  \href{http://dx.doi.org/10.1103/PhysRevD.104.112010}{\emph{Phys. Rev. D} {\bf
  104} (2021) 112010}, [\href{http://arxiv.org/abs/2108.07586}{{\tt
  2108.07586}}].

\bibitem{ATLAS:2022hbt}
{\scshape ATLAS} collaboration, G.~Aad et~al., \emph{{Search for direct pair
  production of sleptons and charginos decaying to two leptons and neutralinos
  with mass splittings near the W-boson mass in $ \sqrt{s} $ = 13 TeV pp
  collisions with the ATLAS detector}},
  \href{http://dx.doi.org/10.1007/JHEP06(2023)031}{\emph{JHEP} {\bf 06} (2023)
  031}, [\href{http://arxiv.org/abs/2209.13935}{{\tt 2209.13935}}].

\bibitem{CMS:2022vpy}
{\scshape CMS} collaboration, A.~Tumasyan et~al., \emph{{Search for higgsinos
  decaying to two Higgs bosons and missing transverse momentum in proton-proton
  collisions at $ \sqrt{s} $ = 13 TeV}},
  \href{http://dx.doi.org/10.1007/JHEP05(2022)014}{\emph{JHEP} {\bf 05} (2022)
  014}, [\href{http://arxiv.org/abs/2201.04206}{{\tt 2201.04206}}].

\bibitem{Hahn:2000kx}
T.~Hahn, \emph{{Generating Feynman diagrams and amplitudes with FeynArts 3}},
  \href{http://dx.doi.org/10.1016/S0010-4655(01)00290-9}{\emph{Comput. Phys.
  Commun.} {\bf 140} (2001) 418--431},
  [\href{http://arxiv.org/abs/hep-ph/0012260}{{\tt hep-ph/0012260}}].

\bibitem{KUBLBECK1990165}
J.~Küblbeck, M.~Böhm and A.~Denner, \emph{Feyn arts — computer-algebraic
  generation of feynman graphs and amplitudes},
  \href{http://dx.doi.org/https://doi.org/10.1016/0010-4655(90)90001-H}{\emph{Computer
  Physics Communications} {\bf 60} (1990) 165--180}.

\bibitem{Hahn:2001rv}
T.~Hahn and C.~Schappacher, \emph{{The Implementation of the minimal
  supersymmetric standard model in FeynArts and FormCalc}},
  \href{http://dx.doi.org/10.1016/S0010-4655(01)00436-2}{\emph{Comput. Phys.
  Commun.} {\bf 143} (2002) 54--68},
  [\href{http://arxiv.org/abs/hep-ph/0105349}{{\tt hep-ph/0105349}}].

\bibitem{Belanger:2001fz}
G.~Belanger, F.~Boudjema, A.~Pukhov and A.~Semenov, \emph{{MicrOMEGAs: A
  Program for calculating the relic density in the MSSM}},
  \href{http://dx.doi.org/10.1016/S0010-4655(02)00596-9}{\emph{Comput. Phys.
  Commun.} {\bf 149} (2002) 103--120},
  [\href{http://arxiv.org/abs/hep-ph/0112278}{{\tt hep-ph/0112278}}].

\bibitem{Belanger:2006is}
G.~Belanger, F.~Boudjema, A.~Pukhov and A.~Semenov, \emph{{MicrOMEGAs 2.0: A
  Program to calculate the relic density of dark matter in a generic model}},
  \href{http://dx.doi.org/10.1016/j.cpc.2006.11.008}{\emph{Comput. Phys.
  Commun.} {\bf 176} (2007) 367--382},
  [\href{http://arxiv.org/abs/hep-ph/0607059}{{\tt hep-ph/0607059}}].

\bibitem{Belanger:2008sj}
G.~Belanger, F.~Boudjema, A.~Pukhov and A.~Semenov, \emph{{Dark matter direct
  detection rate in a generic model with micrOMEGAs 2.2}},
  \href{http://dx.doi.org/10.1016/j.cpc.2008.11.019}{\emph{Comput. Phys.
  Commun.} {\bf 180} (2009) 747--767},
  [\href{http://arxiv.org/abs/0803.2360}{{\tt 0803.2360}}].

\end{thebibliography}\endgroup
\end{document}